\documentclass[12pt]{report}
\usepackage[usenames,dvips]{pstcol}
\usepackage{color}
\usepackage{amssymb}
\usepackage{amsmath}
\usepackage{epsfig}
\usepackage{footnote}
\usepackage{longtable}
\usepackage{subfigure}
\usepackage{xspace}
\usepackage{rotating}
\usepackage{loginov}

\newcommand{\mettsm}{\mbox{\scriptsize ${\rm \not\! E}_{\rm T}$}}
\newcommand{\mettgmh}{\mbox{${\rm \not\! E}_{\rm T}>35$~GeV}}

\def \kfactor {K\kern -0.38em -\kern -0.38em factor}
\def \Et {{\rm E}_{\rm T}}

\newcommand{\Pt}{{\rm p}_{\rm T}}

\newcommand{\met}{\mbox{${\rm \not\! E}_{\rm T}$}}
\newcommand{\Met}{\mbox{${\rm \not\! E}_{\rm T}$}} 
\newcommand{\metvec}{{\not\!\! \vec{E}_T}}
\newcommand{\lepvec}{{\vec{E}_T^\ell}}
\newcommand{\phovec}{{\vec{E}_T^\gamma}}

\def \Ht {{\rm H}_{\rm T}}

\newcommand{\Wenu}{\mbox{W$^\pm \rightarrow e^{\pm}\nu$}}
\newcommand{\wenu}{\mbox{W$^\pm \rightarrow e^{\pm}\nu$}}

\def \dx{\Delta X}

\newcommand{\Zgstar}{Z^0\kern -0.25em/\kern -0.15em\gamma^*}
\def\Zgamma{\Z\kern -0.1em/\kern -0.1em\gamma}

\def\pizero{\pi ^0}
\def\eeggmet{ee\gamma\gamma\Met}
\def\llgg{\ell\ell\gamma\gamma}

\newcommand{\lgX}{\ell\gamma+X}
\newcommand{\egX}{e\gamma+X}
\newcommand{\mugX}{\mu\gamma+X}
\def\lgmet{\ell\gamma\Met}
\def\ljmet{\ell j\Met}
\def\llg{\ell\ell\gamma}
\def\llj{\ell\ell j}
\def\lgg{\ell\gamma\gamma}

\def\ll{\ell\ell}
\def\egmet{e\gamma\Met}
\def\eemet{ee\Met}
\def\ejmet{ej\Met}
\def\mugmet{\mu\gamma\Met}
\def\mujmet{\mu j\Met}

\def\mmggjj{\mu\mu\gamma\gamma jj}

\def\eg{e\gamma}
\def\eeg{ee\gamma}

\def\mug{\mu\gamma}
\def\mgmet{\mu\gamma\Met}
\def\mugmet{\mu\gamma\Met}

\def\mumug{\mu\mu\gamma}
\def\mumuj{\mu\mu j}

\def\lg{\ell\gamma}

\def\Wgg{W\gamma\gamma}
\def\Zgg{Z\gamma\gamma}

\def\gg{\gamma\gamma}
\def\Wg{W\gamma}
\def\Zg{Z\gamma}
\def\Zeeg{Z(ee)\gamma}
\def\Zmumug{Z(\mu\mu)\gamma}
\def\Ztautaug{Z(\tau\tau)\gamma}
\def\Zeegg{Z(ee)\gamma\gamma}
\def\Zmumugg{Z(\mu\mu)\gamma\gamma}
\def\Ztautaugg{Z(\tau\tau)\gamma\gamma}
\def\Wenug{W(e\nu)\gamma}
\def\Wmunug{W(\mu\nu)\gamma}
\def\Wtaunug{W(\tau\nu)\gamma}
\def\Wenugg{W(e\nu)\gamma\gamma}
\def\Wmunugg{W(\mu\nu)\gamma\gamma}
\def\Wtaunugg{W(\tau\nu)\gamma\gamma}

\def\Wj{Wj}
\def\Zj{Zj}

\def\deg{^\circ}

\def\gt{>}
\def\lt{<}

\def\GeV{GeV}

\def\GeVc2{GeV/{c^2}}

\newcommand{\invpb}{pb^{-1}}
\newcommand{\pbinv}{pb^{-1}}

\def\pbarp{{\bar p}p}
\newcommand{\pbar}{\rm{\bar p}} 
\newcommand{\ppbar}{\rm p{\bar p}} 

\def\mumu{\mu\mu\/}

\def\W{W}

\def\Zeg{e+`\gamma`}

\newcommand{\goes}{\kern -0.18em\rightarrow\kern -0.18em}
\newcommand{\plus}{\kern -0.18em +\kern -0.18em}
\def\Z{Z^0}

\def\lum{{\cal L}}
\def\epem{{\rm e^{+}e^{-}}}

\def\tptm{{\tau^{+}\tau^{-}}}
\def\roots{{\sqrt s}}

\def\95cl{95 \%~C.L.}
\def\95CL{95 \%~C.L.}
\def\r#1 {$^{#1}$}

\def\epem{e^+e^-}

\newcommand{\Wtaunu}{\mbox{W$^\pm \rightarrow \tau^{\pm}\nu$}}

\newcommand{\Wmunu}{\mbox{W$^\pm \rightarrow \mu^{\pm}\nu$}}
\newcommand{\wmunu}{\mbox{W$^\pm \rightarrow \mu^{\pm}\nu$}}

\newcommand{\Wlnu}{\mbox{W$^\pm \rightarrow l^{\pm}\nu$}}

\newcommand{\Zmumu}{\mbox{Z$^0 \rightarrow \mu ^+ \mu ^-$}}
\newcommand{\zmumu}{\mbox{Z$^0 \rightarrow \mu ^+ \mu ^-$}}

\newcommand{\Zee}{\mbox{Z$^0 \rightarrow e^+e^-$}}
\newcommand{\zee}{\mbox{$Z^0 \rightarrow e^+e^-$}}
\newcommand{\Zll}{\mbox{Z$^0 \rightarrow \ell^+\ell^-$}}

\newcommand{\xtiny}{\tiny}
\newcounter{question}

\newcommand{\NONE}{\mbox{$\widetilde{\chi}_1^0$}}
\newcommand{\NTWO}{\mbox{$\widetilde{\chi}_2^0$}}

\newcommand{\none}{\NONE}
\newcommand{\ntwo}{\NTWO}

\newcommand{\TeV}{\ensuremath{\mathrm{Te\kern -0.1em V}}\xspace}
\newcommand{\bfTeV}{\ensuremath{\bf{Te\kern -0.1em V}}\xspace}
\newcommand{\rrr}{\rightarrow}
\newcommand{\Etgamma}{\ensuremath{\mathrm{E_T^{\gamma}}}}
\newcommand{\Etlepton}{\ensuremath{\mathrm{E_T^{\ell}}}}

\newcommand{\noflg}{574}

\newcommand{\noflgdphi}{422}

\newcommand{\noflgzlike}{218}

\newcommand{\noflgzunlike}{79}

\newcommand{\noflgtwobody}{204}

\newcommand{\noflgallother}{152}

\newcommand{\noflglgmet}{42}

\newcommand{\noflgmultil}{31}

\newcommand{\noflgmultig}{0}

\input{predictions/total.summary}

\newcommand{\apprluminosity}{305}

\input{predictions/events_pb/events_pb.summary}
\newrgbcolor{maroon}{.45 0.1 0.1}
\begin{document}
\bibliographystyle{unsrt}
\thispagestyle{empty}
{\sf
\vspace{-3cm}

\begin{center}
{State Scientific Center of the Russian Federation \\
INSTITUTE FOR THEORETICAL AND EXPERIMENTAL PHYSICS\\
Moscow, Russia}

\vspace{1cm}
\hfill{\small manuscript}

\vspace{3cm}

{\itshape LOGINOV ANDREY BORISOVICH}

\vspace{0.7cm}

{\Large\bf\boldmath 
Search for Anomalous Production of Events \\ 
with a High Energy Lepton and Photon at the Tevatron}

\vspace{2.5cm}

Speciality 01.04.23 - High Energy Physics

\vspace{0.7cm}

The dissertation submitted in conformity with the requirements\\
for the degree of
Doctor of Philosophy\\
at the \\
Institute for Theoretical and Experimental Physics

\vspace{1.5cm}

\hfill Thesis co-supervisors:

\vspace{0.5cm}

\hfill Professor H.~J.~Frisch

\hfill University of Chicago

\vspace{0.5cm}

\hfill Professor A.~A.~Rostovtsev

\hfill ITEP Moscow

\vspace{1.45cm}

Moscow 2006
\end{center}

}

\newpage
\thispagestyle{empty}
\vspace*{3in}

\newpage
\thispagestyle{empty}

{\sf
\vspace{-3cm}

\begin{center}
{State Scientific Center of the Russian Federation \\
INSTITUTE FOR THEORETICAL AND EXPERIMENTAL PHYSICS\\
Moscow, Russia}

\vspace{1cm}
\hfill{\small manuscript}

\vspace{3cm}

{\itshape LOGINOV ANDREY BORISOVICH}

\vspace{0.7cm}

{\Large\bf\boldmath 
Search for Anomalous Production of Events \\ 
with a High Energy Lepton and Photon at the Tevatron}
\end{center}
}
\vspace{0.2cm}

\begin{center}
{\bf Abstract}
\vspace{-0.3cm}
\end{center}

We present results of a search for for the anomalous production of
events containing a high-transverse momentum charged lepton ($\ell$,
either $e$ or $\mu$) and photon ($\gamma$), accompanied by missing
transverse energy ($\met$), and/or additional leptons and photons, and
jets (X). We use the same kinematic selection criteria as in a
previous CDF search, but with a substantially larger data set,
$\apprluminosity$ $\invpb$, a $\ppbar$ collision energy of 1.96
$\TeV$, and the upgraded CDF II detector. We find $\noflglgmet$
$\lgmet$ events versus a standard model expectation of $\smnoflglgmet
\pm \totdsysnoflglgmet$ events. The level of excess observed in Run I,
16 events with an expectation of 7.6 $\pm$ 0.7 events (corresponding
to a 2.7$\sigma$ effect), is not supported by the new data. In the
signature of $\llg + X$ we observe $\noflgmultil$ events versus an
expectation of $\smnoflgmultil \pm
\totdsysnoflgmultil$ events. In this sample we find no events with an
extra photon or $\met$ and so find no events like the one $\eeggmet$
event observed in Run I.

{\sf
\begin{center}
\vspace{0.4cm}

\hfill Thesis co-supervisors:

\vspace{0.5cm}

\hfill Professor H.~J.~Frisch

\hfill University of Chicago

\vspace{0.5cm}

\hfill Professor A.~A.~Rostovtsev

\hfill ITEP Moscow

\vspace{1.4cm}

Moscow 2006
\end{center}
}

\newpage
\thispagestyle{empty}
\vspace*{4in}

\newpage
\thispagestyle{empty}
\vspace*{4in}

\begin{center}
...to my dear grandmother...
\end{center}

\newpage

\tableofcontents
\newpage
\listoffigures
\newpage
\listoftables

\newpage
\chapter*{Introduction}
\label{introduction}
\addcontentsline{toc}{section}{Introduction}

An important test of the standard model (SM) of particle
physics~\cite{SM} is to measure and understand the properties of the
highest momentum-transfer particle collisions, which correspond to
measurements at the shortest distances. The predicted high energy
behavior of the SM, however, becomes unphysical at an interaction
energy on the order of several TeV. These phenomena beyond the SM may
involve new elementary particles, new fundamental forces, and/or a
modification of space-time geometry. These new phenomena are likely to
show up as an anomalous production rate of a combination of the known
fundamental particles.

The unknown nature of possible new phenomena in the energy range
accessible at the Tevatron is the motivation for a search strategy
that does not focus on a single model of new physics, but presents a
wide net for new phenomena. We compared SM predictions with the rates
measured at the Tevatron with the CDF detector for final states with
at least one high-PT lepton (e or $\mu$) and photon ($\gamma$), plus
other detected objects (leptons, photons, jets, and missing transverse
energy, $\met$). {\it A priori} definition of selection cuts for the
search allows to test Run I anomalies, such as the observation of an
event consistent with the production of two energetic photons, two
energetic electrons, and large $\met$ (the ``$\eeggmet$ event''), in
Run II data.  Another intriguing Run I result that is important to
test is a 2.7$\sigma$ excess above the Standard Model expectations in
the $\lgmet$ signature~\cite{jeff_prd,jeff_prl}.

The Fermilab Tevatron has the highest center-of-mass energy collisions
(per nucleon) of any accelerator to date, and thus has the potential
to discover new physics. The upgraded CDF II detector provides us
better solid angle coverage and particle identification. The
production of two vector gauge bosons, precisely predicted in the
Standard Model, provides a set of signatures in which to search for
the production of new particles which couple to the SM gauge sector
(the top quark being the last new example).

This analysis has been done with 305 $\invpb$ of $\ppbar$ collisions at
$\roots$= 1.96 $\TeV$, collected with CDF detector at the Tevatron,
Fermilab between March 23, 2002 and August 22, 2004. The main results
of this thesis have been published in
~\cite{lgx_prl,newphysics_photons_epj,lgx_crimea}. Standard Model
$\Wg$ and $\Zg$ production CDF Run II results are published
in~\cite{CDF_WZgamma}. The status of the Lepton+Photon+X search has
been presented at the APS Conference (Philadelphia, 2003). 
The
results have been presented at the SUSY 2005 conference (Durham,
2005)~\cite{newphysics_photons_susy2005}, the International School of
Subnuclear Physics (Erice, 2005)~\cite{newphysics_photons_epj}, Lake
Louise Winter Institute (2006)~\cite{henry_lake_louise_institute}, and
also at the CDF Collaboration Meeting (Sitges, 2005) and at the
Exotics, Photon and Very Exotic Phenomena working group meetings.

At the International School of Subnuclear Physics (Erice, 2005) I have
received the ``New Talents'' Award for an Original Work in
Experimental Physics for the talk ``Search for New Physics in Photon
Final states''. The work has been reviewed and approved for
publication in~\cite{lgx_prl,newphysics_photons_epj} by G. t'Hooft,
1999 Nobel Laureate in Physics.

One of the most important tools for a better understanding of the
events that could possibly be New Physics candidates is a CDF Run II
Event Display visualization package~\cite{evd_code,evd_manual}, which
is widely used for offline analysis as well as to monitor online data
taking~\cite{evd_live}. Development and Support of the EventDisplay
package is a responsibility of ITEP (Moscow) group at CDF. I am the
project leader~\cite{evd_team} and responsible for this task.

The thesis consists of an introduction, 13 chapters, and conclusions.
Chapter~\ref{motivation} presents the motivation for the analysis, and
gives an introduction to the Run I results and Signature-Based
searches. Chapter~\ref{cdfattevatron} gives a description of the CDF
experiment at the Tevatron Collider. We describe the CDF coordinate
system, and give information about the tracking, calorimetry, muon and
luminosity systems. We introduce the trigger and data acquisition
systems.

Chapter~\ref{evd} presents a detailed description of the CDF Run II
Event Display (EVD) package and related projects. The EVD is used for
online monitoring, offline analysis and for public relation (PR)
purposes.

Chapter~\ref{lgx_selection} presents the inclusive high-$\Pt$
electron, muon, and photon datasets from which we select $\lgX$
candidates, as well as the time intervals of data-taking, used to test
the stability of the event yields (Section~\ref{datasets}). It also
presents an overview of the kinematic selection criteria for the
$\lgX$ events (Section~\ref{selection}).

The identification criteria for objects and control samples for muon
candidates are described in detail in Chapter~\ref{muonid}, for
electron candidates in Chapter~\ref{electronid}, and for photon
candidates in Chapter~\ref{photonid}. Chapter~\ref{miscid} describes
how missing transverse energy ($\met$) is calculated, and gives the
definition and describes calculation of the total transverse energy
($\Ht$).

Chapter~\ref{predictions} presents the Standard Model expectations
from SM physics processes that give the lepton-photon signature.  The
primary ones are production of $\Wg$, $\Zg$; we include estimates from
the two-photon (3-boson) processes $\Wgg$ and $\Zgg$. For each of
these predictions we have used at least 2 independent Monte Carlo
generators. Backgrounds from SM processes with a `fake' (misidentified
other object, such as a jet) photon or lepton are described in
Chapter~\ref{fakes}. Chapter~\ref{systematics} gives an overview of
the experimental, theoretical and luminosity systematic uncertainties.

Chapter~\ref{photonlepton} presents the topologies of the signatures
we are looking for, and gives the number of events
observed. Section~\ref{comparison} gives the comparison of the
observed event counts with expectations from the sum of SM physics
processes and background. In conclusion we summarize the results and
present future prospects.

Appendix~\ref{a.appendix} presents lists of Lepton-Photon-$\met$ and
Multi-Lepton-Photon events (Section~\ref{list_of_lgx}) and additional
plots for $\lgmet$ and $\llg$ signatures (Sections~\ref{lgmet_plots}
and~\ref{llg_plots}). It also presents the stability of the event
yields for W+jets and Z+jets (Section~\ref{stability_plots_zj_wj}),
distributions of the isolation variables for different muon types
(Section~\ref{cmx_vs_cmup}). Finally, it presents supplementary
information about conversion electrons (Section~\ref{conversions}) and
additional checks for non-Z backgrounds for the $\mumug$ signature
(Section~\ref{mumug_checks}).

\chapter{Motivation}
\label{motivation}

The goal of elementary particle physics is to find the ultimate
constituents of matter and to study the fundamental interactions that
occur among them. To address these questions we need to perform
measurements at the shortest distances, and therefore to study the
properties of the highest momentum-transfer particle
collisions. Particle physics seeks a classification of the elementary
particles and a consistent theoretical description of their
interactions that leads to an accurate description of experimental
observables.

\section{Standard Model, Supersymmetry, or Something Else?}
\label{sm_or_susy}

The Standard Model (SM) is an effective field theory~\cite{SM} that
has so far described the fundamental interactions of elementary
particles remarkably well. 
All of the data from collider experiments, are explained and (in
principle) are calculable in the framework of the SM. However, the SM
does not include gravity and is expected to be an effective low-energy
field theory~\cite{higgs_hunters_guide}. The SM contains no dark
matter candidate(s). The SM higgs boson mass receives quadratically
divergent loop corrections. This results in the well-known hierarchy
problem~\cite{hierarchy} of the SM.

The different approaches to solving the hierarchy problem include
eliminating the Higgs scalar entirely from the theory
(Technicolor~\cite{lane-1995-352}), lowering the cutoff scale (large
extra dimensions~\cite{ledsref}), or embedding the Higgs field in a
multiplet of a symmetry group larger than the 4D Poincare group
(supersymmetry~\cite{susy,allanach-2002-89}).

The existence of supersymmetry (SUSY) would provide solutions to the
fine tuning problem~\cite{susskind}, and possibly the hierarchy
problem, which we currently encounter in the SM. The experimental
signatures of supersymmetry are complex, as all known fermions of the
SM have bosons as supersymmetric partners while all bosons acquire
fermions as superpartners. Due to the large number of free parameters,
it is necessary to make further assumptions in the context of specific
SUSY models~\cite{MSSM} for specific searches.

Part of the SM unified "electroweak" theory of the electromagnetic and
weak forces is based on the exchange of four particles: the photon for
electromagnetic interactions, and two charged W particles and a
neutral Z particle for weak interactions. These particles, $\gamma$,
$Z^0$, $W^{\pm}$, are fundamental in the SM.

In searching for new particles/quantum numbers, the signature of pairs
of gauge bosons ($\Wg$, $\Zg$, $WW$, $\Z\Z$, $\gamma\gamma$) is
natural if pairs of particles with a conserved quantum number are
produced because of flavor conservation in QCD. $\Wg$ and $\Zg$ SM
physics processes lead to inclusive production of events with a
high-energy lepton and a high-energy photon.

\section{The Lepton-Photon Events}
\label{lg_signature}

Besides the specific theoretical models, searching for new physics
with photons has several advantages. For example, the photon is one of
the three SU(2) x U(1) gauge bosons and as such is likely to be a good
probe of new interactions since it might couple to any new gauge
sector. Final-state photons have additional distinct detection
advantages over $W^\pm$ or $Z^0$ bosons since they do not decay. Thus
they do not suffer a sensitivity loss from branching ratios and
momentum sharing between the decay products. The photon is coupled to
electric charge, and thus is radiated by all charged particles,
including the incoming states, which is important for searching for
invisible final states. The photon is a boson and could be produced by
a fermiphobic parent. For the search we also require a lepton: the
events containing high-$E_T$ photon and high-$P_T$ lepton, $\lgX$, are
rare in the SM, and therefore backgrounds are low.

There are many models~\cite{lhc_wkgp} of new physics that could
produce $\lgX$ events. Gauge-mediated models of
supersymmetry~\cite{susy_gauge}, in which the lightest super-partner
(LSP) is a light gravitino, provide a model in which each partner of a
pair of supersymmetric particles produced in a $\ppbar$ interaction
decays in a chain that leads to a produced gravitino, visible as
$\met$. If the next-to-lightest neutralino (NLSP) has a photino
component, each chain also can result in a photon. Models of
supersymmetry in which the symmetry breaking is due to gravity also
can produce decay chains with photons~\cite{susy_gravity}. For
example, if the NLSP is largely photino-like, and the lightest is
largely higgsino, decays of the former to the latter will involve the
emission of a photon~\cite{Ambrosiano}. More generally,
pair-production of selectrons or gauginos can result in final-states
with large $\met$, two photons and two leptons and lead to events like
the Run I $\eeggmet$ candidate event (Section~\ref{runi_results}).

For example, an initial model invoked low-energy supersymmetry with a
neutralino LSP as a possible interpretation of the CDF Run I
$\eeggmet$ event~\cite{hepph9607414} via the process:

$p\overline{p} \rightarrow {\tilde e}^+{\tilde e}^- (+ X)$, ${\tilde
e} \rightarrow \ntwo + e$, $\ntwo \rightarrow \none\gamma$

where ${\tilde e}$ is the selectron (the bosonic partner of the
electron), and $\none$ and $\ntwo$ are the lightest and
next-to-lightest neutralinos, the fermionic partners of the neutral
bosonic states formed by mixing the fermionic partners $W^0$, the $B$,
and the neutral Higgses into mass eigenstates.

Gauge-mediated models in which a photino decays into a gravitino are
also popular choices\cite{susy_gravity}, and have the appealing
feature that they have a natural dark matter candidate.

\begin{figure}[!h]
\centering
\includegraphics[angle=0,width=3.5in,clip]{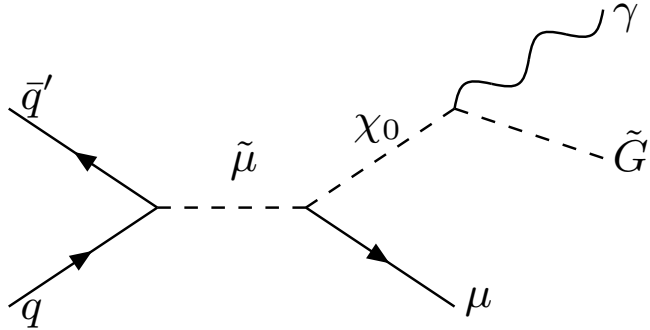}
\caption{Resonant smuon production and subsequent decay, producing the $\mu\gamma\met$ events}
\label{smuon_production.figure}
\end{figure}

Further expanding the space of parameters, a more recent SUSY
interpretation of the CDF { $\mgmet$} events~\cite{allanach-2002-89}
is resonant smuon $\tilde{\mu}$ production with a single dominant
R-parity violating coupling(Figure~\ref{smuon_production.figure}).

The current interest in models of extra dimensions~\cite{ledsref},
which can produce events of interest to the $\lgX$ search, is a good
example of an innovation that was searched for before it was
conceived. These models predict excited states of the known standard
model particles. The production of a pair of excited
electrons~\cite{excited_electron} would provide a natural source for
two photons and two electrons (although not $\met$ unless the pair
were produced with some other, undetected, particle.). As in the case
of supersymmetry, there are many parameters in such models, with a
resulting broad range of possible topologies with multiple gauge
bosons.

However the parameter space of SUSY models is so large, and there are
so many other models beyond SUSY, including ones that have not yet
been thought of, that we have adopted the strategy of testing the SM
predictions in promising signatures. This strategy, the {\it
Signature-Based Search}, is nothing more than testing the
SM~\cite{SM}.

\section{Signature-Based Searches}
\label{signature_based_searches}

While it is good to be guided by theory, one should also remain open
to the unexpected. Therefore we use a quasi-model-independent {\it
Signature-Based Searches} technique, and look for significant
deviations from the SM~\cite{toback_thesis,jeff_thesis,ray_prd}. In
the Run I dataset, no significant evidence for new physics was found,
but there were some hints that the SM may be incomplete
(Section~\ref{runi_results}). CDF has preferred to highlight some
potential anomalies as worth pursuing in Run II, thus setting up
selection criteria in {\it a priori} fashion.

We perform the search by systematically looking at events by their
final state particles. The strategy for the {\it Signature-Based
Search} is to test the SM by looking for an excess over the SM
prediction. The challenge also extends to the theoretical community -
to look for something new we will need to understand the non-new,
i.e. the SM predictions, at an unprecedented level of precision. Some
amount of this can be done with control samples - it is always best to
use data rather than Monte Carlo, but this is not always possible.

\section{Run I Results and Present Analysis}
\label{runi_results}

\subsection{The $\eeggmet$ Candidate Event}
\label{eeggmet.section}

\begin{figure}[!h]
\centering
\resizebox{0.6\textwidth}{!}{%
\includegraphics*{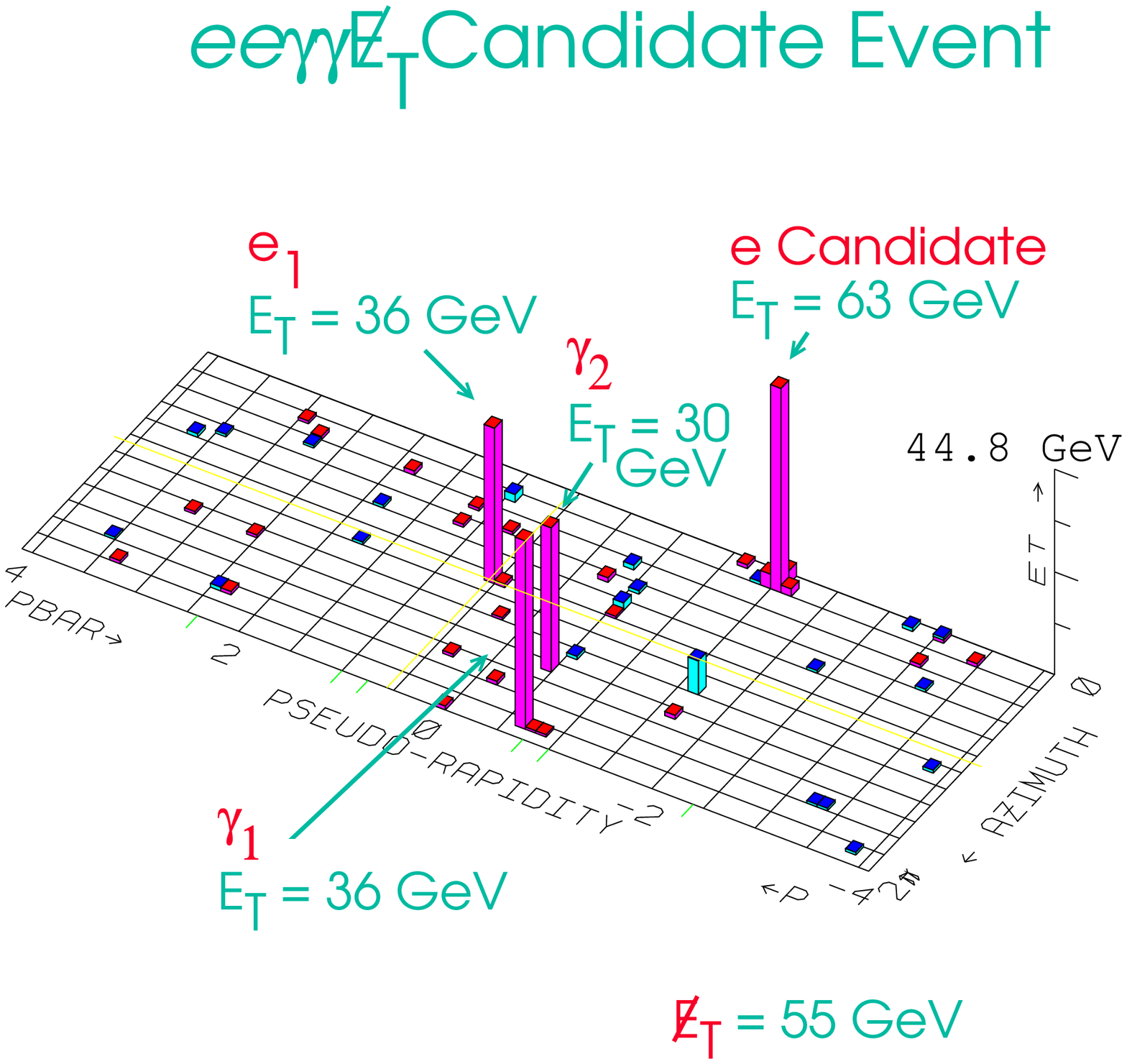}
}
\caption
[The Run I $\eeggmet$ candidate event]
{The Run I $\eeggmet$ candidate event.}
\label{eeggmet.figure}       
\end{figure}

In 1995 the CDF experiment, measuring $\pbarp$ collisions at a
center-of-mass energy of 1.8 TeV at the Fermilab Tevatron, using 85
$\invpb$ of data, observed an event consistent with the production of
two energetic photons, two energetic electrons, and large missing
transverse
energy~\cite{toback_prd,toback_prl,toback_thesis}(Figure~\ref{eeggmet.figure}).

This signature is predicted to be very rare in the Standard Model of
particle physics, with the dominant contribution being from the
WW$\gg$ production: $WW\gamma\gamma\goes (e\nu)(e\nu)\gamma\gamma\goes
ee\gamma\gamma\met$, from which we expect 8$\times$10$^{-7}$ events.
All other sources (mostly due to detector misidentification) lead to
5$\times 10^{-7}$ events. Therefore, we expect (1 $\pm$ 1) $\times$
10$^{-6}$ events, which would give us one $\eeggmet$ candidate event
if we have taken one million times more data than we actually had in
Run I.

The event raised theoretical interest, however, as the two-lepton
two-photon signature is expected in some models of physics `beyond the
Standard Model'~\cite{SM} such as gauge-mediated models
of supersymmetry~\cite{allanach-2002-89,Ambrosiano}. 

\subsection{$\gg$+X Search}
\label{ggx.section}
The detection of this single event led to the development of
`signature-based' inclusive searches in Run I to cast a wider net: in this case
one searches for two photons + X
~\cite{toback_prd,toback_prl,toback_thesis}, where X stands for
anything, with the idea that if pairs of new particles were being
created these inclusive signatures would be sensitive to a range of
decay modes or the creation and decay of different particle types.

In Run I Searches for {\bf$\gg$+X}, all results were consistent with
the SM background expectations with no other exceptions
other than observation of $\eeggmet$ candidate
event(Table~\ref{ggx_runi.table})~\cite{toback_prl}.

\begin{table}[!t]
\begin{center}
\begin{tabular}{lcc}
\hline
Signature (Object) & Obs. & Expected \\
\hline
\mettgmh, $|\Delta\phi_{\mettsm-{\rm jet}}|>10^\circ$           & 1    & 0.5 $\pm$ 0.1       \\
N$_{\rm jet}\ge 4$, ${\rm E}_{\rm T}^{\rm jet}>10$~GeV, 
$|\eta^{\rm jet}|<2.0$                                          & 2    & 1.6 $\pm$ 0.4       \\
$b$-tag, ${\rm E}_{\rm T}^{b}>25$~GeV                           & 2    & 1.3 $\pm$ 0.7       \\
Central $\gamma$, ${\rm E}_{\rm T}^{\gamma_3}>25$~GeV           & 0    & 0.1 $\pm$ 0.1       \\
Central $e$ or $\mu$, ${\rm E}_{\rm T}^{e~{\rm or}~\mu}>25$~GeV & 3    & 0.3 $\pm$ 0.1       \\
Central $\tau$, ${\rm E}_{\rm T}^{\tau}>25$~GeV                 & 1    & 0.2 $\pm$ 0.1       \\
\hline
\end{tabular}
\caption
[Run~I $\gamma\gamma+X$ search]
{Number of observed and expected  $\gamma\gamma$ events with additional
objects in 85 pb$^{-1}${\protect \cite{toback_prl}}}
\label{ggx_runi.table}
\end{center}
\end{table}

\subsection{From $\gg$ to $\lg$: $\lgX$ Search}
\label{lgx.section}

Another `signature-based' inclusive search, motivated by $\eeggmet$
event was for one photon plus one lepton +
X~\cite{jeff_prd,jeff_prl,jeff_thesis}. In general data agrees with
expectations, with the exception for the $\lgmet$ category. We have
observed 16 $\lgmet$ events on a background of 7.6 $\pm$ 0.7
expected. The 16 $\lgmet$ events consist of 11 $\mugmet$ events and 5
$\egmet$ events, versus expectations of 4.2$\pm$0.5 and 3.4$\pm$0.3
events, respectively.  The SM prediction yields the observed rate of
$\ell\gamma\met$ with {\bf 0.7\%} probability (which is equivalent to
{\bf 2.7} standard deviations for a Gaussian distribution). One of the
first SUSY interpretation of the CDF { $\mgmet$}
events~\cite{allanach-2002-89} was resonant smuon $\tilde{\mu}$
production with a single dominant R-parity violating
coupling(Figure~\ref{smuon_production.figure}).

\begin{table}[!t]
\begin{center}
\begin{tabular}{l c c c}
\hline
Category & $\mu_{SM}$ & $N_0$ & P($N\ge N_0|\mu_{SM}$), \% \\
\hline
All $\lg X$           &          --           &           {\bf 77}     &   -- \\
\hline
Z-like $e\gamma$      &          --           &            17     &   -- \\
Two-Body   $\lg    X$ &      24.9$\pm$2.4  &            33     &   9.3 \\
Multi-Body $\lg    X$ &      20.2$\pm$1.7  &            27     &  10.0 \\
\hline
Multi-Body $\llg   X$ &      5.8  $\pm$ 0.6  &             5      &  68.0 \\
Multi-Body $\lgg   X$ &      0.02$\pm$0.02 &             1      &   1.5 \\
 Multi-Body $\lgmet X$ &
{\bf7.6  $\pm$ 0.7}        & 
{\bf16 }                   &
{\bf0.7} \\
\hline
\end{tabular}
\caption
[Run~I $\lgX$ search]
{Run I Photon-Lepton Results: Number of observed and expected  $\ell\gamma$ events with additional
objects in 85 pb$^{-1}${\protect \cite{jeff_prl}}}
\label{lgx_runi.table}
\end{center}
\end{table}

The Run I search was initiated by an anomaly in the data itself, and
as such the 2.7 sigma excess above the SM expectations must be viewed
taking into account the number of such channels that a fluctuation
could have occurred in. The Run I paper concluded: ``However, an
excess of events with 0.7\% likelihood (equivalent to 2.7 standard
deviations for a Gaussian distribution) in one subsample among the
five studied is an interesting result, but it is not a compelling
observation of new physics.  We look forward to more data in the
upcoming run of the Fermilab Tevatron.''~\cite{jeff_prl}.

\chapter{The CDF Experiment at the Tevatron Collider}
\label{cdfattevatron}

An important part of the study of elementary particle physics is to
understand experimental tools - the accelerators, beams and detectors
by means of which particles are accelerated, their trajectories
controlled and their collisions studied.

The Tevatron is currently the world's highest energy particle
accelerator. Protons(p) and anti-protons ($\pbar$) are accelerated to
be brought into collision with a center of mass energy of 1.96
TeV. Two detectors are situated at the B$\O$ and D$\O$ collision
points, the Collider Detector at Fermilab (CDF) and D$\O$.

\section{The Tevatron}
\label{tevatron}
\begin{figure}[!b]
\centering
\includegraphics[width=270pt,angle=0]{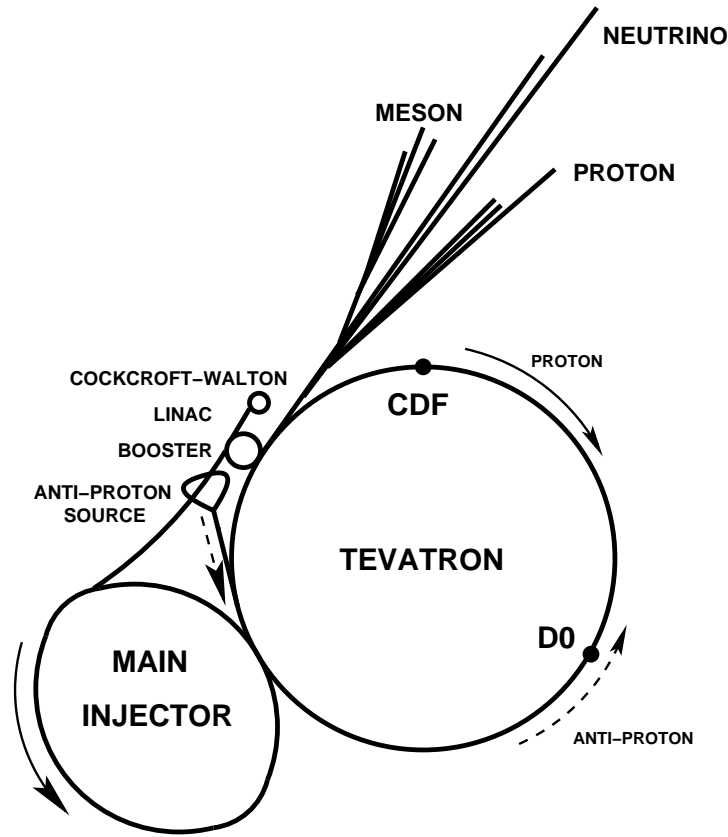}
\caption[Layout of the Fermilab accelerator complex]
{Layout of the Fermilab accelerator complex. Protons (solid arrow) are 
accelerated at the Cockcroft-Walton, Linac, Booster, Main Injector and 
finally at the Tevatron. The anti-protons (dashed arrow) from the anti-proton 
source are first accelerated at the Main Injector and then at the Tevatron.}
\label{fig:acc_complex}
\end{figure}

Fermilab uses a series of accelerators to create the world's most
energetic particle beams. The diagram in Figure~\ref{fig:acc_complex}
shows the paths taken by p and $\pbar$ from initial acceleration to
collision in the Tevatron. In the first stage of acceleration H$^{-}$
ions are created from the ionization of the hydrogen gas and
accelerated to a kinetic energy of 750 KeV in the Cockcroft-Walton
pre-accelerator~\cite{cockcroft:web}.
The H$^-$ ions enter a linear accelerator
(Linac)~\cite{Schmidt:1993fz}, where they are accelerated to 400
MeV. The acceleration in the Linac is done by a series of ``kicks''
from Radio Frequency (RF) cavities. The oscillating electric field of
the RF cavities groups the ions into bunches.
Before entering the next stage, a carbon foil removes the electrons
from the H$^-$ ions at injection, leaving only the protons. The 400
MeV protons are then injected into the circular synchrotron
(``Booster''). The protons travel around the Booster to a final energy
of 8~GeV.

Protons are then extracted from the Booster into the Main Injector
~\cite{main_injector}, where they are accelerated from 8~GeV to
150~GeV before the injection into the Tevatron. The Main Injector also
produces 120~GeV protons. These protons are extracted and collide with
a nickel target, producing a wide spectrum of secondary particles,
including $\pbar$. In the collisions, about 20 $\pbar$ are produced
per one million protons. The $\pbar$ are collected, focused, and then
stored in the Accumulator ring. Once a sufficient number of $\pbar$
are collected, they are sent to the Main Injector and accelerated to
150~GeV.

Finally, both the p and $\pbar$ are injected into the Tevatron. The
Tevatron, the last stage of Fermilab's accelerator chain, receives
150~GeV p and $\pbar$ from the Main Injector and accelerates them to
980~GeV. The p and $\pbar$ travel around the Tevatron in opposite
directions. The beams are brought to collision at the center of the
two detectors, CDF II and D$\O$ II (see Figure~\ref{fig:acc_complex}).

\section{The CDF Detector}
\label{cdfdetector}

A discovery will rely heavily on a thorough understanding of the
detector.  Two aspects are critical: the identification of objects
that make up each signature, and the understanding of the calibration
and resolution of the detector. The objects for which we have a good
understanding of the efficiencies and fake-rates are those for which
tracking is essential: electrons, muons, and photons (i.e. a high
confidence of the absence of a track), all in the central
region. Similarly, the energy scale and resolutions of the
calorimeters are well understood in the central region, where the
magnetic spectrometer is used to calibrate the calorimeters.

\begin{figure}[!t]
   \begin{center}
      \parbox{4.4in}{\epsfxsize=\hsize\epsffile{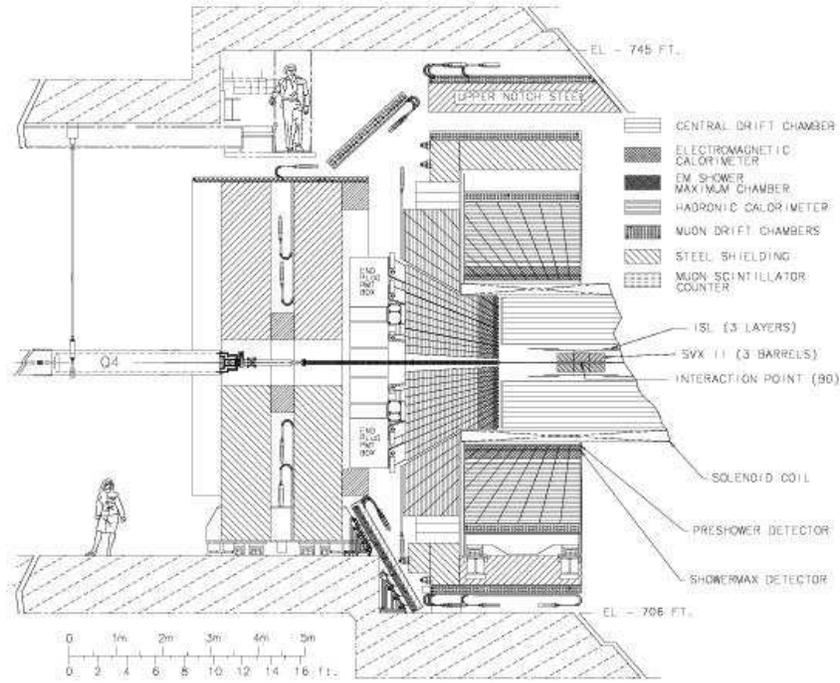}}
\caption 
[Cross-section r-z view of the CDF]
{Cross-section r-z view of the CDF 
(see Section~\ref{detector_coordinate_system}).}
      \label{fig:cdfelev}
   \end{center}
\end{figure}

The CDF II detector is a cylindrically-symmetric spectrometer designed
to study $\ppbar$ collisions at the Fermilab Tevatron based on the
same solenoidal magnet and central calorimeters as the CDF I
detector~\cite{CDFI}. A cross-section of one half of the detector is
shown in Figure~\ref{fig:cdfelev}.

Because the analysis described here is intended to repeat the Run I
search as closely as possible, we note especially the differences from
the CDF I detector relevant to the detection of leptons, photons, and
$\met$. The tracking systems (Section~\ref{detector_tracking}) used to
measure the momenta of charged particles have been replaced with a
central outer tracker (COT) that has smaller drift cells~\cite{COT},
and an enhanced system of silicon microstrip detectors~\cite{SVX}. The
calorimeters in the regions (Section~\ref{detector_coordinate_system})
with pseudorapidity $|\eta| \gt 1$ have been replaced with a more
compact scintillator-based design, retaining the projective geometry
(Section~\ref{detector_calorimetry}).  The coverage in $\varphi$ of
the central upgrade muon detector (CMP) and central extension muon
detector (CMX) systems (Section~\ref{detector_muon}) has been
extended; the central muon detector (CMU) system is unchanged.

The main upgrades to the CDF detector from Run~I to Run~II, relevant
to the analysis, can be summarized as follows:

\begin{itemize}
\setlength{\itemsep}{-0.07in}
\item {Fully digital DAQ system designed for 132 ns bunch crossing times}
\item {Significantly upgraded silicon detector:}
\subitem{-- 707,000 channels compared with 46,000 in Run~I}
\subitem{-- Axial, stereo, and 90$^\circ$ strip readout}
\subitem{-- Full coverage over the luminous region along the beam axis}
\subitem{-- Radial coverage from 1.35 to 28 cm for $|\eta| < 2$}
\subitem{-- Innermost silicon layer(``L00'') on the beampipe with 6 $\mu$m axial hit resolution}
\item {Outer drift chamber capable of 132 ns maximum drift}
\subitem{-- 30,240 sense wires, 44-132 cm radius, 96 d$E$/d$x$ samples possible per track}
\item {Fast scintillator-based calorimetry out to $|\eta| \simeq 3$}
\item {Expanded muon coverage}
\item {Improved trigger capabilities}
\subitem{-- Drift chamber tracks with high precision at Level-1}
\subitem{-- Silicon tracks for detached vertex triggers at Level-2}
\item {Expanded particle identification via time-of-flight and d$E$/d$x$}
\end{itemize}

\subsection{CDF Coordinate System}
\label{detector_coordinate_system}

The CDF detector uses a right-handed coordinate system. The horizontal
direction pointing out of the ring of the Tevatron is the positive
$x$-axis. The vertical direction pointing upwards is the positive
$y$-axis. The proton beam direction, pointing to the east, is the
positive $z$-axis.

A spherical coordinate system is also used. The radius $r$ is measured
from the center of the beamline. The polar angle $\theta$ is taken
from the positive $z$-axis. The azimuthal angle $\varphi$ is taken
counter-clockwise from the positive $x$-axis.

At a $p\bar{p}$ collider, the production of any process starts from a
parton-parton interaction which has an unknown boost along the
$z$-axis, but no significant momentum in the plane perpendicular to
the $z$-axis, i.e. the transverse plane. This makes the transverse
plane an important plane in a $p\bar{p}$ collision. Momentum
conservation requires the vector sum of the transverse energy and
momentum of all of the final particles to be zero. The transverse
energy $\Et$ and transverse momentum $\Pt$ for a particle produced in
a $\ppbar$ collision are defined by $\Et = E\times\sin\theta$ and $\Pt
= p\times\sin\theta$. We use the convention that ``momentum'' refers
to $pc$ and ``mass'' to $mc^2$.

Another quantity invariant under Lorentz boosts along the beamline is
Rapidity, which is defined as
$y={{1}\over{2}}log({{E+P_L}\over{E-P_L}})$, where $P_L$ is the
longitudinal momentum along the beamline and E is the energy.

Pseudorapidity $\eta$ is used by high energy physicists and is defined
as $\eta = -\ln\tan\frac{\theta}{2}$. For massless particles
$\eta\equiv y$.

Hard $p\bar{p}$ head-on collisions produce significant momentum in the
transverse plane. The CDF detector has been optimized to measure these
events. Typically, particles in a $p\bar{p}$ collision event tend to
be more in the forward and backward regions than in the central region
because there is usually a boost along the $z$-axis. The derivative of
$\eta$ is $d\eta = -\frac{d\theta}{\sin\theta}$.

A constant $\eta$ slice corresponds to variant $\theta$ slice which is
smaller in the forward and backward regions than in the central
region. This makes the $\eta$ occupancy more uniform than $\theta$
occupancy. Therefore, for example, calorimeters are constructed in
$\eta$ slices instead of $\theta$ slices.

\subsection{Tracking}
\label{detector_tracking}

\begin{figure}
   \begin{center}
      \parbox{4.7in}{\epsfxsize=\hsize\epsffile{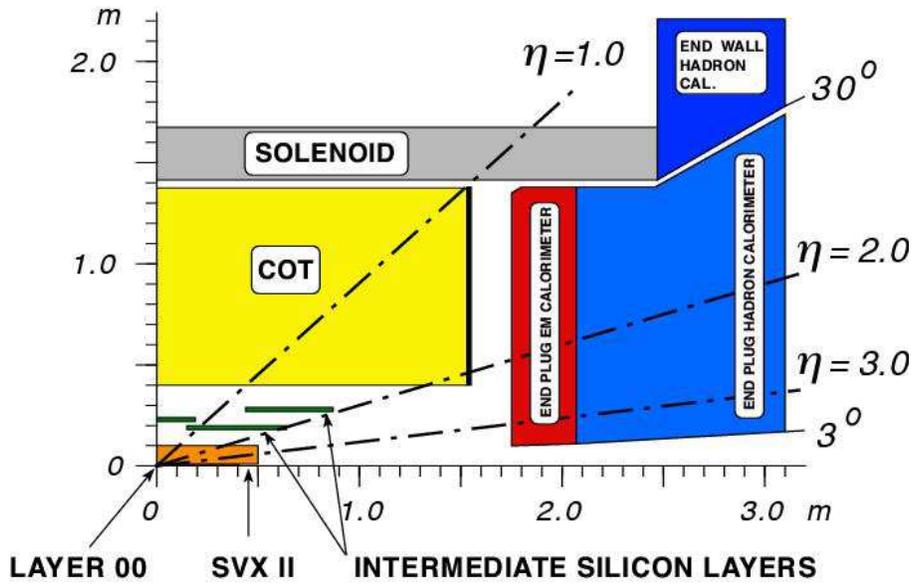}}
      \caption[CDF II tracking volume]
              {CDF II tracking volume.}
      \label{fig:cdfii_tracker_quad}
   \end{center}
\end{figure}

The CDF detector features excellent charged particle tracking and good
electron and muon identification in the central region.  The detector
is built around a 3 m diameter, 5 m long superconducting solenoid
operated at 1.4 T. The tracking volume is surrounded by the solenoid
magnet and the endplug calorimeters as shown in
Figure~\ref{fig:cdfii_tracker_quad}.

The CDF tracking system includes a central outer drift chamber (COT)
and the silicon tracker. The main parameters of the CDF tracking
system are summarized in Table~\ref{tb_tracking}.

\begin{table}[htbp]
\begin{center}
\begin {tabular} {l c}
\hline \hline
COT                         &    \\ \hline
Radial coverage             & 44 to 132 cm \\
Number of superlayers       &  8  \\
Measurements per superlayer & 12  \\
Maximum drift distance      & 0.88 cm \\
Resolution per measurement  & 180 $\mu$m \\
Rapidity coverage           &  $|\eta| \leq 1.0$ \\
Number of channels          & 30,240 \\
\hline \hline
\end{tabular}
\end{center}
\vspace*{.2cm}
\caption[CDF tracking systems]{Design parameters of the CDF tracking systems}
\label{tb_tracking}
\end{table}

{\bf The Silicon Tracker }

\begin{figure}[!h]
\begin{center}
   \begin{minipage}[b]{2.5in}
   \begin{center}
      \parbox{2.5in}{\epsfxsize=\hsize\epsffile{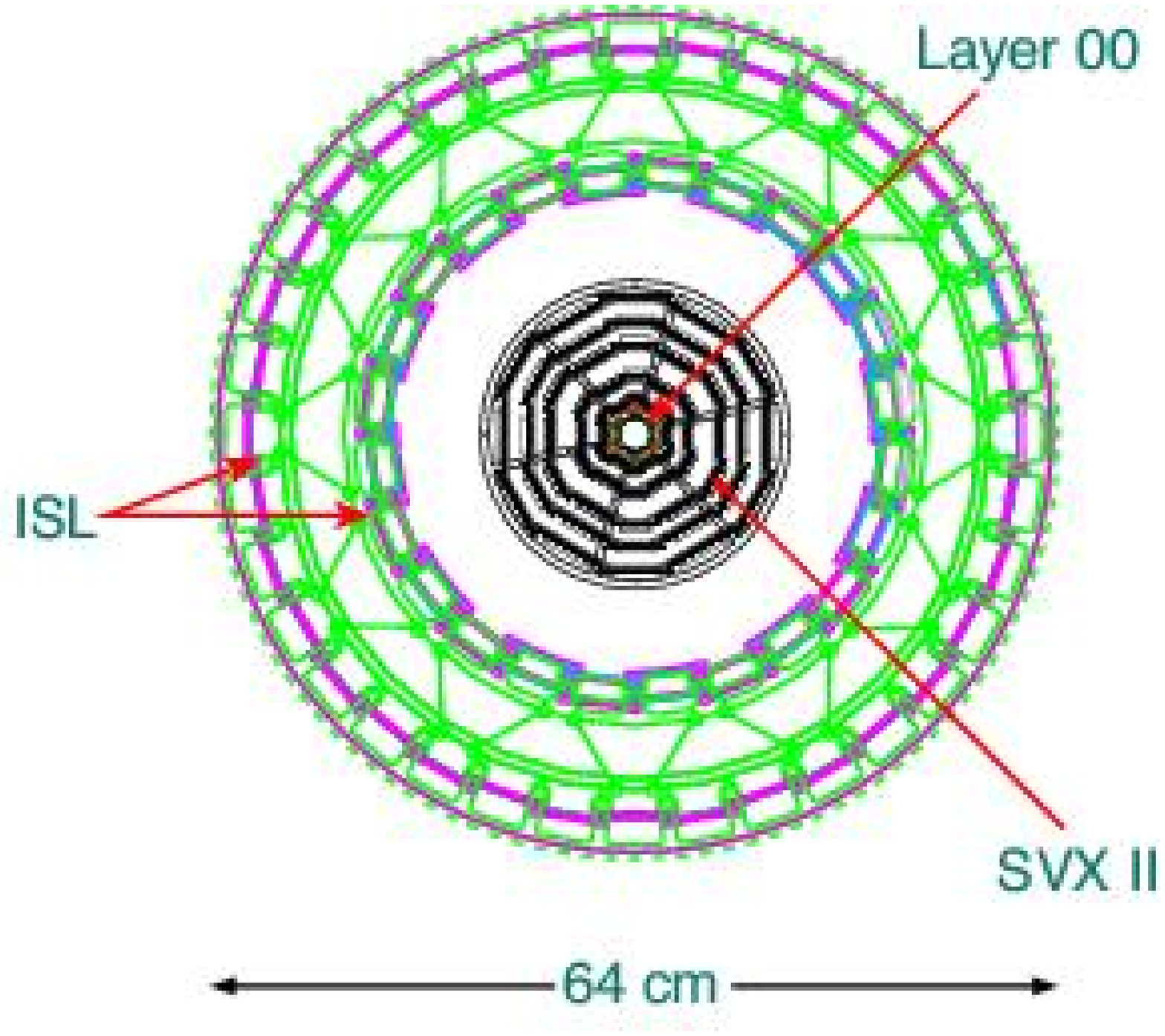}}
      \caption[Silicon system]
              {Silicon system.}
      \label{fig:cdf_silicon_endview}
   \end{center}
   \end{minipage}
\hspace{0.4in} 
   \begin{minipage}[b]{2.5in}
   \begin{center}
      \parbox{2.5in}{\epsfxsize=\hsize\epsffile{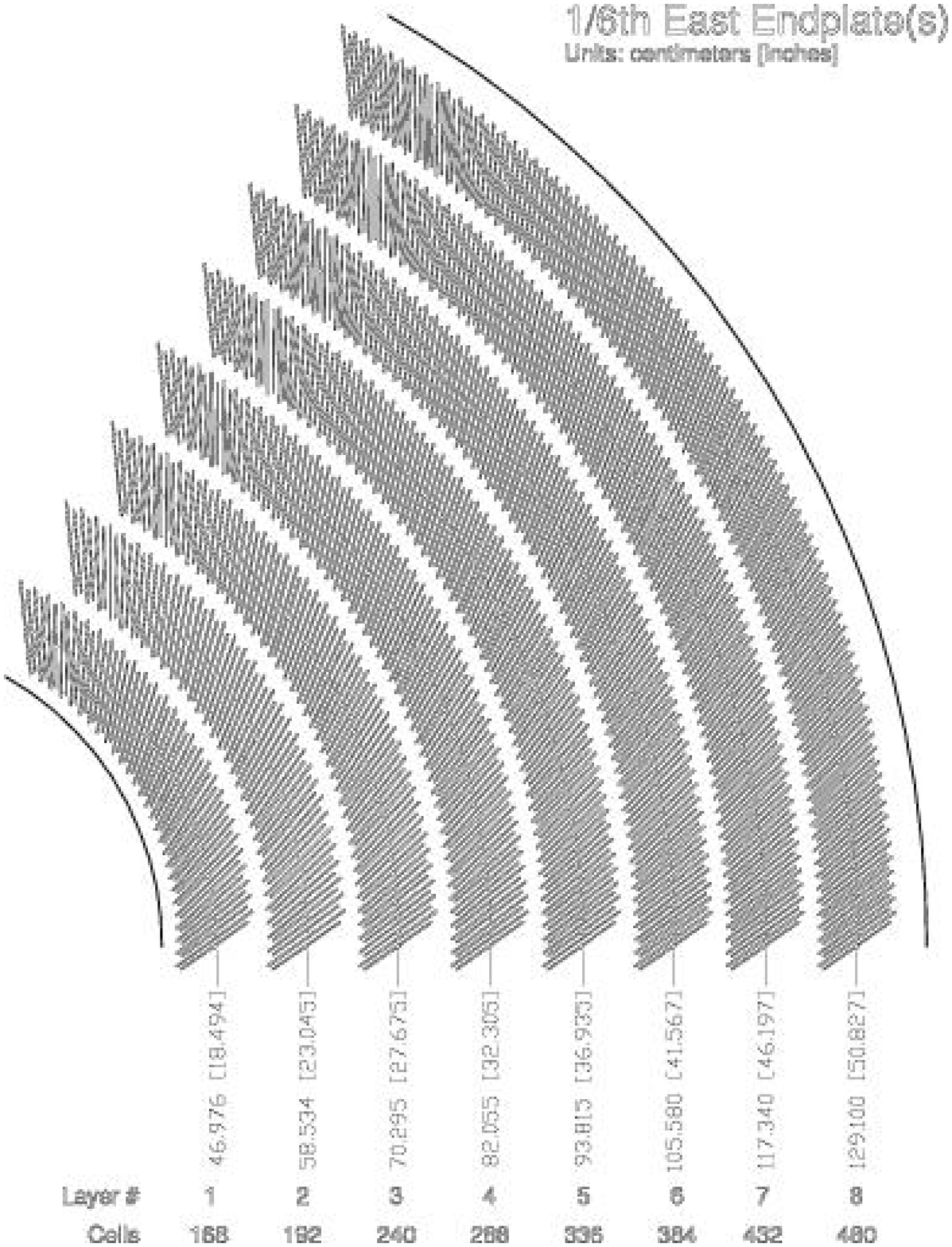}}
      \caption[COT superlayers]
              {COT superlayers.}
      \label{fig:cot_plate}
   \end{center}
   \end{minipage}
\end{center}
\end{figure}

Enhanced system of silicon microstrip detectors~\cite{SVX} consists of
three components: Layer~00, the Silicon VerteX detector II (SVX~II),
and the Intermediate Silicon Layers (ISL). $r-\phi$ view of the
silicon tracker is shown in Figure~\ref{fig:cdf_silicon_endview}.

A single layer rad-hard Layer~00 detector is mounted on and supported
by the beam pipe. The Layer~00 single-sided sensors provide 6 $\mu$m
axial hit resolution.

The next five layers compose the SVX II and are double-sided
detectors.  The axial side of each layer is used for $r$-$\phi$
measurements. The stereo side of each layer is used for $r$-$z$
measurements.

The two outer layers compose the ISL and are double-sided
detectors. This entire system allows charged particle track
reconstruction in three dimensions. The impact parameter resolution of
SVX~II~+~ISL is 40~$\mu$m including 30~$\mu$m contribution from the
beamline.  The~$z_0$ resolution of SVX~II~+~ISL is 70~$\mu$m. The main
parameters of the silicon tracker are summarized in
Table~\ref{tb_tracking}.

{\bf The Central Outer Tracker (COT) }

The COT~\cite{COT} is a multi-wire open-cell drift chamber for charged
particle reconstruction, occupying the radial region from 44 to 132 cm
and $|z|<$155 cm. The COT replaced the Central Tracking Chamber (CTC),
which, in addition to aging problems observed during Run I, would also
suffer from degraded performance at $\lum\ge 1\times 10^{32}
cm^{-2}s^{-1}$. The major problem with the CTC would be its maximum
drift time (800 ns) relative to the expected bunch crossing time in
Run II (132 ns).

To address this, the COT uses small drift cells ($\sim$2 cm wide, a
factor of 4 smaller than the CTC) and a fast gas to limit drift times
to less than 130 ns. Each cell consists of 12 sense wires oriented in
a plane, tilted with respect to the radial
(Figure~\ref{fig:cot_plate}). A group of such cells at a given radius
is called a superlayer. There are eight alternating superlayers
(Figure~\ref{fig:cot_plate}) of stereo (nominal angle of 2$^\circ$,
used for $r$-$z$ measurement) and axial (used for $r$-$\phi$
measurement) wire planes. The main parameters of the COT are
summarized in Table~\ref{tb_tracking}.

The COT is filled with a mixture of Argon:Ethane~=~50:50 which
determines the drift velocity. A charged particle travels through the
gas mixture and produces ionization electrons. The electrons drift
toward the sense wires in the electric field created by cathode field
panels and potential wires of the cell. In the crossed magnetic and
electric fields electrons originally at rest move in the plane
perpendicular to the magnetic field at an angle $\alpha$ with respect
to the electric field lines. The value of $\alpha$ depends on the
magnitude of both magnetic and electric fields and the properties of
the gas mixture. In the COT $\alpha\approx 35^\circ$.

The optimal situation for the resolution is when the drift direction
is perpendicular to that of the track, which is true for high $\Pt$
tracks because they are almost radial. To make the ionization
electrons drift in the $\phi$ direction all COT cells are tilted by
35$^\circ$ with respect to the radial.

When an electron gets near a sense wire the local 1/r field
accelerates them causing more ionization and thus forms an "avalanche"
producing a signal (hit) on the sense wire. By measuring the arrival
time of ``first'' electrons (drift time) at sense wire relative to
collision time, the distance of the hit is calculated.

COT tracks above 1.5 GeV are available for triggering at Level 1 (XFT
tracks). SVX layers 0-3 are combined with XFT tracks at Level-2 (SVT
tracks) (see Section~\ref{detector_trigger} for the description of CDF
Run II trigger system).

\subsection{Calorimetry}
\label{detector_calorimetry}

The energy measurement is done by sampling calorimeters, which are
absorber and sampling scintillator sandwiches with phototube readout.
Outside the solenoid, Pb-scintillator electromagnetic (EM) and
Fe-scintillator hadronic (HAD) calorimeters cover the range $|\eta| <
3.6$. The central calorimeter systems have been retained from Run I,
but the plug calorimeters are new detectors for Run II.

Both the central ($|\eta| < 1.1$) and plug ($1.1 < |\eta| < 3.6$)
electromagnetic calorimeters have fine grained shower profile
detectors at electron shower maximum, and preshower pulse height
detectors at approximately $1 X_o$ depth. Electron identification is
accomplished using $E/p$ from the EM calorimeter; using $HAD/EM\sim
0$; and using shower shape and position matching in the shower max
detectors. The calorimeter cell segmentation is summarized in
Table~\ref{calseg} and shown in Figure~\ref{evd_lego_display}. A
comparison of the central and plug calorimeters is given in
Table~\ref{plugcomp}.

\begin{table}[!h]
\begin{center}
\begin {tabular} {l|c|c}
\hline \hline
$|\eta|$ Range & $\Delta\phi$ & $\Delta\eta$ \\
\hline
0. - 1.1 (1.2 h) & $15^{\circ}$ &  $\sim0.1$ \\
1.1 (1.2 h) - 1.8 & $7.5^{\circ}$ & $\sim0.1$ \\
1.8 - 2.1 & $7.5^{\circ}$ & $\sim0.16$ \\
2.1 - 3.64 & $15^{\circ}$ & 0.2 $-$ 0.6 \\
\hline \hline
\end{tabular}
\vspace*{.2cm}
\caption[CDF calorimeter segmentation]{CDF II Calorimeter Segmentation}
\label{calseg}
\end{center}
\end{table}

The region 0.77 $< \eta <$ 1.0, 75 $^\circ < \phi <$ 90$^\circ$ is
uninstrumented to allow for cryogenic utilities servicing the
solenoid.

\begin{table}[!h]
\begin{center}
\begin {tabular} {l|c|c}
\hline \hline
 & Central  & Plug \\
\hline
EM: & &\\
Thickness & $19 X_0, 1 \lambda$ & $21 X_0, 1 \lambda$ \\
Sample (Pb) & $0.6 X_0$ & $0.8 X_0$  \\
Sample (scint.) & 5 mm & 4.5 mm \\
Stoch. res. & $14\%/\sqrt{E_T}$ & $16\%/\sqrt{E}$ \\
Shower Max. seg. (cm)& 1.4$ \phi \times $(1.6-2.0) Z& $0.5\times0.5$ UV\\
\hline
Hadron: &&\\
Thickness& $4.5 \lambda$  & $7 \lambda$\\
Sample (Fe)&1 to 2 in.& 2 in.\\
Sample (scint.)& 10 mm & 6 mm\\
\hline \hline
\end{tabular}
\vspace*{.2cm}
\caption[CDF central/plug calorimeters]{Central and Plug Upgraded Calorimeter
Comparison}
\label{plugcomp}
\end{center}
\end{table}

Any high energy electron or photon passing through the electromagnetic
calorimeters, will undergo pair production ($\gamma\rrr e^+e^-$) and
bremsstrahlung ($e^\pm\rrr \gamma e^\pm$) thus producing an
electromagnetic shower. The point at which the electromagnetic shower
consists of the largest amount of particles is known as the shower
maximum. At this point the average energy per particle becomes low
enough to prevent further multiplication. After the shower maximum,
the shower decays slowly through ionization losses for the electrons
and positrons or by Compton scattering for the photons. The
calorimeters measure the energy deposited by these showers, and hence
the energy of the incident particle. Electromagnetic calorimeters are
designed to fully contain showers from electrons and photons.

Hadrons lose energy by nuclear interaction cascades which can have
pions, protons, kaons, neutrons, neutrinos, muons, photons, etc. This
is significantly more complicated than an electromagnetic cascade and
thus results in a large fluctuation in energy measurement.

\subsubsection{Central Calorimeters}

The central calorimeters consist of the central electromagnetic
calorimeter (CEM)~\cite{Balka:1987ty}, the central hadronic
calorimeter (CHA)~\cite{Bertolucci:1987zn}, and the end wall hadronic
calorimeter (WHA). 

The CEM and CHA are constructed in wedges which span 15$^{\circ}$ in
azimuth and extend about 250~cm in the positive and negative $z$
direction, shown in Figure~\ref{fig:CCAL}. There are thus 24 wedges on
both the $+z$ and $-z$ sides of the detector, for a total of 48.  A
wedge contains ten towers, each of which covers a range 0.11 in
pseudorapidity.  Thus each tower subtends $0.11\times15^{\circ}$ in
$\eta\times\phi$. The CEM covers $0<|\eta|<1.1$, the CHA covers
$0<|\eta|<0.9$, and the WHA covers $0.7<|\eta|<1.3$.

The CEM uses lead sheets interspersed with polysterene scintillator as
the active medium and employs phototube readout, approximately 19$X_0$
in depth, and has an energy resolution of
$13.5\%/\sqrt{\Et}\oplus2\%$\footnote{$\oplus$ denotes addition in
quadrature}.

To provide more accurate information on the position of the
electromagnetic shower inside the calorimeter, the Central
Electromagnetic Shower (CES)~\cite{Balka:1987ty} detector is embedded
inside the CEM at the shower maximum, at a depth of approximately 6
radiation lengths. The CES detector is a proportional strip and wire
chamber situated at a radius of 184 cm from the beamline. In the
azimuthal direction, cathode strips are used to provide the z position
and in the $\phi$ direction, anode wires are used. These wires can
effectively measure the transverse shower profile to distinguish
between a single shower from a prompt photon and two showers from a
decay of a neutral meson to two photons, e.g. $\pi^0\rrr\gg$, with a
position resolution of 2 mm at 50 GeV.

In order to help particle identification, specifically between
electromagnetic and hadronic showers the central preradiator detector
(CPR) is mounted on the front of the calorimeter wedges, at a radius
of 168 cm from the beamline, and uses the solenoid and tracking
detectors as a radiator. It uses proportional chambers to sample the
early development of the shower to measure conversions in the coil,
helping to distinguish prompt photons and electrons from photons
originating from $\pi^0$ decays and electrons from conversions. A
prompt photon has a 60\% probability of converting, while the
conversion probability of at least one photon from $\pi^0\rrr\gg$ is
about 80\%~\cite{hayward_thesis}.

The CHA and WHA use steel absorber interspersed with acrylic
scintillator as the active medium.  They are approximately
4.5$\lambda$ in depth, and have an energy resolution of
$75\%/\sqrt{\Et}\oplus3\%$, as measured on the test beam for single
pions~\cite{Bertolucci:1987zn}.

\subsubsection{Plug Calorimeters}

\begin{figure}[!h]
\begin{center}
   \begin{minipage}[b]{2.8in}
   \begin{center}
      \parbox{2.2in}{\epsfxsize=\hsize\epsffile{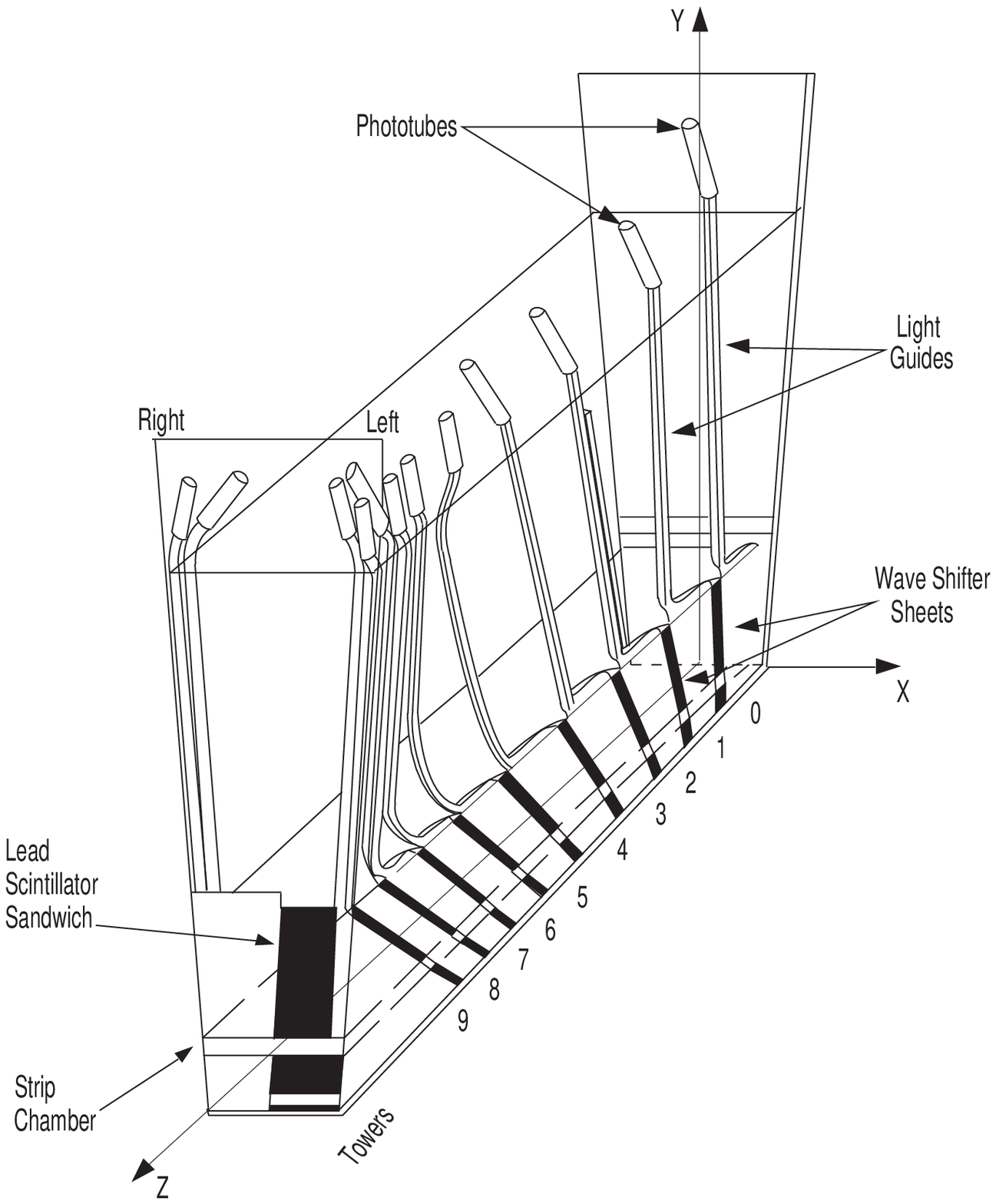}}
      \caption[CEM/CES/CHA wedge]
              {CEM/CES/CHA wedge.}
      \label{fig:CCAL}
   \end{center}
   \end{minipage}
\hspace{0.4in}
   \begin{minipage}[b]{2.2in}
   \begin{center}
      \parbox{2.2in}{\epsfxsize=\hsize\epsffile{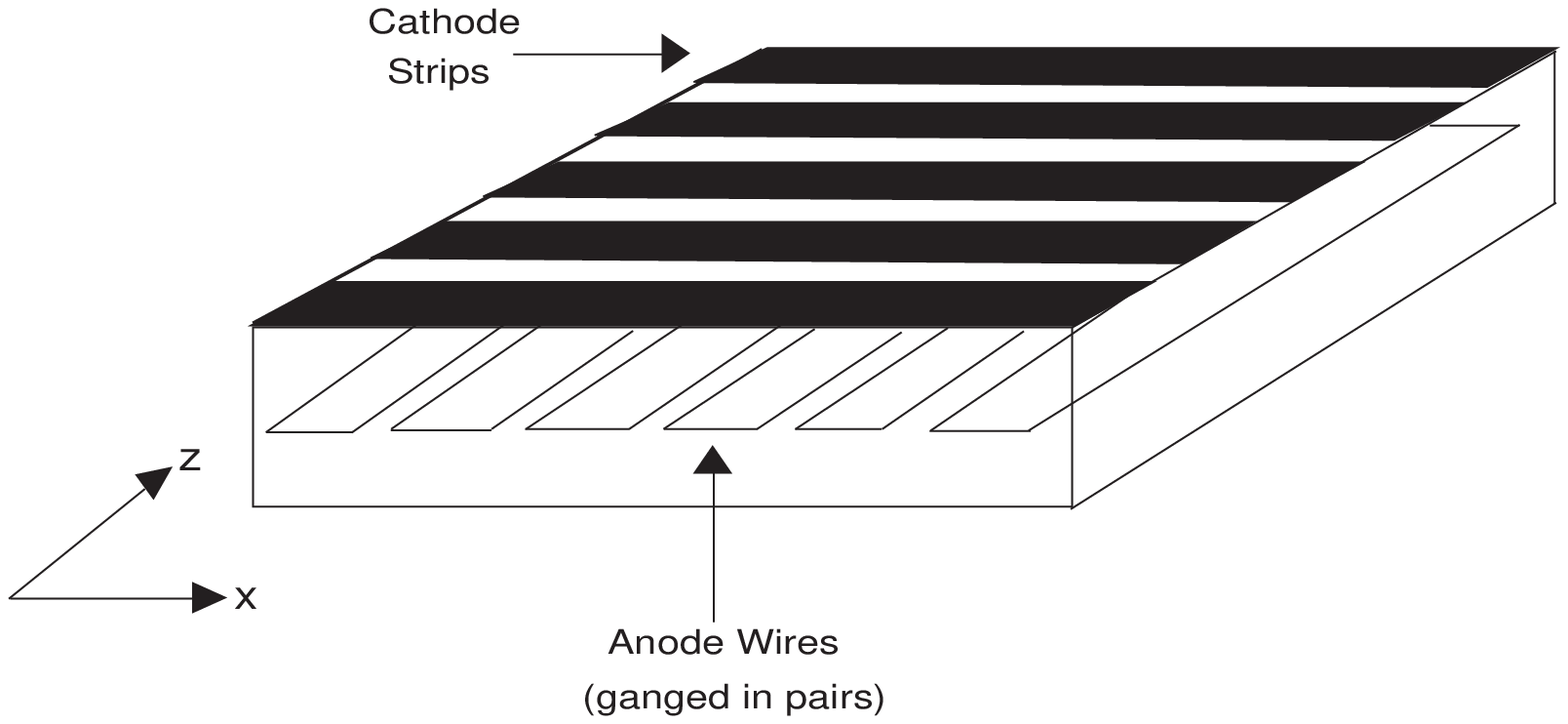}}
      \caption[CES strip and wire]
              {CES strip and wire.}
      \label{fig:CES}
   \end{center}
   \end{minipage}
\\
\vspace{0.3in}
   \begin{minipage}[b]{2.8in}
   \begin{center}
      \parbox{2.8in}{\epsfxsize=\hsize\epsffile{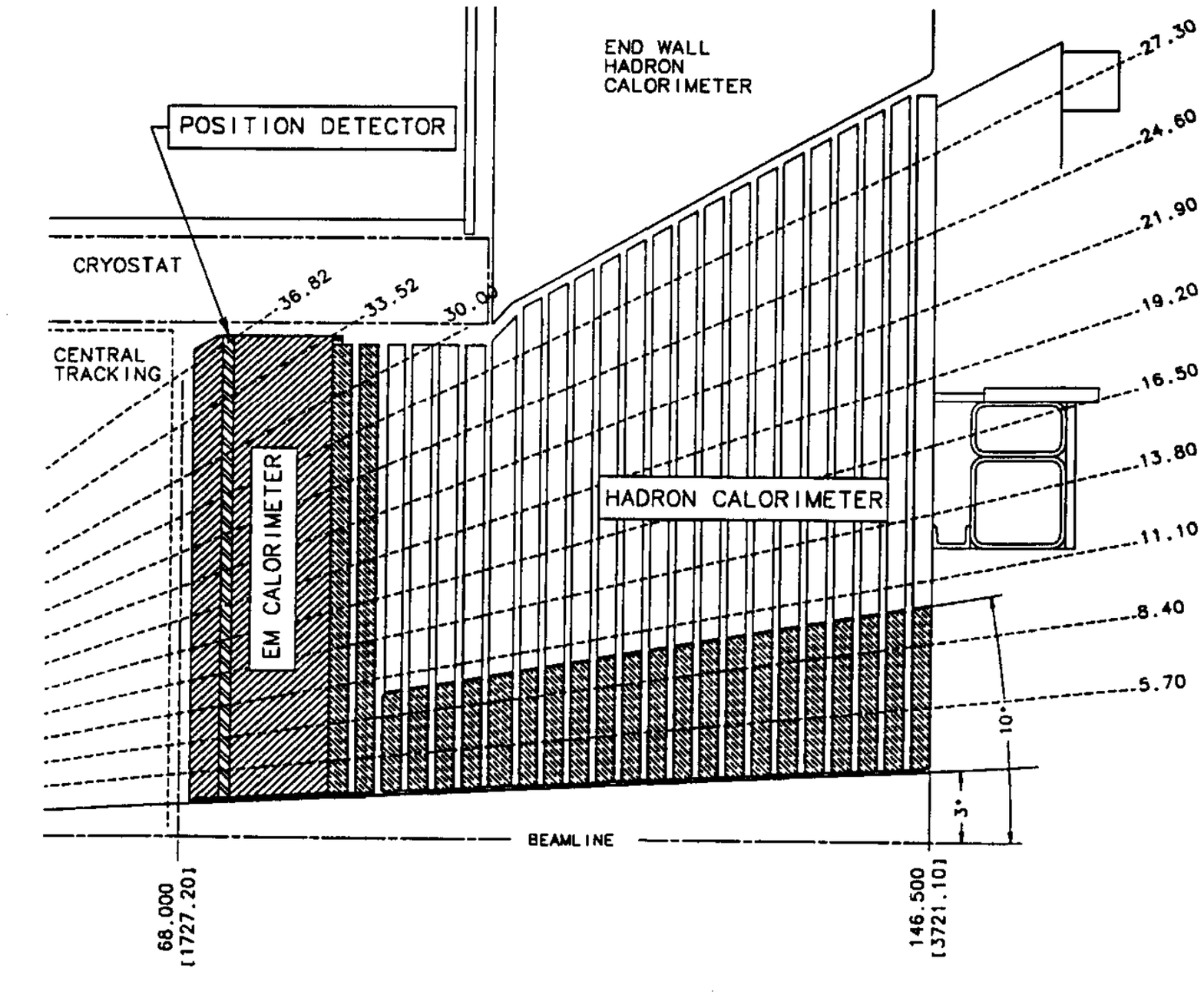}}
      \caption[PEM/PES/PHA layout]
              {PEM/PES/PHA layout.}
      \label{fig:PCAL}
   \end{center}
   \end{minipage}
\hspace{0.4in}
   \begin{minipage}[b]{2.2in}
   \begin{center}
      \parbox{1.8in}{\epsfxsize=\hsize\epsffile{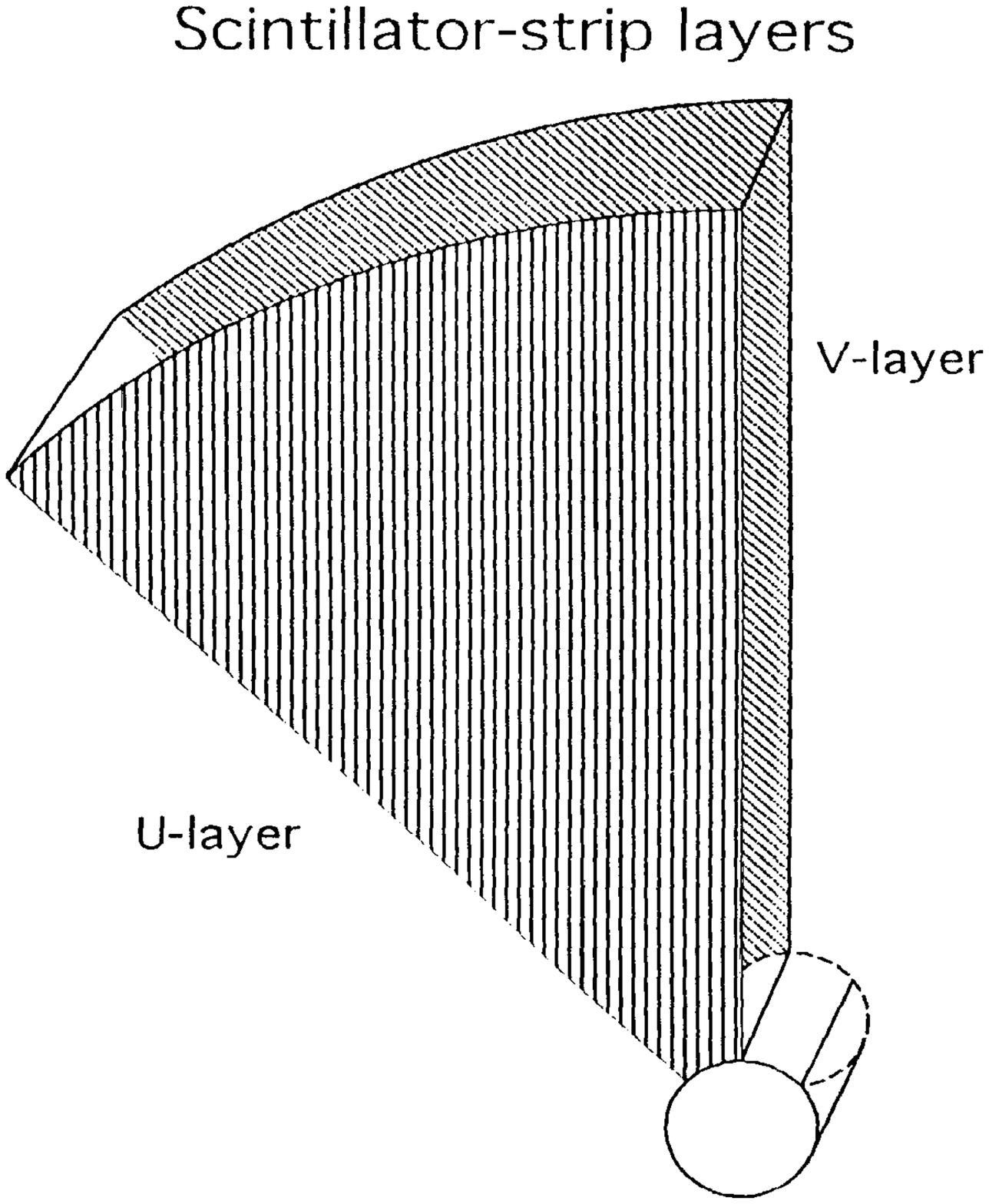}}
      \caption[PES U and V layers]
              {PES U and V layers.}
      \label{fig:PES}
   \end{center}
   \end{minipage}
\end{center}
\end{figure}

The plug calorimeters consist of the plug electromagnetic calorimeter
(PEM)~\cite{Albrow:2001jw}, and the plug hadronic calorimeter (PHA).
At approximately 6$X_0$ in depth in PEM is the plug shower maximum
detector (PES). Figure~\ref{fig:PCAL} shows the layout of the detector
and coverage in polar angle $36.8^{\circ}>\theta>3^{\circ}$
($1.1<|\eta|<3.64$).  Each plug wedge spans 15$^{\circ}$ in azimuth,
however in the range $36.8^{\circ}>\theta>13.8^{\circ}$
($1.1<|\eta|<2.11$) the segmentation in azimuth is doubled and each
tower spans only 7.5$^{\circ}$.

The PEM is a lead-scintillator sampling calorimeter.  It is
approximately 21$X_0$ in depth, and has an energy resolution of
$16\%/\sqrt{E}\oplus1\%$.  The PES consists of two layers of
scintillating strips: U and V layers offset from the radial direction
by $+22.5^{\circ}$ and $-22.5^{\circ}$ respectively, as shown in
Figure~\ref{fig:PES}.  The position resolution of the PES is about
1~mm. The PHA is a steel-scintillator sampling calorimeter.  It is
approximately 7$\lambda$ in depth, and has an energy resolution of
$74\%/\sqrt{E}\oplus4\%$, as measured on the test beam for single
pions~\cite{Bertolucci:1987zn}.

\subsection{Muon Systems}
\label{detector_muon}

The muon is a minimum ionizing particle which loses very little energy
in detector materials. The muon's lifetime, 2.2 $\mu$s, is long enough
for the muon to pass through all the detector components, reach the
muon chambers and decay outside.

There are four independent muon systems: the central muon detector
(CMU)~\cite{RunII_TDR}, the central muon upgrade
(CMP)~\cite{muon_chambers}, the central muon extension
(CMX)~\cite{cite_cmx}, and the intermediate muon detector (IMU). The
calorimeter steel serves as a filter for muon detection in the CMU and
CMX, over the range $|\eta| < 1$, $p_T > 1.4$ GeV. Additional iron
shielding, including the magnet yoke, provides a muon filter for the
CMP in the range $|\eta| < 0.6$, $p_T >2.2$ GeV.  The (non-energized)
forward toroids from Run~I provide muon filters for IMU in the range
$1.0 < |\eta| < 1.5$ for $p_T > 2$ GeV. Scintillators for triggering
are included in CMP, CMX, and IMU.

Muon identification is accomplished by matching track segments in the
muon chambers with COT/SVX tracks; matching is available in $r-\phi$
for all detectors and in $z$ in CMU and CMX. The parameters for the
muon systems are summarized in Table~\ref{tab:overcmu}. The IMU, which
provides coverage in the forward region will not be discussed in
detail, as it is not used for this analysis.

\begin{figure}[!h]
\begin{center}
   \begin{minipage}[b]{3.0in}
   \begin{center}
      \parbox{3.0in}{\epsfxsize=\hsize\epsffile{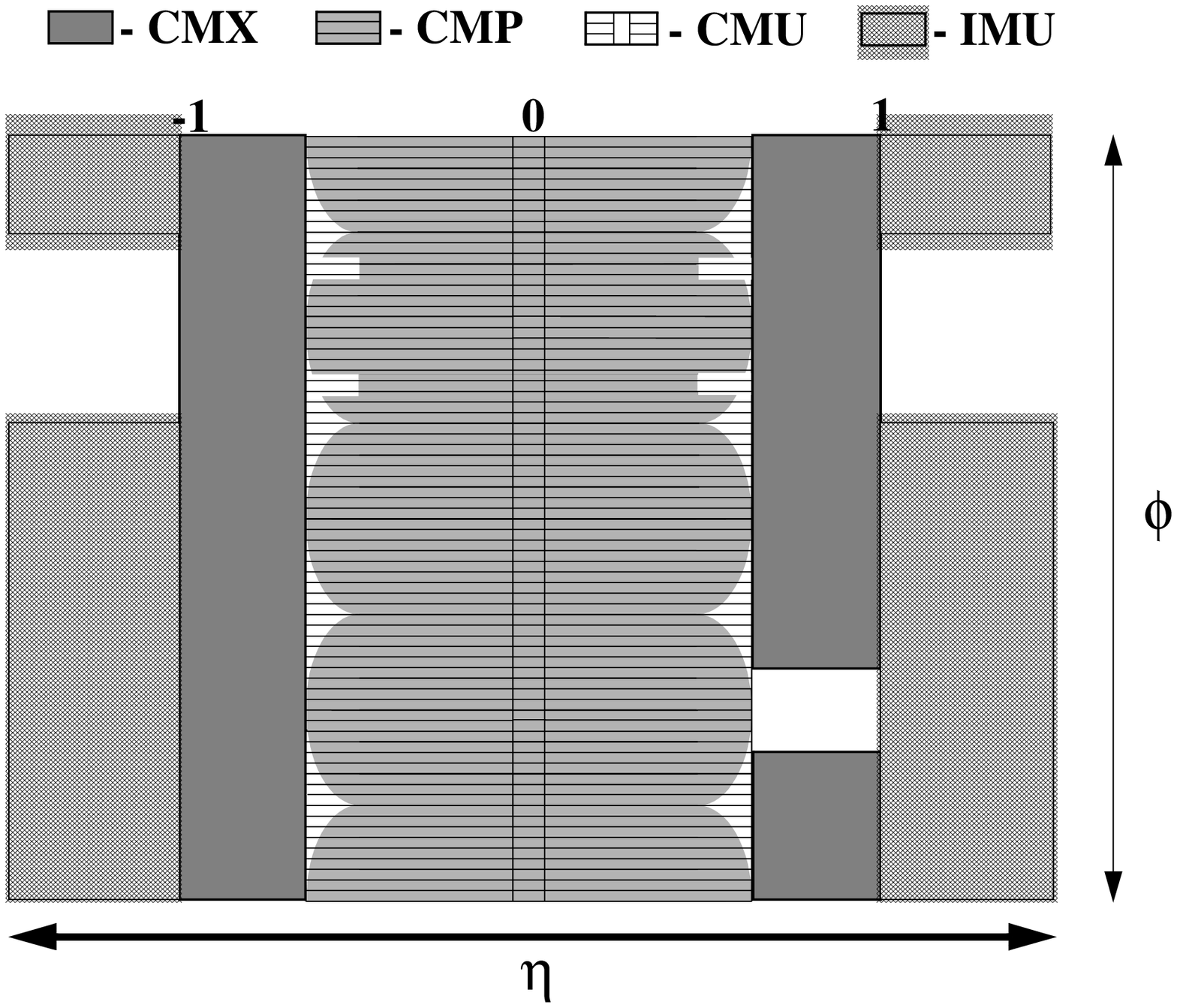}}
\caption 
[Location of the muon detectors in $\phi$ and $\eta$]
{Location of the muon detectors in $\phi$ and $\eta$. On the
east side, there is a gap in coverage in the CMX of 30$^\circ$ in
azimuth, due to the location of the cryogenic utilities servicing the
solenoid.}
\label{newetaphimu.figure}
   \end{center}
   \end{minipage}
\hspace{0.4in}
   \begin{minipage}[b]{3.0in}
   \begin{center}
      \parbox{3.0in}{\epsfxsize=\hsize\epsffile{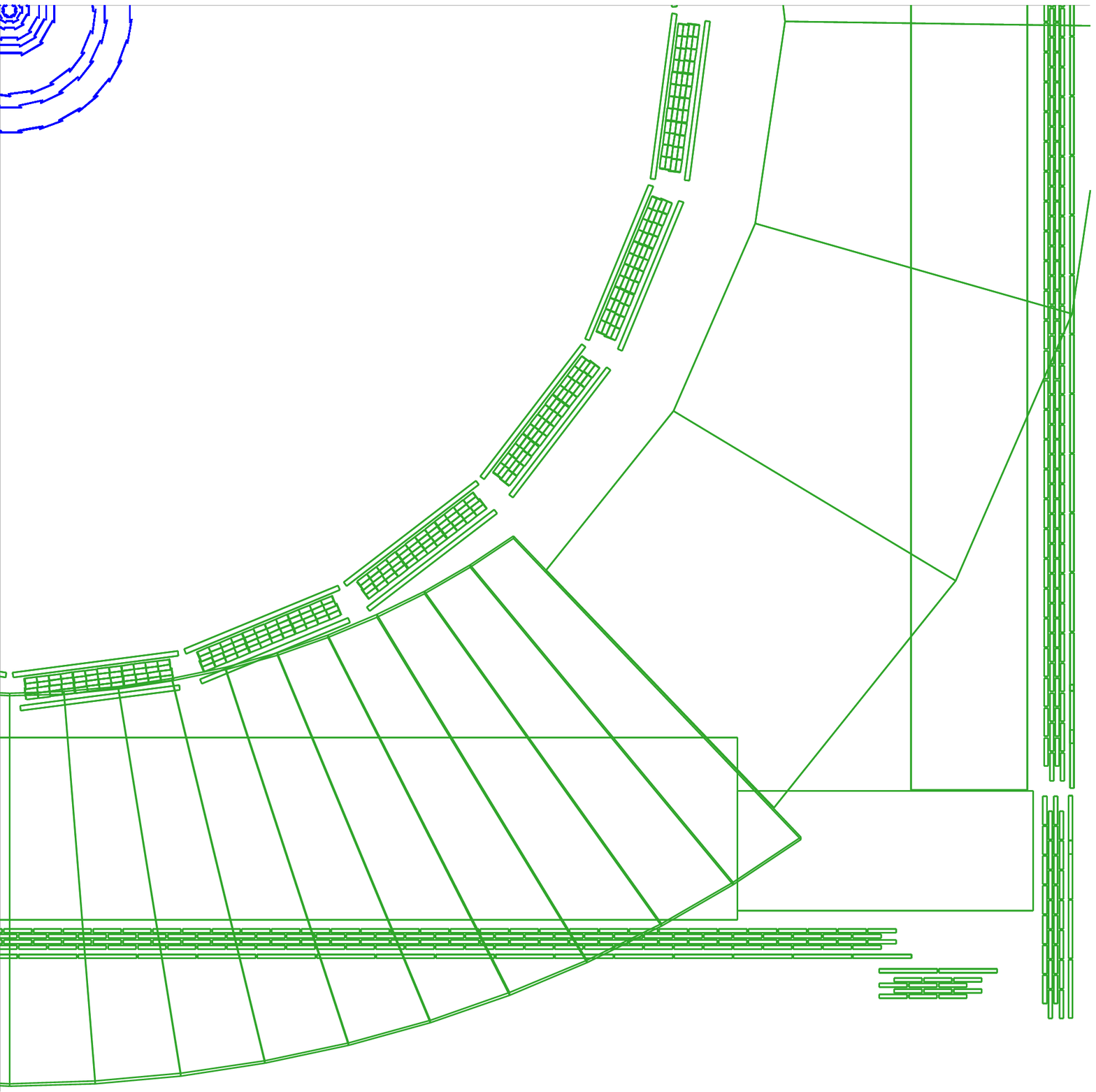}}
      \caption[Central muon systems: CMU, CMP, CMX]
              {Central muon systems: CMU, CMP, CMX.}
      \label{central_muon_system.figure}
   \end{center}
\vskip2.0in
   \end{minipage}
\end{center}
\end{figure}

The coverage for the muon systems in $\eta-\phi$ space is shown in
Figure~\ref{newetaphimu.figure}. CMU, CMP and CMX muon systems are
also shown in Figure~\ref{cdf3d_opengl.figure}.

\begin{table}[t]
\begin{center}
\begin{tabular}{lcccc}
\hline \hline
 & CMU   & CMP   & CMX & IMU \\ \hline
Pseudo-rapidity coverage
  & $|\eta|\leq 0.6$ & $|\eta|\leq 0.6$
  & $0.6 \leq |\eta|\leq 1.0 $
  & $1.0 \leq |\eta|\leq 1.5 $ \\
\hline
\multicolumn{5}{c}{Drift tubes}\\
\hline
Cross-section, cm
 & 2.68 x 6.35 & 2.5 x 15 & 2.5 x 15 & 2.5 x 8.4 \\
Length, cm   & 226   & 640  & 180 & 363 \\
Max drift time, $\mu {\rm s}$
 & 0.8 & 1.4 & 1.4 &  0.8 \\
\hline
\multicolumn{5}{c}{Scintillation counters}\\
\hline
Thickness, cm &      & 2.5  & 1.5 & 2.5 \\
Width    , cm &      & 30   & 30-40 & 17 \\
Length   , cm &      &  320 & 180 & 180 \\
\hline
Minimum muon $p_T$, ~GeV
  & 1.4 & 2.2 & 1.4 & 1.4-2.0\\
\hline
\hline
\end{tabular}
\vspace*{.2cm}
\caption[CDF muon detectors]{Design Parameters of the CDF II Muon Detectors.}
\label{tab:overcmu}
\end{center}
\end{table}

\begin{figure}
   \begin{center}
      \parbox{5.0in}{\epsfxsize=\hsize\epsffile{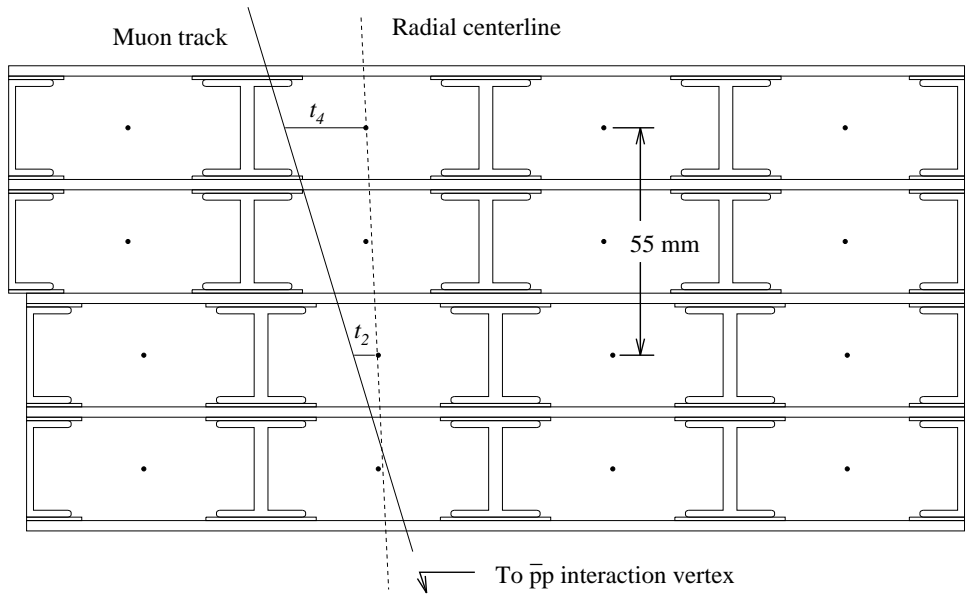}}
      \caption
[CMU module in the r-$\phi$ plane]
{A CMU module in the r-$\phi$ plane with 4 layers of drift
chambers. The drift times t1 and t2 are used to calculate muon
momentum for triggering (Section~\ref{detector_trigger}). 
The sense wires connected to the readout are shown as a black circles.}
      \label{CMU_tower.figure}
   \end{center}
\end{figure}

A muon chamber contains a stacked array of drift tubes and operates
with a gas mixture of Argon:Ethane~=~50:50.  The basic drift principle
is the same as that of the COT, but the COT is a multi-wire chamber,
whereas at the center of a muon drift tube there is only a single
sense wire. The sense wire is connected to a positive high voltage
(HV) while the wall of the tube is connected to a negative HV to
produce a roughly uniform time-to-distance relationship throughout the
tube. The drift time of a single hit gives the distance to the sense
wire, and the charge division at each end of a sense wire can in
principle be used to measure the longitudinal coordinate along the
sense wire. The hits in the muon chamber are linked together to form a
short track segment called a muon stub
(Figure~\ref{CMU_tower.figure}). If a muon stub is matched to an
extrapolated track in the tracking system
(Section~\ref{detector_tracking}), a muon is reconstructed.

\subsubsection{CMU and CMP}

The CMU is unchanged from Run I. It is located behind the towers of
the CHA and divided into wedges covering 12.6$^\circ$ in azimuth for
$\eta<$ 0.6. Only muons with a $\Pt >$ 1.4 GeV reach the CMU. Each
wedge has three towers, each comprised of four layers of four drift
tubes. The second and fourth layers are offset by 2 mm in $\phi$
direction from the first and third as shown in
Figure~\ref{CMU_tower.figure}. Six CMU wedges and their relative
location with respect to CMX and CMP (outer box) subsystems are shown
in Figure~\ref{central_muon_system.figure}.

A 50 $\mu$m diameter stainless steel resistive sense wire is located
in the center of each cell. The wires in the cells in the first and
third (second and fourth) layers are connected in the readout. Each
wire pair is instrumented with a time-to-digital converter (TDC) to
measure the $\phi$-position of the muon and an analogue-to-digital
converter (ADC) on each end to measure z position via charge
division. The position resolution of the detector is 250 $\mu$m in the
drift direction (r-$\phi$) and 1.2 mm in the sense wire direction (z).

Approximately 0.5\% of high energy hadrons produced will pass through
the CMU creating an irreducible fake muon background. In order to
reduce this effect, an additional muon chamber (CMP) is installed
behind 60 cm of steel.

The CMP consists of a four-sided box placed on the outside of the CDF
detector. Muons with $\Pt >$ 2.2 GeV can reach the CMP. The
rectangular form of the CMP detector means that its $\eta$ varies in
azimuth (Figure~\ref{newetaphimu.figure}). The CMP covers
$|\eta|<0.6$.

The maximum drift time of the CMU is longer than the $\ppbar$ bunch
crossing separation, which can cause an ambiguity in the Level~1
trigger (Section~\ref{detector_trigger}). To resolve the ambiguity
scintillation counters are used. The scintillation counters (CSP) are
installed on the outer surface of the CMP.

\subsubsection{Central Muon Extension (CMX)}

The CMX has eight layers and extends the $\eta$ coverage to
$0.6<|\eta|<1.0$. It consists of two 120$^\circ$ arches located at
each end of the central detector, as shown in
Figure~\ref{newetaphimu.figure}. The uninstrumented regions have been
filled by the insertion of a 30$^\circ$ keystone at the top, and two
90$^\circ$ miniskirts for the lower gaps. There is a gap in the
coverage on the east side due to cryogenic utilities servicing the
solenoid as shown in Figure~\ref{newetaphimu.figure}, known as
the "chimney".

A layer of scintillation counters (the CSX) is installed on both the
inside and the outside surfaces of the CMX. No additional steel was
added for this detector because the large angle through the hadron
calorimeter, magnet yoke, and steel of the detector end support
structure provides more absorber material than in the central muon
detectors.

\subsection{Time of Flight System}
\label{detector_tof}

The Time of Flight detector (TOF)~\cite{Acosta:2004kc} measures the
time taken by a particle to travel from the interaction point to the
detector, and has a particle timing resolution of 100 ps. This
information can be used to differentiate between different particle
types (e.g. kaons, pions) and also to help tag cosmic ray events.

The TOF is situated between the COT and the solenoid. It consists of
216 scintillator bars with dimensions 4$\times$4$\times$276 cm. At
each end of the scintillator bars a photomultiplier tube is mounted

\subsection{Cherenkov Luminosity Counters}
\label{detector_clc}

Luminosity ($\lum$) is a measure of particle interaction, specifically
the chance that a proton will collide with an antiproton. The rate of
inelastic scattering in $\ppbar$ interactions can be used to determine
the $\lum$ .

A gas Cherenkov Luminosity Counter (CLC)~\cite{clc} measures the
number of interactions per beam crossing to determine the luminosity
of the Tevatron. There are two CLCs positioned between the beam-pipe
and the plug calorimeters, covering the region 3.7 $<|\eta| <$
4.7. Each CLC consists of 48 thin, conical gas-filled Cherenkov
counters. They are arranged in three concentric circles, each
consisting of 16 counters (see, for example,
Figure~\ref{evd_online.figure}).

The luminosity of a $\ppbar$ collider can be estimated using the
equation:

\begin{equation}
\lum = {{f \times \mu}\over{\sigma}}
\end{equation}

where f is the frequency of bunch crossing, $\mu$ is the average
number of interactions per beam crossing, given by the CLC hit rate
(about 5-6), and $\sigma$ is the inelastic cross-section of $\ppbar$
scattering. The average of the inelastic cross-sections as measured by
CDF Run I and the E811~\cite{e811} is 60.7$\pm$2.3
mb~\cite{inelastic_x_section}.

A total systematic uncertainty of 6\% is quoted for all luminosity
measurements. This includes a 4.4\% contribution from the acceptance
and operation of the luminosity monitor and 4.0\% from the theoretical
uncertainty on the calculation of the total $\ppbar$
cross-section~\cite{inelastic_x_section}.

\subsection{Trigger and Data Acquisition}
\label{detector_trigger}

Many interesting physics processes have cross sections which are many
orders of magnitude smaller than the total inelastic cross
section. The collision rate at the Tevatron is much higher than the
rate at which data can be recorded. Therefore, the trigger needs to be
fast and accurate to record as many interesting events as possible,
while rejecting uninteresting events.

\begin{figure}
   \begin{center}
      \parbox{3.8in}{\epsfxsize=\hsize\epsffile{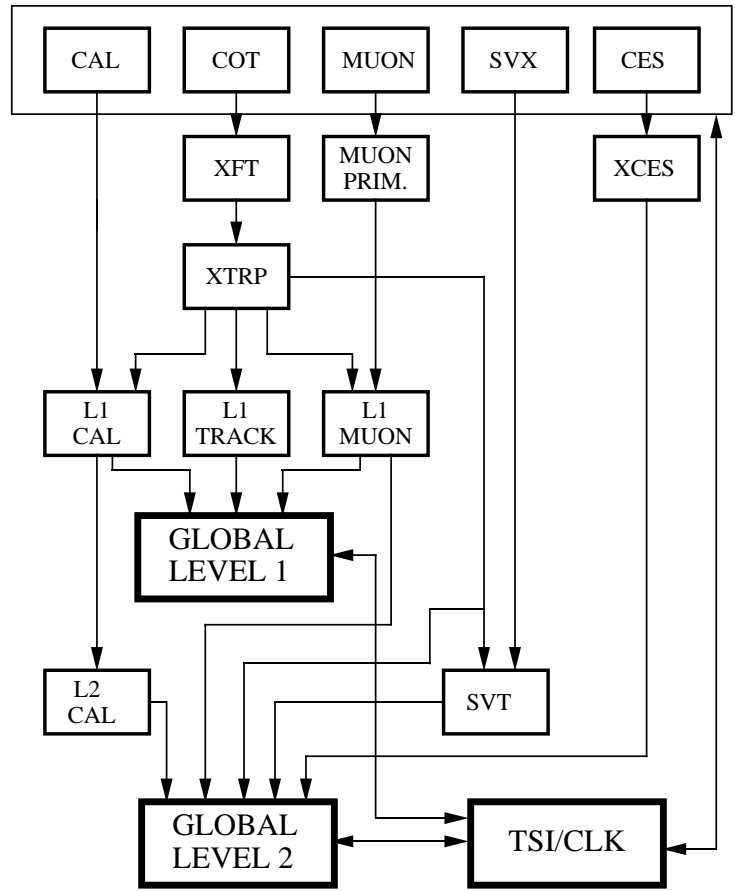}}
\caption[Trigger system block diagram]
        {Trigger system block diagram: flow of data through L1 and L2
        of trigger system. Silicon Vertex Tracker (SVT), based on the
        track impact parameter of displaced tracks, is used by L2. The
        data volume is reduced at each level which allows more refined
        filtering at subsequent levels with minimal deadtime. L3 is
        purely a software trigger consisting of the event builder
        running on a large PC farm.
        } \label{fig:trigger_system} \end{center}
\end{figure}

To accomplish this, the CDF trigger system has a three-level
architecture: Level 1 (L1), Level 2 (L2), and Level 3 (L3). The data
volume is reduced at each level, which allows more refined filtering at
subsequent levels with minimal deadtime. Each sub-detector generates
primitives which can be used in the trigger system to select
events. The trigger system block diagram is shown in
Figure~\ref{fig:trigger_system}.

At L1 axial layers of the COT are used by eXtreme Fast Tracker (XFT)
to reconstruct $\phi$ and $\Pt$ for the tracks. Based on the XFT
tracks and a ratio of the hadronic energy to the electromagnetic
energy of a calorimeter tower (HAD/EM ratio) electrons and photons are
then reconstructed. Muons are reconstructed by matching XFT and muon
hits. Jets are reconstructed based on a sum of the electromagnetic and
hadronic energies for a tower. $\met$ and $\sum\Et$ (a scalar sum of
the energies of all of the calorimeter towers) are also reconstructed
at L1.

L1 is a synchronous hardware trigger and it makes a decision within 4
$\mu$s. This trigger reduces the event rate from 7.6 MHz to 50 KHz,
which is limited by the L2 processing time. Accepted events are then
passed to the L2 hardware.

At L2 SVX layers 0-3 are combined with XFT tracks
(Figure~\ref{fig:trigger_system}.). The L2 uses jet clustering as well
as improved momentum resolution for tracks, finer angular matching
between muon stubs and central tracks and data from the CES for
improved identification of electrons and photons.

The L2 decision time is about 20 $\mu$s. L2 is a combination of
hardware and software triggers and is asynchronous. If an event is
accepted by L1, the front-end electronics moves the data to one of the
four onboard L2 buffers. This is sufficient to process a L1 50 KHz
accept rate and to average out the rate fluctuations. The L2 accept
rate is about 300 Hz which is limited by the speed of the event
builder in L3.

L3 is a purely software trigger consisting of the event builder
running on a large PC farm. Data which passes one of the specified
trigger paths is reconstructed at L3 using full detector information
and the latest calibrations. The L3 accept rate of about 75 Hz is
limited by the speed of writing data to tape for the permanent
storage.

As soon as an event passes L3 it is delivered to the data-logger
subsystem which sends the event out to permanent storage for offline
reprocess, and to online monitors which verify that the entire
detector and trigger systems are working properly. One of the online
monitors, CDF Run II Event Display, is described in detail in
Section~\ref{evd}.

\chapter{CDF Run II Event Display}
\label{evd}

Event displays are indispensable tools for data analysis in high
energy physics experiments. They help to understand the physics of a
recorded interaction, to diagnose the apparatus, to make the detector
geometry imaginable and to illustrate the whole matter to general
audiences.

CDF Run II Event Display (EVD)~\cite{evd_manual} is a major
contribution to the commissioning and operation of CDF II detector.
Development and support of the EVD package~\cite{evd_code} is one of
the responsibilities of the ITEP(Moscow) group at CDF~\cite{evd_team}.

The data from the collider experiment is a stream of signals from
subdetectors. The detector ''sees'' these signals as sequences of
impulses, distributed over many channels of different subdetectors
(see Section~\ref{cdfdetector}). The signals are analyzed by powerful
pattern recognition and analysis programs, which create more
sophisticated objects like clusters, segments, stubs and then tracks,
muons, electrons/photons, jets, etc.

Typically physics results are based on statistical analysis of {\it
many} events. The standard forms of presentation are histograms,
graphs and tables. However, a graphical representation of a {\it
single} event a powerful tool for checking the validity of
reconstruction or analysis algorithms. For a quick assessments of
error conditions as well as for public presentations the visual
representation is the most efficient way to transfer information to a
human brain.

Higher event multiplicities and higher momenta of outgoing particles
are matched by detectors with a growing number of subunits of
increasing granularity, resolution and precision. As a consequence
pictures of detectors and events are getting more complicated and may
even get incomprehensible. This leads to a question if the
presentation of data via visual techniques is useful for complicated
events at the Tevatron.

\section{Overview}
\label{evd_overview}

The aim of the EVD is to enable visual representation of the objects
existing in the CDF Run II software. EVD interacts both with the data
and with the simulation and reconstruction packages. For simulated
data EVD visualizes the Decay Tree, which is constructed from HEPG
information.

It is natural to define three kinds of objects: {\it Real Objects}
(Section~\ref{evd_real_objects}), {\it Graphical Objects}
(Section~\ref{evd_graphical_objects}), and {\it Views}
(Section~\ref{evd_views}). To visualize {\it Real Objects} and to
access information about the event, {\it Operations}
(Section~\ref{evd_operations}) are performed on the {\it Graphical
Objects} and {\it Views}.

\subsection{Real Objects}
\label{evd_real_objects}

A {\it Real Object} is information from a subdetector or combined
information from several subdetectors. For example, to identify
electron one needs calorimeter and tracking information. Some of the
{\it Real Objects} in CDF are listed below:

\begin{itemize}
\setlength{\itemsep}{-0.07in}
\item 
Tracking information (Section~\ref{detector_tracking}) \\
-- axial and stereo hits from the COT\\
-- hits from the silicon tracker\\
-- tracks reconstructed with different tracking algoritms
\item 
Calorimeter information (Section~\ref{detector_calorimetry})\\
-- Central and Plug Shower Chambers information\\
-- Central and Plug Preradiator information
\item 
Hits from TOF system (Section~\ref{detector_tof})
\item 
Information from the muon systems (Section~\ref{detector_muon})\\
-- Hits from the muon systems \\
-- Track segments reconstructed in the muon chambers
\item 
Information from East and West CLC subdetectors, which are used to
monitor luminosity and to identify diffractive events
(Section~\ref{detector_clc})
\item 
Information from Beam Shower Counters~\cite{bsc}, used to identify diffractive events
\item
Information from CDF Run II L1/L2/L3 Trigger System (Section~\ref{detector_trigger})
\item 
EM Timing information~\cite{emtiming}
\item
Pre-reconstructed objects, such as Muon, Electron, Photon, Jet
Candidates, $\met$
\item
Full information about raw and reconstructed data\\ 
-- access to banks and collections in the event
\end{itemize}

A major requirement has been made to keep analysis in the EVD to a
minimum, as objects are identified differently in different analyses
(i.e. loose electron for one analysis may be a jet in some other
analysis). However, EVD has a functionality to clean complicated
events by selective presentation of parts of the data and the
detector.  (Section~\ref{evd_operations}). For example, EVD can show
or hide tracks depending on their properties ($\phi$, $\eta$, $\Et$,
$\Pt$, number of hits in a subsystem etc.).

\subsection{Graphical Objects}
\label{evd_graphical_objects}

{\it Real Objects} in the EVD are mapped to their visual
representation into {\it Graphical Objects} of different types,
corresponding to different {\it Views}. The graphical objects
correspond to the stored real objects (for example, to the list of
hits and tracks) and other real objects (for example, to electron,
muon and photon candidates) created from them.

The properties of {\it Real Objects} are used to display {\it
Graphical Objects}. For instance, information from the calorimeter is
shown as towers with a size proportional to the deposited energy. We
use graphics libraries available in the ROOT package~\cite{root}.

\subsection{Views}
\label{evd_views}

{\it View} is a method of visualizing a set of graphical objects. For
a user, view is generally a window with a defined way of displaying
{\it Graphical Objects}. We define three categories of views for the
CDF Run II Event Display:

\begin{enumerate}
\item 
{\bf Realistic Views} are obtained by either a sequence of rotations,
linear scaling, and projections of a geometry of detector/identified
objects. Realistic views are understood intuitively, although the
pictures may become too complex (for example, see
Figures~\ref{evd_rz_display} and~\ref{cdf3d_opengl.figure}).

Hits density as well as detector precision grow towards the
interaction point. Therefore, an ability to hide parts of the detector
obscuring the picture is crucial. Another feature of the EVD is to
show or hide Graphical objects depending on the properties of Real
Objects (number of hits, drift time etc.)

\item 
{\bf Schematic Views} can be obtained from realistic views by changing
the {\it aspect ratio} or {\it focal length} for a subdetector. For
example, the scale may be decreased with increasing radius, so that
the inner subdetectors appear enlarged. This emphasizes the commonly
used construction principle of detectors, namely that precision and
sampling distance decrease, when stepping from the inner to the outer
detectors.

Schematic Views do not necessarily represent the detector in its real
proportions or shapes.  For example a box that changes color depending
on error conditions may serve as a schematic view of any detector
component. In many cases the schematic views are relatively easy to
understand and very efficient to use (for example, see
Figure~\ref{evd_cot_display}).

\item 
{\bf Abstract Views} have little resemblance to the detector image in
cartesian space. These views are not intuitively understood but given
some training may be the most powerful. Examples of abstract views are
angular projections, histograms, LEGO plots, etc. (for example, see
Figure~\ref{evd_lego_display})
\end{enumerate}

\subsection{Operations}
\label{evd_operations}

{\bf Operations on Graphical Objects}:\\
These operations only change the visual properties of the graphical
objects, the real objects are not changed. Visual operations can be
performed on the graphical objects or on the views. The following
operations on graphical objects are supported:
\begin{itemize}
\setlength{\itemsep}{-0.05in}
	\item rotation, zooming, translation, scaling
	\item viewpoint changing
        \item object hiding/unhiding
        \item sub-view creation
        \item redefinition of visual properties of graphical objects
\end{itemize}
 
{\parindent=0mm \bf Operations on Real Objects}: \\
Operations on real objects can access information about the objects or
change the state of these objects. In addition, visual properties of
the corresponding graphics objects can also be changed as a result of
changing real objects. The following operations on real objects are
supported:

\begin{itemize}
\setlength{\itemsep}{-0.05in}
	\item access to public member functions for an object from the EVD
        \item obtaining the detailed information on an object properties (for
        example, for tracks one can access information on number/type of
        hits in COT/Silicon detectors, dE/dx, $\chi^2$, $d_0$, $z_0$
        etc.)
        \item access to event record information (data banks and collections)
        \item application of identification cuts
	\item interface to a histogramming package
\end{itemize}

\section{CDF Run II Application Framework}
\label{evd_module_structure}

The design of the EVD is based on the belief that both requirements
and graphics software abilities will be very broad at any time and
will constantly evolve. The EVD is designed to accommodate that
diversity and change. This can be accomplished only by sufficient
flexibility and modularity of the core control
structure~\cite{framework}.

An application framework, in the context of a high energy physics
experiment, is a "system" that allows physicist-developed code to be
combined with code developed by other people and to be used in both
the online and offline environments. Either real or simulated data can
be used as the input and the output can include (modified) copy of the
input as well as additional reconstructed quantities. This output then
forms the input in the next stage of a multi-stage data reduction
environment. CDF Run II Application Framework (AC++) is based on the ROOT
object oriented analysis framework~\cite{root}, which is incorporated
in CDF Run II C++ software.

\begin{figure}[!b]
  \begin{center} 
\mbox{
\subfigure[A screen shot of the manager window for the EVD.]
{\epsfig{file=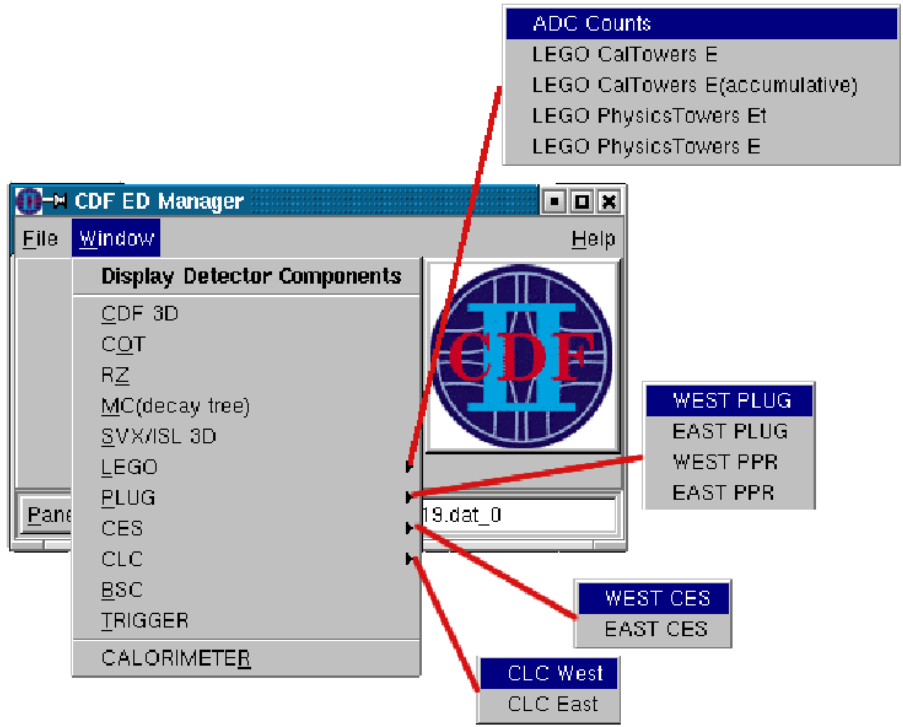, ,width=5.5in}} \quad
     }
\mbox{
\subfigure[The EVD control panel with the {\tt Loop} tab selected.]
{\epsfig{file=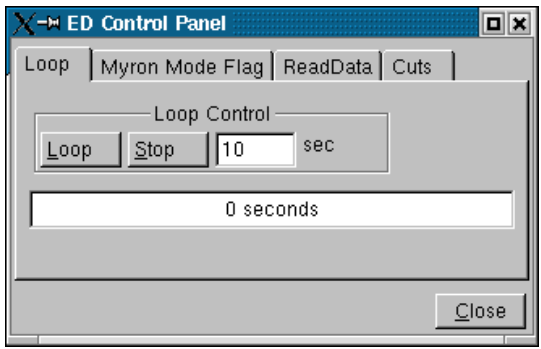,width=2.8in}} \quad
\subfigure[Data Menu to select which data to read.]
{\epsfig{file=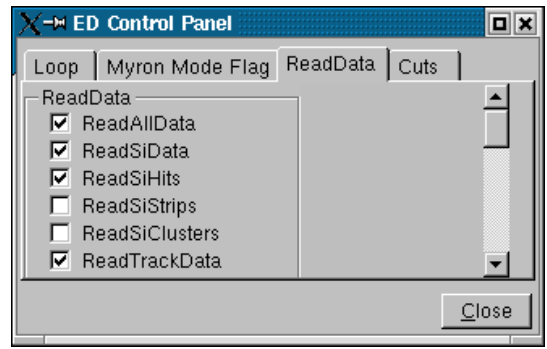,width=2.8in}} \quad
     }
\caption 
[CDF Run II Event Display Graphical User Interface]
{CDF Run II Event Display Graphical User Interface. The start of the
EVD is signalled by the appearance of the manager window (a). Clicking
the {\tt Panel} button brings up the EVD control panel (b). The
control panel allows you to configure the behavior of the EVD. For
example, user may select which data to read using the DataMenu
(c).}
\label{gui.figure} \end{center}
\end{figure}

The AC++ provides a unified framework for event reconstruction,
post-production analysis and online monitoring, triggering and
calibration. The goals of the framework is to provide a simple and
straightforward means of combining any number of independent classes,
called modules, into a single executable and to provide a flexible
system for specifying (either interactively or in a batch mode) how
these modules are run.

Therefore, EVD has the flexibility to enable/disable different
modules, so that one can work with unprocessed data for the immediate
feedback. Alternatively, one can also run reconstruction modules
inside the EVD with user-defined parameters for debugging purposes.

\section{Graphical User Interface}
\label{gui}

The start of the EVD is signalled by the appearance of the manager
window (Figure~\ref{gui.figure},a). From the manager window user
select one of the event displays, and control automatic looping
through events, as is done in the CDF Run II control room at
$B\O$. The name of the input file is displayed at the bottom.

\begin{figure}[!b]
  \begin{center}
    \includegraphics[width=0.8\textwidth]{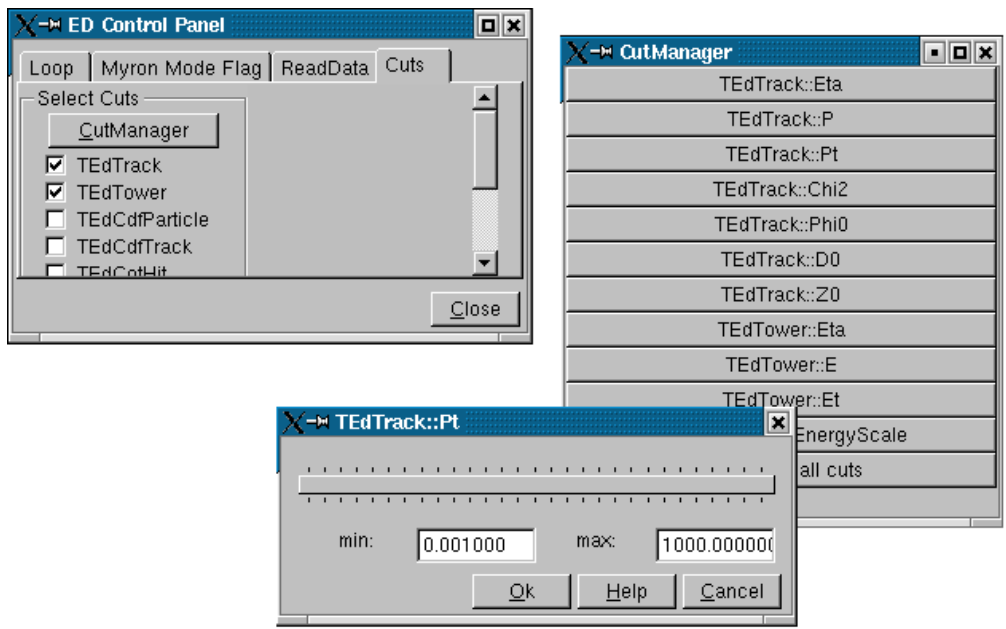}    
    \caption
[Cut Manager]
{Cut Manager. Complicated events might be cleaned by selective
presentation of the data. User can specify cuts on physical quantities
of the displayed objects, such as the $\eta$, $\phi$, $\Pt$ of tracks
or the energies deposited in the calorimeter towers.}
    \label{cut_manager.figure}
  \end{center}
\end{figure}

Clicking the {\tt Panel} button brings up the EVD control panel
(Figure~\ref{gui.figure}, b). The control panel allows user to
configure the behavior of the EVD. User may select which data to read
using the DataMenu (Figure~\ref{gui.figure}, c)

Complicated events might be cleaned by selective presentation of parts
of the data and the detector. User can specify cuts on physical
quantities of the displayed objects, such as the $\eta$, $\phi$, $\Pt$
of tracks or the energies deposited in the calorimeter towers. This
helps to clean up the event by removing low-$\Pt$ tracks and low-$\Et$
towers, or to debug reconstruction problems by requiring EVD to show
objects which pass some cuts (e.g. number of COT or Silicon hits, or
tracks with large $d_0$ etc.). Figure~\ref{cut_manager.figure} shows
sample Cut Manager session.

\section{Displays}
\label{evd_displays}

\subsection{$r-\phi$ and $r-z$ Views}
\label{cot_display}

\begin{figure}[!b]
  \begin{center}
   \includegraphics[width=0.99\textwidth]{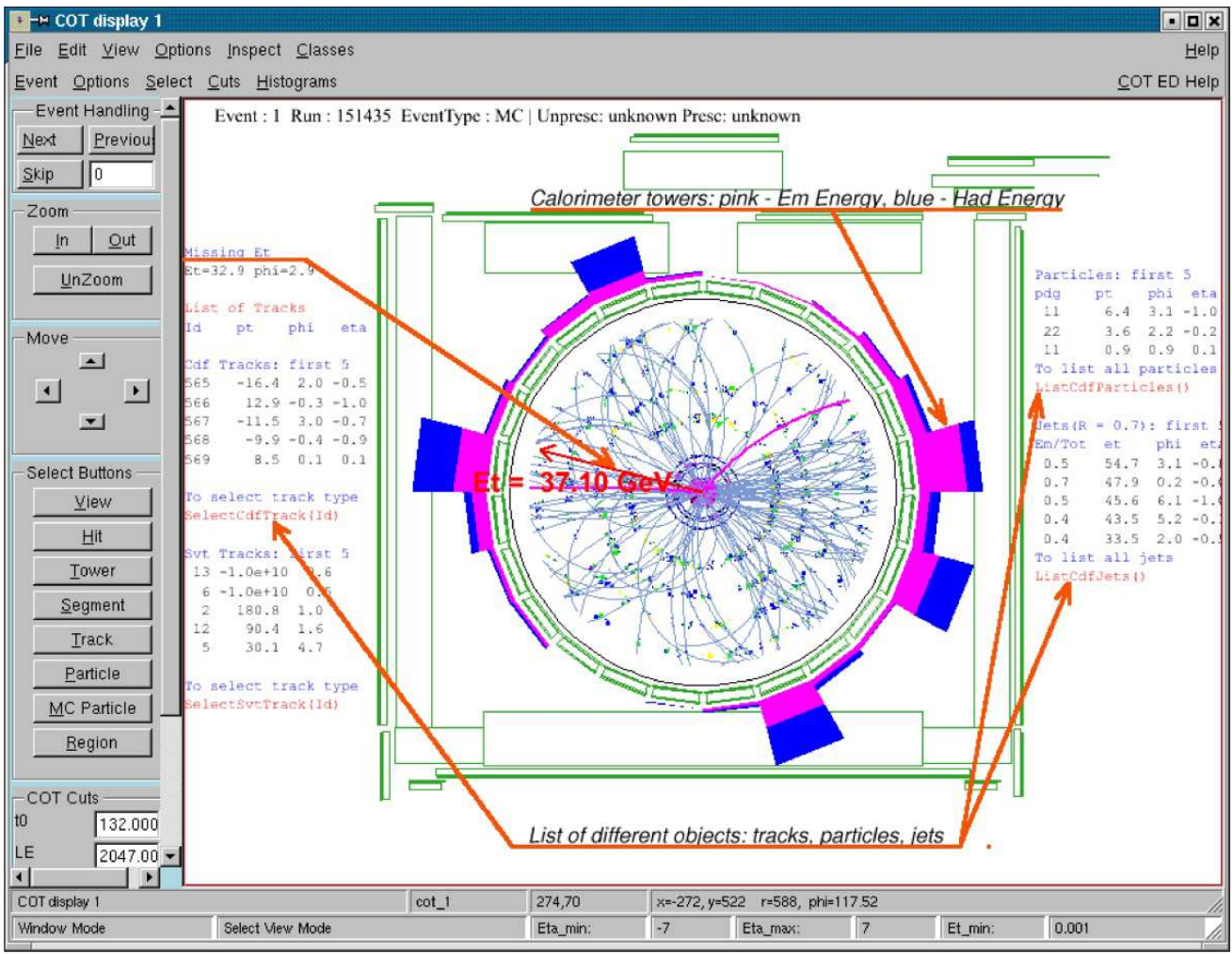}
    \caption
[COT Display ($r-\phi$ view)]
{COT Display ($r-\phi$ view).}
    \label{evd_cot_display}
  \end{center}
\end{figure}

In layered projections, each geometry object acquires a 2-dimensional
shape which can be different in each projection. For example, the
drift chamber outline is a circle in the $r-\phi$ view and a rectangle
in the $r-z$ view.

The $r-\phi$ view (COT Display, Figure~\ref{evd_cot_display}) shows
hits in the subdetectors (COT, Silicon trackers, Muon Chambers, TOF,
XTRP etc.), energies in the central EM and HAD calorimeter towers
(summed over $\eta$), missing $\Et$ information, information about CDF
reconstructed objects, such as $e$, $\mu$, $\gamma$, jet
candidates. It also gives the information about run/event number, as
well as trigger information.

Figure~\ref{evd_cot_display} is an example of a schematic view
(Section~\ref{evd_views}), and it is obtained from a realistic
$r-\phi$ projection by applying a {\it fish-eye} transformation to a
COT volume.

A fish-eye view introduces a nonlinear transformation of radius in the
layered $r-\phi$ projection, with the aim to enable simultaneous
inspection of tracking chambers, calorimetry and muon system within
the same picture (Figure~\ref{evd_cot_display}).

\begin{figure}[htbp]
  \begin{center}
   \includegraphics[width=0.99\textwidth]{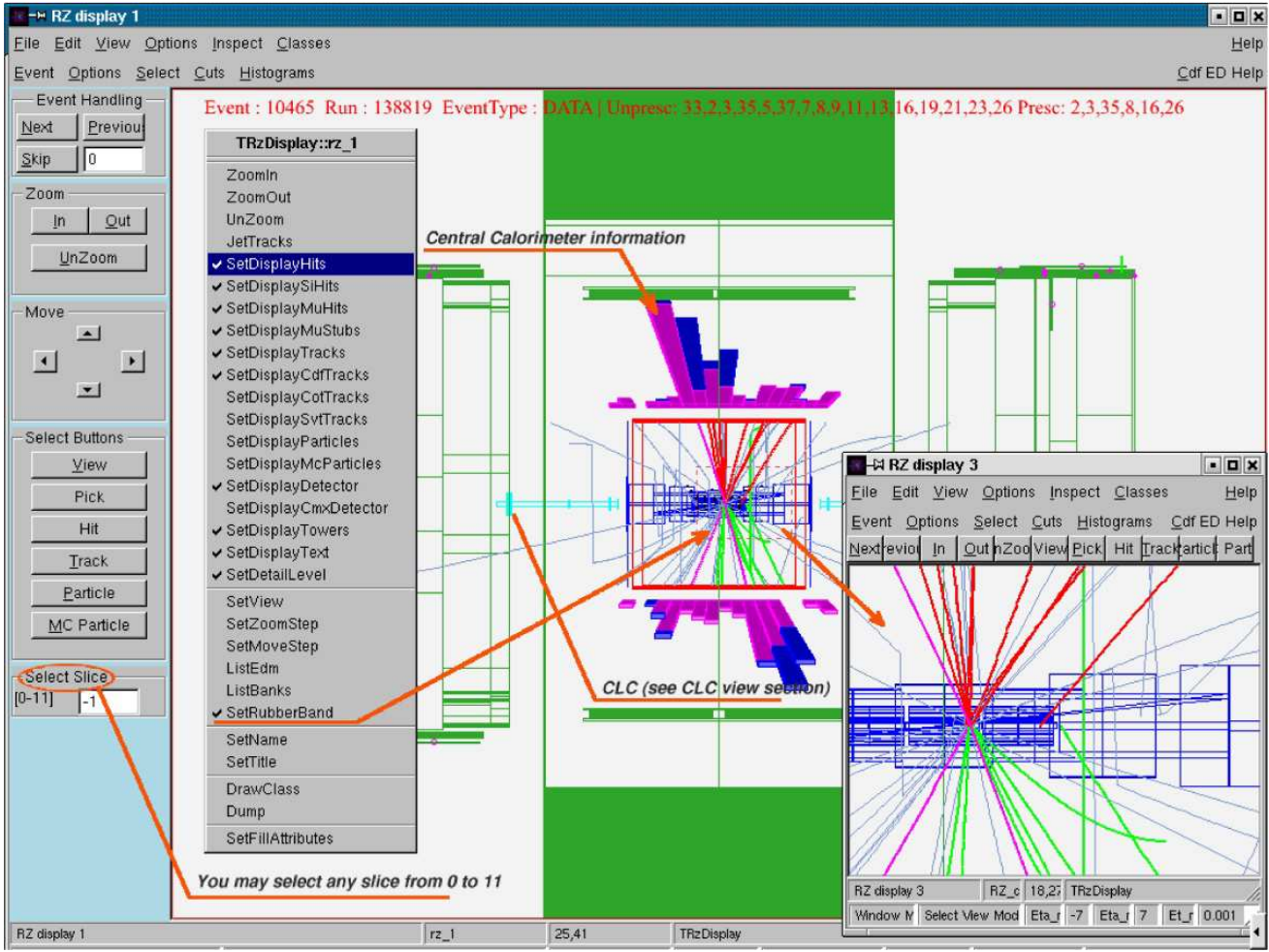}
    \caption
[RZ Display ($r-z$ view)]
{RZ Display ($r-z$ view).}
    \label{evd_rz_display}
  \end{center}
\end{figure}

The $r-z$ view (Figure~\ref{evd_rz_display}) is designed to show the
same information as $r-\phi$ view, but in $r-z$ projection. The only
exception are COT hits, which do not have z coordinate and therefore
are displayed in $r-\phi$ projection only. Energies in the central EM
and HAD calorimeter towers are summed over
$\phi$. Figure~\ref{evd_rz_display} is an example of a realistic view
(Section~\ref{evd_views}).

User can ``slice'' the detector by selectin pair of opposite wedges
("Select Slice" option at the bottom of the RZ display menu,
Figure~\ref{evd_rz_display}) to reduce the amount of information
displayed. In this case one will have only those wedges' tracks, hits,
bits, and calorimeter towers. There are 24 15$^\circ$ wedges, which
are displayed in 12 opposite pairs. The default slice value is -1,
which folds all upper wedges onto the top and all lower wedges onto
the bottom, which is the way the RZ display used to work in CDF Run I.

\subsection{Lego Displays}
\label{lego_displays}

\begin{figure}[!b]
  \begin{center}
   \includegraphics[width=0.99\textwidth]{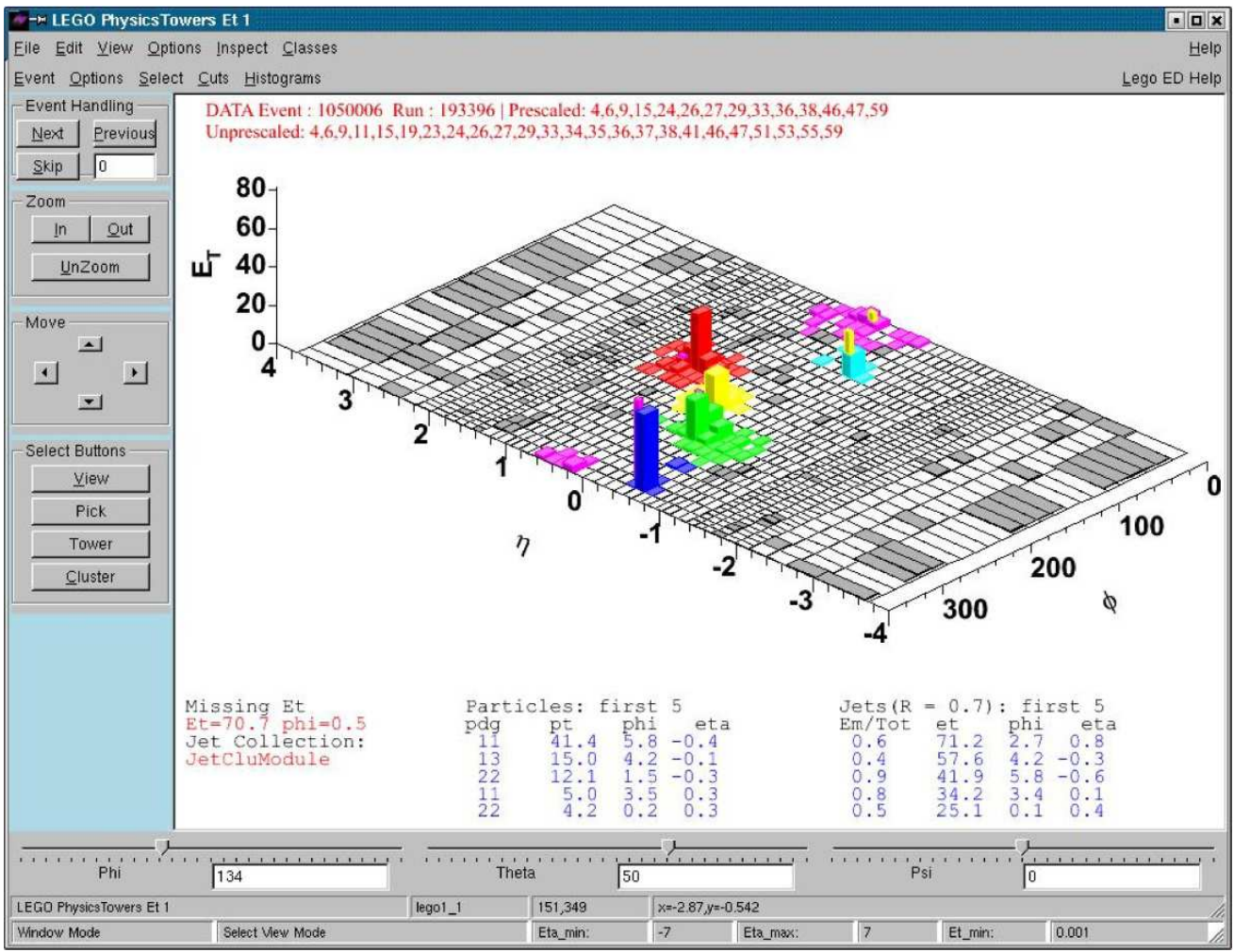}
    \caption[Lego Display: $\Et$ vs. $\eta-\phi$]{The Lego Display shows the variable $\Et$ as a function of eta-phi.}
    \label{evd_lego_display}
  \end{center}
\end{figure}

These are generic LEGO plot windows (Figure~\ref{evd_lego_display})
showing a variable (E, $\Et$, ADC counts) as a function of
eta-phi. Figure~\ref{evd_lego_display} is an example of an abstract
view (Section~\ref{evd_views}). $\eta-\phi$ grid corresponds to
$\eta-\phi$ segmentation of CDF calorimeter system
(Section~\ref{detector_calorimetry}); each bin on the lego display
represents a tower. Towers corresponding to CDF Run II particle
candidates (e, $\mu$, $\gamma$, jets) have been added to LEGO views to
improve the display (see Figure~\ref{evd_lego_display}). User can
interactively rotate LEGO display/change default settings to obtain a
better view of an event.

There also many other LEGO-based views, used to monitor/debug specific
subdetectors, such as the PLUG/PPR Views to visualize Plug Calorimeter
information, and the CES View to show the sums of energies for CES
Strips and Wires. In addition to the views itself there is an
interface to a histogramming package to obtain specific
distributions. For example, one can see distributions of measurements
in strips/wires of the CES. Views are designed to give a general idea
about an event, and histograms allow to see more detailed picture for
a part of a subdetector.

\subsection{3D Displays}
\label{lego_displays}

\begin{figure}
\includegraphics[width=0.9\textwidth]{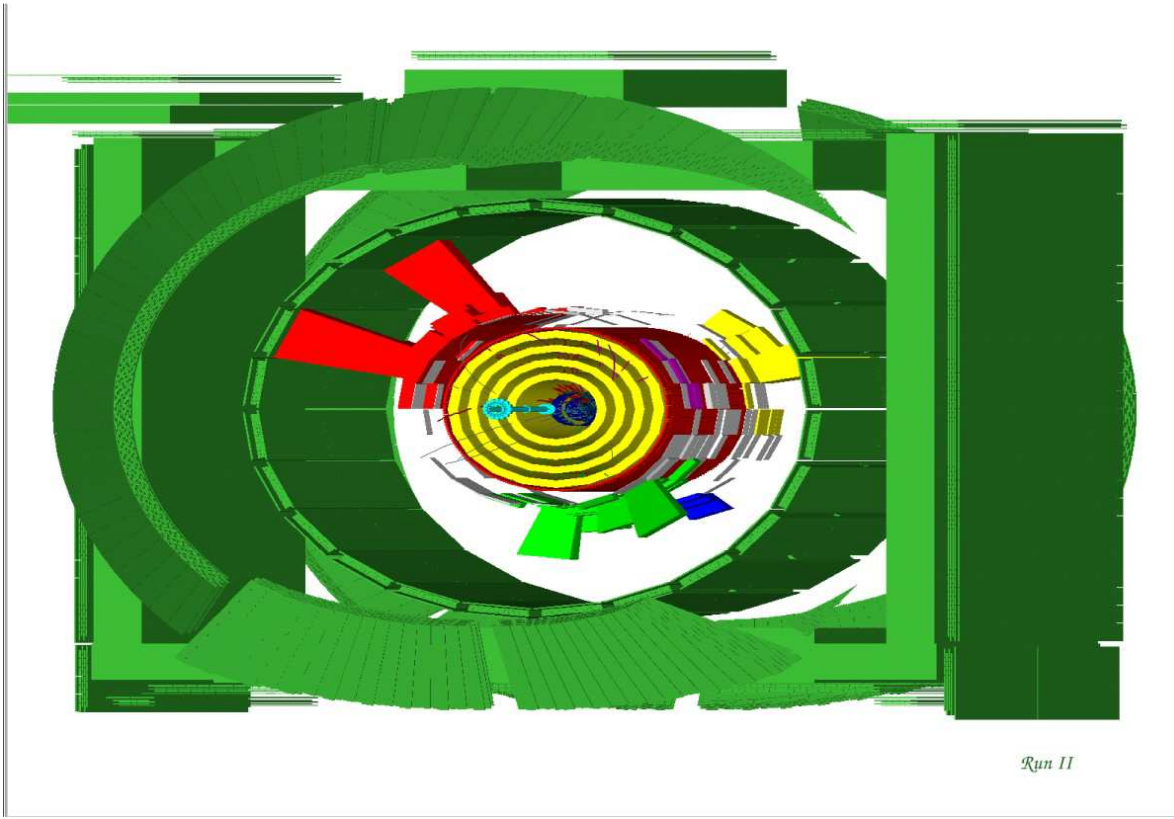}
\caption
[3D OpenGL view of CDF detector]
{3-dimensional OpenGL view of detector geometry with hidden lines removal.}
\label{cdf3d_opengl.figure}
\end{figure}

Perspective 3-dimensional views with hidden lines and hidden surface
removal are very useful for understanding detector geometry, and
provide attractive pictures for public relation (PR) purposes
(Figure~\ref{cdf3d_opengl.figure}). Analyzing the event itself is
often less successful in this mode, since the complicated geometry
tends to obscure the tracks and hits. 

Figure~\ref{cdf3d_opengl.figure} is an example of a realistic view
(Section~\ref{evd_views}). There are several available 3D views which
one can use separately for specific needs or combine to obtain a
complicated view.

\begin{itemize}
\setlength{\itemsep}{-0.11in}
\item 
CDF 3D display \\
-- 3-dimensional view of the CDF detector with tracks/silicon hits/muon hits and stubs)
\item
3D calorimeter display \\
-- central and plug calorimeter towers together
with tracks
\item 
SVX 3D display \\
-- dedicated silicon detector display, which shows the
silicon hits/strips together with tracks). SVX 3D display is designed
to obtain detailed information for all silicon hits associated with a
given track, which is helpful for debugging purposes.
\end{itemize}

Three-dimensional Views with hidden lines and surface removal are
possible through the special OpenGL viewer which is integrated in
view. Figure~\ref{cdf3d_opengl.figure} has been obtained using
OpenGL. In this mode live rotations are possible given a suitable
hardware acceleration for the instantaneous response.

\subsection{Other Displays}
\label{other_displays}

\begin{figure}[!t]
  \begin{center}
   \includegraphics[width=0.8\textwidth]{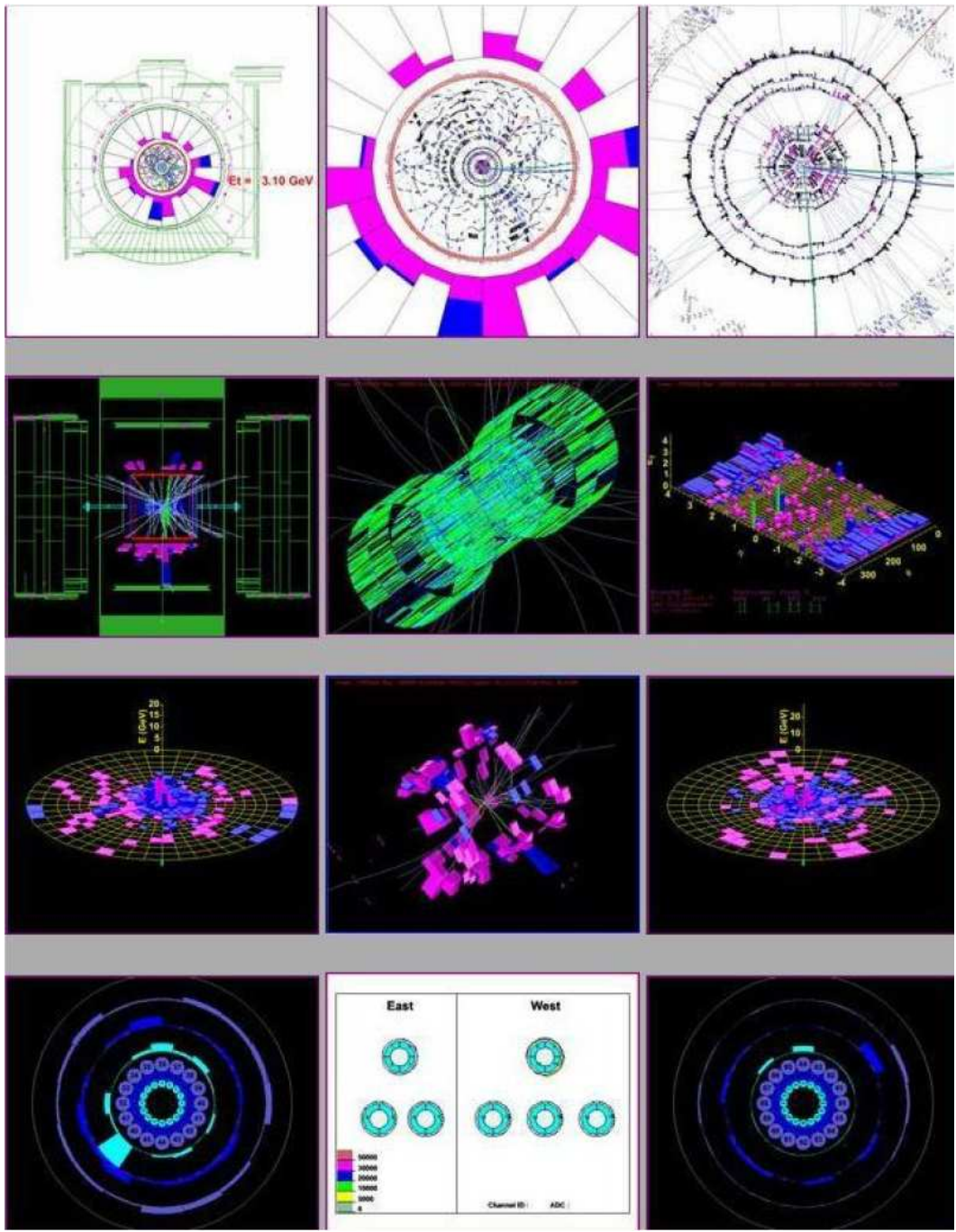}
\caption
[CDF "Live Events" public page]
{CDF "Live Events" public page.}
\label{evd_online.figure}
  \end{center}
\end{figure}

There many other views (see Figure~\ref{evd_online.figure}), used to
monitor different CDF Run II subsystems and to analyze real and
simulated data:

\begin{itemize}
\setlength{\itemsep}{-0.11in}
\item 
Wedge Display \\
-- CES/CPR information together with central calorimeter information
\item 
Trigger Display \\
-- Level 1, Level 2, Level 3 trigger bits and corresponding names
\item 
BSC Display \\
-- shows ADC counts from beam shower counter subdetectors
\item 
CLC Display \\
-- shows ADC counts from Cherenkov Luminosity Counters
\item 
MC Decay Tree Display \\
-- decay tree constructed from HEPG information for the simulated data
\end{itemize}

\section{Live Events}

The Live Events page has been designed to provide attractive pictures
for PR purposes~\cite{evd_live}. The Views displayed on
Figure~\ref{evd_online.figure} from the top to the bottom are as
follows:

\begin{itemize}
\setlength{\itemsep}{-0.11in}
\item {\it 1st row:}\\
COT Display. COT Display zoomed to access central outer tracker
information. COT Display zoomed to access silicon tracker information
\item {\it 2nd row:}\\
RZ Display. SVX 3D Display. Calorimeter LEGO Display
\item {\it 3rd row:}\\
East Plug LEGO Display. Calorimeter 3D Display. West Plug LEGO Display
\item {\it 4th row:}\\
East CLC Display. BSC Display. West CLC Display
\end{itemize}

\section{Conclusions}

From the very beginning of Run II data taking EVD is in continuous use
for the online monitoring and for the analyses. This answers the
question whether a presentation of data via visual techniques is
possible for complicated events at the Tevatron. More importantly,
features of the EVD make it to be one of the most important tools for
a better understanding of the events which could possibly be New
Physics candidates in a CDF Run II data.

\clearpage

\chapter{$\lgX$ Selection}
\label{lgx_selection}
\section{Datasets}
\label{datasets}

The data presented in the thesis was taken between March 21, 2002, and
August 22, 2004 and represent $\apprluminosity\invpb$ for which the
silicon detector (Section~\ref{detector_tracking})~\cite{SVX}, and all
three central muon systems (CMP, CMU and CMX), described in
Section~\ref{detector_muon} were operational.

The $\mug$ candidates are taken from a logical `OR' of the inclusive
high-$\Pt$ muon sample and the inclusive high-$\Et$ photon sample;
this was done to ensure a high and stable trigger efficiency for the
muons. For consistency, $\eg$ candidates are also obtained from a
logical 'OR' of the inclusive high-$\Et$ electron sample and the
inclusive high-$\Et$ photon sample. Each of the samples\footnote{We
used {\it bhel0d} as high-$\Et$ electron sample; {\it bhmu0d} as
high-$\Pt$ muon sample; and {\it cph10d} as high-$\Et$ photon sample}
was ntupled using the UC flat ntuple~\cite{UCNTUPLE_all}.

To accept an event from the inclusive high-$\Pt$ lepton sample we
require the event to have a loose lepton and a photon, or two leptons
(either tight or loose), or a tight lepton and $\met\gt 15 GeV$ (see
Tables~\ref{muon_cuts.table},~\ref{electron_cuts.table},~\ref{plugelectron_cuts.table}
and~\ref{photon_cuts.table}).

To accept an event from the inclusive high-$\Pt$ photon sample we
require the event to have a tight photon (see
Table~\ref{photon_cuts.table}) and a loose lepton. The muon selection
criteria are listed in Table~\ref{muon_cuts.table}; the electron
selection criteria are listed in Tables~\ref{electron_cuts.table} and
\ref{plugelectron_cuts.table}; 

\begin{figure}[!t]
  \begin{center}
    \mbox{
\subfigure[Luminosity in stability bins, $\invpb$]
{\epsfig{file=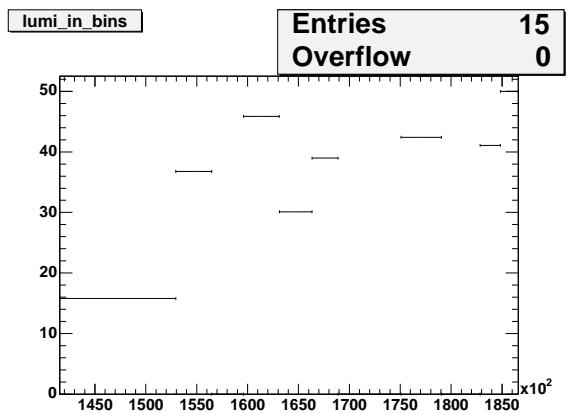,width=3in}}
         } 
\caption 
[The luminosity in each of the 8 time intervals, used to check stability of event gains]
{The luminosity in each of the 8 time intervals defined in
Table~\ref{lumi_stability.table}, used to check the stability of
control event yields during the run. The boundaries of the intervals
have been chosen to correspond to shutdowns or to major changes in the
trigger table. The total luminosity is $\apprluminosity~\invpb$. The
6\% luminosity uncertainty (Chapter~\ref{systematics}) is not shown.}
\label{lumi_stability.figure} 
\end{center}
\end{figure}

We check data integrity during the run by plotting the stability of
the event yields for the control samples.
We use the 8 time intervals defined in
Table~\ref{lumi_stability.table}~\cite{lumi_stability}.  The
boundaries of the intervals have been chosen to correspond to
shutdowns or to major changes in the trigger table. The luminosity in
each bin is plotted in Figure~\ref{lumi_stability.figure}.

\begin{table}[!h]
\small
 \begin{center}
  \begin{tabular} {|p{1.2cm}|p{4.5cm}|p{2.2cm}|p{2.5cm}|p{2cm}|}
\hline
Run      &     Table                       &  Date       & $\lum$,     $\invpb$ & Comment   \\
\hline
141544   &     PHYSICS\_1\_01 4\_275       &  2002.03.23 &                      & Start	   \\
152949   &     PHYSICS\_1\_02 7\_175\_323  &  2002.10.16 &  15.8 $\pm$ 0.9      & TrigTable \\
\hline
152593   &     PHYSICS\_1\_03 1\_185\_325  &  2002.10.16 &                      & TrigTable \\
156487   &     PHYSICS\_1\_03 2\_194\_329  &  2003.01.12 &  36.8 $\pm$ 2.2      & Shutdown  \\
\hline
159603   &     PHYSICS\_1\_04 4\_255\_357  &  2003.02.28 &                      & Startup   \\
163113   &     PHYSICS\_1\_04 9\_288\_373  &  2003.05.19 &  45.9 $\pm$ 2.8      & TrigTable \\
\hline
163130   &     PHYSICS\_1\_05 1\_290\_375  &  2003.05.19 &                      & TrigTable \\
\end{tabular}
 \end{center}
\caption[The time intervals used to check the stability of W
          and Z event yields versus time]
{ The intervals in run number used to check the stability of W
          and Z event yields versus time. The boundaries of the
          intervals have been chosen to correspond to shutdowns or to
          major changes in the trigger table. The offline luminosity
          is shown; the luminosity scale factor of 1.019 has not been
          applied. The systematic error of 6\% (Chapter~\ref{systematics}) in
          the $\lum$ is shown. }
\label{lumi_stability.table}
\end{table}

To summarize, in the resulting inclusive electron and muon samples
every event contains either a $l\gamma$ candidate, or a candidate for
the $W$ and $\Z$ control samples, described in detail below in
Chapters~\ref{muonid} and~\ref{electronid}.

\section{Selection Overview}
\label{selection}

Events with a high transverse momentum ($\Pt$) lepton or photon are
selected by a three-level trigger~\cite{RunII_TDR} that requires an
event to have either a lepton with $\Pt > 18~\GeV$ or a photon with
$\Et > 25~\GeV$ within the central region, $|\eta|
\lesssim 1.0$. The trigger system selects photon and electron
candidates from clusters of energy in the central electromagnetic
calorimeter. Electrons are further distinguished from photons by
requiring the presence of a COT track pointing at the cluster. The
muon trigger requires a COT track that extrapolates to a reconstructed
track segment (``stub'') in the muon drift chambers.

We use the same kinematic event selection as in the Run I analysis:
inclusive $\lg$ events are selected by requiring a central photon
candidate with $\Etgamma>25$ $\GeV$, a central lepton candidate ($e$
or $\mu$) with $\Etlepton>25$ $\GeV$ passing the ``tight'' criteria
listed below, and a point of origin along the beam-line not more than
60 cm from the center of the detector.

A muon candidate (Chapter~\ref{muonid}) passing the ``tight'' cuts has
the following properties: a) a well-measured track in the COT; b)
energies deposited in the electromagnetic and hadron compartments of
the calorimeter consistent with expectations; c) a muon ``stub'' track
in the CMX detector or in both the CMU and CMP detectors~\cite{RunII_TDR}
consistent with the extrapolated position of the COT track; and d) COT
timing measurements consistent with a track from a $\ppbar$ collision
and not from a cosmic ray.

An electron candidate (Chapter~\ref{electronid}) passing the ``tight''
selection has the following properties: a) a high-quality track with
$\Pt$ of at least half the shower energy, unless the $\Et > 100$
$\GeV$, in which case the $\Pt$ threshold is set to 25 $\GeV$; b) a
transverse shower profile consistent with an electron shower shape and
that matches the extrapolated track position; c) a lateral sharing of
energy in the two calorimeter towers containing the electron shower
consistent with that expected; and d) minimal leakage into the hadron
calorimeter.

Photon candidates (Chapter~\ref{photonid}) are required to have no
track with $\Pt>1$ $\GeV$, and at most one track with $\Pt<1$ $\GeV$,
pointing at the calorimeter cluster; good profiles in both transverse
dimensions at shower maximum; and minimal leakage into the hadron
calorimeter.

To reduce background from photons or leptons from the decays of
hadrons produced in jets, both the photon and the lepton in each event
are required to be ``isolated''. The $\Et$ deposited in the
calorimeter towers in a cone in $\eta-\varphi$ space of radius $R=0.4$
around the photon or lepton position is summed, and the $\Et$ due to
the photon or lepton is subtracted. The remaining $\Et$ in the cone is
required to be less than $2.0~\GeV+0.02\times(\Et-20~\GeV)$ for a
photon, or less than 10\% of the $\Et$ for electrons or $\Pt$ for
muons. In addition, for photons the sum of the $\Pt$ of all COT tracks
in the cone must be less than $2.0~\GeV+0.005\times\Et$.

Missing transverse energy $\met$ (Section~\ref{metid}) is calculated
from the calorimeter tower energies in the region $|\eta| <
3.6$. Corrections are then made to the $\met$ for non-uniform
calorimeter response~\cite{jet_corr} for jets with uncorrected $\Et >
15$ $\GeV$ and $\eta < 2.0$, and for muons with $\Pt > 20$ $\GeV$.

We use $W^\pm$ and $Z^0$ production as control samples (see
Section~\ref{muon_control_sample} for the details for the muon channel
and Section~\ref{electron_control_sample} for the electron channel) to
ensure that the efficiencies for high-$\Pt$ electrons and muons, as
well as for $\met$, are well understood. The photon control sample is
constructed from the events in which one of the electrons radiates a
high-$\Et$ photon, with an additional requirement that the $\eg$
invariant mass be within 10 $\GeV$ of the $Z^0$ mass.

The first search we perform is in the $\lgmet+X$ subsample, defined by
requiring that an event contain $\met> 25~\GeV$ (Section~\ref{metid})
in addition to the photon and ``tight'' lepton.

A second search, for the $\llg+X$ signature, is constructed by
requiring another muon (Chapter~\ref{muonid}) or electron
(Chapter~\ref{electronid}) in addition to the ``tight'' lepton and the
photon. The additional muons are required to have $\Pt > 20~\GeV$ and
to satisfy at least one of two different sets of criteria: the same as
those above for ``tight'' muons but with fewer hits required on the
track, or a more stringent cut on track quality but no requirement
that there be a matching ``stub'' in the muon systems. Additional
central electrons are required to have $\Et > 20~\GeV$ and to satisfy
the same criteria as tight central electrons but with a track
requirement of only $\Pt>10$ $\GeV$ (rather than 0.5$\times\Et$), and
no requirement on a shower maximum measurement or lateral energy
sharing between calorimeter towers. Electrons in the end-plug
calorimeters (Section~\ref{plug_electrons}), $1.2 < |\eta| < 2.0$, are
required to have $\Et> 15~\GeV$, minimal leakage into the hadron
calorimeter, a ``track'' containing at least 3 hits in the silicon
tracking system, and a shower transverse shape consistent with that
expected, with a centroid close to the extrapolated position of the
track~\cite{wenu_asymmetry_paper}.

The analysis includes a search for $e\mu\gamma$ events, for which the
estimated SM expectation is of order of 0.2 events. We also search for
$\lgg$ events by requiring another photon with $\Et>$25 GeV in
addition to the ``tight'' lepton and the photon. The additional
photons are required to pass standard photon cuts, described in
Chapter~\ref{photonid}.

\chapter{Muon Identification and Control Samples}
\label{muonid}

This chapter describes the selection of muon objects that are used
both in the searches and for the control samples. We require at least
one `tight central muon' in an event for it to be classified as a
$\mug$ event. In both $\eg$ and $\mug$ events we search for additional
muons using a definition of `loose central muon'. In this chapter we
describe these two sets of cuts and the numbers of muon objects
passing each cut below. As this is a chapter on object identification,
the tables show the number of objects passing each cut.

The summary on the number of events in the Muon Sample is shown in
Table~\ref{muon_summary.table}. The counting experiments based on
event counts rather than object counts for the $\mug$ candidates are
described in Chapter~\ref{photonlepton}. This chapter also describes
the control samples of $\Wmunu$ and $\Zmumu$ decays used to check the
temporal stability, and also the product of acceptance and efficiency.

\section{Muon Selection Criteria}
\begin{table}[h]
 \begin{center}
\begin{tabular} {|p{5.0cm}|p{3.4cm}|p{3.4cm}|p{3.4cm}|}
\hline
Variable & Tight & Loose & Stubless \\ 
\hline
Track $P_t$ &
$~>~25~GeV$ & 
$~>~20~GeV$ & 
$~>~20~GeV$ \\
\hline
Track quality cuts &
3x3SLx5~hits &
3x2SLx5~hits &
3x3SLx5~hits \\
\hline
Track $|z0|$  & 
$~<~60~cm$    & 
$~<~60~cm$    & 
$~<~60~cm$    \\
\hline
Calorimeter Energy (Em) &
$<$ 2 + sliding & 
$<$ 2 + sliding & 
$<$ 2 + sliding \\
\hline
Calorimeter Energy (Had)& 
$<$ 6 + sliding & 
$<$ 6 + sliding & 
$<$ 6 + sliding \\
\hline
Fractional Calorimeter Isolation $E_T$ & 
$~<~0.1$ & 
$~<~0.1$ & 
$~<~0.1$ \\
\hline
Cosmic      & 
False  & 
False  & 
False  \\
\hline
Chi2/(N of COT hits-5)&
-       &
-       &
$<$ 3   \\
\hline
Cal.Energy (EM+Had) &
-       &
-       &
$>$ 0.1 \\
\hline
CMUP muons cuts &
yes                 &
yes                 &
no                  \\
\hline
CMX  muons cuts &
yes                 &
yes                 &
no                 \\
\hline
\end{tabular}
 \end{center}
\caption
[Muon identification and isolation cuts]
{Muon identification and isolation cuts for Tight, Loose, and Stubless
muons. Tight and Loose are further subdivided into CMUP and CMX
categories. CMUP muon cuts are: $|\Delta X(CMU)|~<~3~cm$, $|\Delta
X(CMP)|~<~5~cm$. CMX muon cuts are: $|\Delta X(CMX)|~<~6~cm$, COT exit
radius of the muon track $\rho$(COT) $>$ 140 cm.}
\label{muon_cuts.table}
\end{table}

The muon selection cuts are similar to the standard CDF Run II cuts
~\cite{muons}. We describe the selection of the tight muons in
Section~\ref{tight_muons}. The selection of loose muons is described
in Sections~\ref{loose_muons} and~\ref{stubless_muons}.

\subsection{Tight Central CMUP and CMX Muons}
\label{tight_muons}

The muon selection criteria for a tight central muon are listed in
Tables~\ref{muon_cuts.table} and are described below.  Tight central
muons are identified by extrapolating tracks in the COT through the
calorimeters, and the extrapolation is required to match to a stub
either in both the CMU and CMP muon detectors (`CMUP' muon) or in the
CMX system ('CMX' muon), see Table~\ref{muon.table}). Tight central
muons are required to have a track-stub matching distance less than 3
cm for CMU, less than 5 cm for CMP, and less than 6 cm for CMX. For a
CMX muon we also require COT exit radius of the muon track $\rho$(COT)
to be greater than 140 cm to ensure that the track is well-measured.

The CMUP and CMX muon identifications require a muon object with the
requisite muon stubs. There are 355105 such objects in the
$\nentriesmu$ events of the muon sample. These are then divided into
CMUP muon candidates and CMX muon candidates. Stubless muon candidates
are treated separately in Section~\ref{stubless_muons}; there are
55346 stubless muon objects in the $\nentriesmu$ events.

\begin{table}[!t]
 \begin{center}
  \begin{tabular} {|p{5.5cm}|p{6.5cm}|l|l|}
\hline
Variable & Cut & Cumulative & This Cut \\ 
\hline
$\mu$ Candidates                &  & 
355105 & 355105 \\ \hline
Track $P_T$                     & $\gt$ 25 GeV & 
309679 & 309679 \\ \hline
Track quality cuts(loose)       & 3x2SLx5~hits & 
292208 & 314742 \\ \hline
Track quality cuts(tight)       & 3x3SLx5~hits & 
284759 & 304392 \\ \hline
Track $|z_0|$                   & $\lt$ 60 cm & 
280969 & 348881 \\ \hline
Calorimeter Energy (Em)         & $\lt$ 2+max (0, 0.0115$\times$(p-100)) GeV & 
278614 & 341205 \\ \hline
Calorimeter Energy (Had)        & $\lt$ 6+max (0, 0.028$\times$(p-100))  GeV & 
277925 & 346820 \\ \hline
Fractional Calorimeter \newline Isolation $E_T$   & $\lt$ 0.1 & 
277508 & 321987 \\ \hline
Cosmic      & FALSE  & 
194038 & 248861 \\ \hline
\hline
CMU stub                      & TRUE  & 
113369 & 201714 \\ \hline
$|\Delta X~CMU|$             & $\lt$ 3 cm & 
113057 & 347140 \\ \hline
CMP stub                      & TRUE  & 
111498 & 203653 \\ \hline
$|\Delta X~CMP|$             & $\lt$ 5 cm & 
111407 & 347649 \\ \hline
Region is OK                  & TRUE  & 
111407 & 349193 \\ \hline
Track $|d_0|$                   & $\lt$ 0.02 cm (si hits) OR \newline $\lt$ 0.2 cm (no si hits) & 
101933 & 211425 \\ \hline
\hline
CMX stub                      & TRUE  & 
76066 & 89334 \\ \hline
$|\Delta X~CMX|$             & $\lt$ 6 cm & 
75989 & 351303 \\ \hline
Region is OK                  & TRUE  & 
73367 & 349193 \\ \hline
Track $|d_0|$                   & $\lt$ 0.02 cm (si hits) OR \newline $\lt$ 0.2 cm (no si hits) & 
60554 & 211425 \\ \hline
\end{tabular}
 \end{center}
\caption
[Tight CMUP and CMX muon identification]
{Tight CMUP and CMX muon identification.  The cumulative totals of
tight central CMUP and CMX muons, showing the behavior as the cuts are
applied.  The initial entries in the table, before the rows in which
the stub requirement is applied, start with the total number of muon
objects with muon stubs in either the CMUP or CMX systems. Each entry
corresponds to a muon in the CDF Muon Collection in an event. The
``Region is OK'' cut for the CMUP muons requires $|\Delta
X(CMU)|~<~3~cm$, $|\Delta X(CMP)|~<~5~cm$. The ``Region is OK'' cut
for the CMX muons requires $|\Delta X(CMX)|~<~6~cm$, COT exit radius
$\rho$(COT) $>$ 140 cm. The $d_0$ cut is not applied to select muons,
but is tabulated for reference and to see its effect. The column
labeled `Cumulative' gives the effect of each successive cut on the
number of muon candidates in the $\nentriesmu$ events in the muon
subsample; The heading `This Cut' represents the effect of applying
only the cut listed.}
\label{muon.table}
\end{table}

The impact parameter calculation uses the default muon track rather
than the parent COT track, and a tighter impact parameter cut is
applied if the track does in fact contain silicon hits. Instead, we
have tabulated this $d_0$ cut for reference but we do not use it to
select tight or loose muons. The muon tracks used in the initial
selection for this analysis are beam-constrained
COT-only~\cite{muons}. For default muon tracks that contain silicon we
link backwards to the COT-only parent track and use that track for all
subsequent analysis. This technique, while losing valuable information
from the silicon at this stage, puts all prompt COT tracks on the same
footing.

\begin{figure}[htbp]
  \begin{center}
    \mbox{
\subfigure[$\eta-\phi$ distribution for the muons which pass loose muon identification cuts]{\epsfig{file=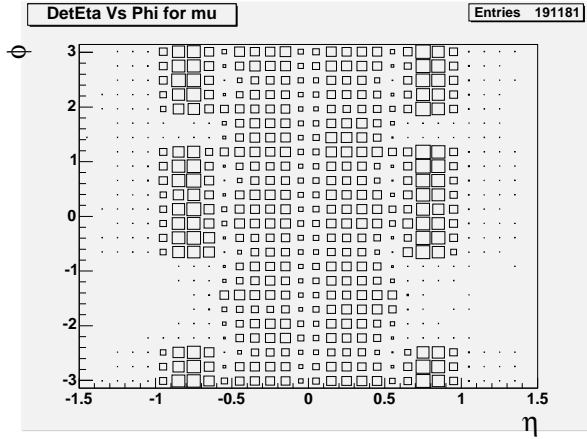,width=3in}} \quad
      } 
\caption 
[$\eta-\phi$ distribution for the muons which pass muon ID cuts]
{ (a) The $\eta-\phi$ distribution for the muons which pass 
muon identification cuts,
listed in Table~\ref{muon_cuts.table}.}
      \label{muonetaphi.figure}
  \end{center}
\end{figure}

The $\eta-\phi$ distributions are shown in
Figure~\ref{muonetaphi.figure} for the muons that pass the loose muon
identification cuts (Table~\ref{muon_cuts.table}). Muon candidates
which have stubs reconstructed from hits in either the 'bluebeam',
'miniskirt' or 'keystone' regions~\cite{muons} of the detector are
rejected. These sections of the detector were not fully operational
for the entire data sample.

All central muons are required to have $|z_0| < 60~cm$ so that the
collision is well-contained within the CDF detector. In order to be
well-measured, the muon track is required to have minimum of 3 axial
and 3 stereo superlayers with at least 5 hits in each superlayer.

High energy muons are typically isolated `minimum-ionizing' particles
that have limited calorimeter energy. A muon traversing the central
electromagnetic calorimeter(CEM) deposits an average energy of $\sim$
0.3 GeV. Therefore we require muon candidates to deposit less than 2
GeV total in the CEM towers (we take into account two towers in
the CEM) the muon track intersects.  Similarly, muons transversing the
central hadronic calorimeter(CHA) deposit an average energy of $\sim$
2 GeV; we consequently require muon candidates to deposit a total
energy less than $\sim$ 6 GeV, also increasing with muon momentum, in
the CHA towers intersected by the track extrapolation.  To take into
account the (slow) growth of energy loss with momentum, for very high
energy muons ($p > 100~GeV$) we require the measured CEM energy to be
less than $2.0 + 0.0115\times(p-100)~GeV$ and CHA energy to be less than
$6.0+ 0.028*(p-100)~GeV$.

To suppress hadrons and decay muons created from hadrons in jets we
require the total transverse energy deposited in the calorimeters in a
cone of R=0.4 around the muon track direction(known as the fractional
calorimeter isolation $E_T$) to be less than 0.1 of the muon track pt.

The COT cosmic finder by itself is essentially fully efficient.
Therefore, to suppress cosmic rays we use the COT-based cosmic
rejection~\cite{cosmics} and reject events which it tagged as cosmic
ray muons.

\subsection{Loose Central CMUP and CMX Muons}
\label{loose_muons}

While each $\mug$ event has to contain at least one tight CMUP or CMX
muon, both $\eg$ and $\mug$ events are searched for additional
high-$\Pt$ muons that could come from the decays of heavy particles.
There are two types of secondary muons we accept: `Loose' CMUP and CMX
muons, described here, and stubless muons (see
Section~\ref{stubless_muons}).

Loose muons are muon objects with either CMUP or CMX stubs, but with
looser COT cuts than the tight CMUP or CMX muons (see
Table~\ref{muon_cuts.table}). We require 3 axial and 2 stereo COT
super layers with at least 5 hits each for loose CMUP and CMX muons.

\subsection{Loose Central Stubless Muons}
\label{stubless_muons}

\begin{table}[!t]
 \begin{center}
  \begin{tabular} {|p{5.5cm}|p{6.5cm}|l|l|}
\hline
Variable & Cut & Cumulative & This Cut \\ 
\hline
$\mu$ Candidates                &  & 
74888 & 74888 \\ \hline
Track $P_T$                     & $\gt$ 20 GeV & 
42701 & 42701 \\ \hline
Track quality cuts(loose)       & 3x2SLx5~hits & 
25328 & 39120 \\ \hline
Track quality cuts(tight)       & 3x3SLx5~hits & 
22362 & 34366 \\ \hline
Track $|z_0|$                   & $\lt$ 60 cm & 
21041 & 72372 \\ \hline
Calorimeter Energy (Em)         & $\lt$ 2+max (0, 0.0115$\times$(p-100)) GeV & 
18634 & 62363 \\ \hline
Calorimeter Energy (Had)        & $\lt$ 6+max (0, 0.028$\times$(p-100))  GeV & 
18337 & 67502 \\ \hline
Fractional Calorimeter \newline Isolation $E_T$   & $\lt$ 0.1 & 
18115 & 46783 \\ \hline
Cosmicu     & FALSE  & 
6427  & 55172 \\ \hline
Calorimeter Energy (Em+Had)   & $\gt$ 0.1 GeV & 
6175  & 56551 \\ \hline
Track quality cuts            & $\chi^2$/(N of COT hits-5) $\lt$ 3 & 
6052  & 33472 \\ \hline
Region is OK                  & TRUE  & 
6052  & 74888 \\ \hline
Track $|d_0|$                   & $\lt$ 0.02 cm (si hits) OR \newline $\lt$ 0.2 cm (no si hits) & 
6024  & 35160 \\ \hline
\end{tabular}
 \end{center}
\caption
[Stubless muon identification and isolation cuts]
{Stubless muon identification and isolation cuts. Each entry
corresponds to a 'Stubless' muon in the CDF Muon Collection. The $d_0$
cut is not applied to select muons, but is tabulated for reference and
to see its effect. The column labeled `Cumulative' gives the effect
of each successive cut on the number of stubless muon candidates in
the $\nentriesmu$ events in the muon subsample. The heading `This Cut'
represents the effect of applying only the cut listed. }
\label{stubless.table}
\end{table}

The cuts for the Stubless muons, described in
Table~\ref{stubless.table}, are looser than the tight cuts, and in
particular do not require a stub in the muon chambers. There are three
types of `Stubless` muons:
\vspace*{-0.1in}
\begin{itemize}
\setlength{\itemsep}{-0.05in}
\item CMU muons  (muon track matches the CMU stub only)
\item CMP muons  (muon track matches a stub in the CMP only)
\item CMIO muons (muon track does not match a stub in CMU, CMP or CMX)
\end{itemize}

To identify stubless muons, we require at least some energy in the
calorimeter towers that the muon extrapolates to, Calorimeter Energy
(Em+Had) $\gt$ 0.1 GeV, and a good fit to the COT track,
$\chi^2$/(Number of COT hits-5)$\lt$3~\cite{muons}. These two cuts are
used to reject charged kaon decays in flight in which a low-momentum
kaon ($\sim 5$ GeV, typically) decays inside the COT with the kaon and
decay-muon tracks forming a `seagull' pattern which is reconstructed
as a single high-momentum track.

The pattern-finding algorithm often removes a complete stereo layer in
order to get a good fit, and so these tracks are badly
mis-reconstructed in polar angle. They consequently often are recorded
to leave zero energy in the extrapolated traversed calorimeter
towers~\cite{lauren}.

\section{Muon Control Samples}
\label{muon_control_sample}

The $\W$ and $\Z$ provide control samples for the $\lg$ samples.  We
use data triggered by the high-$\Pt$ muon trigger (MUON\_CMUP\_18 or
MUON\_CMX\_18) for both $\Z$ and $\W$ candidates, where tight muons
are the muons that pass the cuts in Table~\ref{muon.table}.

For comparisons with data we used the $\Zmumu$ Monte Carlo
sample~\cite{top_mc}. We applied the trigger efficiencies and scale
factors~\cite{muons}, listed in Table~\ref{muon_sf_and_te.table}.

\begin{table}[!h]
 \begin{center}
  \begin{tabular} {|p{6cm}|l|}
\hline
\multicolumn{2}{|c|} {
Trigger efficiencies
}\\
\hline
$\epsilon^{trigger}_{CMUP}$  & (90.78 $\pm$ 0.47)\%  \\
\hline
$\epsilon^{trigger}_{CMX}$   & (96.49 $\pm$ 0.40)\%  \\
\hline
\multicolumn{2}{|c|} {
Scale factors
}\\
\hline
tight CMUP                   & 0.8738 $\pm$ 0.0086    \\
\hline
tight CMX                    & 0.9889 $\pm$ 0.0063    \\
\hline
loose CMUP                   & 0.8921 $\pm$ 0.0088    \\
\hline
loose CMX                    & 0.9990 $\pm$ 0.0060    \\
\hline
Stubless                     & 0.9760 $\pm$ 0.0026    \\
\hline
\end{tabular}
 \end{center}
\caption
[Scale factors and trigger efficiencies for the muons]
{Scale factors and trigger efficiencies for the muons, applied
to $\Zmumu$ MC sample.}
\label{muon_sf_and_te.table}
\end{table}

\subsection{The $\zmumu$ Control Sample}

\begin{figure}[!h]
  \begin{center} \mbox{
\subfigure[Invariant Mass]{\epsfig{file=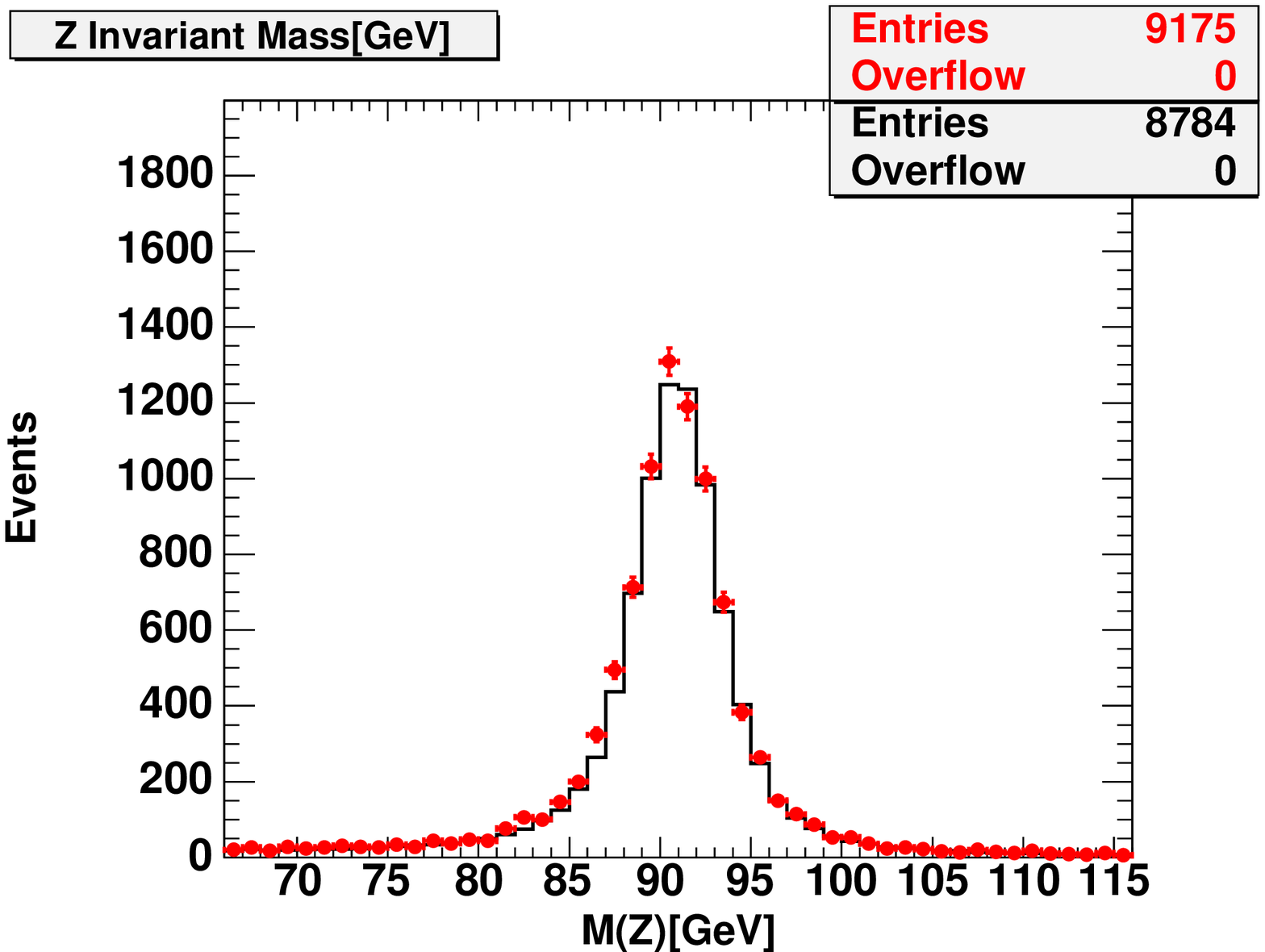,width=2.8in}} \quad
\subfigure[Invariant Mass (log)]{\epsfig{file=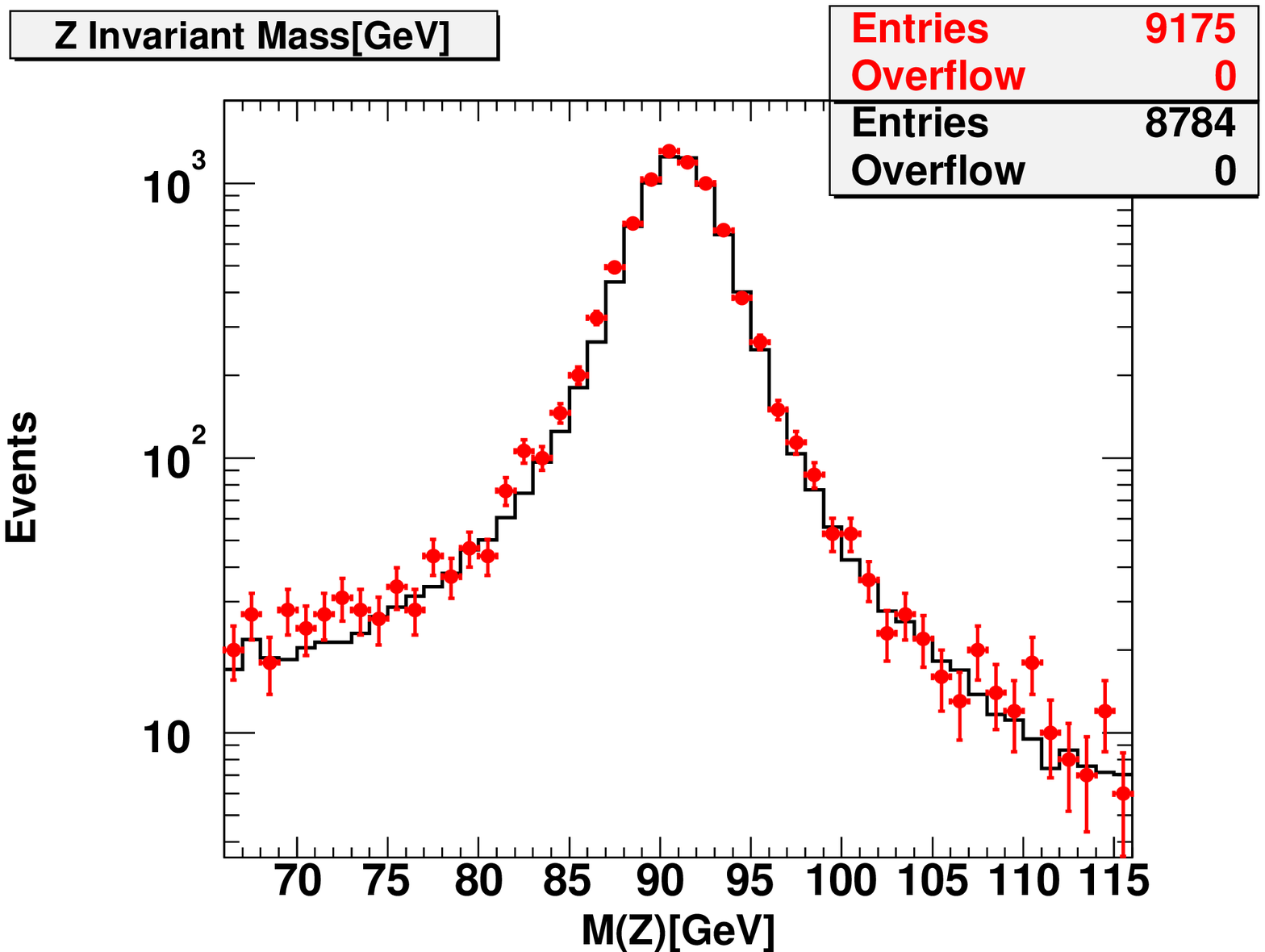,width=2.8in}} \quad
      }
    \mbox{
\subfigure[$P_T$ of $\Zmumu$]{\epsfig{file=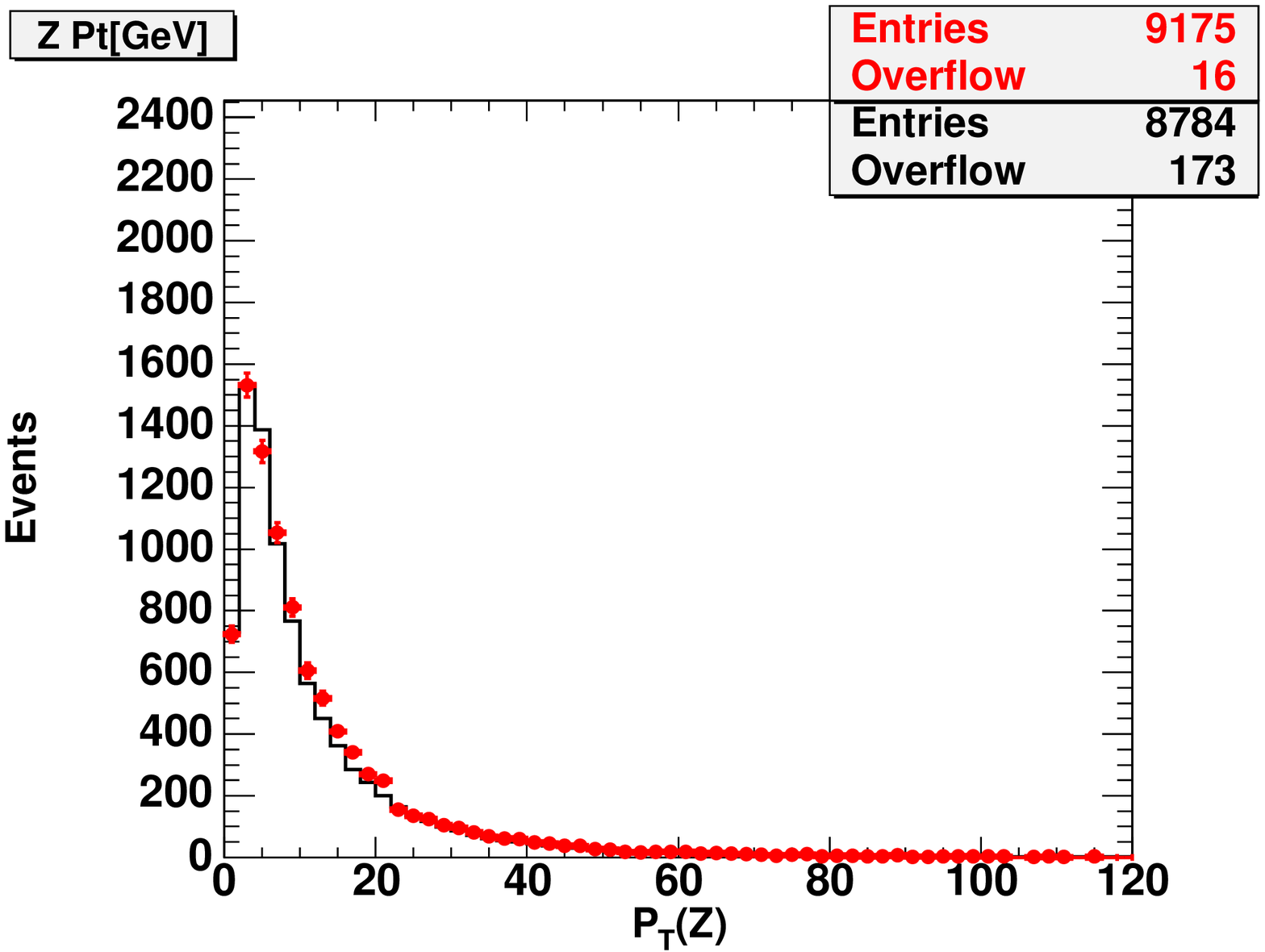,width=2.8in}} \quad
\subfigure[$P_T$ of $\Zmumu$ (log)]{\epsfig{file=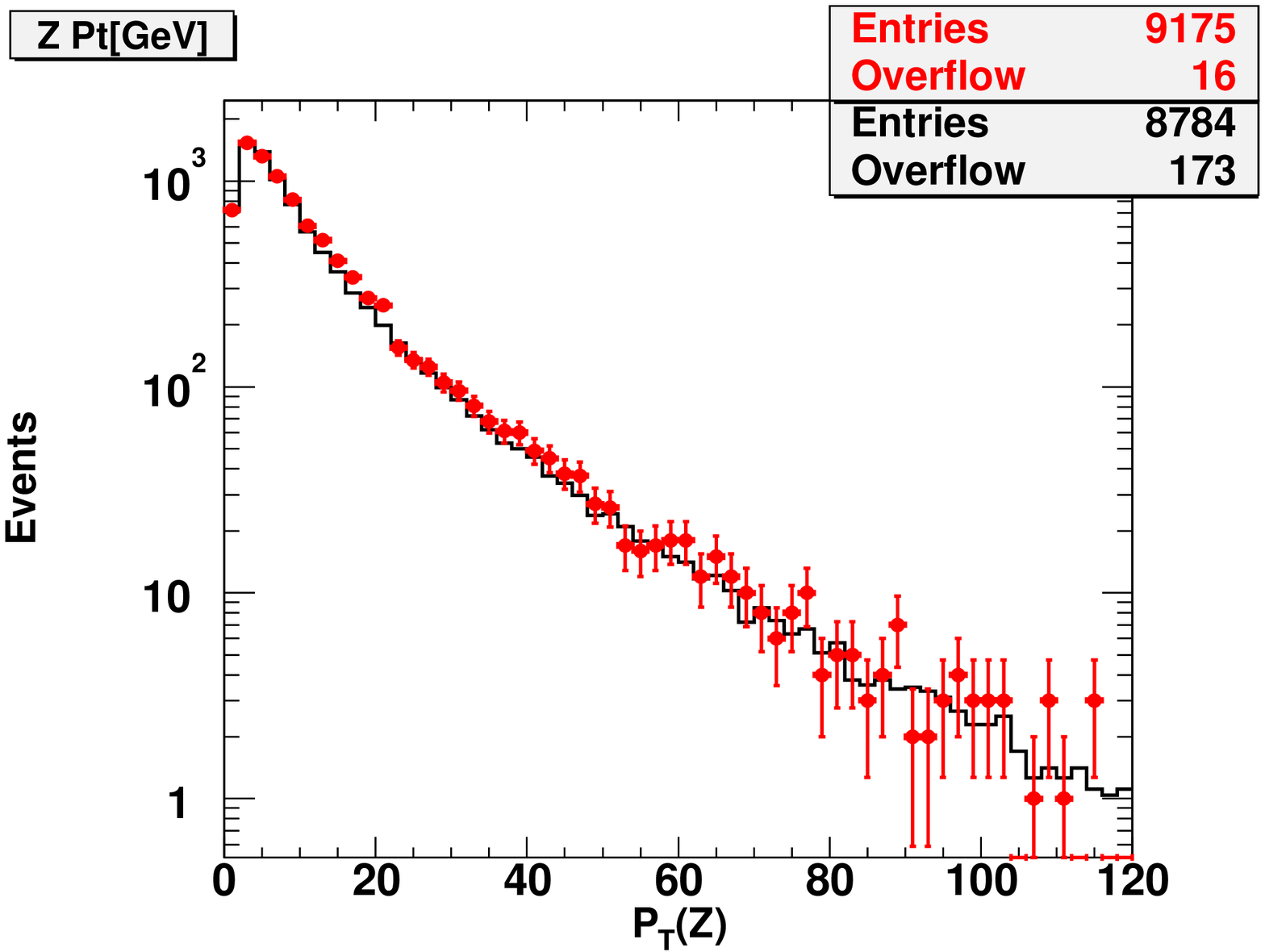,width=2.8in}} \quad
      }

\caption 
[The distributions for $\Zmumu$: data vs MC]
{The $\Zmumu$ control sample `sanity-check' plots.
    The distributions for Invariant Mass of $\Zmumu$, $P_T$ of
    $\Zmumu$, linear plots(a, c), log plots(b, d).  The histogram is
    the prediction from the $\Zmumu$ MC sample; the points
    are the $\Zmumu$ candidates from the data. Background estimates
    are not included.}
\label{zmumu_datamc.figure}
\end{center}

\end{figure}

The selection criteria for the $\zmumu$ control sample are listed in
Table~\ref{zmumu.table}. We require two muons in the event. One must
pass the tight cuts (Table~\ref{muon.table}), and another must pass
the loose cuts (Table~\ref{muon_cuts.table}). We further require the
two muons to have opposite charge, and require the difference of the
$z_0$ beam-line coordinates of the muon tracks to be less than 4
cm. The last requirement for counting $\Z$ events is that the
invariant mass of the muon pair of the $\Z$ candidate should be
between 66 GeV and 116 GeV.  We find $\nofzmumu$ $\Z$ events. We also
find $\nofzmumusamesign$ same-sign `$\Z$' events, indicative of the
maximum level of track reconstruction problems. 
The selection variable
distributions for the $\zmumu$ control sample are shown in
Figure~\ref{zmumu_datamc.figure}.
\begin{table}[!h]
\begin{center}
\begin{tabular} {|p{4cm}|p{12cm}|}
\hline
Variable              &  Cut                                                 \\
\hline
\hline
Tight Muon            &  Cuts are listed in Table~\ref{muon.table}           \\
\hline
Loose Muon            &  Cuts are listed in Tables~\ref{muon.table}
                         with looser COT cuts, 
                         Table~\ref{stubless.table}                          \\
\hline
Muon tracks must be      
of opposite charge    &  $Q^{\mu1} + Q^{\mu2}$ = 0                           \\
\hline
Delta Z cut           &  $|Z_{TRACK}^{\mu1}$ - $Z_{TRACK}^{\mu2}| < $ 4 cm   \\
\hline
Mass Window cut       &  $M > $ 66 GeV and $ < 116 $ GeV                     \\
\hline
Trigger               &  MUON\_CMUP\_18 or MUON\_CMX\_18                     \\
\hline
\end{tabular}
\end{center}
\caption
[The selection cuts for the $\zmumu$ control sample]
{The selection cuts for the $\zmumu$ control sample. The superscripts
$\mu1$ and $\mu2$ stand for the 2 muons in the event.}
\label{zmumu.table}
\end{table}

To normalize the $\Zmumu$ MC sample, we used the measured
$\sigma$($\Zmumu$)$\times$BR~\cite{CDF_WZ}. The comparison of data vs MC for
different muon types is shown in Table~\ref{zw_muons.table}. The
$\Zmumu$ event yields are shown in
Figure~\ref{zmumuwmunustability.figure}.

\subsection{The $\wmunu$ Control Sample}
\label{wmunu_control_sample}

The selection criteria for the $\wmunu$ control sample are listed in
Table~\ref{wmunu.table}. We require one tight muon
($\Pt\gt\cutmutightpt$ GeV, see Table~\ref{muon.table}), and
$\met>\cutmumet$ GeV (Section~\ref{metid}). We require the transverse
mass of the $\W$ candidate to be in the mass window 20-140 GeV.

To reject $\zmumu$ events in which one muon is not identified, events
with a second track with $\Pt>10$ GeV and associated EM and HAD
calorimeter energies less than 3 and 9 GeV, respectively, are
rejected.  We find $\nofwmunu$ $\wmunu$ events, $\nofwmunup$ positive
and $\nofwmunum$ negative $\W$'s. The $\Wmunu$ event yields are shown
in Figure~\ref{zmumuwmunustability.figure}.

\begin{table}[!h]
\begin{center}
\begin{tabular} {|p{4cm}|p{12cm}|}
\hline
Variable              &  Cut                                                 \\
\hline
\hline
Tight Muon            &  Cuts are listed in Table~\ref{muon.table}           \\
\hline
Missing $E_T$ Cut     &  $E_T > $ 25 GeV                                     \\
\hline
Z-Rejection           &  No 2nd Track with $P_T>10$ GeV, $E_{EM}< 3$ GeV, 
                         $E_{HAD}<9$ GeV                                     \\
\hline
Mass Window Cut       &  $M_T > 20$ GeV and $< 140$ GeV                      \\
\hline
Trigger               &  MUON\_CMUP\_18 or MUON\_CMX\_18                     \\
\hline
\end{tabular}
\end{center}
\caption
[The selection cuts for the $\wmunu$ control sample]
{The selection cuts for the $\wmunu$ control sample.}
\label{wmunu.table}
\end{table}

\subsection{Summary of the Muon Control Sample Event Counts. Stability Plots}

\begin{figure}[!h]
\begin{center}
    \mbox{
\subfigure[$\Zmumu$]{\epsfig{file=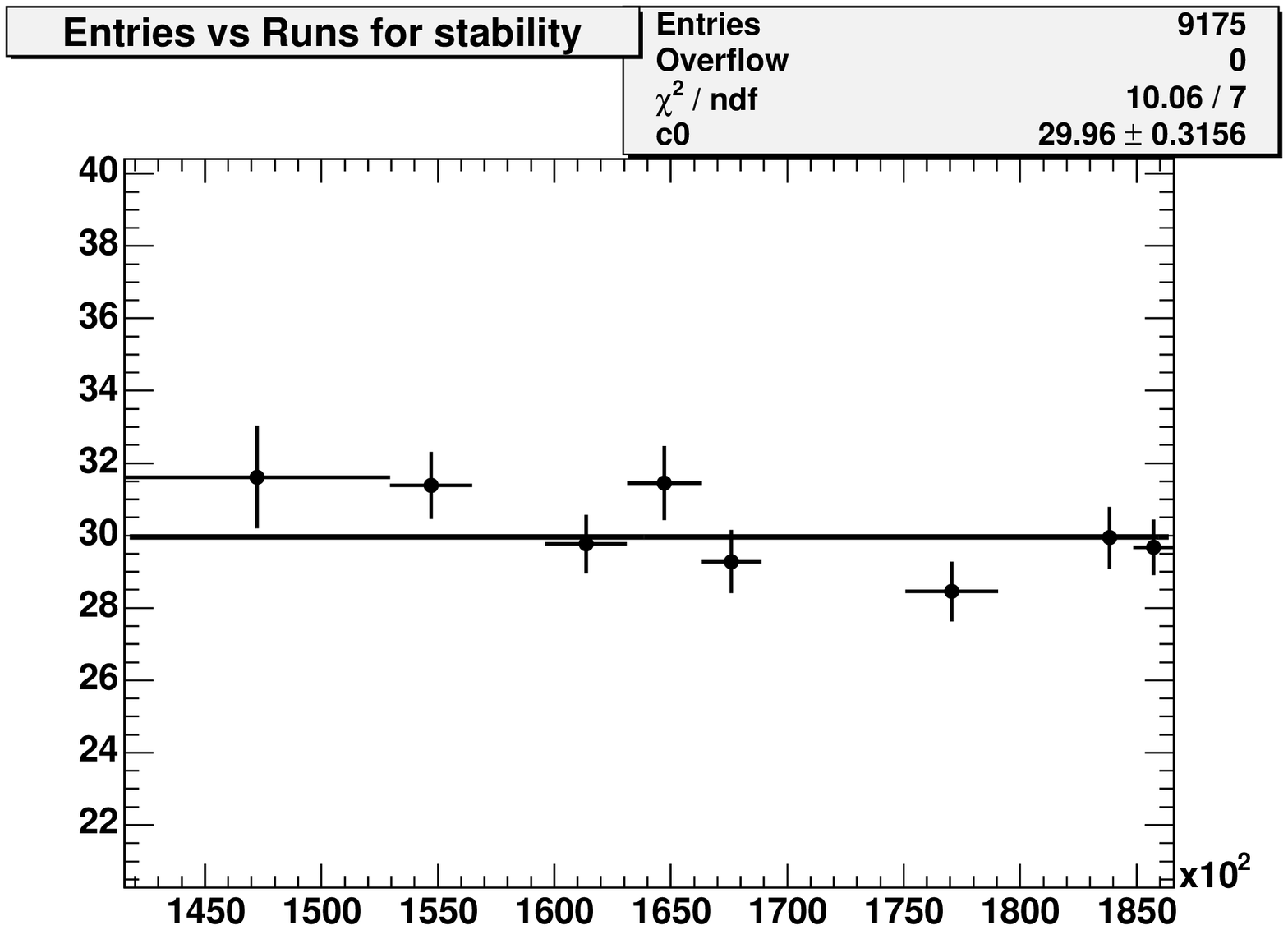,width=3in}} \quad
\subfigure[$\Wmunu$]{\epsfig{file=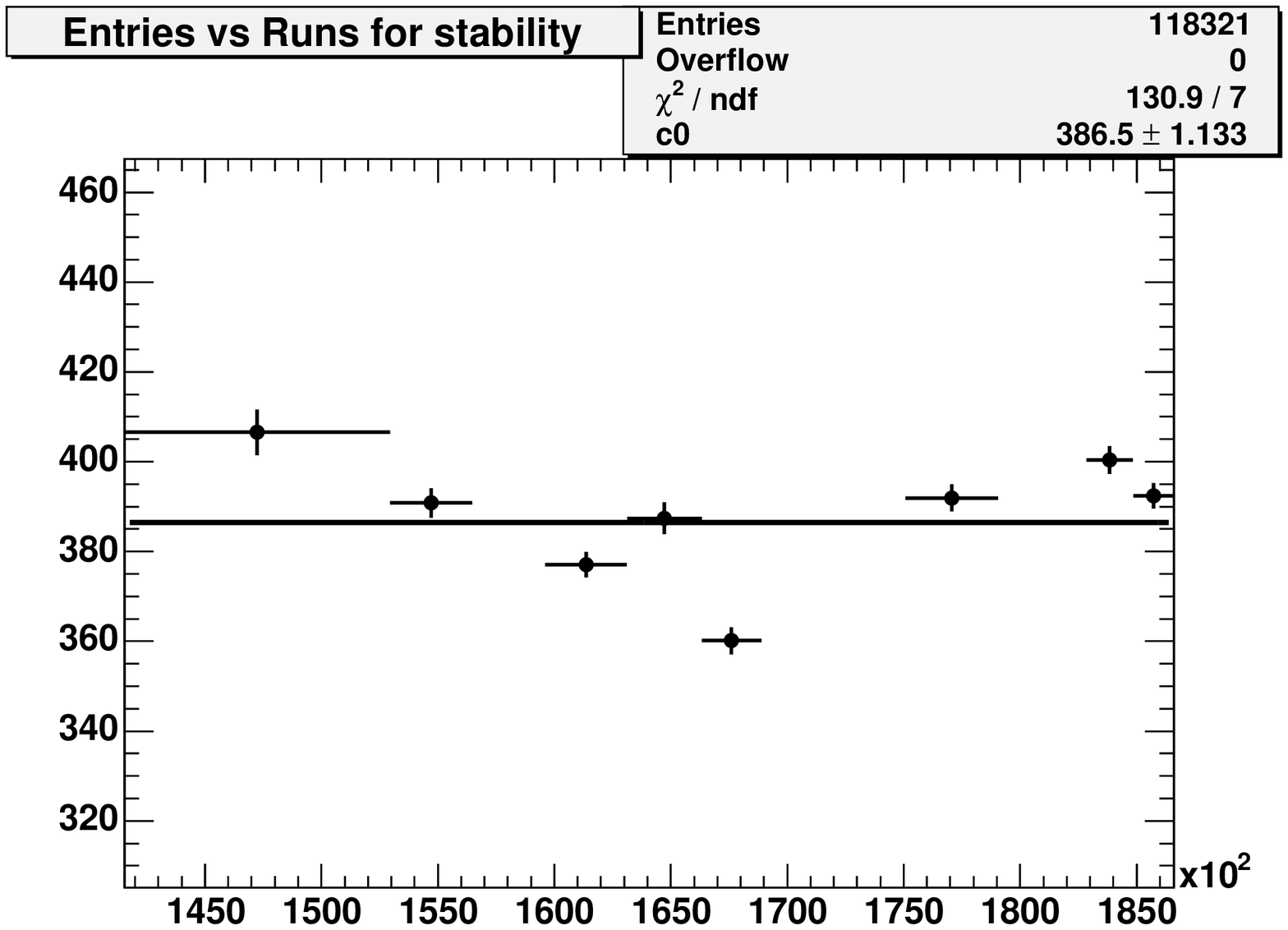,width=3in}} \quad
         }
\caption 
[Stability plots for $\Zmumu$ and $\Wmunu$]
{Stability plots of: (a) the $\Zmumu$, and (b) $\Wmunu$ control sample
cross sections versus run number. The bins are those of
Table~\ref{lumi_stability.table}. Shown are the statistical errors, the
luminosity systematic error of 6\% (Chapter~\ref{systematics}) is not
included. The trends in $\Zmumu$ ($\Wmunu$) are similar to
$\Zee$($\Wenu$), see Figure~\ref{zeewenustability.figure}. We attribute
this to common effects (luminosity, trigger, COT).}
\label{zmumuwmunustability.figure} 
\end{center}
\end{figure}

We use the control samples of $\Wmunu$ and $\Zmumu$ decays to check
temporal stability of the event yields
(Figure~\ref{zmumuwmunustability.figure}). The summary on the number
of events in the Muon Sample is shown in
Table~\ref{muon_summary.table}.

\begin{table}[!h]
 \begin{center}
\begin{tabular}{|l|p{3cm}|p{3cm}|}
\hline
                            & DATA                & MC                  \\ 
\hline
\multicolumn{3}{|c|} {
$\Zmumu$	     }\\
\hline
$\Zmumu_{ALL~Tight}      $  & 3465            & 3403           \\ 
$\Zmumu_{CMUP-CMUP~Tight}$  & 1462    & 1428   \\
$\Zmumu_{CMUP-CMX~Tight} $  & 1533	    & 1529    \\
$\Zmumu_{CMX-CMX~Tight}  $  & 470	    & 446     \\
$\Zmumu_{ALL}            $  & 9175              & 8784             \\ 
\hline
\hline
$\Zmumu_{CMUP-STUBLESS}  $  & 3594  & 3464 \\
$\Zmumu_{CMX-STUBLESS}   $  & 1929   & 1868  \\
\hline
\end{tabular}
\caption
[$\Zmumu$ summary: data vs $\Zmumu$ MC]
{$\Zmumu$ summary: data vs $\Zmumu$ MC. Background estimates are not included.}
\label{zw_muons.table}
 \end{center}
\end{table}

\begin{table}[htbp]
 \begin{center}
  \begin{tabular} {|p{10cm}|r|}
\hline
Category & N \\ 
\hline
\hline
Events with at least one Tight Muon		&  \nofmuevents \\
\hline
Tight Muons            				&  \nofmu \\
\hline
$\Wmunu$(triggered+$M_W$ window cut)		&  \nofwmunu \\
\hline
\mbox{Tight + Loose Muons}			&  \nofzmumurawcandidates \\
\hline
$\Zmumu$(triggered+$M_Z$ window cut)		&  \nofzmumu \\
\hline
\end{tabular}

 \end{center}
\caption
[The numbers of events for the muon control samples]
{The numbers of events for the muon control samples. The muon selection
cuts are given in Table~\ref{muon_cuts.table}, the $\zmumu$ selection
cuts in Table~\ref{zmumu.table}, and the $\wmunu$ cuts in
Table~\ref{wmunu.table}.
}
\label{muon_summary.table}
\end{table}

\clearpage
\newpage
\chapter{Electron Identification and Control Samples}
\label{electronid}

We require at least one `tight central electron' in an event for it to
be classified as an $\eg$ event. In both $\eg$ and $\mug$ events we
search for additional 'loose' electrons in the central and end-plug
electromagnetic calorimeters. We describe the tight central and loose
central and plug cuts below. The counting experiments based on event
counts rather than object counts for the $\eg$ candidates are
described in Chapter~\ref{photonlepton}. This chapter also describes
the control samples of $\Wenu$ and $\Zee$ decays used to check the
temporal stability, and also the product of acceptance and efficiency.

\section{Electron Selection Criteria}
\begin{table}[!b]
 \begin{center}
\input{local_tables/electron_cuts.table}
 \end{center}
\caption
[Central electron identification and isolation cuts]
{Central electron identification and isolation cuts.}
\label{electron_cuts.table}
\end{table}

The electron selection cuts are similar to the standard CDF Run II
cuts~\cite{electrons}. We describe the selection of the tight central
electrons in Section~\ref{tight_electrons}. The selection of loose
central electrons is described in Section~\ref{loose_electrons}; the
selection of electrons in end-plug is presented in
Section~\ref{plug_electrons}. The selection cuts are
standard~\cite{electrons} with the exception that the CES fiducial
requirement (see Section~\ref{tight_electrons}) and the conversion cut
(see Sections~\ref{conversions} and~\ref{nonwz}) are not applied.

\begin{table}[!t]
 \begin{center}
\begin{tabular} {|p{7.0cm}|p{4cm}|p{4cm}|}
\hline
Variable & Tight & Phoenix Tight \\ 
\hline
$E_T$, $\GeV$ &
$\gt$ 15  & 
$\gt$ 15  \\
\hline
Had/Em     & 
$\lt$ 0.05    & 
$\lt$ 0.05    \\
\hline
Fractional Calorimeter Isolation $E_T$ &
$\lt$ 0.1  & 
$\lt$ 0.1  \\
\hline
Chi2 Strips & 
$\lt$ 10     & 
$\lt$ 10     \\
\hline
$\Delta R_{xy}$, cm     & 
$\lt$ 3.0    & 
$\lt$ 3.0   \\
\hline
PES 5by9 U and V & 
$\gt$ 0.65  & 
$\gt$ 0.65  \\
\hline
PEM $|\eta|$ &
2.0 $\lt |\eta| \lt$ 1.2  & 
2.0 $\lt |\eta| \lt$ 1.2  \\
\hline
PhxMatch        &
-               &
TRUE            \\ 
\hline
Number of Silicon Hits &
-                      &
$\ge$ 3                \\ 
\hline
$|$Z(Phoenix)$|$, cm       &
-                      &
$\lt$ 60            \\ 
\hline
\end{tabular}
 \end{center}
\caption
[Plug electron identification and isolation cuts]
{Identification and isolation cuts for the electrons in end-plug. We
are using the ``Phoenix Tight'' selection~\cite{CDF6278}.}
\label{plugelectron_cuts.table}
\end{table}

\subsection{Tight Central Electrons}
\label{tight_electrons}

\begin{table}[!h]
 \begin{center}
  \begin{tabular} {|p{5.5cm}|p{6.5cm}|l|l|}
\hline
Variable & Cut & Cumulative & This Cut \\ 
\hline
$e$ Candidates                         &  & 
470379 & 470379 \\ \hline
$E_T$, $\GeV$                                   & $\gt$ 25 & 
323506 & 323506 \\ \hline
Track $P_T$, $\GeV$                             & $\gt$ 10; \newline $E_T\gt$ 100 GeV: $\gt$ 25 & 
316943 & 324739 \\ \hline
Track $|z_0|$, cm                           & $\lt$ 60 & 
316930 & 468731 \\ \hline
Had/Em                                  & $\lt$ 0.055 + 0.00045$\times$E & 
315752 & 420940 \\ \hline
E/P (for $E_T \lt$ 100 GeV only)       & $\lt$ 2.0 & 
308932 & 365229 \\ \hline
Lshr                                    & $\lt$ 0.2 (for $E_T \lt$ 100 GeV only) & 
307955 & 362783 \\ \hline
Chi2 Strips                             & $\lt$ 10  (for $E_T \lt$ 100 GeV only) & 
306883 & 433201 \\ \hline
$\Delta X$, cm                          & -3.0 $\lt Q_{trk}\times\Delta X \lt$ 1.5; \newline $E_T\gt$ 100 GeV: $|\Delta X|\lt$ 3.0 & 
306768 & 362872 \\ \hline
$|\Delta Z|$, cm                        & $\lt$ 3.0; \newline $E_T \gt$ 100 GeV: $\lt$ 5.0 & 
306660 & 361208 \\ \hline
Fractional Calorimeter \newline Isolation $E_T$ & $\lt$ 0.1 & 
269624 & 289848 \\ \hline
Track quality cuts                      & 3 Axial, 2 Stereo SL $\times$ 5 hits & 
269621 & 398424 \\ \hline
Conversion Flag                         & $\neq$ 1 & 
252900 & 439187 \\ \hline
Fiducial                                &  based on shower max & 
222844 & 367024 \\ \hline
\end{tabular}
 \end{center}
\caption
[Tight central electron identification and isolation cuts]
{Tight central electron identification and isolation cuts. Each entry
corresponds to an CDF Central EM object in an event. The column
labeled `Cumulative' gives the effect of each successive cut on the
number of electron candidates in the $\nentriese$ events in the
electron subsample. The heading `This Cut' represents the effect of
applying only the cut listed. We do not apply 'Conversion Flag' and
'Fiducial' cuts to select electrons, they are tabulated for reference
and to see their effects.}
\label{electron.table}
\end{table}

The selection criteria for tight central electrons are listed in
Table~\ref{electron_cuts.table} and are described below.

Electrons are identified in the CEM by matching high momentum tracks
to high-energy CEM clusters. The electron track is the highest
momentum track which intersects one of two towers in the CEM cluster.
The electron tracks that we use in this analysis are beam-constrained
COT-only. We apply the same corrections to the electron tracks as we
do to the muon tracks.

Fiduciality is a variable, which can have following values:
\begin{itemize}
\setlength{\itemsep}{-0.08in}
   \item[{\bf -1~~:}]  error (null strip/wire cluster)
   \item[{\bf  0~~:}]  not fiducial in central or plug
   \item[{\bf  1~~:}]  fiducial in central or plug
   \item[{\bf  2~~:}]  fiducial, but in CEM Tower 9
   \item[{\bf  3~~:}]  fiducial, but  in Chimney wedge tower 7
   \item[{\bf  4~~:}]  fiducial in CEM using max pt track extrapolated to plane of CES
   \item[{\bf  5~~:}]  fiducial in PEM using PES
   \item[{\bf  6~~:}]  fiducial in PEM using max $\Pt$ track extrapolated to plane of PES
\end{itemize}

An electron candidate is required to have tracking momentum (P) which
exceeds half of its calorimeter energy (E). The electron track is
required to have a minimum of 3 axial and 2 stereo SL segments
containing at least 5 hits each. In order that the momentum resolution
does not make for inefficiencies for very high-energy electrons, for
$E_T > 100~GeV$ the E/P cut is not applied (leaving only the $\Pt>25$
GeV cut as the requirement on the track).  The electrons are required
to have the track extrapolate to the beam line within $|z_0| < 60~cm$
so that CDF detector contains the collision well.

The position of the track extrapolated to the CES radius must satisfy
the following requirements: it must fall within charge-signed CES
shower position of the cluster in the r-phi view 
-3.0 cm $\lt Q_{trk}\times\Delta X \lt$ 1.5 cm
and it must fall within 3 cm of
the CES shower position in the Z-direction($\Delta Z$).

The CEM shower characteristics should be consistent with that of a
single charged particle. We require the ratio of the total energy of
the CHA towers located behind the CEM towers in the electron cluster
to that of the electron itself to be less than than
0.055+0.00045$\times$E GeV. A comparison of the lateral shower sharing
with neighboring towers in the CEM cluster with test-beam data is
parameterized by a dimensionless quantity,
$L_{shr}$~\cite{Balka:1987ty}, which must have a value less than 0.2.

\begin{figure}[!b]
  \begin{center}
    \mbox{
\subfigure[CES X]
{\epsfig{file=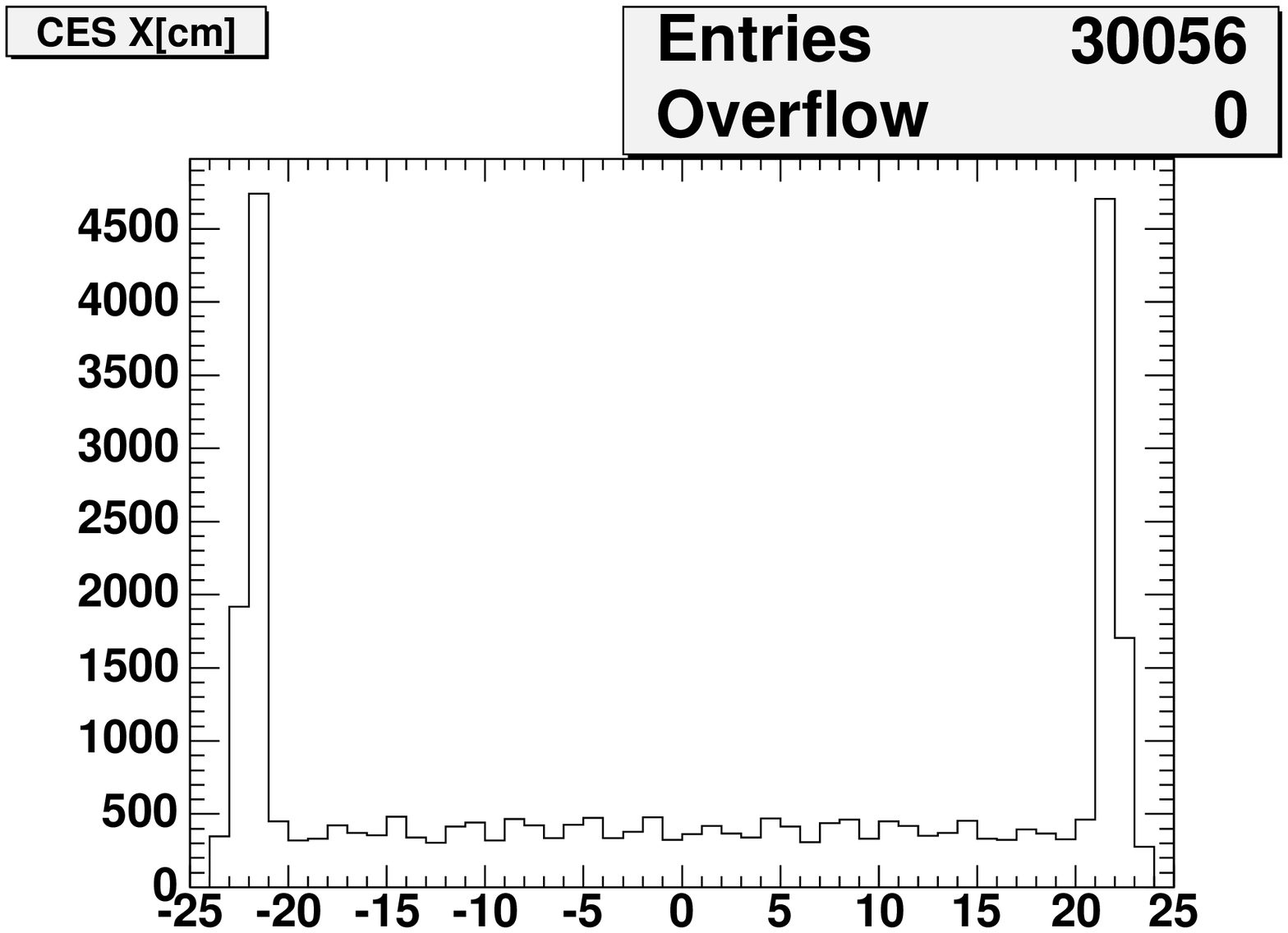,width=2.1in,clip=}} \quad
\subfigure[CES Z]
{\epsfig{file=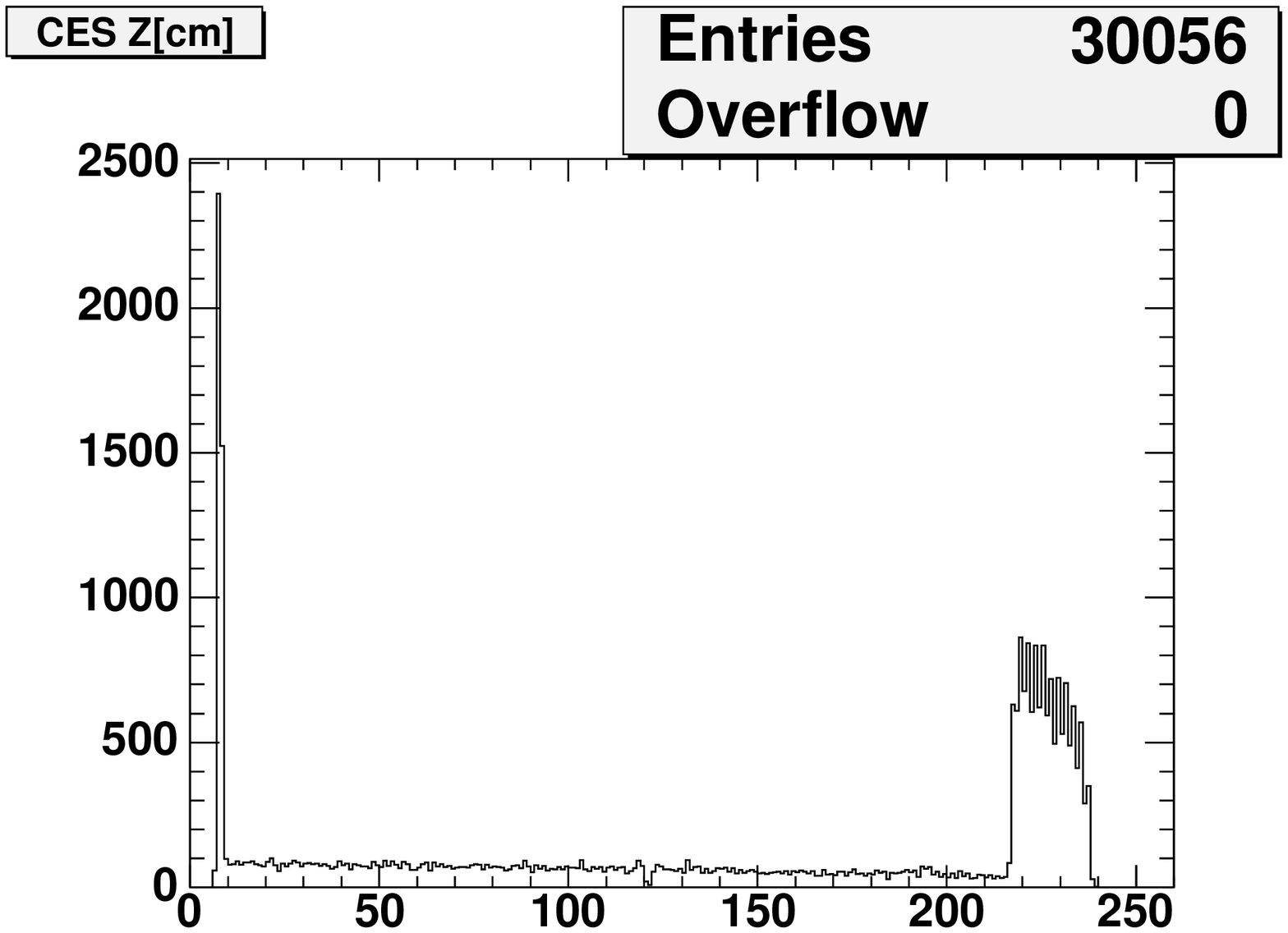,width=2.1in,clip=}} \quad
\subfigure[Detector $\eta$ vs. $\phi$]
{\epsfig{file=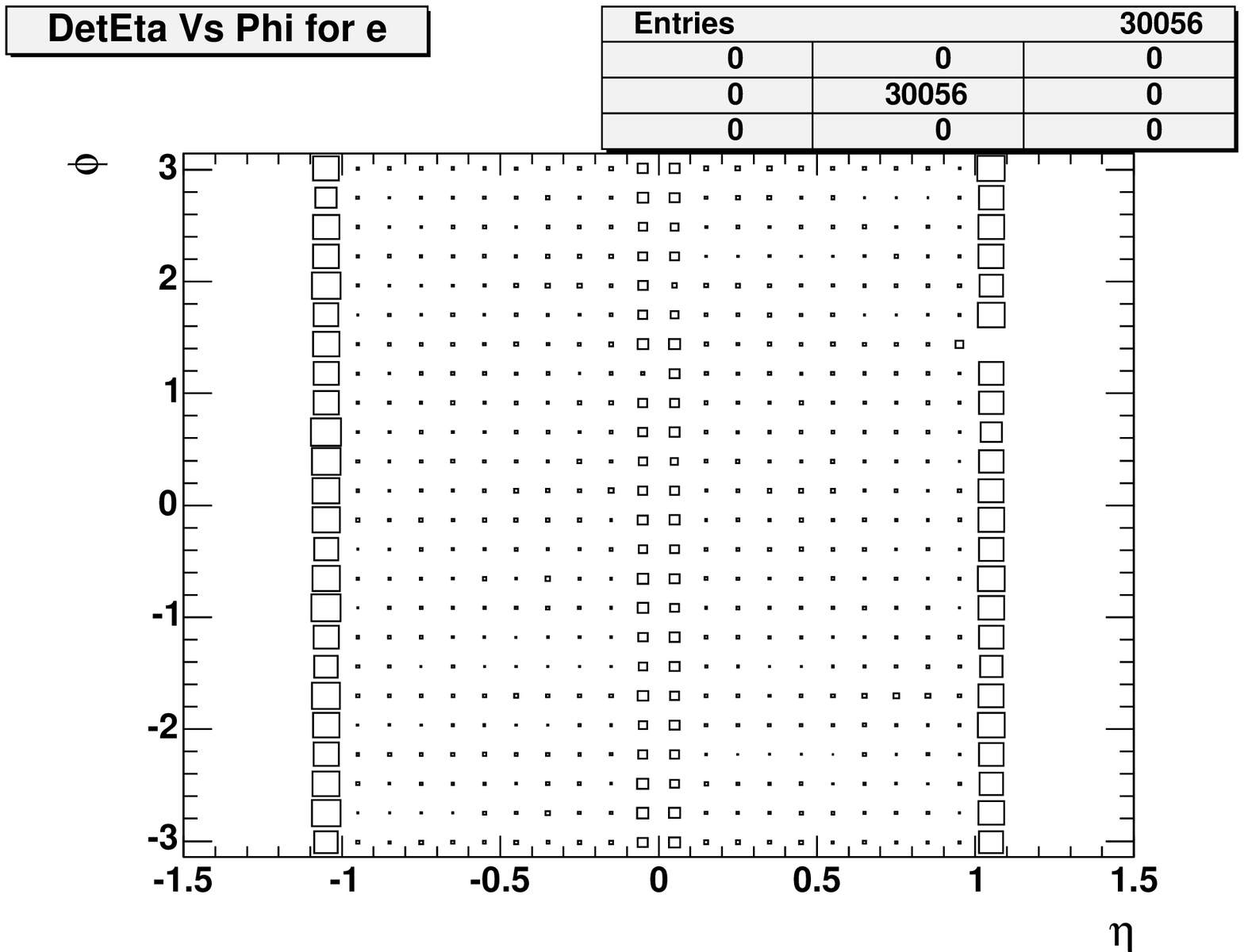,width=2.1in,clip=}} \quad
      }
    \caption
[Electrons which pass all cuts but the CES fiducial requirement]
{Plots for electrons which pass cuts all but CES fiducial requirement.}
   \label{nonfid_electrons.figure}
  \end{center}
\end{figure}

We require the $\chi^2$ for the profile of energy deposited in the CES
strips compared to that expected from test beam
data~\cite{Balka:1987ty} to be less than 10. No $\chi^2$ cut is made
on the profile in the CES wires as bremsstrahlung will separate from
the electron in the $r\phi$ view.

As an additional isolation requirement, the total transverse energy
deposited in the calorimeter in a cone R=0.4 around the electron
track, must be less than 0.1 of the $E_T$ of the electron. The
isolation is corrected via the standard algorithm~\cite{CDF4170}, for
leakage, but not the number of vertices.

We do not apply 'Conversion Flag' and 'Fiducial' cuts to select
electrons, they are tabulated for reference and to see their effects.

The acceptance gain by removing the fiduciality requirement is
approximately 14\%. Figure~\ref{nonfid_electrons.figure} shows the
distributions for the electrons which pass all cuts but fiducial
requirement.

\subsection{Loose Central Electrons}
\label{loose_electrons}

While each $\eg$ event has to contain at least one tight
electron, both $\eg$ and $\mug$ events are searched for
additional high-$\Pt$ electrons that could come from the decays of
heavy particles.  The cuts for these additional electrons are described
in Table~\ref{electron_cuts.table}. These cuts are looser than the tight
cuts, and in particular do not require any of the CES variables,
i.e. no track-cluster match in $\Delta X$ or $\Delta Z$ and no cut on
strip $\chi^2$, and also no cut on $L_{shr}$~\cite{Balka:1987ty}.

\subsection{Plug Electrons}
\label{plug_electrons}

Additional isolated electrons in the plug calorimeter with $\Et> 15$
GeV are identified for measured PEM rapidities of $1.2 < |\eta| <
2.0$. The cuts used for plug electron identification are given in
Table~\ref{plugelectron.table}. We require minimal leakage or activity
in the hadron calorimeter, Had/Em $\lt$ 0.05, a fractional isolation
(isolation energy over the electron energy) less than 0.1, and the
shower shape to satisfy the PEM 3x3 $\chi^2$ and PES 5by9 5-strip to 9
strip ratio cuts. These cuts are similar to standard cuts~\cite{electrons}.

We apply face corrections to the PEM energy of the plug electron
candidate, add the PPR energy and scale resulting number by
1.0315~\cite{veramendi}, as shown in Equation~\ref{pem.equation}.

\begin{equation}
  \begin{array}{l}
E_{plug~electron} = (E_{pem}^{cor} + E_{ppr})\times 1.0315
 \label{pem.equation}
 \end{array}
\end{equation}

\begin{table}[!b]
 \begin{center}
  \begin{tabular} {|p{5.5cm}|p{6.5cm}|l|l|}
\hline
Variable & Cut & Cumulative & This Cut \\ 
\hline
$e$ Candidates                         &  & 
54247 & 54247 \\ \hline
$E_T$                                   & $\gt$ 15 GeV  & 
15341 & 15341 \\ \hline
Had/Em                                  & $\lt$ 0.05 & 
11971 & 37048 \\ \hline
Fractional Calorimeter \newline Isolation $E_T$ & $\lt$ 0.1 & 
10274 & 21053 \\ \hline
Chi2 Strips                             & $\lt$ 10 & 
9931  & 21708 \\ \hline
Delta R                                 & $\lt$ 3.0 cm & 
9886  & 47213 \\ \hline
PES 5by9 U and V                        & $\gt$ 0.65 & 
9478  & 25977 \\ \hline
PEM $|\eta|$                           & 1.2 $\lt\eta\lt$ 2.0 & 
9057  & 33376 \\ \hline
PhxMatch                                & TRUE & 
9028  & 13654 \\ \hline
Number of Silicon Hits                  & $\ge$ 3  & 
8836  & 11647 \\ \hline
$|$Z(Phoenix)$|$                        & $\lt$ 60 cm & 
8836  & 11733 \\ \hline
\end{tabular}
 \end{center}
\caption
[Identification and isolation cuts for additional plug electrons]
{Identification and isolation cuts for additional plug electrons.}
\label{plugelectron.table}
\end{table}
\section{Electron Control Samples}
\label{electron_control_sample}

As in the muon case, the W and $\Z$ provide control samples for the
$\eg$ samples. We require at least one tight electron pass the
high-$\Et$ electron trigger (ELECTRON\_CENTRAL\_18) for both W and
$\Z$ candidates, where tight electrons are the electrons that pass the
cuts in Table~\ref{electron_cuts.table}.

For comparisons with data we used the $\Zee$ MC
sample~\cite{top_mc}. We applied trigger efficiencies~\cite{CDF7401}
and scale factors~\cite{electrons}, which are listed in
table~\ref{electron_sf_and_te.table}. The CEM Trigger efficiency for
high-$E_T$ electrons is 1-2.784*exp(-1.749*($E_T$ -
17.86))~\cite{CDF7401}. The total Track Trigger efficiency is
$\approx$ 96-98\%, depending on the run number and silicon/non-silicon
list of good runs~\cite{CDF7401}.We apply this trigger efficiency to
our run-dependent $\Zee$ MC sample.

\begin{table}[!t]
 \begin{center}
  \begin{tabular} {|p{6cm}|l|}
\hline
\multicolumn{2}{|c|} {
Trigger efficiencies
}\\
\hline
$\epsilon^{trigger}_{CEM}$    & 1-2.784*exp[-1.749*($E_T$ - 17.86)]  \\
\hline
$\epsilon^{trigger}_{Track }$ & $\approx$0.96-0.98  \\
\hline
\multicolumn{2}{|c|} {
Scale factors
}\\
\hline
tight CEM              & 0.999 $\pm$ 0.006    \\
\hline
loose CEM              & 1.005 $\pm$ 0.005    \\
\hline
tight PHX              & 0.95  $\pm$ 0.01     \\
\hline
\end{tabular}
 \end{center}
\caption
[Scale factors and trigger efficiencies for the electrons]
{Scale factors and trigger efficiencies for the electrons, applied 
to the $\Zee$ MC samples.}
\label{electron_sf_and_te.table}
\end{table}

\subsection{The $\zee$ Central-Central Control Sample}

The selection criteria for the $\zee$ central-central control sample
are listed in Table~\ref{zee.table}.  For this sample a $\Z$ candidate
is required to have two central electrons, one passing the tight cuts,
and the other the loose cuts (Table~\ref{electron_cuts.table}).  The
mass of the Z candidate is required to be within the window 66 GeV to
116 GeV. The difference in the $z_0$ coordinates of the two electron
tracks must be less than 4 cm. We find $\nofzee$ opposite-sign events
and $\nofzeesamesign$ same-sign central-central $\Z$ events satisfying
these criteria (the large number of same-sign events in the electron
sample but not in the muon sample is largely due to photon conversions -
see Chapter~\ref{fakes}).

\begin{figure}[!t]
  \begin{center} \mbox{
\subfigure[Invariant Mass]{\epsfig{file=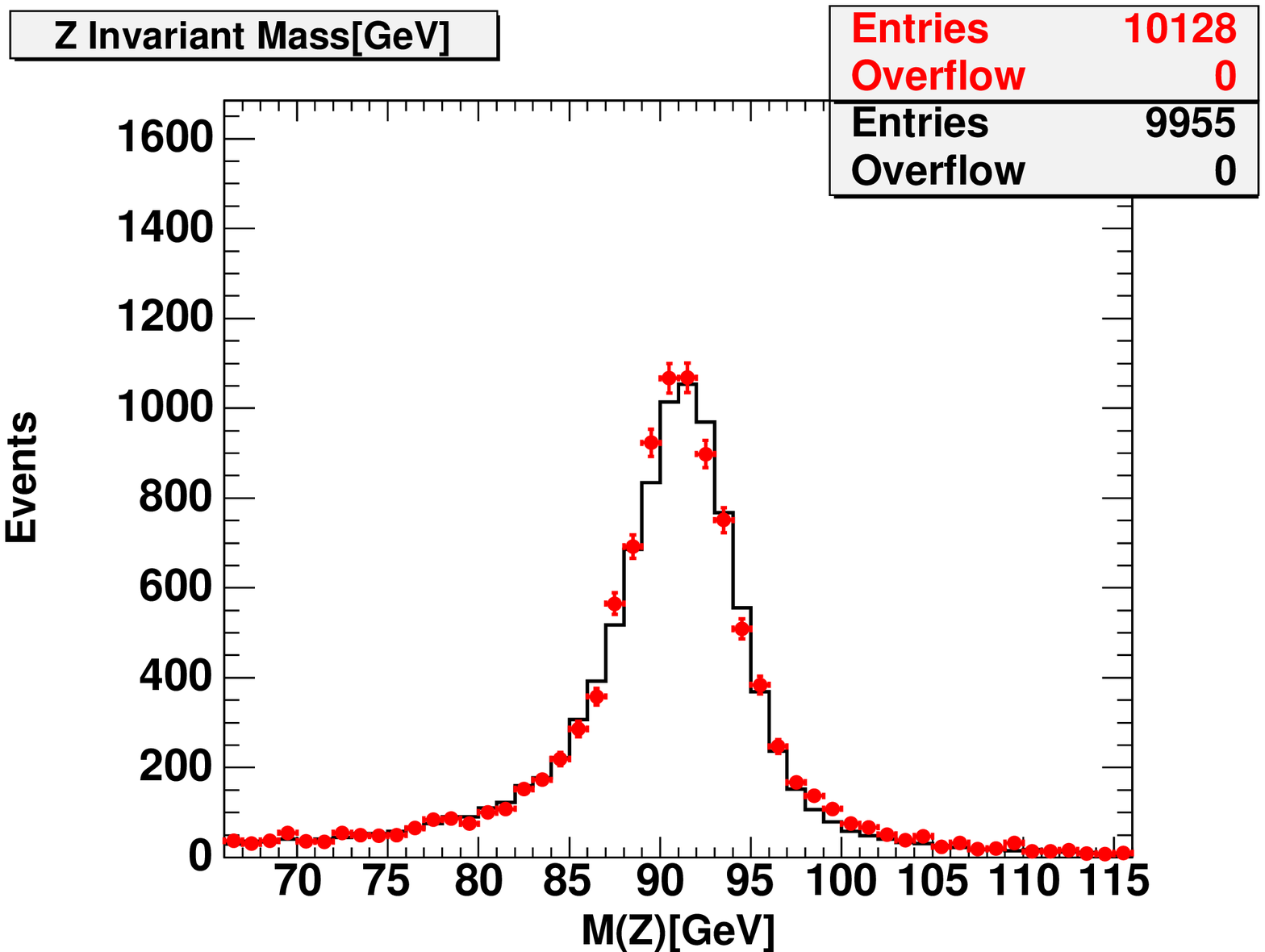,width=2.8in}} \quad
\subfigure[Invariant Mass (log)]{\epsfig{file=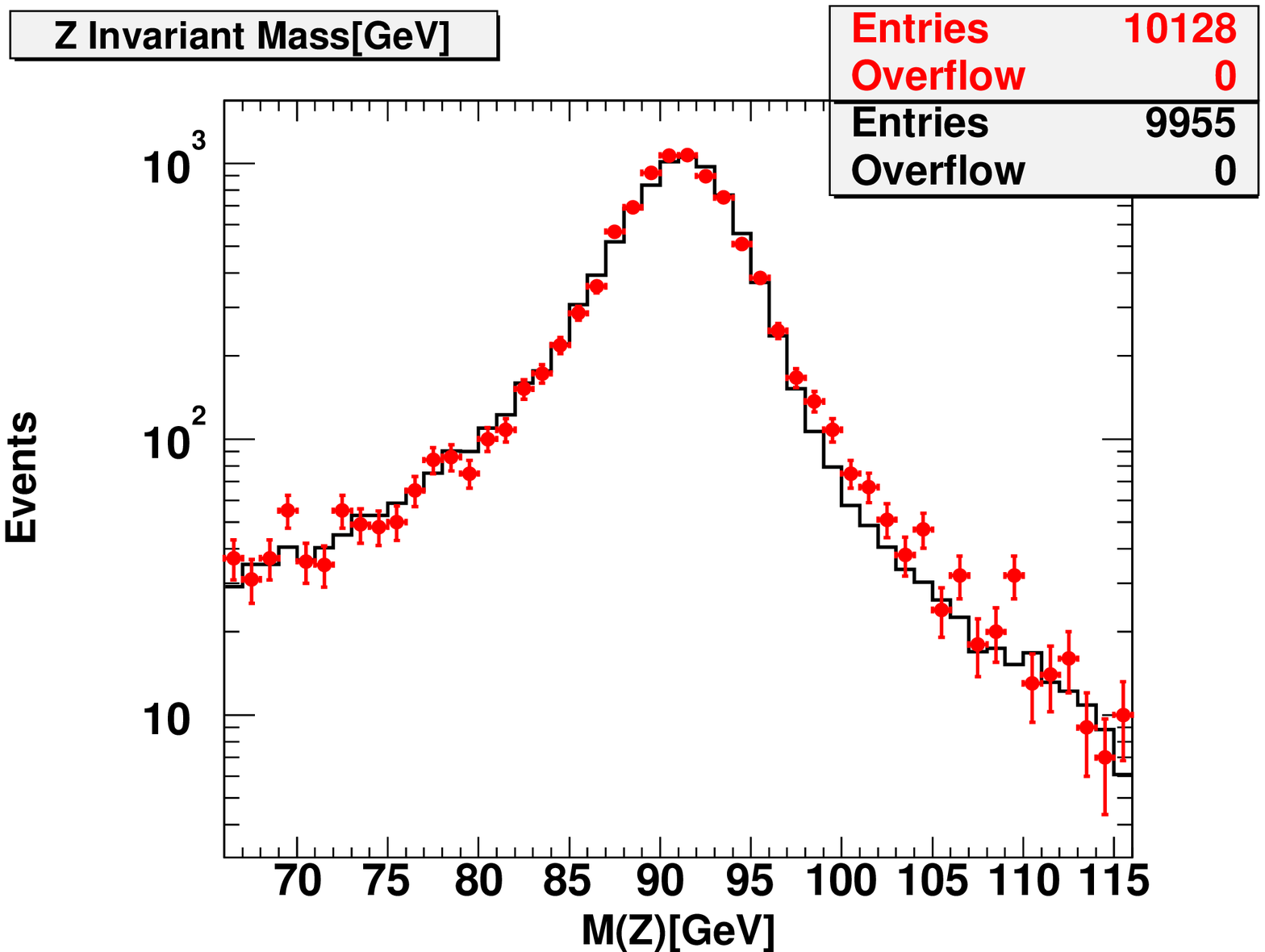,width=2.8in}} \quad
      }
    \mbox{
\subfigure[$P_T$ of $\Zee$]{\epsfig{file=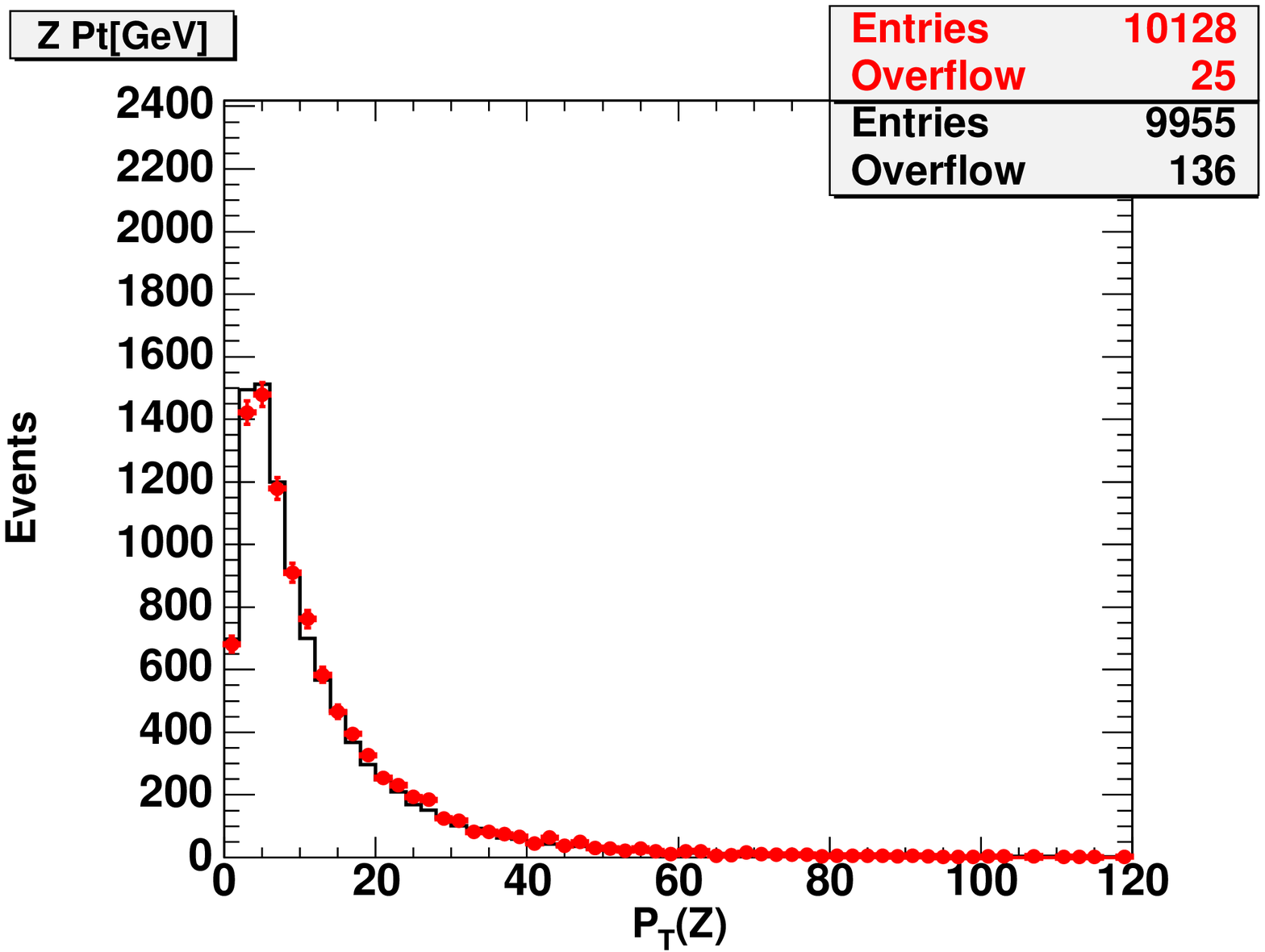,width=2.8in}} \quad
\subfigure[$P_T$ of $\Zee$ (log)]{\epsfig{file=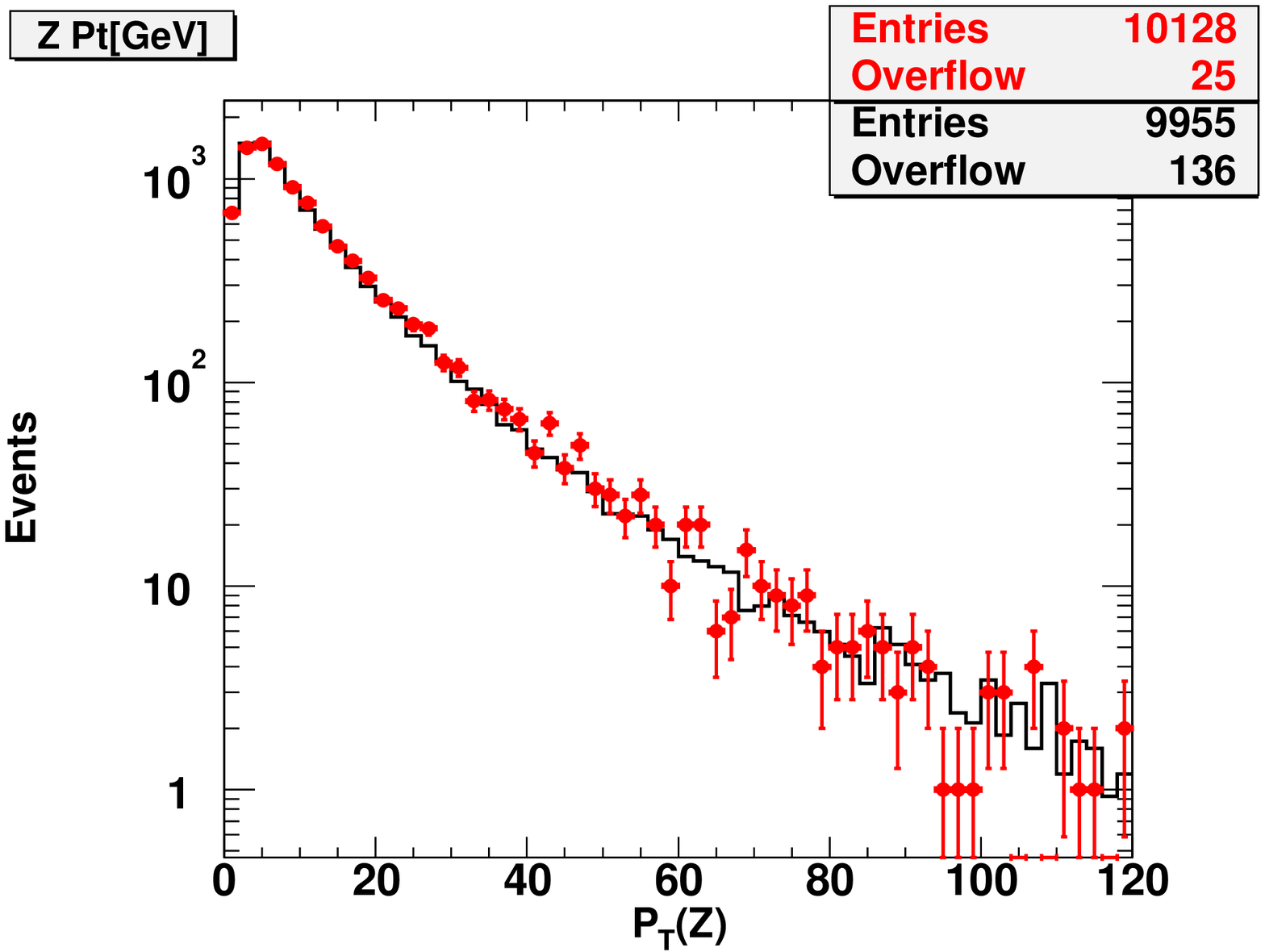,width=2.8in}} \quad
      }
\caption 
[The distributions for $\Zee$: data vs MC]
{ The distributions for Invariant Mass of $\Zee$,
    $P_T$ of $\Zee$, linear plots(a, c), log plots(b, d). The
    histogram is the prediction from the $\Zee$ MC sample (ztop2i),
    the points are $\Zee$ candidates in the data. Background estimates
    are not included.}
\label{zee_datamc.figure}
\end{center}
\end{figure}

\begin{table}[!t]
\begin{center}
\begin{tabular} {|p{4cm}|p{10cm}|}
\hline
Variable              &  Cut                                                 \\
\hline
\hline
Tight Electron        &  Cuts are listed in Table~\ref{electron.table}       \\
\hline
Loose Electron        &  Cuts are listed in Table~\ref{electron_cuts.table}  \\
\hline
Electron tracks must be      
of opposite charge    &  $Q^{e1} + Q^{e2}$ = 0                               \\
\hline
Delta Z cut           &  $|Z_{TRACK}^{e1}$ - $Z_{TRACK}^{e2}| < $ 4 cm       \\
\hline
Mass Window cut       &  $M > $ 66 GeV and $ < 116 $ GeV                     \\
\hline
Trigger               &  ELECTRON\_CENTRAL\_18                               \\
\hline
\end{tabular}
\end{center}
\caption
[The selection cuts for the central-central and central-plug $\zee$ control samples]
{The selection cuts for the CC and CP $\zee$ control samples. 
The superscripts $e1$ and $e2$ stand for electrons in the event.}
\label{zee.table}
\end{table}
The distributions in mass, $\Pt$ of the pair, and $\Delta\phi$ are
shown in Figure~\ref{zee_datamc.figure}.  To normalize the $\Zee$ MC
sample, we used the measured
$\sigma$($\Zmumu$)$\times$BR~\cite{CDF_WZ}.

\subsection{The $\zee$ Central-Plug Control Sample}

We also form a central-plug dielectron $\Z$ control sample to monitor
the identification performance for the electrons in end-plug
calorimeters. We require a tight central electron, and a plug electron
passing the cuts of Table~\ref{plugelectron.table}. The mass of the Z
candidate is required to be within the window 66 GeV to 116 GeV.  We
find $\nofzeece$ ($\nofzeecw$) central-plug $\Z$ events satisfying
these criteria with the plug electron in the East (West) calorimeter.
\begin{figure}[htbp]
  \begin{center}
    \mbox{
\subfigure[Invariant Mass]{\epsfig{file=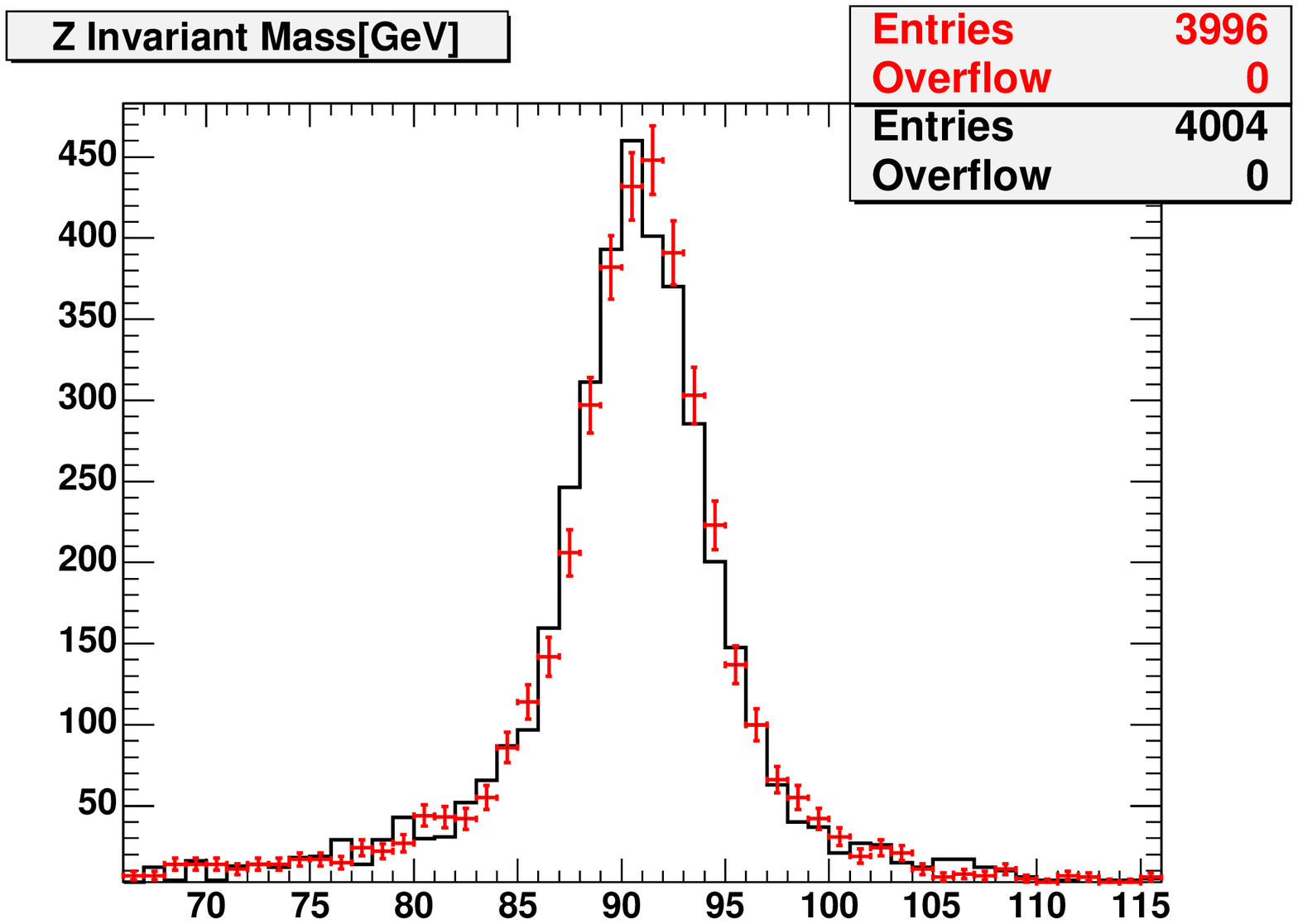,width=2.7in}} \quad
\subfigure[Invariant Mass (log)]{\epsfig{file=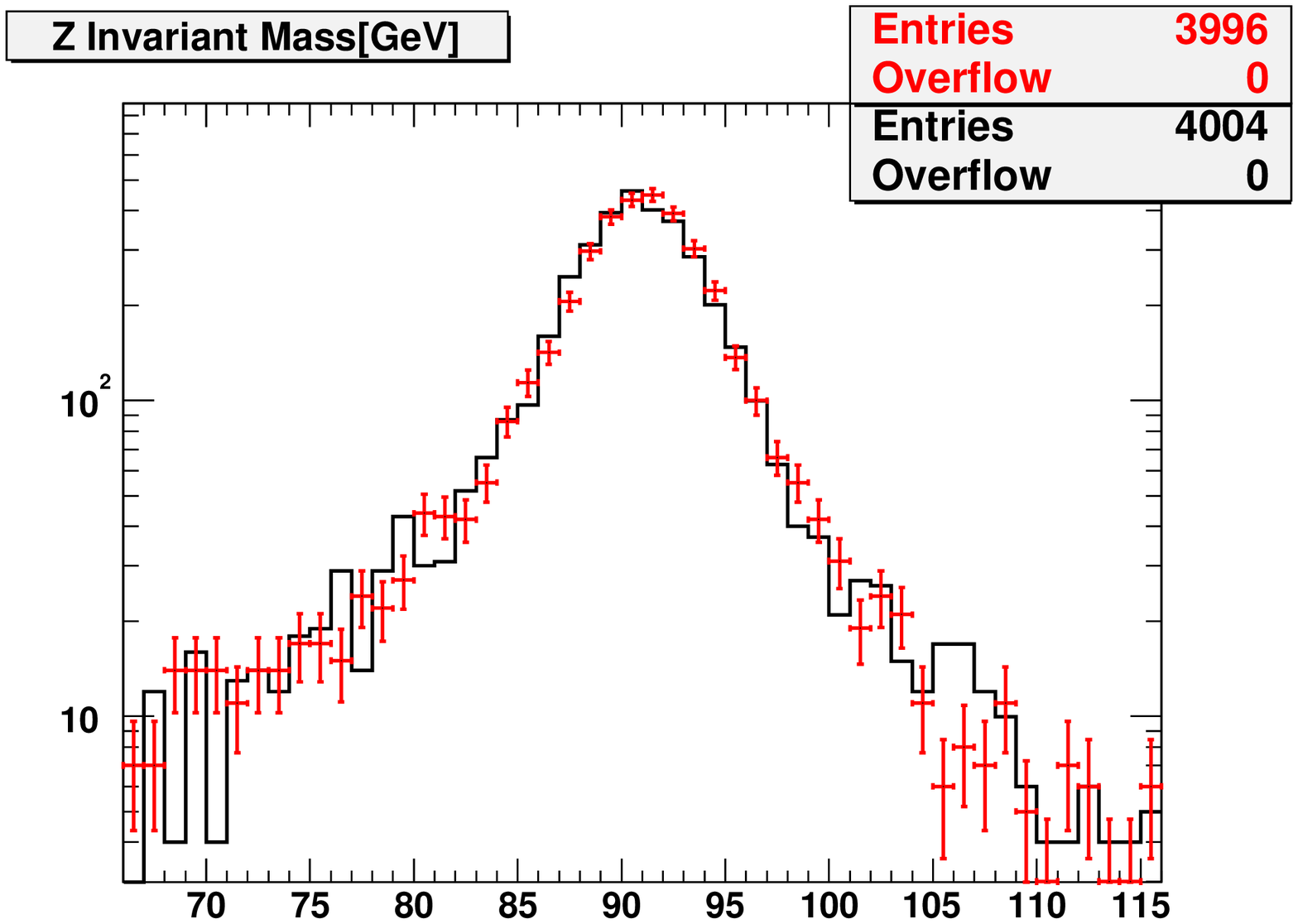,width=2.7in}} \quad
      }
    \mbox{
\subfigure[$P_T$ of $\Zee$]{\epsfig{file=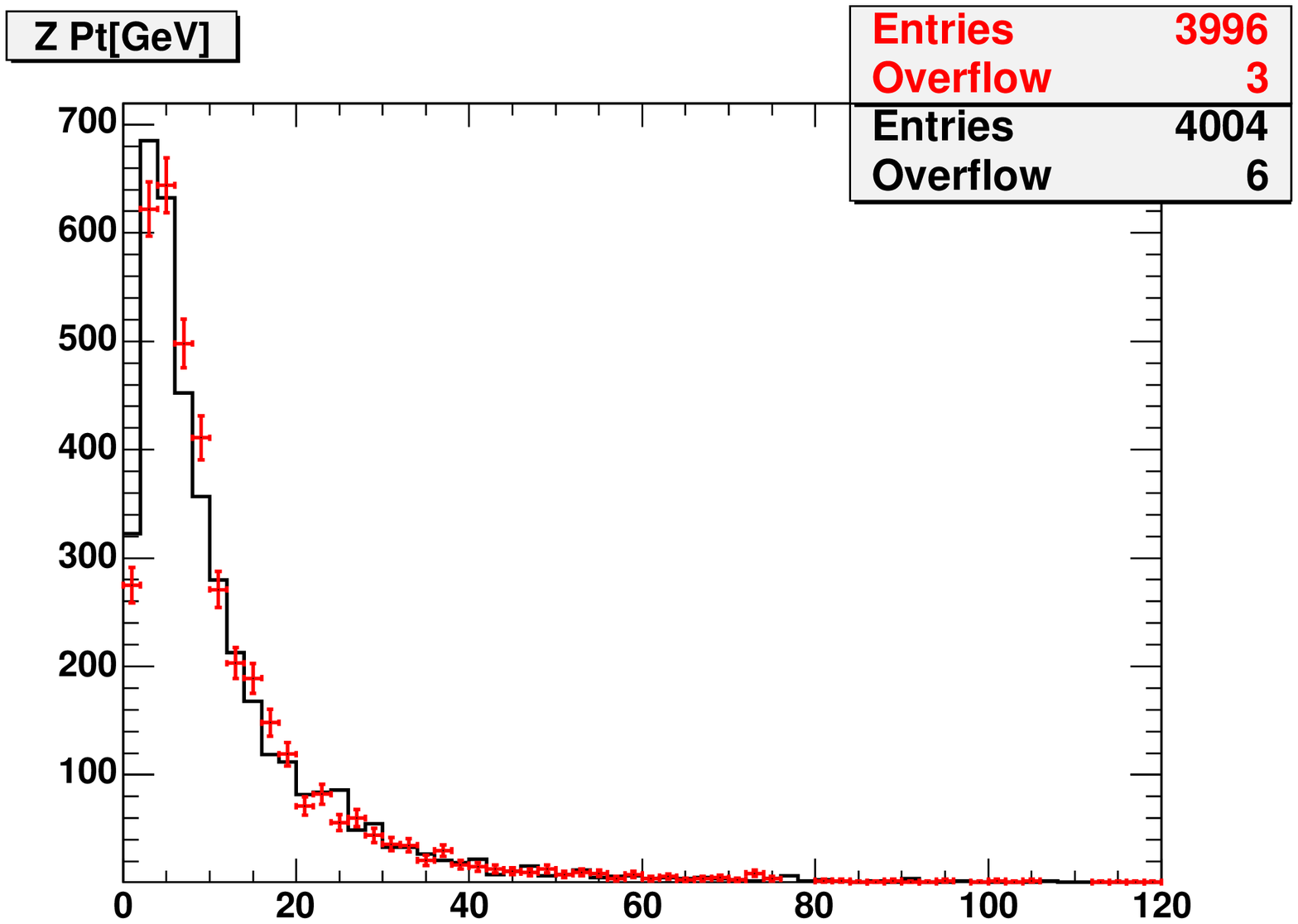,width=2.7in}} \quad
\subfigure[$P_T$ of $\Zee$ (log)]{\epsfig{file=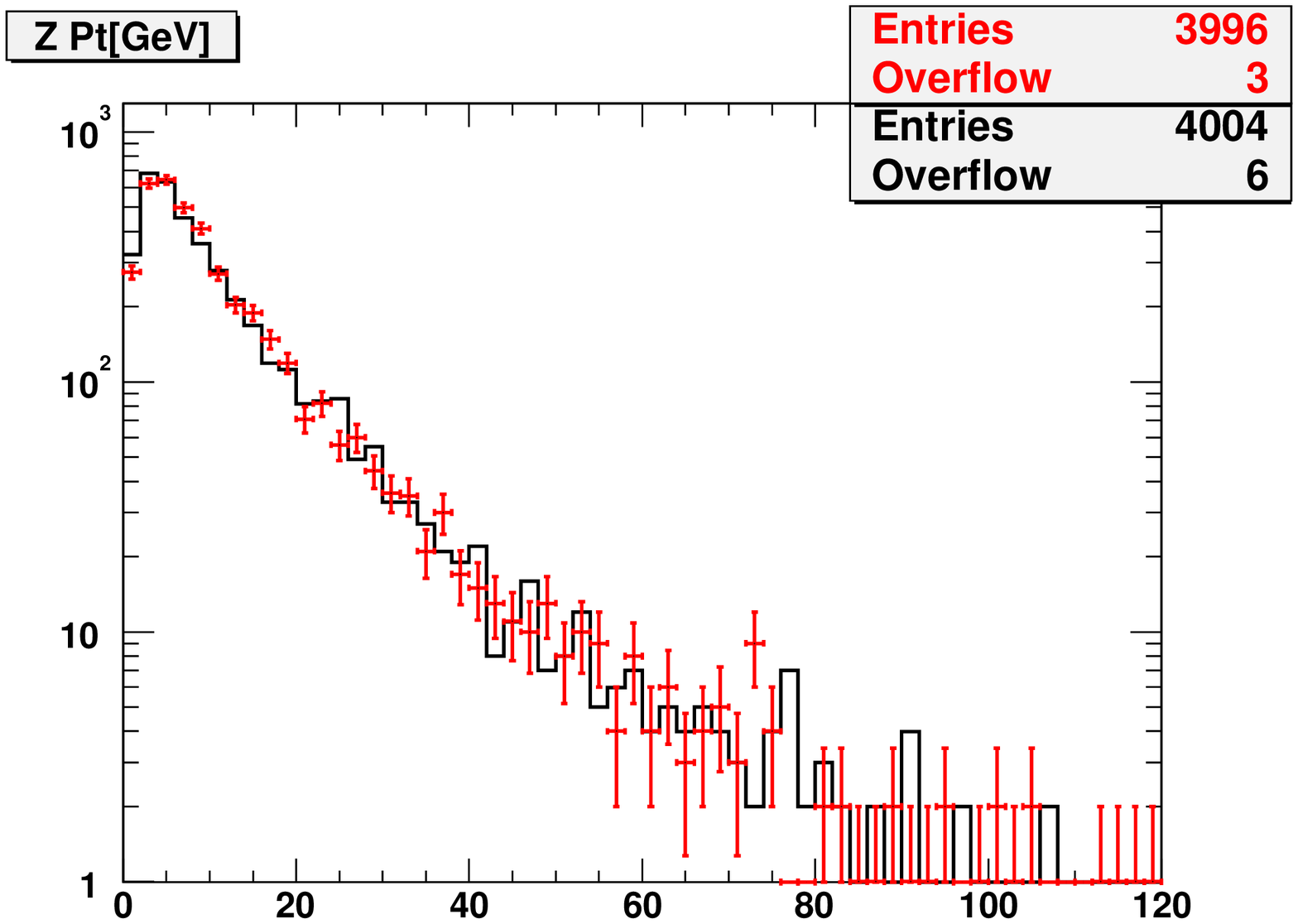,width=2.7in}} \quad
      }
    \caption 
[Central-plug dielectron $\Zee$ control sample]
{Central-plug dielectron $\Zee$ control sample to monitor the plug
electron identification performance. We require a tight central
electron(Table~\ref{electron.table}), and a plug electron passing the cuts of
Table~\ref{plugelectron.table}.  The distributions for
Invariant Mass of $\Zee$, $P_T$ of $\Zee$,
linear plots(a, c), log plots(b, d). 
We find $\nofzeece$ ($\nofzeecw$) central-plug $\Z$ events satisfying these 
criteria with the plug electron in the East (West) calorimeter.
The points are Central-East Plug $\Zee$ candidates; the histogram is 
Central-West Plug $\Zee$ candidates.
             }
    \label{zee_plug.figure}
  \end{center}
\end{figure}

The distributions in mass, $\Pt$ of the pair, and $\Delta\phi$ are
shown in Figure~\ref{zee_plug.figure}. The comparison of data vs MC is
shown in Table~\ref{zw_electrons.table}. The $\Zmumu$ event yields are
shown in Figure~\ref{zeewenustability.figure}.

\subsection{The $\wenu$ Control Sample}

The selection criteria for the $\wenu$ control sample are listed in
Table~\ref{wenu.table}. We require one tight central electron
(i.e. $E_T>\cutetightpt$ GeV), and corrected $\met$
(Section~\ref{metid}) greater than $\cutemet$ GeV.  We require the
transverse mass of the W candidate to be in the mass window 20-140
GeV. We find $\nofwenu$ $\wenu$ events, $\nofwenup$ positive and
$\nofwenum$ negative W's. The $\Wenu$ event yields are shown in
Figure~\ref{zeewenustability.figure}.

\begin{table}[!h]
\begin{center}
\begin{tabular} {|p{4cm}|p{10cm}|}
\hline
Variable              &  Cut                                                 \\
\hline
\hline
Tight Electron        &  Cuts are listed in Table~\ref{electron.table}       \\
\hline
Missing $E_T$ Cut     &  $E_T > $ 25 GeV                                     \\
\hline
Mass Window Cut       &  $M_T > 20$ GeV and $< 140$ GeV                      \\
\hline
Trigger               &  ELECTRON\_CENTRAL\_18                               \\
\hline
\end{tabular}
\end{center}
\caption
[$\wenu$ selection cuts]
{$\wenu$ selection cuts.}
\label{wenu.table}
\end{table}

\subsection{Summary of the Electron Control Sample Event Counts and
    Stability Plots}

\begin{figure}[!h]
  \begin{center}
    \mbox{
\subfigure[$\Zee$]{\epsfig{file=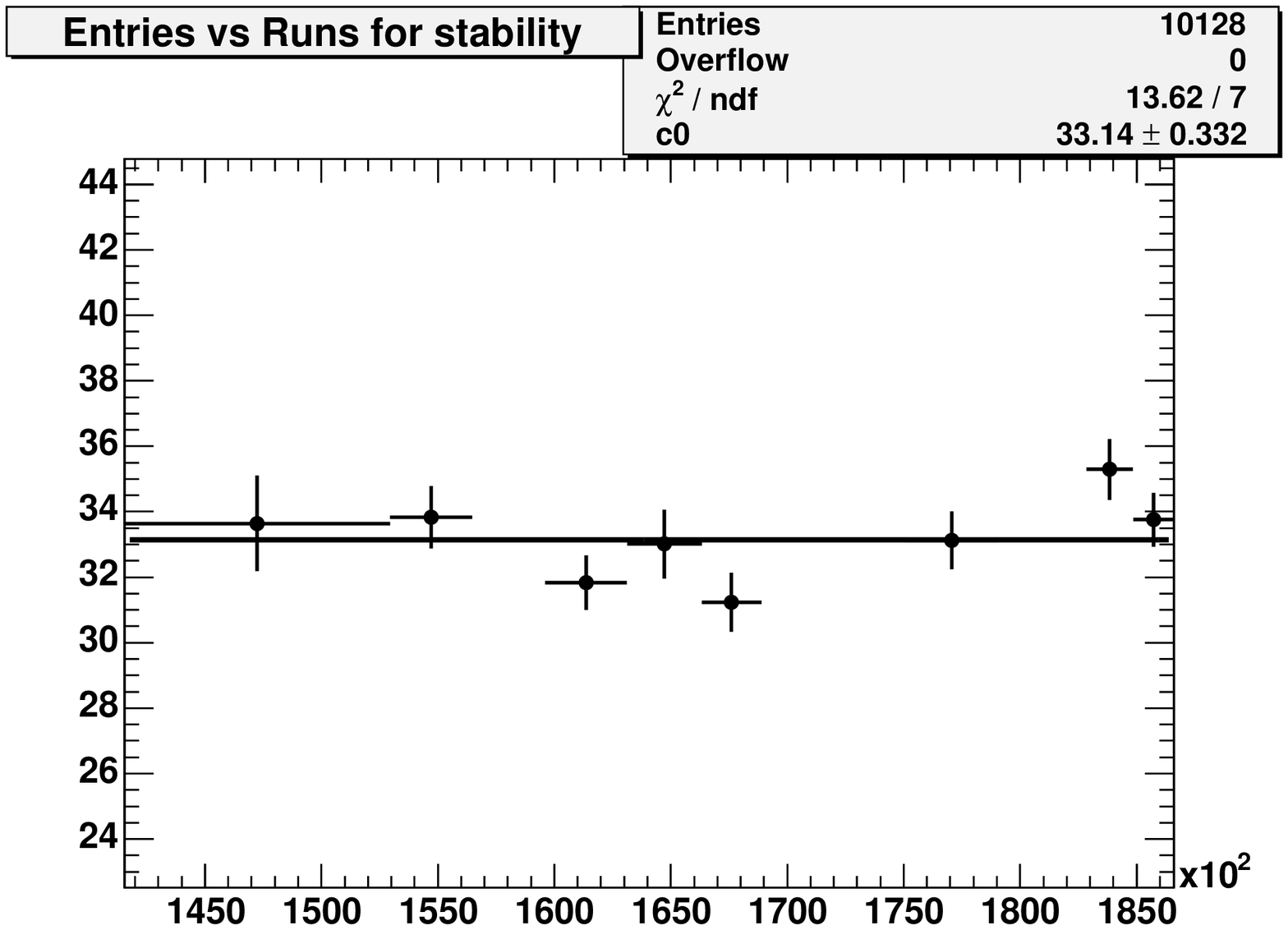,width=3in}} \quad
\subfigure[$\Wenu$]{\epsfig{file=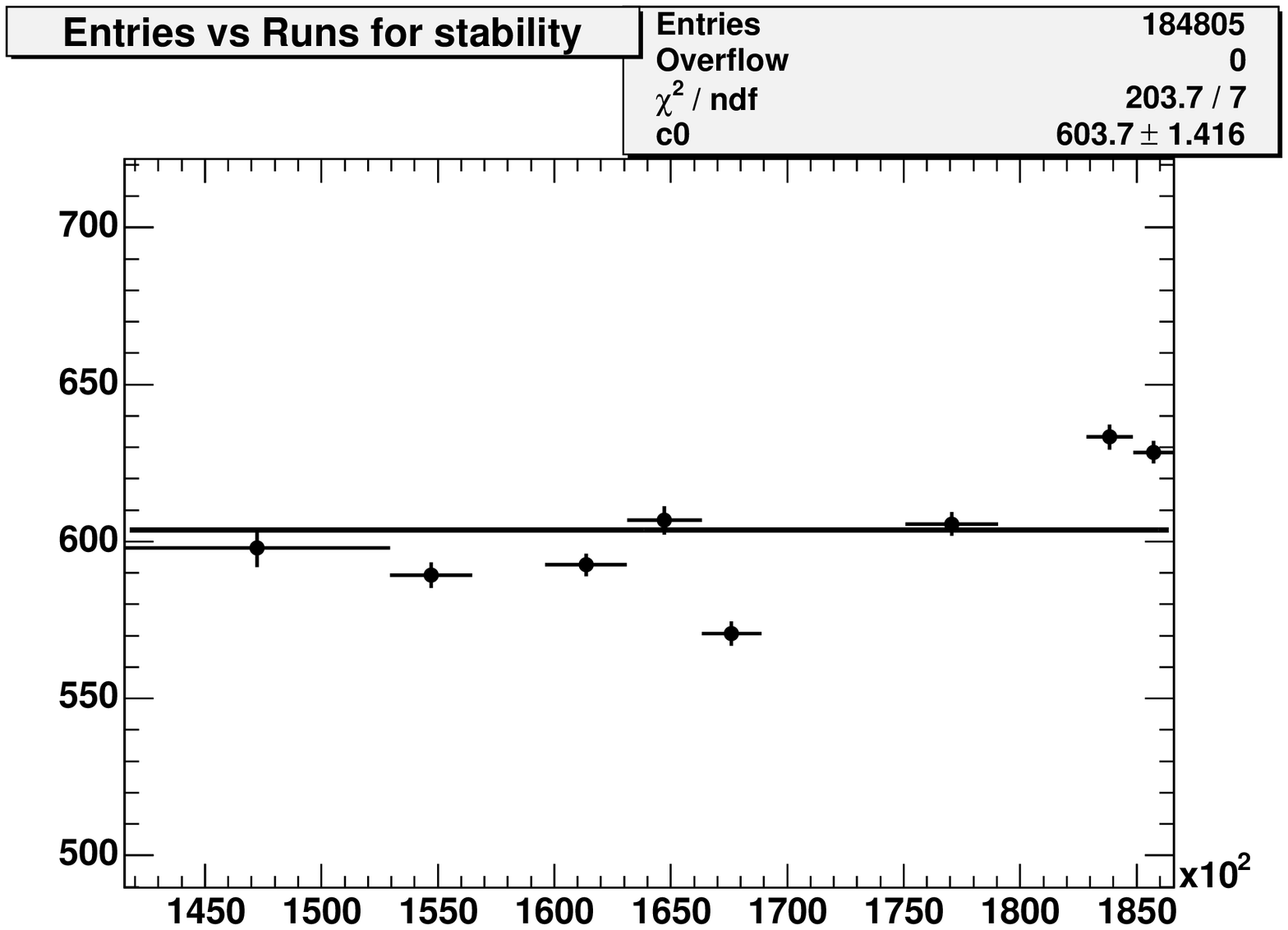,width=3in}} \quad
      }
    \mbox{
\subfigure[$\Zee$ Central-East]{\epsfig{file=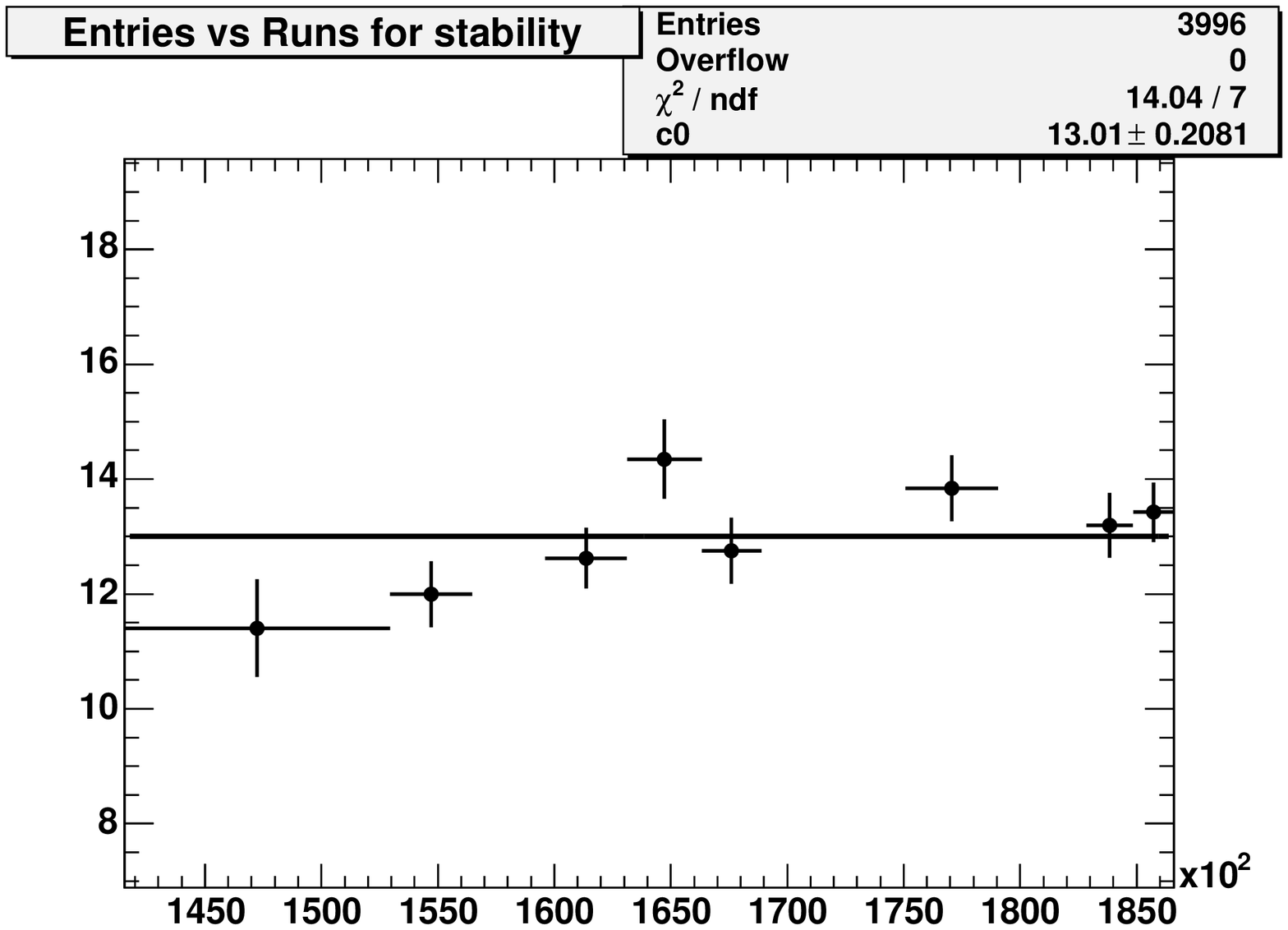,width=3in}} \quad
\subfigure[$\Zee$ Central-West]{\epsfig{file=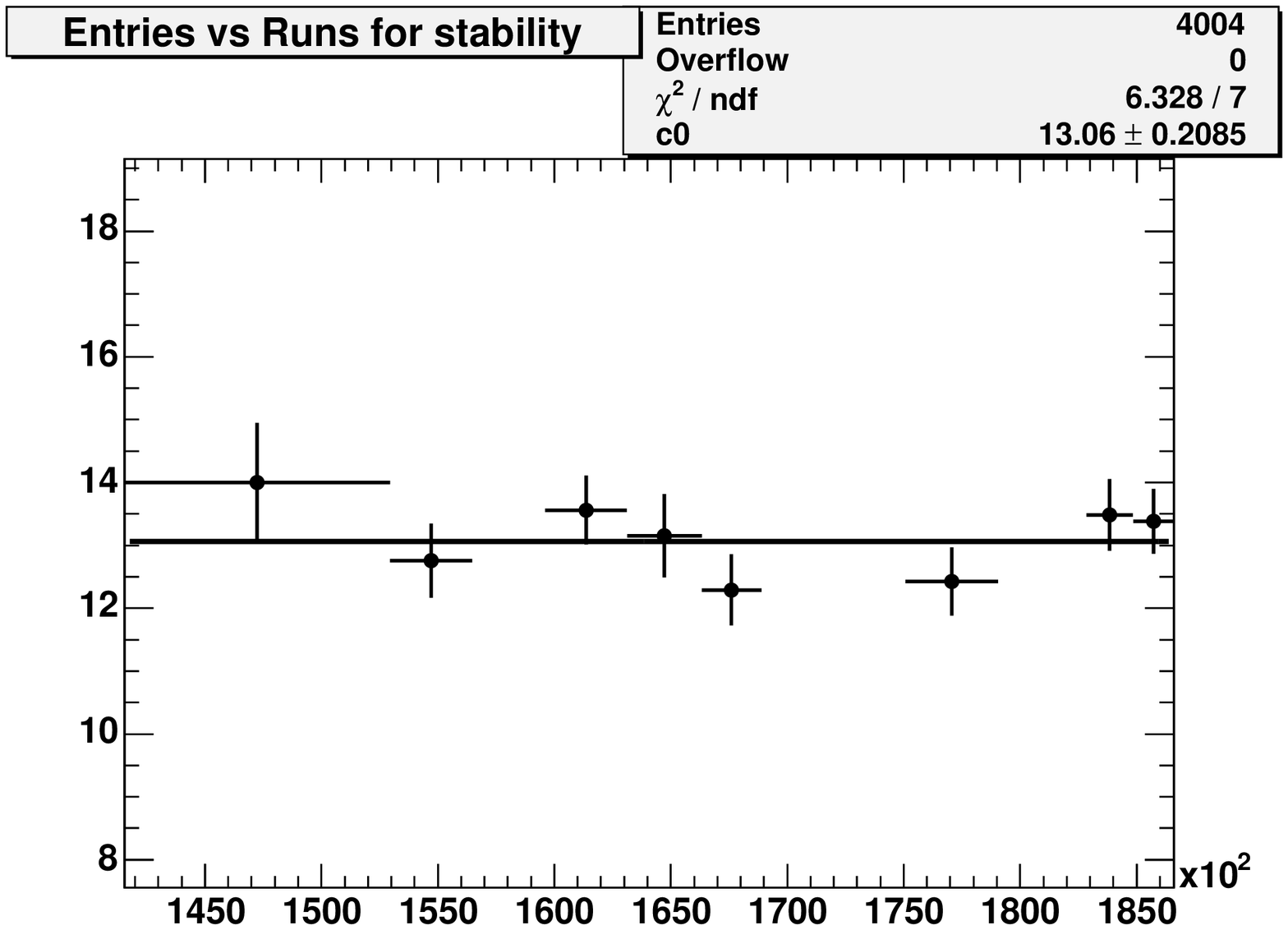,width=3in}} \quad
      }
\caption 
[Stability plots for $\Zee$ and $\Wenu$]
{Stability plots for $\Zee$(a), $\Wenu$(b), $\Zee$ Central-East(c),
$\Zee$ Central-West(d). The bins are those of
Table~\ref{lumi_stability.table}. Shown are the statistical errors,
the luminosity systematic error of 6\% (Chapter~\ref{systematics}) is
not included. The trends in $\Zee$($\Wenu$) are similar to
$\Zmumu$ ($\Wmunu$), see Figure~\ref{zmumuwmunustability.figure}. We
attribute this to some common effects (luminosity, trigger, COT).}
\label{zeewenustability.figure}
  \end{center}
\end{figure}

We use the control samples of $\Wenu$ and $\Zee$ decays to check
temporal stability of the event yields
(Figure~\ref{zeewenustability.figure}). Table~\ref{zw_electrons.table}
presents summary of the electron control sample event counts.

\begin{table}[!h]
 \begin{center}
\begin{tabular}{|p{5cm}|p{3cm}|p{3cm}|}
\hline
                           & DATA                & MC                  \\ 
\hline
\multicolumn{3}{|c|} {
$\Zee$	     }\\
\hline
$\Zee_{CC}           $  & 10128              & 9955             \\ 
$\Zee_{CC Same-Sign} $  & 199            & 127           \\ 
$\Zee_{CE}           $  & 3996              & 4257             \\ 
$\Zee_{CW}           $  & 4004              & 4131             \\ 
\hline
\end{tabular}
\caption 
[$\Zee$ summary: data vs $\Zee$ MC]
{$\Zee$ summary: data vs $\Zee$ MC. The material is underestimated in
MC, so we estimate $e\rightarrow\gamma$ fake rate from data.}
\label{zw_electrons.table}
 \end{center}
\end{table}

\begin{table}[!h]
 \begin{center}
  \begin{tabular} {|p{10cm}|r|}
\hline
Category & N \\ 
\hline
\hline
Events with at least one Tight Electron		&  \nofeevents \\
\hline
Tight Electrons					&  \nofe \\
\hline
$\Wenu$(triggered+$M_W$)			&  \nofwenu \\
\hline
\mbox{Tight + Loose Central Electron}		&  \nofzeerawcandidatescc \\
\hline
Tight + East Plug Electron			&  \nofzeerawcandidatesce \\
\hline
Tight + West Plug Electron			&  \nofzeerawcandidatescw \\
\hline
\mbox{CC $\Zee$(triggered+$M_Z$)}		&  \nofzee \\
\hline
\mbox{CE Plug $\Zee$(triggered+$M_Z$)}		&  \nofzeece \\
\hline
\mbox{CW Plug $\Zee$(triggered+$M_Z$)}		&  \nofzeecw \\
\hline
\end{tabular}


 \end{center}
\caption
[The numbers of events for the electron control samples]
{The numbers of events for the electron control samples. 
The electron selection cuts are given in Table~\ref{electron.table}, 
the $\zee$ selection cuts in Table~\ref{zee.table},
and the $\wenu$ selection cuts in Table~\ref{wenu.table}.
CC refers to Central-Central events; 
CE refers to Central-East Plug, CW refers to Central-West Plug.
}
\label{electron_summary.table}
\end{table}

\clearpage
\newpage
\chapter{Photon Identification}
\label{photonid}

The photon selection criteria are identical for photons in both the
muon and electron samples; the photon cuts, and the number of events
passing in each sample, are enumerated in
Table~\ref{photon_lepton.table} and are described below.

A sample of photons from $\Zee$ events used to measure the probability
of an electron radiating an energetic photon is also introduced.
  
\section{Photon Selection Criteria}

\begin{table}[!b]
 \begin{center}
\begin{tabular} {|p{6.5cm}|p{8cm}|}
\hline
Variable & Cut \\ 
\hline
$E_T^{corr}$                            & $\gt$ 25 GeV                                   \\
\hline
Had/Em                                  & $\lt$ 0.125 or  $\lt$ 0.055 + 0.00045$\times E^{corr}$  \\
\hline
$\chi^2$  (Strips+Wires)/2.0            & $\lt$ 20                                       \\
\hline
N Tracks                                & $\le$ 1                                       \\ 
\hline
Track $P_T$                             & $\lt$ 1+0.005$\times E_T^{corr}$ GeV            \\
\hline
Cone 0.4 Iso$E_T^{corr}$                & $\lt$ 2.0+0.02$\times(E_T^{corr}-20)$ GeV                 \\
\hline
Cone 0.4 TrackIso                       & $\lt$ 2.0+0.005$\times E_T^{corr}$ GeV               \\
\hline
2nd CES Cluster (Strip and Wire)        & $\lt$ 2.4+0.01$\times E_T^{corr}$ GeV                     \\
\hline
Fiducial                                & Ces$|X|\lt$ 21 cm, 9 cm $\lt$ Ces $|Z|\lt$ 230 cm    \\
\hline
\end{tabular}
 \end{center}
\caption
[Photon identification and isolation cuts]
{Photon identification and isolation cuts.}
\label{photon_cuts.table}
\end{table}

A photon candidate is required to have corrected transverse energy
greater than 25 GeV.  For photons or electrons the CES shower position
is determined by the energy-weighted centroid of the highest energy
clusters of those strips and wires in the CES which correspond to the
seed tower.  The direction of the photon is determined by the line
connecting the primary event vertex to the shower position in the CES.
To ensure that events are well-measured the shower position of the
photon is required to fall within the fiducial region of the CES so
that the shower is fully contained in the active region.

Photon candidates are required to have characteristics consistent with
those of a neutral electromagnetically-interacting particle. No COT
track with $\Pt\gt$ 1 GeV may point at the photon cluster. One track
with $\Pt\lt$ 1 GeV may point at the cluster.

The variable `Iso$E_T^{corr}$' is the Run I cone
$R_{\eta-\varphi}=0.4$ isolation energy with the Run I correction to
isolation energy due to phi-crack leakage~\cite{CDF4170}. The tracking
isolation variable `Track Isolation' is the sum of the $\Pt$ of tracks
in a in a cone in $\eta-\varphi$ space of radius $R=0.4$ surrounding
the photon, measured in GeV.

Table~\ref{photon_lepton.table} summarizes the selection of photons in
the muon and electron subsamples (Section~\ref{datasets}).

\begin{table}[!t]
 \begin{center}
  \begin{tabular} {|p{4.2cm}|p{4.2cm}|l|l||l|l|}
\hline
\multicolumn{2}{|c|}{Subsample} & \multicolumn{2}{|c||}{Muon} & \multicolumn{2}{|c|}{Electron}\\
\hline
\hline
Variable & Cut      & Cumulative & This & Cumulative & This \\ 
\hline
$\gamma$ Candidates       &              & 
75026 & 75026 & 524626 & 524626 \\ \hline
$E_T^{corr}$, $\GeV$                     & $\gt$ 25 & 
4567  & 4567  & 333650 & 333650 \\ \hline
HAD/EM                                     & $\lt$0.125$||\newline $0.055+0.00045$\times E$ & 
4169  & 68819 & 330675 & 499963 \\ \hline
$\chi^2$ × CES                             & $\lt$ 20 & 
2978  & 39468 & 279701 & 389644 \\ \hline
N Tracks                                   & $\le$ 1 & 
2500  & 58325 & 210366 & 371793 \\ \hline
Track $P_T$, $\GeV$                        & $\lt$ 1  +0.005$\times E_T$      & 
1787  & 43186 & 4073   & 96410  \\ \hline
Cone 0.4 Iso$E_T^{corr}$, $\GeV$           & $\lt$ 2.0+0.02$\times(E_T-20)$  & 
1647  & 19068 & 3818   & 302862 \\ \hline
Track Isolation, $\GeV$                  & $\lt$ 2.0+0.005$\times E_T$     & 
1610  & 36352 & 3771   & 64776  \\ \hline
E (2nd CES Cluster) \newline (Strip and Wire), $\GeV$    & $\lt$ 2.4+0.01$\times E_T$      & 
1604  & 67048 & 3762   & 471879 \\ \hline
Fiducial: \newline Ces$|X|$, Ces$|Z|$, cm                    & Ces$|X|\lt$ 21, \newline 9 $\lt$ Ces$|Z|\lt$ 230 & 
1598  & 40879 & 3735   & 409371 \\ \hline
\end{tabular}
 \end{center}
\caption
[Selection of photons in the muon and electron subsamples]
{Photon identification and isolation cuts. Each entry corresponds to
one CDF EM Object, Central or Plug. The column labeled `Cumulative'
gives the effect of each successive cut on the number of photon
candidates. The heading `This' represents the effect of applying only
the one cut listed.  After a final cut requiring a Tight muon to be in
the event with the photon we find a total of $\nofmug$ $\mug$
candidate events. After a final cut requiring a Tight electron to be
in the event with the photon we find a total of $\nofeg$ $\eg$
candidate events. }
\label{photon_lepton.table}	
\end{table}

For the muon subsample we require the event to be triggered by either
a high-$\Et$ muon trigger (which is MUON\_CMUP\_18 or MUON\_CMX\_18)
or by a high-$\Et$ photon trigger (PHOTON\_25\_ISO). After requiring a
Tight muon (Table~\ref{muon_cuts.table}) to be in the event with the
photon (Table~\ref{photon_cuts.table}) we find a total of $\nofmug$
$\mug$ candidate events.

For the electron subsample we require event to be triggered by either
high-$\Et$ electron trigger (ELECTRON\_CENTRAL\_18) or by high-$\Et$
photon trigger (PHOTON\_25\_ISO). After requiring a tight
electron(Table~\ref{electron_cuts.table}) to be in the event with the
photon(Table~\ref{photon_cuts.table}) we find a total of $\nofeg$
$\eg$ candidate events. The disparity between this number and the
$\nofmug$ $\mug$ events is due to a number of causes, in particular
hard photon bremsstrahlung of an electron, as discussed later in
Section~\ref{photon_control_sample}.

\section{Introducing the Photon Control Sample} 
\label{photon_control_sample}

Unlike for the electron or muon, finding a pure sample of high-Pt
photons is difficult; the `Compton' sample of $\gamma$-jet events has
QCD fake backgrounds, for example. We describe here a control sample
of high-Pt photons derived from the $\Z$.

\begin{figure}[!t]
  \begin{center}
    \mbox{
\subfigure[Invariant Mass, GeV]{\epsfig{file=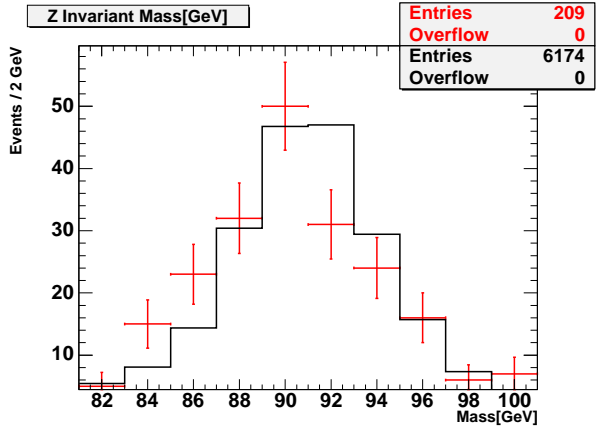,width=3in}} \quad
\subfigure[$p_T$ of $\Z$, GeV]{\epsfig{file=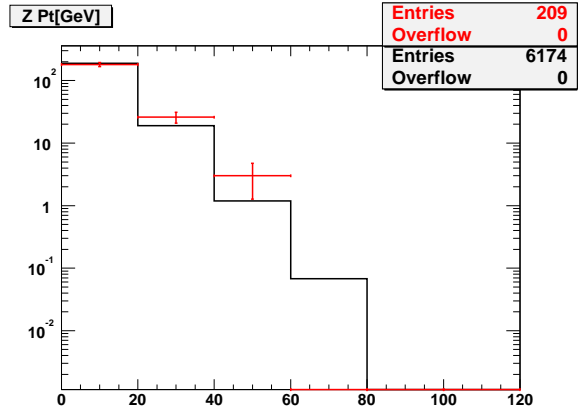
,width=3in}} \quad
      }
    \mbox{
\subfigure[$\Delta \phi$]{\epsfig{file=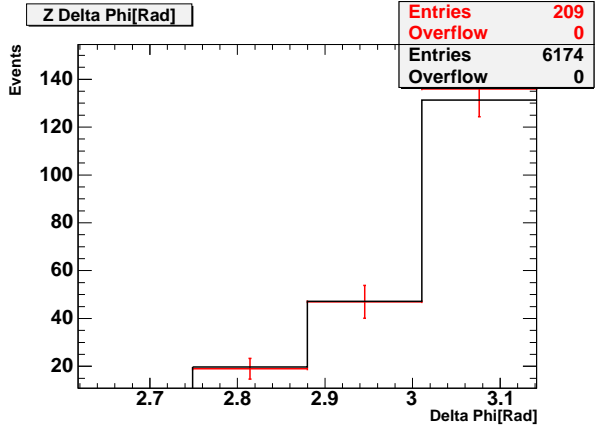,width=3in}} 
\subfigure[$\Delta R$]{\epsfig{file=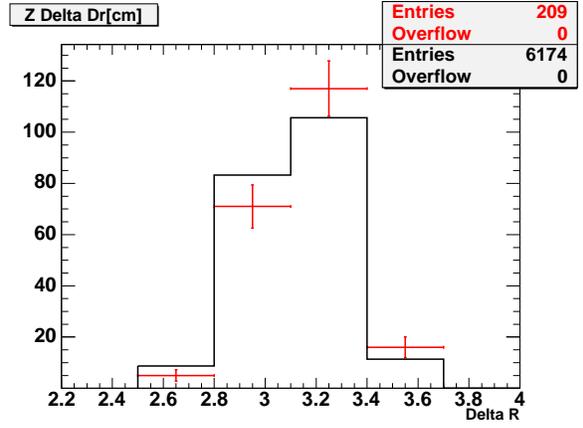,width=3in}} 
      }
    \caption 
[The distributions for $\eg$ control sample]
{The distributions for: (a) Invariant Mass of $\Zeg$, (b) $p_T$ of
$\Zeg$, (c) $\Delta \phi$ of $\Zeg$, (d) $\Delta R$ of $\Zeg$.  The
points are the $Z^0$-like $\eg$ sample; the shaded histogram
represents electron-electron events from data with the same
kinematic cuts, normalized to the number of events in the control
sample.  Invariant mass for $\Zeg$ is slightly shifted to the left,
because energy of the radiated photon is less than energy of the
original electron. }

    \label{zeezeg.figure}
  \end{center}
\end{figure}

In looking for photons in the electron sample, one has to take into
account that the dominant source of fake background for $\eg$ events
is $Z^{0}\rightarrow e^+e^-$ production, wherein one of the electrons
undergoes hard photon bremsstrahlung in the detector material, or the
COT fails to detect one of the electron tracks, and that electron
subsequently passes all of the photon cuts.  There are approximately
$\nofzeetightleptons$ tight central electron pairs in the CDF data, so
a photon fake rate as low as $1\%$ will give rise to
$\nofzeebackground$ $\eg$ background events, which would be
unacceptably high for finding sources of new physics comparable to
$W/Z^{0}+\gamma$ production. The Run II detector has significantly
more material inside the outer tracking chamber than the Run I
detector had, and so the number of `$e\gamma$` events from $\Zee$
production will be significantly higher. We measure this fake rate
directly from the data (the material in simulated data is
underestimated, see Table~\ref{zw_electrons.table}).

A control sample of $Z^{0}$-like events, $\Zeg$, is selected from the
$\nofegdphi$ two-body $\eg$ candidates (see
Chapter~\ref{photonlepton}) by requiring that the invariant mass of
the $\eg$ pair, $M_{e\gamma}$, be within 10 GeV of the $Z^{0}$ mass
(91 GeV).  There are $\nofegzlike$ such events in the CDF data. It is
observed that the shapes of the distributions of the two samples
($\zee$ and $\Zeg$) are similar to each
other(Figure~\ref{zeezeg.figure}).

\clearpage
\newpage
\chapter{Calculating the Missing Transverse Energy and $H_T$}
\label{miscid}

This chapter describes how missing transverse energy ($\met$) is
calculated, and gives the definition and describes calculation of the
total transverse energy.

\section{Calculating the $\met$}
\label{metid}

Missing transverse energy ($\met$) is associated with particles that
escape detection. For example, $\met$ is the signature of weakly
interacting neutral particles such as neutrinos, or possible new
particles such as the gravitino or LSP. It also can come from
mismeasurement of the true $E_T$ of objects, or from backgrounds such
as cosmic rays or beam halo.

Missing $\rm E_T$ ($\metvec$) is defined by $\metvec = -\sum_{i} E_T^i
\hat{n}_i$, where i is the calorimeter tower number for $|\eta| <$ 3.6, 
and $\hat{n}_i$ is a unit vector perpendicular to the beam axis and
pointing at the i$^{th}$ tower. We define the magnitude
$\met=|\metvec|$. 

Corrections are made to the $\met$ for non-uniform calorimeter
response~\cite{jet_corr} for jets with uncorrected $\Et > 15$ $\GeV$
and $\eta < 2.0$, and for muons with $\Pt > 20$ $\GeV$:

\begin{itemize}
\setlength{\itemsep}{-0.05in} 

  \item Muons: \\
-- correct for $E_T-P_T$, where $E_T$ is a transverse energy
  deposited in electromagnetic and hadron calorimeters, and $P_T$ is a
  transverse momentum of a muon track. We correct $\met$ for all
  muons with $\Et\gt\cuteloosept$ GeV.

  \item Jets: \\
-- correct for $E_T-E_T^{corr}$, where $E_T$ is a
  transverse energy of an uncorrected jet, and $E_T^{corr}$ is a
  transverse energy of a jet, corrected for non-uniform calorimeter
  response. We correct for jets with $\Et^{corr}\gt 15$ GeV.

\end{itemize}
 
 When identifying jets we check that jet object does not have any of
 the objects identified in the current analysis close to it (within
 $\Delta R \lt$ 0.5).

 For the $\lgmet$ analysis we set the cut on $\met$ to be
 $\met>\cutemet$ GeV.

\section{Calculating the Total Transverse Energy}
\label{htid}

Total transverse energy $\Ht$ is defined for each event as the sum of
the transverse energies of the leptons, photons, jets, and $\met$ that
pass the analysis selection criteria. To calculate $H_T$ we use Tight
and Loose Central Electrons (Table~\ref{electron_cuts.table}), Tight
Phoenix Electrons (Table~\ref{plugelectron_cuts.table}), Tight and
Loose CMUP and CMX muons, Stubless muons (Table~\ref{muon_cuts.table}),
$\met$, and jets in the event with $|\eta|\lt 2$ and
$E_T^{corr}\gt\cutejetptformetcorr$.

\chapter{Standard Model Predictions}
\label{predictions}

\begin{figure}[!h]
\centering
\includegraphics[angle=0,width=7in,clip]{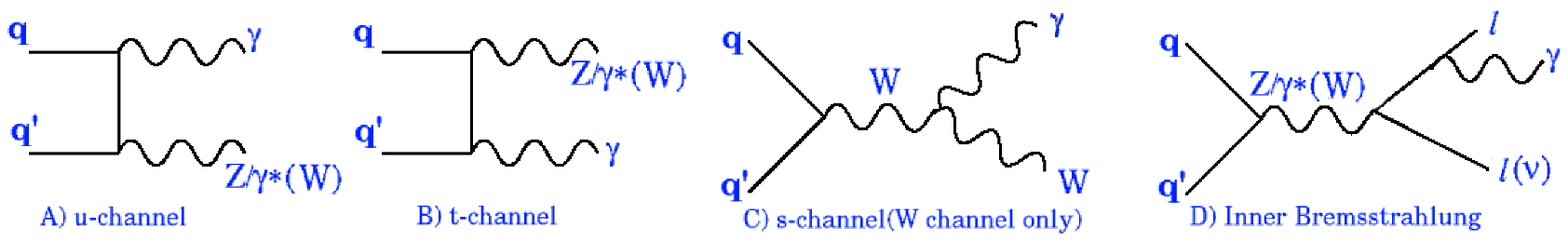}
\caption
[Tree-level diagrams for $\Zg$ and $\Wg$ production]
{Tree-level diagrams for $\Zg$ and $\Wg$ production.}
\label{zgwg_diagrams.figure}
\end{figure}

The dominant source of $\lg$ events at the Tevatron is electroweak
diboson production (Figure~\ref{zgwg_diagrams.figure}), in which an
electroweak boson ($W$ or $\Z$) decays leptonically ($\ell \nu$ or
$\ell\ell$) and a photon is radiated from either an initial state
quark, a charged electroweak boson ($W$), or a final state lepton.
The number of $\lg$ events from electroweak diboson production is
estimated using several leading-order (LO) Monte Carlo (MC) event
generator programs. These programs are MadGraph~\cite{madgraph},
CompHep~\cite{comphep}, and Baur~\cite{diboson_mc,baur}.

These programs output 4-vectors and helicities of particles emanating
from a diboson production event in an ASCII format. In addition the
information on how the particles are produced (``mother'' and
``daughter'') is recorded, including the energy scale and other parameters
used for the matrix element calculation.

These files are then fed into the LesHouchesModule~\cite{CDF6333},
which runs Pythia~\cite{Pythia} to add parton fragmentation and
final-state radiation and initial-state radiation (both QED and QCD) ,
and then writes out the events in CDF HEPG format. These files are
then used as input to the CDF detector simulation program.  This
program outputs simulated data in a format identical to that of an
actual CDF Run II event.  Simulated $\lg$ event rates are then
estimated in a manner identical to that of CDF data.

\section{$\Wg$, $\Zg$, $\Wgg$ and $\Zgg$ MC Sets}

The details on the generator level settings of the Baur~\cite{baur} MC
$\Wg$, $\Zg$ datasets can be found in Ref.~\cite{CDF6619}.

The MadGraph $\Wg$, $\Zg$, $\Wgg$, $\Zgg$ datasets created for this
analysis are listed in Table~\ref{madgraph_samples.table}.  Details on
the MadGraph and CompHep MC samples can be found
in Ref.~\cite{CDF6057,CDF7523}.

The kinematic cuts used for the generation of $\Wg$, $\Zg$, $\Wgg$ and
$\Zgg$ MC Sets are listed in
Table~\ref{madgraph_comphep_baur_cuts.table}.

\begin{table}[!h]
 \begin{center}
  \begin{tabular} {|p{4.5cm}|p{2.5cm}|p{2.5cm}|p{2.5cm}|}
\hline
Object & \multicolumn{3}{|c|}{Cuts} \\
\hline
\multicolumn{4}{|c|}{MadGraph and CompHep Samples} \\
\hline
 & $\Et$, $\GeV$ & $\eta$ & $\Delta R$ \\
\hline
`First' Lepton      & 6.0 & 4.0  & 0.2 \\
Additional Leptons  & 6.0 & 4.0  & 0.2 \\
Neutrinos           & 1.0 & 10.0 & 0   \\
`First' Photon      & 6.0 & 4.0  & 0.2 \\
Additional Photons  & 6.0 & 4.0  & 0.2 \\
\hline
\hline
\multicolumn{4}{|c|}{Baur Samples} \\
\hline
 & $\Et$, $\GeV$ & $\eta$ & $\Delta R$ \\
\hline
`First' Lepton      & 0.0 & 10.0 & 0.2 \\
Additional Leptons  & 0.0 & 10.0 & 0.2 \\
Neutrinos           & 0.0 & 10.0 & 0   \\
`First' Photon      & 5.0 & 10.0 & 0.2 \\
\hline
minimum m($\ll$),  $\GeV$ & \multicolumn{3}{|c|}{$>$20} \\
minimum m($\llg$), $\GeV$ & \multicolumn{3}{|c|}{$>$20} \\
\hline
\end{tabular}
\caption
[Generator level cuts used to produce the $\Zg$, $\Wg$, $\Zgg$
and $\Wgg$ CompHep, MadGraph and Baur samples]
{The cuts used at generator level to produce the $\Zg$, $\Wg$, $\Zgg$
and $\Wgg$ samples for CompHep, MadGraph and Baur datasets.}
\label{madgraph_comphep_baur_cuts.table}
 \end{center}
\end{table}

\begin{table}[!h]
 \begin{center}
  \begin{tabular}{|l|r|c|}
\hline \hline
~~~~~~~~~DataSet Name  & Events & Cross section (pb) \\
\hline
\multicolumn{3}{|c|}{ $\Zg$, $\Wg$}\\
\hline
 $\Zeeg$       &       395482  &  6.855990  \\
 $\Zmumug$     &       395482  &  6.855990  \\
 $\Ztautaug$   &       199047  &  6.423524  \\
\hline
 $\Wenug$      &       199831  & 27.2  \\
 $\Wmunug$   . &       199850  & 27.2  \\
 $\Wtaunug$    &       199891  & 24.0  \\ 
\hline
\multicolumn{3}{|c|}{ $\Zgg$, $\Wgg$}\\
\hline
\hline
 $\Zeegg$      &       198830  &  0.089137  \\
 $\Zmumugg$    &       198830  &  0.089137  \\
 $\Ztautaugg$  &       198700  &  0.078612  \\
\hline
 $\Wenugg$     &       199351  & 0.126   \\
 $\Wmunugg$    &       199351  & 0.126   \\
 $\Wtaunugg$   &       198910  & 0.0939  \\
\hline
\end{tabular}
  \caption
[The $\Wg$, $\Zg$, $\Wgg$ and $\Zgg$ MadGraph datasets]
{The $\Wg$, $\Zg$, $\Wgg$ and $\Zgg$ MadGraph datasets.} 
  \label{madgraph_samples.table}
 \end{center}
\end{table}

\begin{table}[!h]
 \begin{center}
  \begin{tabular}{|l|r|c|}
\hline \hline
~~~~~~~~~DataSet Name  & Events & Cross section (pb) \\
\hline
\multicolumn{3}{|c|}{ $\Zg$, $\Wg$}\\
\hline
 $\Zeeg$       &        429979 &  8.62  \\
 $\Zmumug$     &        438468 &  8.61  \\
\hline
 $\Wenug$      &       140130  & 31.9   \\
 $\Wmunug$   . &       164732  & 31.9   \\
\hline
\end{tabular}

  \caption
[The $\Wg$ and $\Zg$ Baur datasets]
{The $\Wg$ and $\Zg$ Baur datasets.} 
  \label{madgraph_samples.table}
 \end{center}
\end{table}

To account for NLO corrections to the $\Wg$, $\Zg$, $\Wgg$, $\Zgg$
processes we use the $E_T$-dependent
K-factor=${{\sigma_{NLO}}\over{\sigma_{LO}}}$ obtained using Baur's
NLO calculation programs~\cite{Baur_NLO}.

We apply these corrections
to our LO MC(MadGraph and Baur). 
For $\Wgg$ and
$\Zgg$ we used the same K-factor formulas as for $\Wg$ and $\Zg$,
respectively.

Since the Baur NLO program only considers the $s$, $t$ and $u$ channel
contributions and not the bremstrahlung off the lepton lines(in this
case on the generator level $M_W \le 76.0$, $M_Z \le 86.0$) we apply
the inclusive W cross-section K-factor of 1.36 to the W+photon
processes.

The K-factor applied to $\Wg$ and $\Wgg$ MadGraph MC samples is shown
in Equation~\ref{kfactor_wgwgg.equation}. The K-factor applied to
$\Zg$ and $\Zgg$ MadGraph MC samples is shown in
Equation~\ref{kfactor_zgzgg.equation}. In the
Equations~\ref{kfactor_wgwgg.equation}
and~\ref{kfactor_zgzgg.equation} $M_W$($M_Z$) is the mass of the
generated W(Z) system , and $P_T^\gamma$ is a (generated) photon
transverse energy.

\begin{equation}
  \begin{array}{l}
  M_W \le 76.0 \Rightarrow \kfactor = 1.36 \\
  M_W \gt 76.0 \Rightarrow \kfactor = 1.62 + 0.00001 \times P_T^\gamma-  0.386 \times exp(-0.100 \times P_T^\gamma) \\
 \label{kfactor_wgwgg.equation}
 \end{array}
\end{equation}

\begin{equation}
  \begin{array}{l}
  M_Z \le 86.0 \Rightarrow \kfactor = 1.36 \\
  M_Z \gt 86.0 \Rightarrow \kfactor = 1.46 - 0.000728 \times P_T^\gamma- 0.125 \times exp(-0.0615 \times P_T^\gamma) \\
 \label{kfactor_zgzgg.equation}
 \end{array}
\end{equation}

Every MC event is weighted with the appropriate K-factor.

\section{Checks}
To be highly confident in the SM predictions, we compared the
predictions from the three independent LO matrix-element generators at
generator and HEPG level.

There is excellent agreement (within 10\% or within statistics)
between MadGraph and CompHep in all channels. As the two generators
are really different in technique (a helicity amplitude calculation in
MadGraph, as opposed to the symbolic evaluation of squared matrix
element in CompHep), this gives us confidence in the predictions.

\begin{figure}[!t]
\begin{center} 
  \mbox{
  \subfigure[$e\nu\gamma$]
	{\epsfig{file=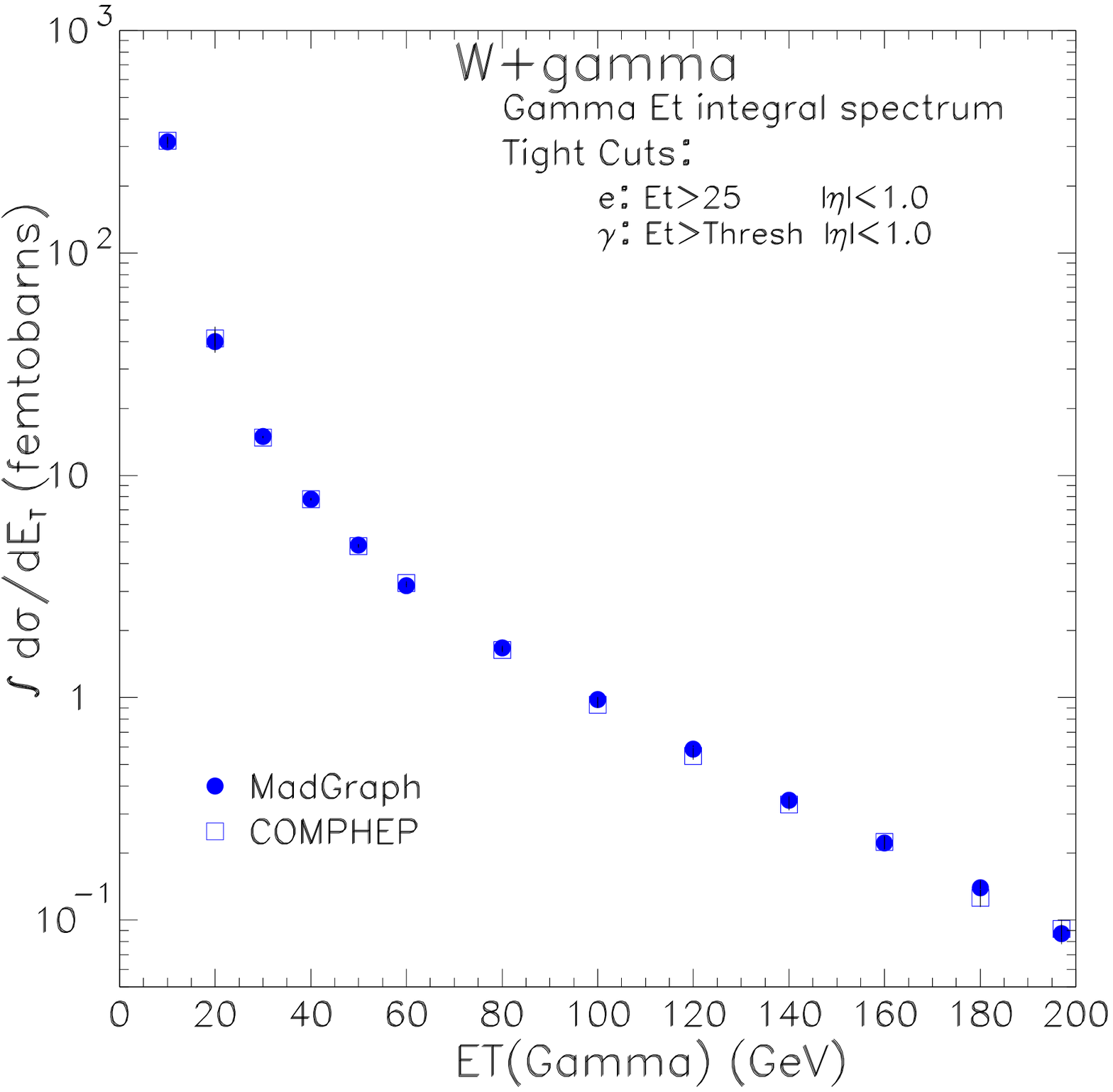, width=2.7in}} \quad
  \subfigure[$e^+e^-\gamma$]
	{\epsfig{file=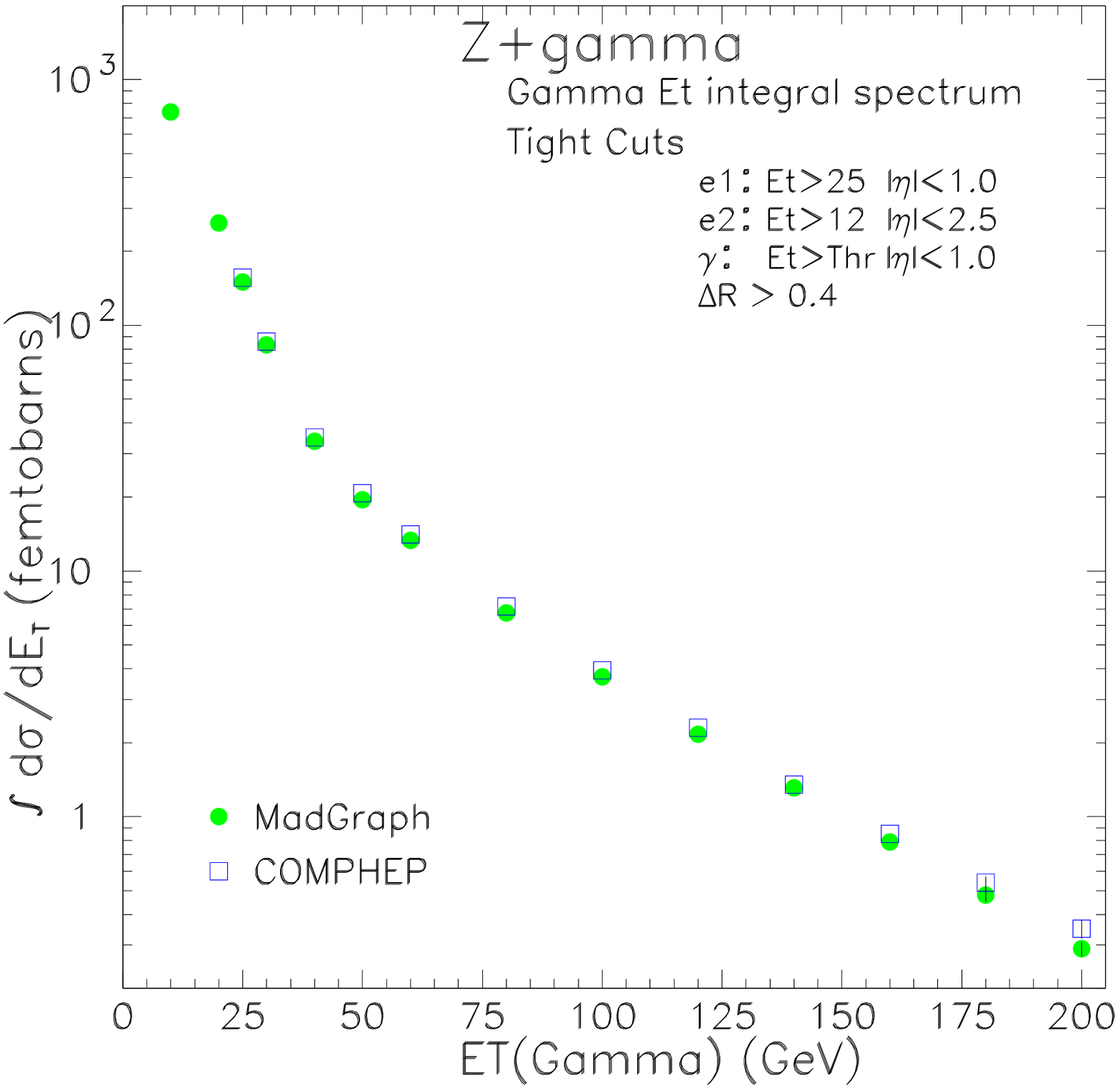, width=2.7in}} \quad
      }
 \caption
[MadGraph vs Baur, generator level: $e\nu\gamma$ and $e^+e^-\gamma$ production]
{The integral cross-sections in fb from MadGraph and COMPHEP
 at 1.96 GeV versus the gamma $\Et$ for (a) $e\nu\gamma$ production,
 (b) $e^+e^-\gamma$ production~\cite{CDF6057,CDF7523}.}
    \label{wgzg_madgraph_vs_comphep.figure} 
\end{center}
\end{figure}

\begin{figure}[!t]
\begin{center} 
  \mbox{
  \subfigure[$e\nu\gg$]
	{\epsfig{file=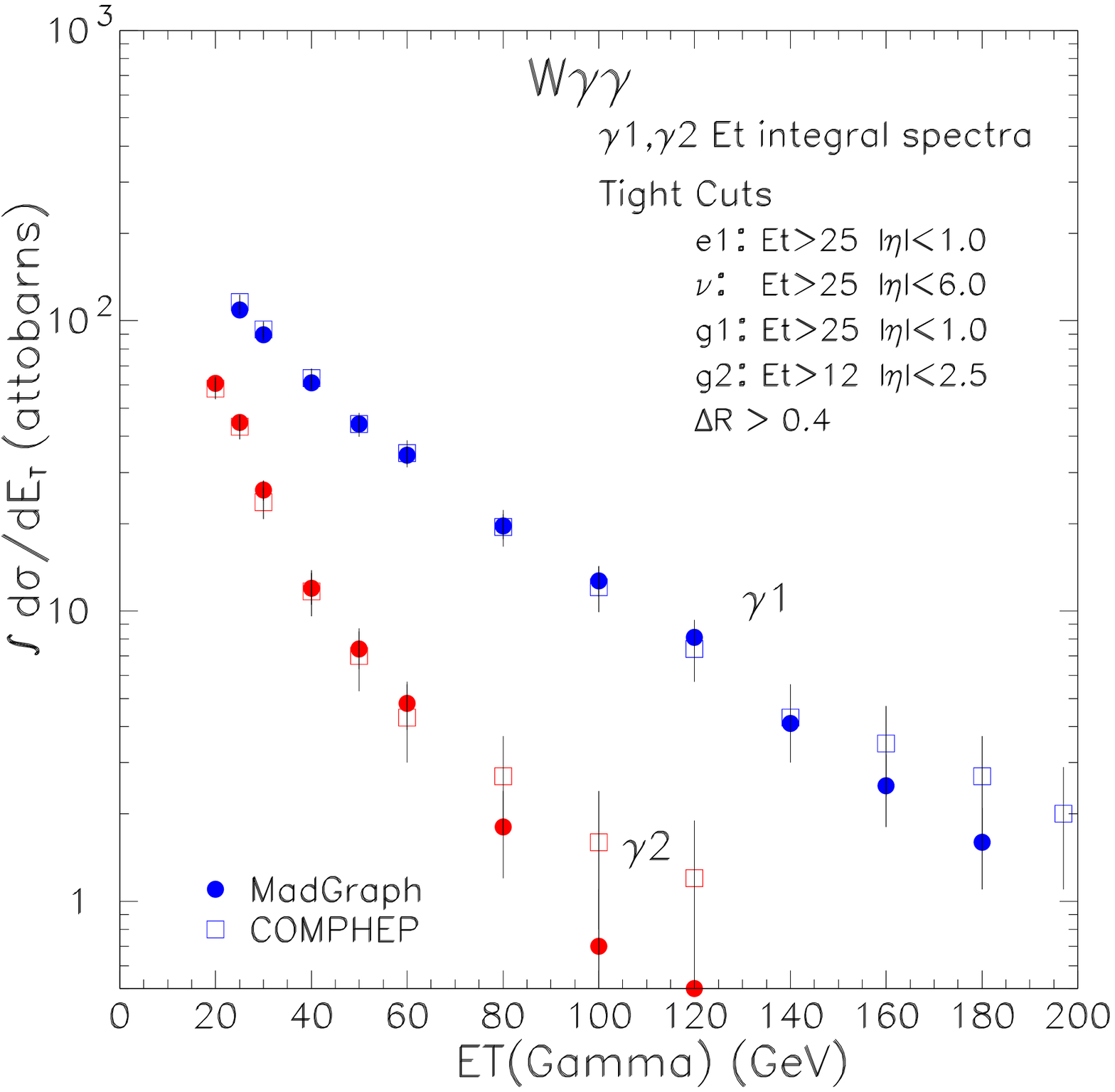, width=2.7in}} \quad
  \subfigure[$e^+e^-\gg$]
	{\epsfig{file=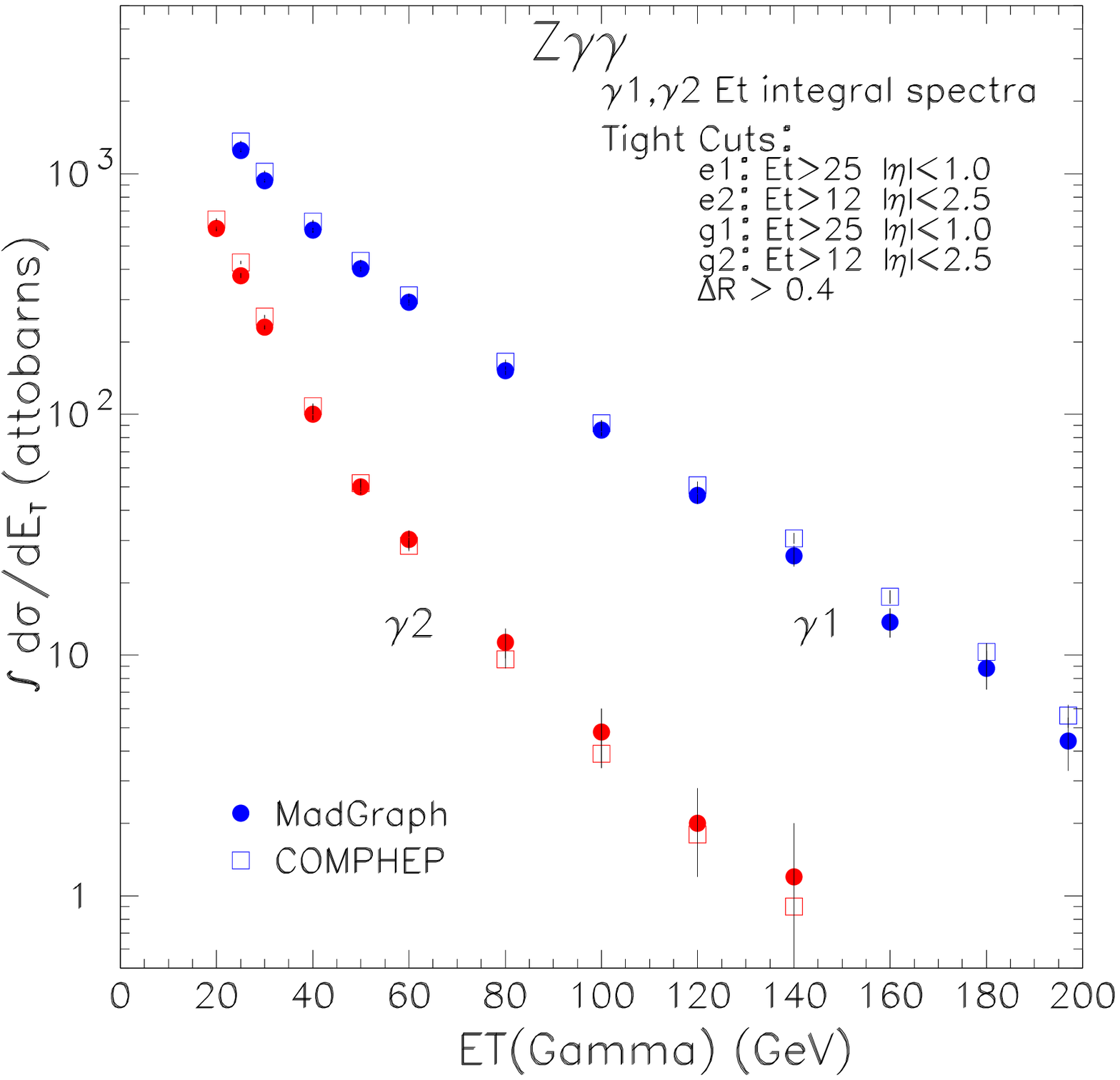, width=2.7in}} \quad
      }
 \caption
[MadGraph vs Baur, generator level: $\Wgg$ and $\Zgg$ production]
{The integrated cross section in attobarns ($10^{-3}$ fb or
 $10^{-6}$ pb) versus the gamma $\Et$ for for (a) $\Wgg$ production
 and (b) $\Zgg$ production. The cross section for the highest-$\Et$
 photon to be above the threshold is in blue; and the 2nd photon is in
 red~\cite{CDF6057,CDF7523}.}
    \label{wggzgg_madgraph_vs_comphep.figure} 
\end{center}
\end{figure}

Figure~\ref{wgzg_madgraph_vs_comphep.figure} shows the integrated
cross section versus $\Et$ of the photon for the MadGraph and CompHep
$\Wg$ and $\Zg$
samples~\cite{CDF6057,CDF7523}. Figure~\ref{wggzgg_madgraph_vs_comphep.figure}
shows the integrated cross section versus $\Et$ of the photon for the
MadGraph and CompHep $\Wgg$ and $\Zgg$ samples~\cite{CDF6057,CDF7523}.

\begin{figure}[!h]
  \begin{center} 
\vspace{-0.2in}
\mbox{
\hspace{-0.2in}
\subfigure[First Lepton $E_T$ GeV]
{\epsfig{file=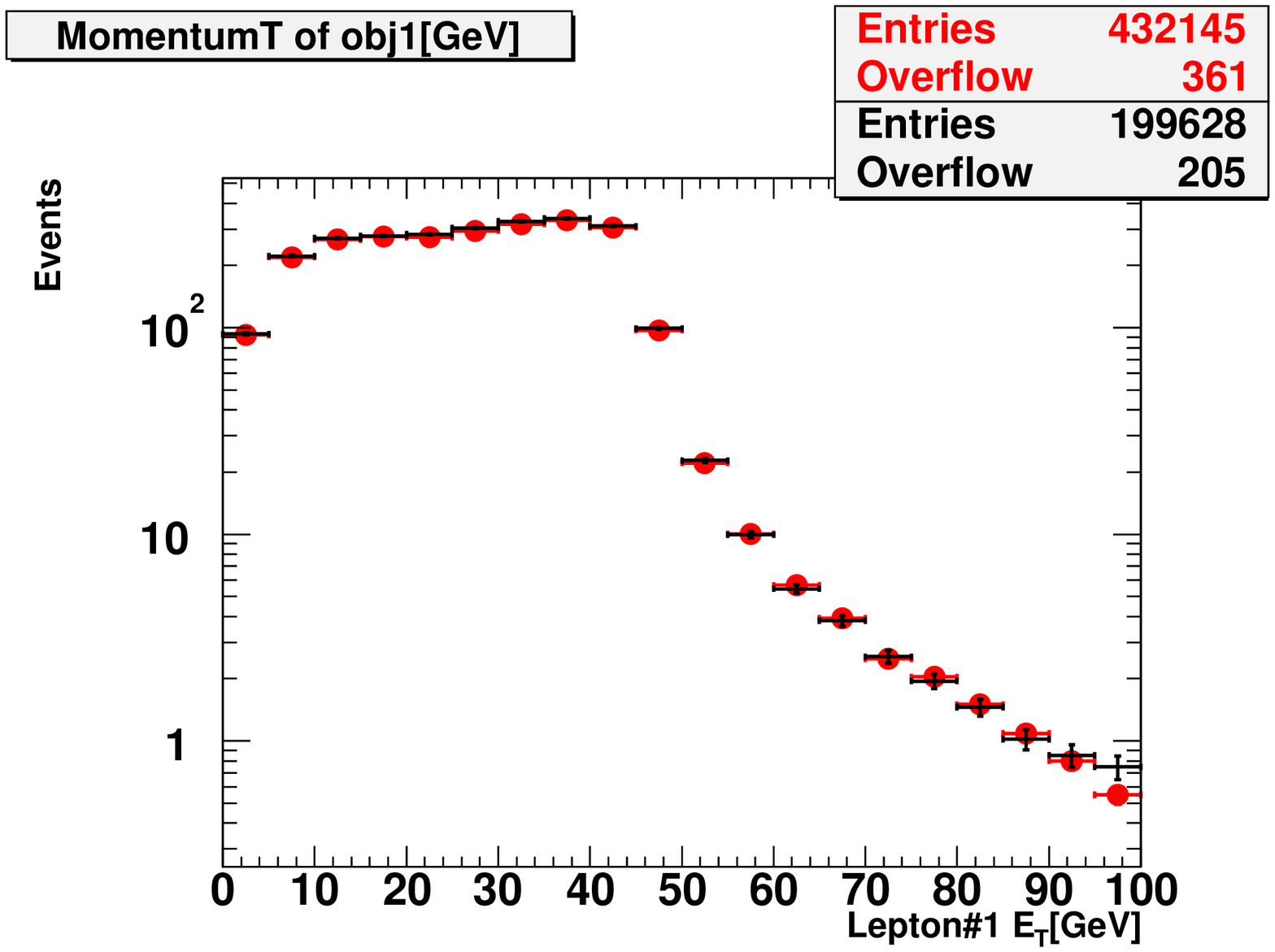,width=2.3in}} \quad
\hspace{-0.2in}
\subfigure[Second Lepton $E_T$ GeV]
{\epsfig{file=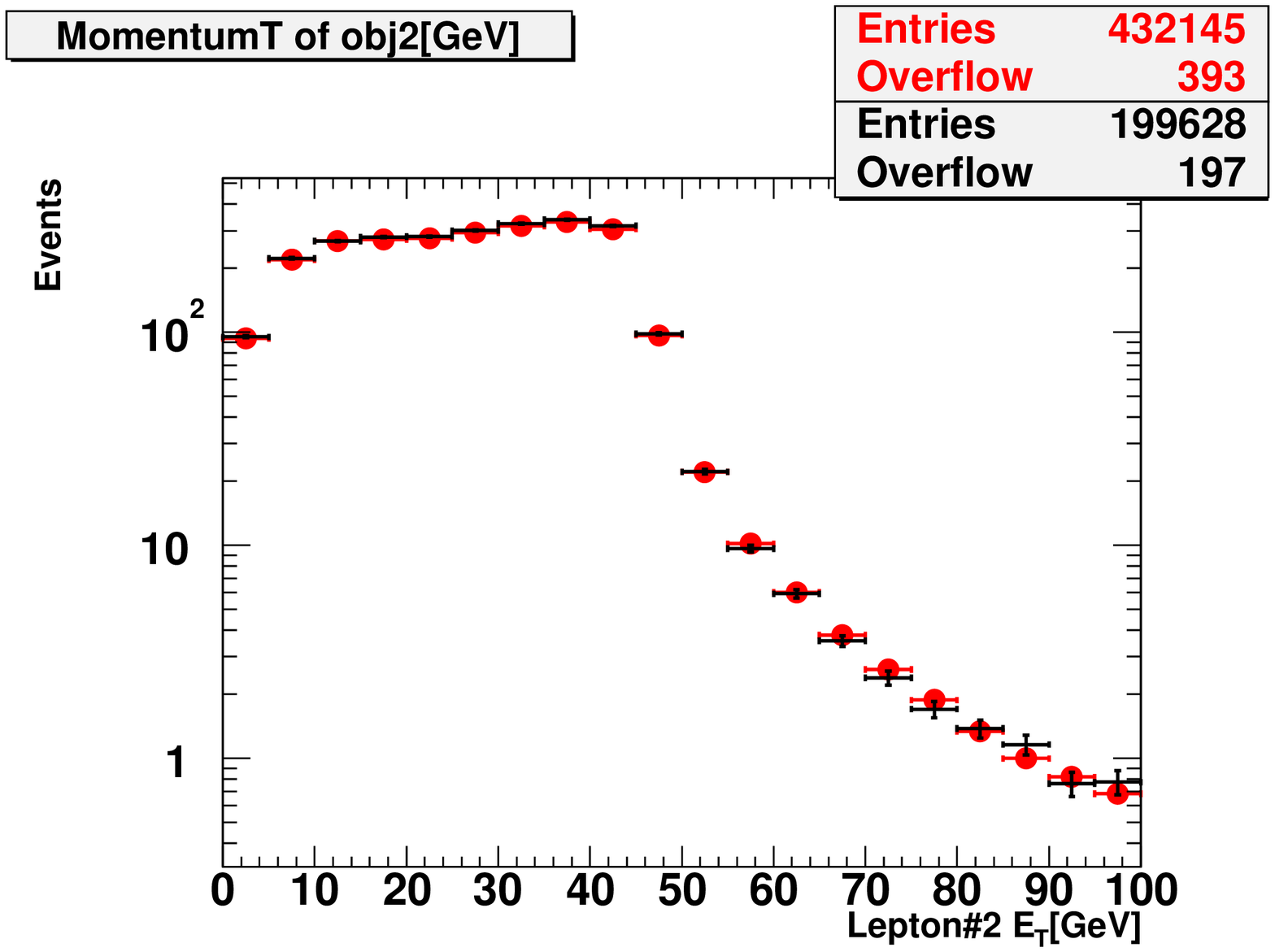,width=2.3in}} \quad
\hspace{-0.2in}
\subfigure[Photon $E_T$ GeV]
{\epsfig{file=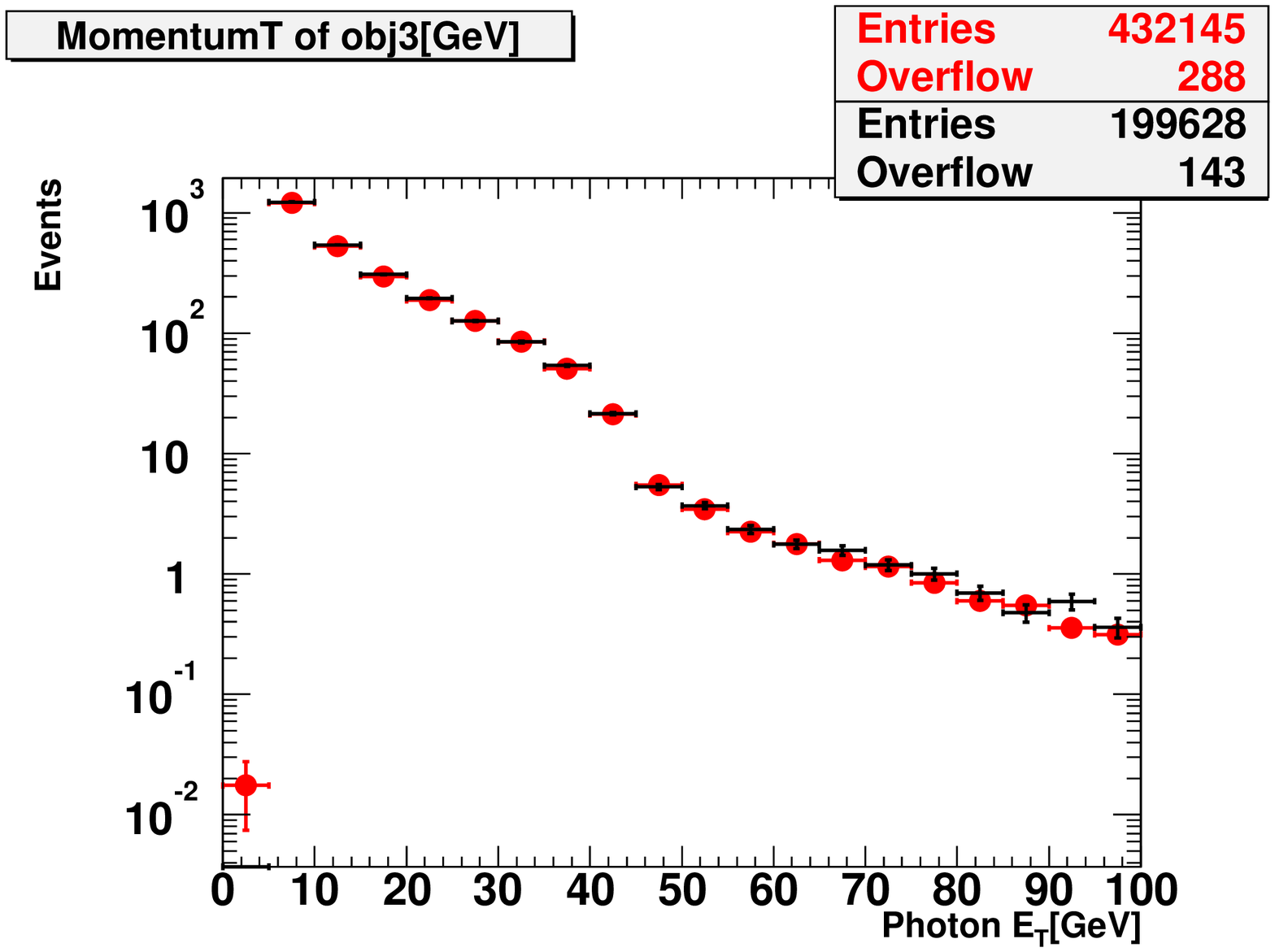,width=2.3in}} \quad
}
\vspace{-0.2in}
    \mbox{
\hspace{-0.2in}
\subfigure[First Lepton $\eta$]
{\epsfig{file=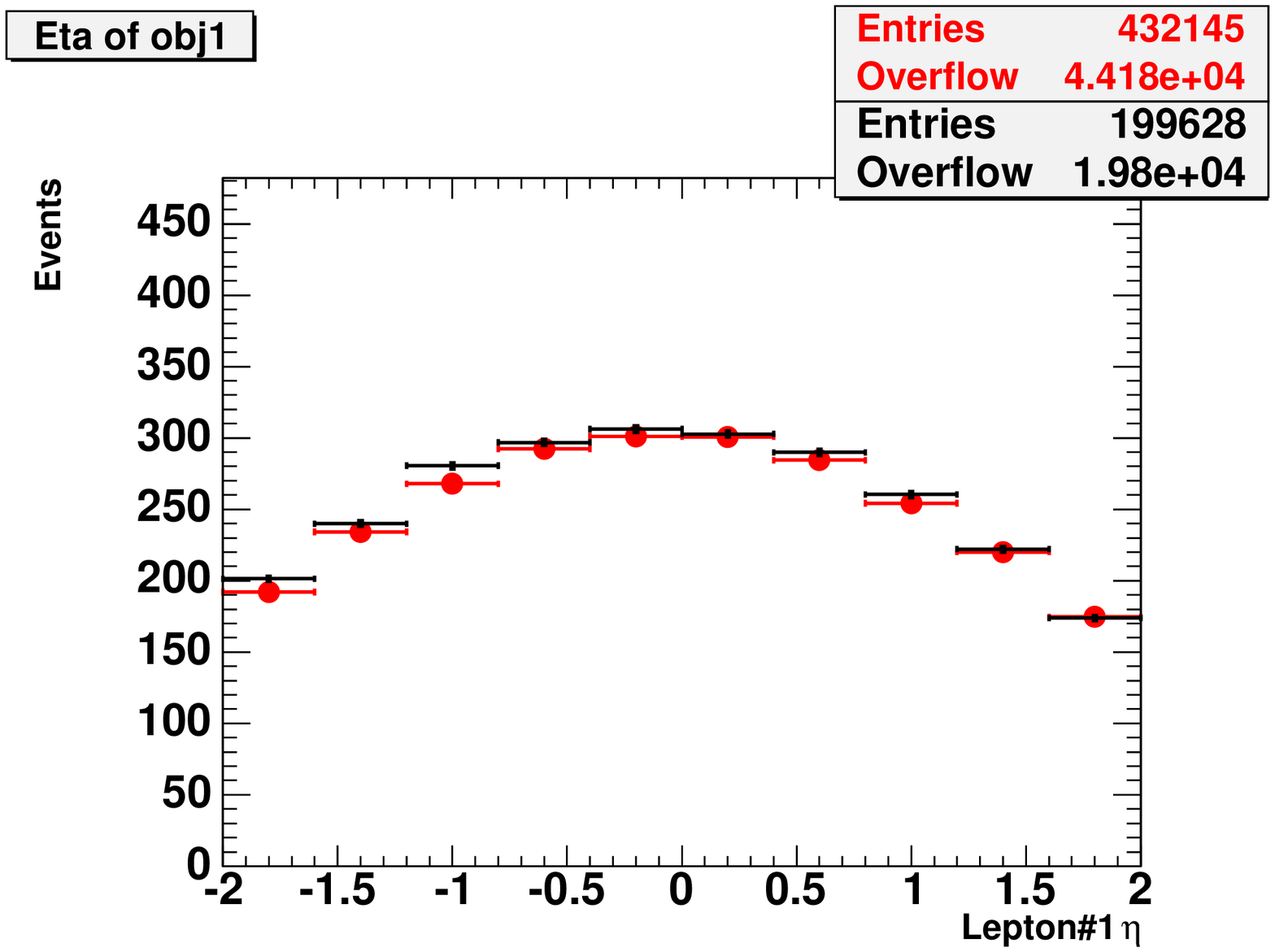,width=2.3in}} \quad
\hspace{-0.2in}
\subfigure[Second Lepton $\eta$]
{\epsfig{file=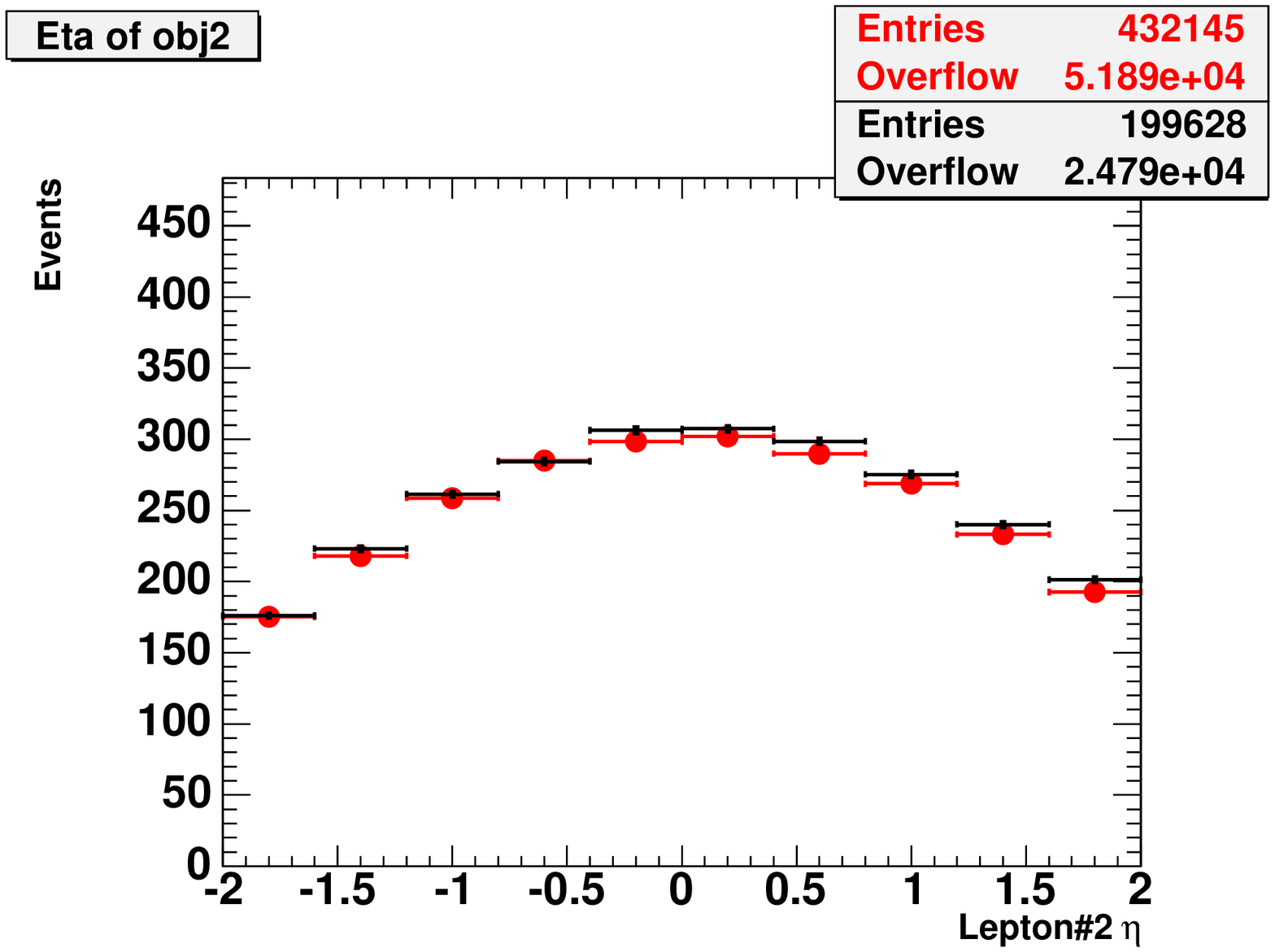,width=2.3in}} \quad
\hspace{-0.2in}
\subfigure[Photon $\eta$]
{\epsfig{file=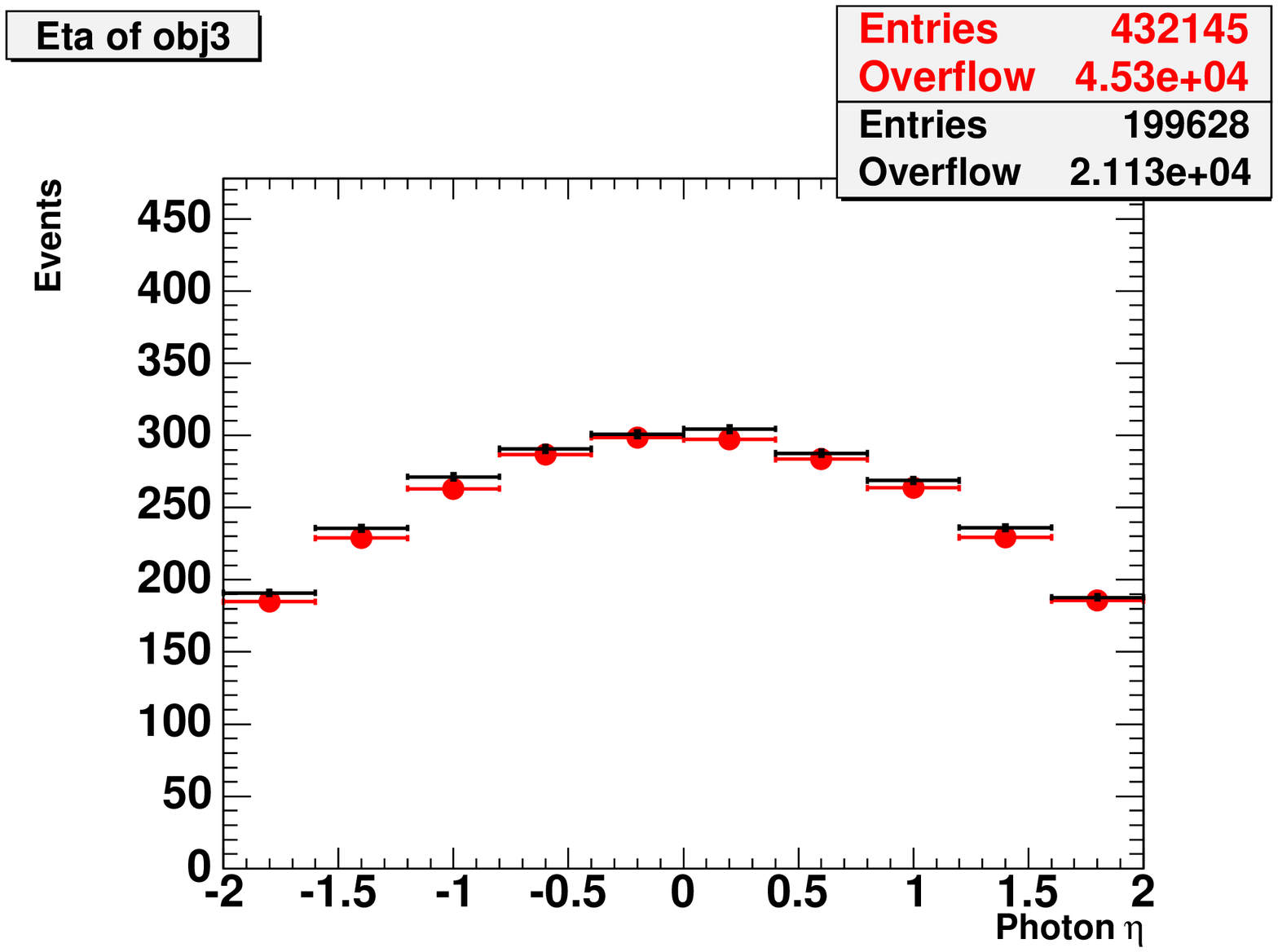,width=2.3in}} \quad
      }
\vspace{-0.2in}
    \mbox{
\hspace{-0.2in}
\subfigure[M(Z) GeV]
{\epsfig{file=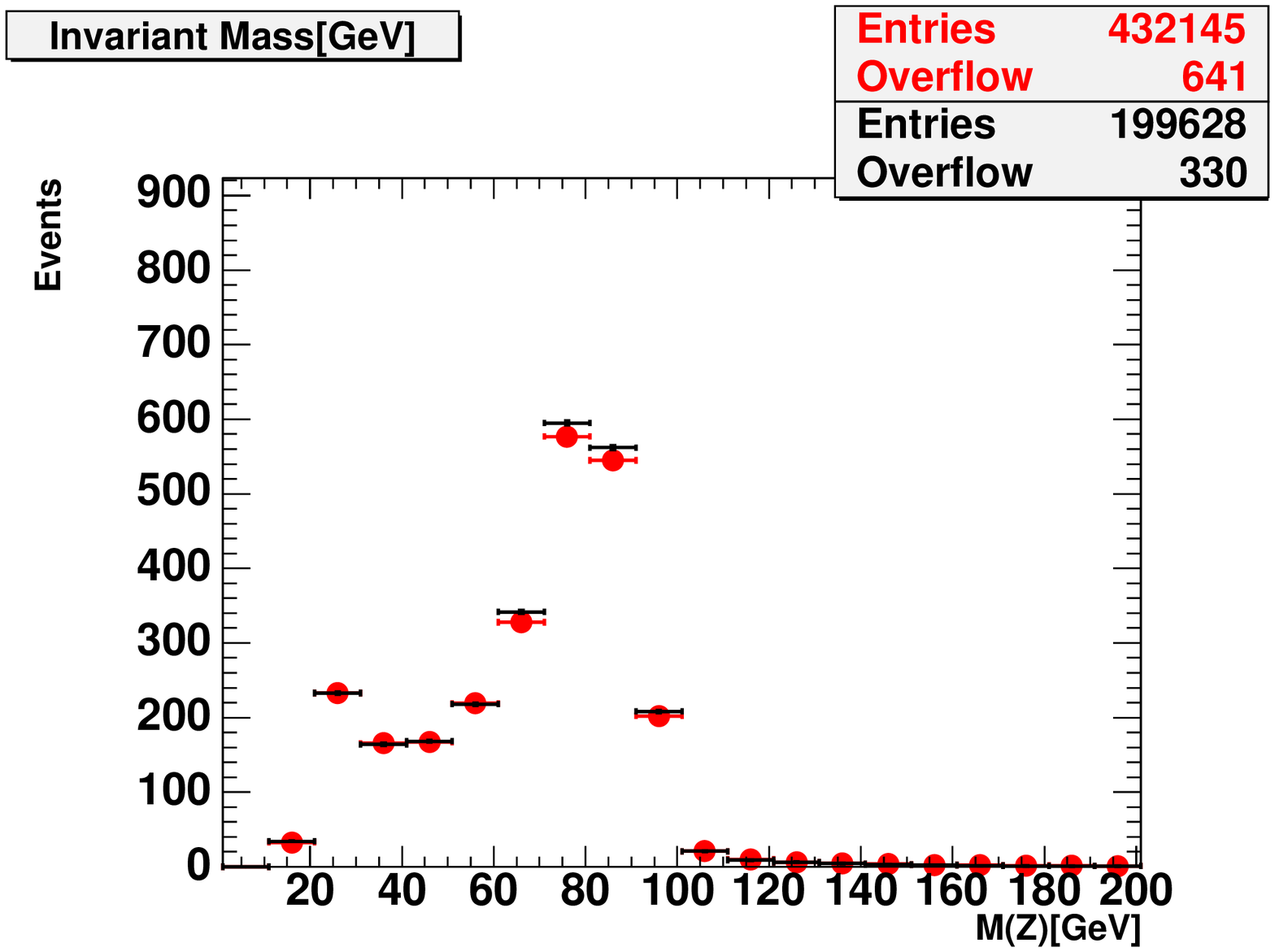,width=2.3in}} \quad
\hspace{-0.2in}
\subfigure[$\Delta\phi (ll)$]
{\epsfig{file=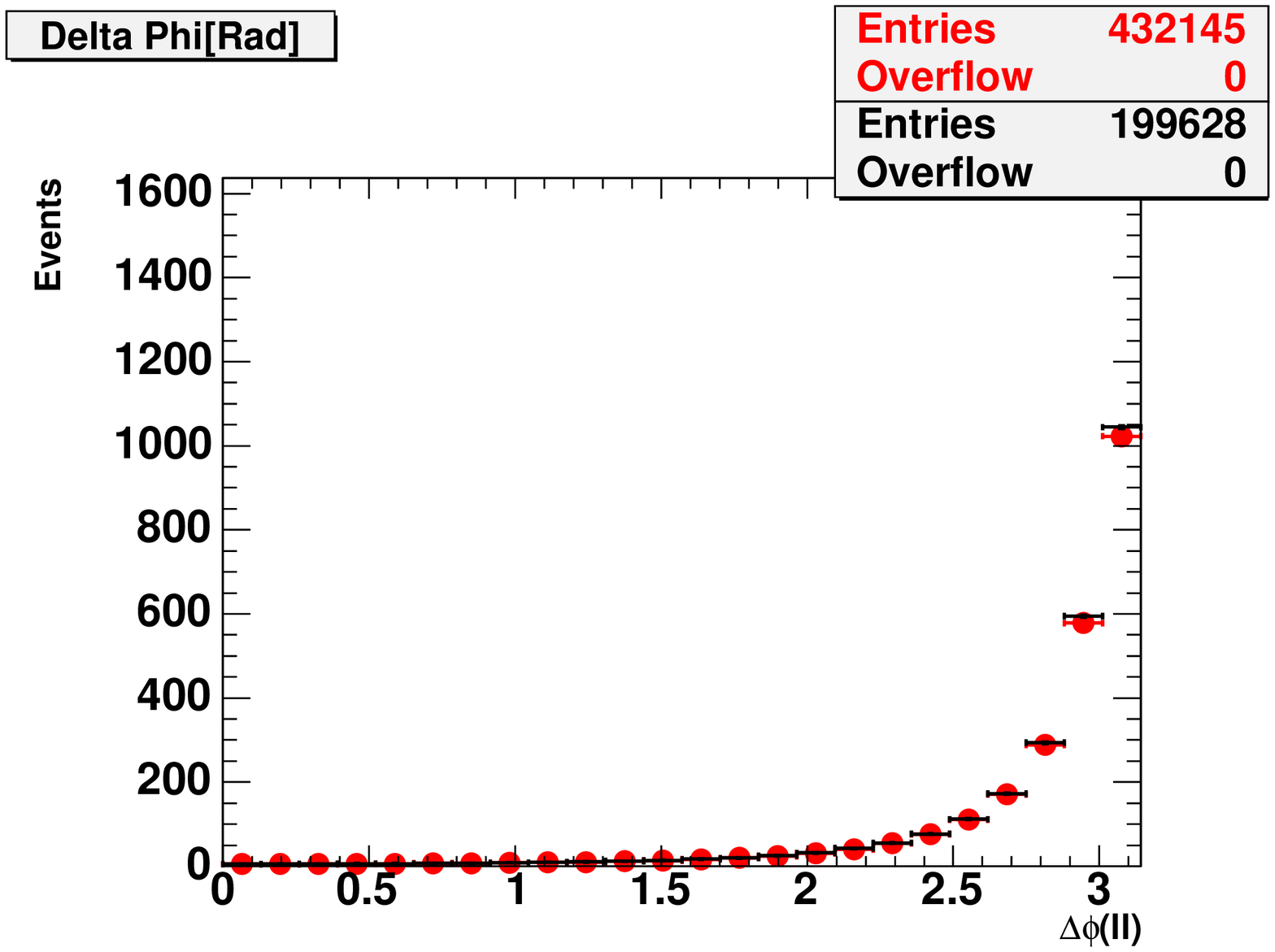,width=2.3in}} \quad
\hspace{-0.2in}
\subfigure[M($\Zg$) GeV]
{\epsfig{file=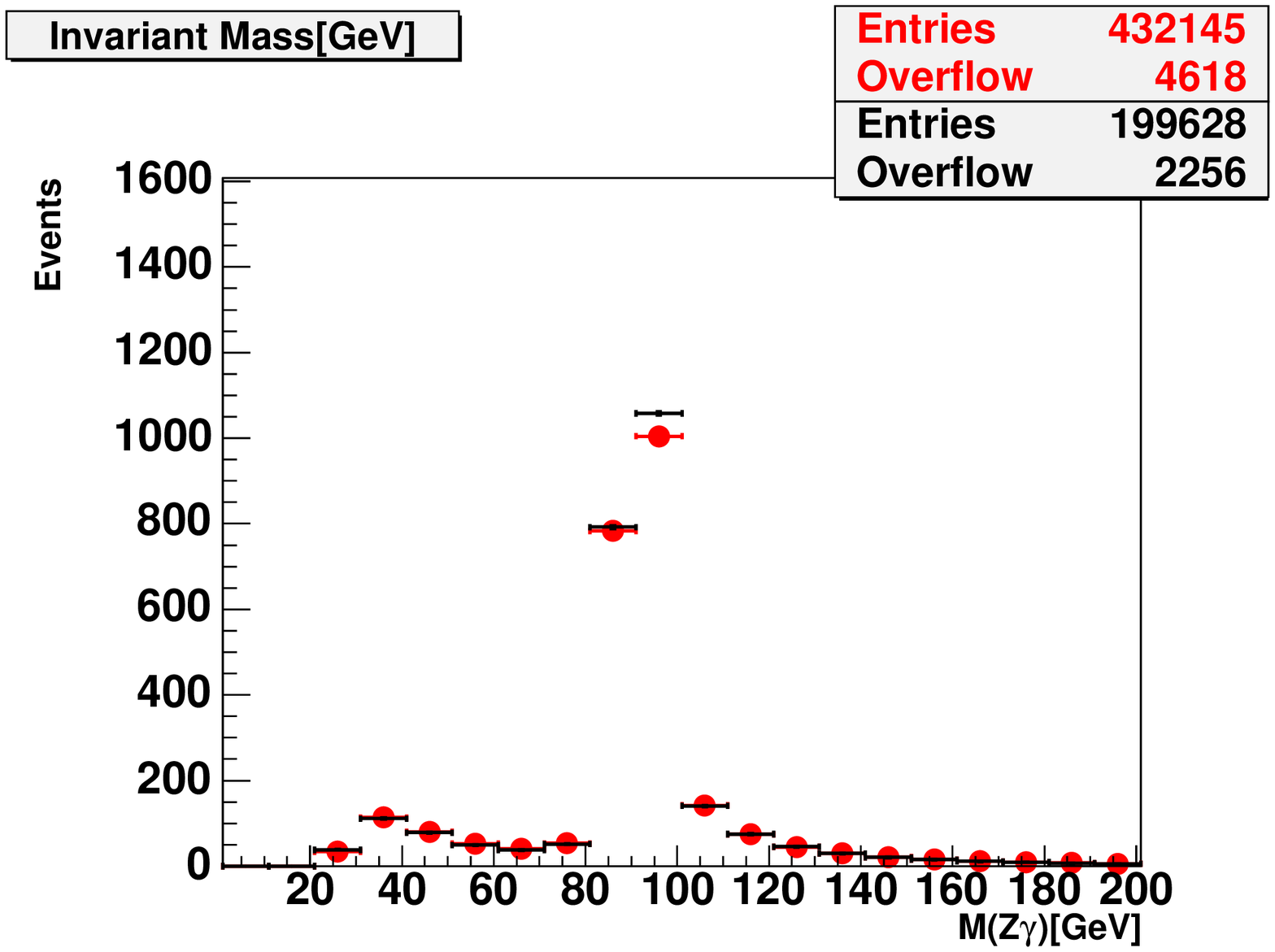,width=2.3in}} \quad
      }
\caption 
[MadGraph vs Baur after fragmentation]
{Comparison of distributions for {\bf MadGraph} (black) vs
{\color{red}Baur(red dots)} for $\Zg$ ($\mu$ channel) after
fragmentation (HEPG level). See~\cite{CDF8116} for details.}
\label{comparison_gen_madbaur.figure}
    \end{center}
\end{figure}

We have compared MadGraph, CompHep and MadGraph samples at
GENERATOR~\cite{CDF6057} and HEPG~\cite{CDF7040}. The more detailed
study for MadGraph and Baur $\Zg$ and $\Wg$ samples have been
performed at Ref.~\cite{CDF8116,CDF8126}. For example, the
distributions for the muon channel is shown in
Figure~\ref{comparison_gen_madbaur.figure}.

\section{The SM Diboson $\Wg$ and $\Zg$ Processes as Sources of
    Lepton-Photon Events}
The $\Wg$ channel was the main SM contributor to the $\ell\gamma\met$
signature with the Run I cuts~\cite{jeff_thesis}. In the Run I analysis,
$W\gamma$ was expected to contribute $1.93\pm0.26$ events to the
$e\gamma\met$ channel, out of a total of $3.41 \pm0.34$, and
$1.99\pm0.27$, out of a total of $4.23\pm0.46$ for the $\mu\gamma\met$
channel. Having a reliable prediction of this signature is crucial.

\begin{table}[!h]
 \begin{center}
\begin{tabular} {|p{2.0cm}||p{2.5cm}|p{2.5cm}|p{2.5cm}|p{2.5cm}|p{2.5cm}|}
\hline
\rotatebox{75}{Categories} &
\rotatebox{75}{All Other } &
\rotatebox{75}{1$\ell$, 1$\gamma$, $\Delta\phi_{\lg}\lt$150} &
\rotatebox{75}{$\lgmet$ } &
\rotatebox{75}{Multi-Lepton } &
\rotatebox{75}{Multi-Photon } \\
\hline
\multicolumn{6}{|c|} { $\Wg$, electron channel }\\
\hline
MadGraph
& 15.3$\pm$0.99 & 1.61$\pm$0.32 & 13.6$\pm$0.93 & 0$\pm$0.042 & 0$\pm$0.042\\
\hline
Baur
& 15.9$\pm$1.31 & 2.05$\pm$0.47 & 13.8$\pm$1.21 & 0$\pm$0.069 & 0$\pm$0.069\\
\hline
\hline
Average & 15.60$\pm$0.82{\tiny (stat)} & 1.83$\pm$0.28{\tiny (stat)} & 13.70$\pm$0.76{\tiny (stat)} & 0.0$\pm$0.040{\tiny (stat)} & 0.0$\pm$0.040{\tiny (stat)}\\
 & $\pm$1.70{\tiny (sys)} & $\pm$0.48{\tiny (sys)} & $\pm$1.41{\tiny (sys)} & $\pm$0.0{\tiny (sys)} & $\pm$0.0{\tiny (sys)}\\
\hline
\multicolumn{6}{|c|} { $\Wg$, muon channel }\\
\hline
MadGraph
& 10.4$\pm$0.77 & 1.97$\pm$0.34 & 8.44$\pm$0.70 & 0$\pm$0.042 & 0$\pm$0.042\\
\hline
Baur
& 10.8$\pm$0.95 & 1.56$\pm$0.36 & 9.25$\pm$0.88 & 0$\pm$0.059 & 0$\pm$0.059\\
\hline
\hline
Average & 10.60$\pm$0.61{\tiny (stat)} & 1.77$\pm$0.25{\tiny (stat)} & 8.84$\pm$0.56{\tiny (stat)} & 0.0$\pm$0.036{\tiny (stat)} & 0.0$\pm$0.036{\tiny (stat)}\\
 & $\pm$1.18{\tiny (sys)} & $\pm$0.45{\tiny (sys)} & $\pm$1.23{\tiny (sys)} & $\pm$0.0{\tiny (sys)} & $\pm$0.0{\tiny (sys)}\\
\hline
\end{tabular}
\caption
[The SM contributions from the $\Wg$ to the analysis categories]
{The SM contributions from the $\Wg$ channel to the analysis categories. The LO
calculations from Baur and MadGraph have been corrected by the
K-factor and CDF efficiencies. The difference between the two LO
generators has been included as a systematic uncertainty (see text);
the uncertainty on the average includes both the statistical
uncertainties and this systematic uncertainty. }
\label{mg_baur_wa.table}
 \end{center}
\end{table}

The photon can be radiated from the incoming quarks, from the outgoing
electron, or the intermediate W
(Figure~\ref{zgwg_diagrams.figure}). More detail on the kinematic
distributions and the cross sections is available in
Ref~\cite{CDF6057}. Initial-state radiation is simulated by the {\sc
pythia} MC program~\cite{Pythia} tuned to reproduce the underlying
event. The generated particles are then passed through a full detector
simulation, and these events are then reconstructed with the same code
used for the data.

The predicted numbers of detected $\eg$ and $\mug$ events in
$\apprluminosity\pbinv$ from SM $\W\gamma$ production satisfying the
analysis cuts from MadGraph are given in Table~\ref{mg_baur_wa.table}.

The uncertainties on the SM contributions include those from parton
distribution functions (5\%), factorization scale (2\%), K-factor
(3\%), a comparison of different MC generators ($\sim$ 5\%), and the
luminosity (6\%) (Chapter~\ref{systematics}).

We have studied predictions from MadGraph, CompHep and Baur generators
for data and MC reconstructed in Ref.~\cite{CDF7040}, and these studies
showed good agreement in predicted rates.

The process $\ppbar \goes \Zgstar + \gamma$ is also one of the major
SM backgrounds for the lepton+ photon+X searches. It is the largest
contributor to the inclusive multi-body category in the Run I search,
with expected contributions in 86 $\pbinv$ of 5.01 $\pm$ 0.54 events
in the $e\gamma$ mode and 4.60$\pm$0.54 in the $\mu\gamma$
mode~\cite{jeff_prd}. It is also significant in the $\ell\gamma\met$
channel, especially for muons, as one muon can be missed inducing
$\Met$. The expected contributions for the electron and muon channels
in Run I were $0.32\pm0.5$ and $0.96\pm0.15$ events, respectively,
smaller than the $W\gamma$ contributions, but still significant, with
a total ($e+\mu$) of $1.27\pm0.17$ events out of the $7.64\pm0.71$
events expected in 86 $\pbinv$.

The photon can be radiated from the incoming quarks or from the
outgoing electron (Figure~\ref{zgwg_diagrams.figure}). More detail on
the kinematic distributions and the cross sections is available in
Ref~\cite{CDF6057}.

The predicted numbers of detected (inclusive) $\eg$ and $\mug$ events
in $\apprluminosity\pbinv$ from SM $Z\gamma$ production satisfying the
analysis cuts from MadGraph are given in Table~\ref{mg_baur_za.table}.
The uncertainties on the SM contributions include those from parton
distribution functions (7\%), a comparison of different MC generators
($\sim$ 5\%), and the luminosity (6\%) (Chapter~\ref{systematics}).

\begin{table}[!h]
 \begin{center}
\begin{tabular} {|p{2.0cm}||p{2.5cm}|p{2.5cm}|p{2.5cm}|p{2.5cm}|p{2.5cm}|}
\hline

\rotatebox{75}{Categories} &
\rotatebox{75}{All Other } &
\rotatebox{75}{1$\ell$, 1$\gamma$, $\Delta\phi_{\lg}\lt$150} &
\rotatebox{75}{$\lgmet$ } &
\rotatebox{75}{Multi-Lepton } &
\rotatebox{75}{Multi-Photon } \\

\hline
\multicolumn{6}{|c|} { $\Zg$, electron channel }\\
\hline
MadGraph
& 21$\pm$0.39 & 7.77$\pm$0.24 & 0.97$\pm$0.084 & 12.1$\pm$0.29 & 0.12$\pm$0.030\\
\hline
Baur
& 22.7$\pm$0.43 & 8.31$\pm$0.26 & 1.34$\pm$0.11 & 12.9$\pm$0.33 & 0.17$\pm$0.038\\
\hline
\hline
Average & 21.85$\pm$0.29{\tiny (stat)} & 8.04$\pm$0.18{\tiny (stat)} & 1.16$\pm$0.069{\tiny (stat)} & 12.50$\pm$0.22{\tiny (stat)} & 0.15$\pm$0.024{\tiny (stat)}\\
 & $\pm$2.80{\tiny (sys)} & $\pm$0.98{\tiny (sys)} & $\pm$0.39{\tiny (sys)} & $\pm$1.51{\tiny (sys)} & $\pm$0.052{\tiny (sys)}\\
\hline
\multicolumn{6}{|c|} { $\Zg$, muon channel }\\
\hline
MadGraph
& 15.7$\pm$0.32 & 3.46$\pm$0.15 & 4.28$\pm$0.17 & 7.96$\pm$0.22 & 0.026$\pm$0.013\\
\hline
Baur
& 16.6$\pm$0.35 & 4.18$\pm$0.18 & 4.7$\pm$0.19 & 7.67$\pm$0.23 & 0.022$\pm$0.013\\
\hline
\hline
Average & 16.15$\pm$0.24{\tiny (stat)} & 3.82$\pm$0.12{\tiny (stat)} & 4.49$\pm$0.13{\tiny (stat)} & 7.81$\pm$0.16{\tiny (stat)} & 0.024$\pm$0.01{\tiny (stat)}\\
 & $\pm$1.92{\tiny (sys)} & $\pm$0.82{\tiny (sys)} & $\pm$0.63{\tiny (sys)} & $\pm$0.87{\tiny (sys)} & $\pm$0.0047{\tiny (sys)}\\
\hline
\end{tabular}
\caption
[The SM contributions from the $\Zg$ to the analysis categories]
{The SM contributions from the $Z\gamma$ electron channel to the
analysis categories. The LO calculations from Baur and MadGraph have
been corrected by the K-factor and CDF efficiencies.  The difference
between the two LO generators has been included as a systematic
uncertainty (see text); the uncertainty on the average includes both
the statistical uncertainties and this systematic uncertainty.
}
\label{mg_baur_za.table}
 \end{center}
\end{table}

\section
{The SM Triboson $W\gamma\gamma$ and $Z\gamma\gamma$ Processes as Sources
of Lepton-Photon Events}

While small, the $W\gamma\gamma$, $Z\gamma\gamma$ processes are the
largest true SM sources of a signature of a high-$\Pt$ lepton plus two
photons. The observation of several such events has motivated a
careful study of the cross-sections for these
sources~\cite{CDF6057}. In this study we have used both MadGraph and
Comphep; we get excellent agreement between the two generators
(Figure~\ref{wggzgg_madgraph_vs_comphep.figure}), giving us confidence
in the predictions.

The final state of one lepton and two photons, $\ell\nu\gg$, is
produced in the SM through an intermediate W, with radiation off of
any of the charged lines in the diagrams. We denote the final state of
$\ell\nu\gg$ as `$W\gamma\gamma$' for convenience, although the $W$ is
virtual and the kinematics of the final state are more complicated
than the name would suggest. Note that this is process has three
spin-one bosons in the final state; MadGraph treats the helicities
correctly and writes them into the output file as input to the next
steps.

The predicted numbers of detected $\eg$ and $\mug$ events in
$\apprluminosity\pbinv$ from SM $\Wgg$ production satisfying the
analysis cuts from MadGraph are given in
Table~\ref{mg_waa_zaa.table}. The agreement on the generator level for
$\Wgg$ between the two monte carlos, CompHep and MadGraph is good
(Figure~\ref{wggzgg_madgraph_vs_comphep.figure}), and we will use
MadGraph $\Wgg$ to get predicted rates. More details, including
kinematic distributions, can be found in Ref.~\cite{CDF6057}.

\begin{table}[!h]
\footnotesize
 \begin{center}
\begin{tabular} {|p{2.0cm}||p{2.5cm}|p{2.5cm}|p{2.5cm}|p{2.5cm}|p{2.5cm}|}
\hline

\rotatebox{75}{Categories} &
\rotatebox{75}{All Other } &
\rotatebox{75}{1$\ell$, 1$\gamma$, $\Delta\phi_{\lg}\lt$150} &
\rotatebox{75}{$\lgmet$ } &
\rotatebox{75}{Multi-Lepton } &
\rotatebox{75}{Multi-Photon } \\
\hline
\multicolumn{6}{|c|} { $\Wgg$, electron channel }\\
\hline
MadGraph & 0.14$\pm$ & 0.019$\pm$ & 0.11$\pm$ & 0.0052$\pm$ & 0.0067$\pm$\\
 & 0.0064{\tiny (stat)} & 0.0023{\tiny (stat)} & 0.0057{\tiny (stat)} & 0.0012{\tiny (stat)} & 0.0014{\tiny (stat)}\\
 & $\pm$0.017{\tiny (sys)} & $\pm$0.0023{\tiny (sys)} & $\pm$0.013{\tiny (sys)} & $\pm$6.2e-04{\tiny (sys)} & $\pm$8.0e-04{\tiny (sys)}\\
\hline
\multicolumn{6}{|c|} { $\Wgg$, muon channel }\\
\hline
MadGraph & 0.069$\pm$ & 0.0100$\pm$ & 0.055$\pm$ & 0.0$\pm$ & 0.0037$\pm$\\
 & 0.0042{\tiny (stat)} & 0.0016{\tiny (stat)} & 0.0037{\tiny (stat)} & 1.9e-04{\tiny (stat)} & 9.5e-04{\tiny (stat)}\\
 & $\pm$0.0083{\tiny (sys)} & $\pm$0.0012{\tiny (sys)} & $\pm$0.0066{\tiny (sys)} & $\pm$0.0{\tiny (sys)} & $\pm$4.4e-04{\tiny (sys)}\\
\hline
\multicolumn{6}{|c|} { $\Zgg$, electron channel }\\
\hline
MadGraph & 0.54$\pm$ & 0.26$\pm$ & 0.029$\pm$ & 0.24$\pm$ & 0.015$\pm$\\
 & 0.0100{\tiny (stat)} & 0.0070{\tiny (stat)} & 0.0023{\tiny (stat)} & 0.0067{\tiny (stat)} & 0.0017{\tiny (stat)}\\
 & $\pm$0.065{\tiny (sys)} & $\pm$0.031{\tiny (sys)} & $\pm$0.0035{\tiny (sys)} & $\pm$0.029{\tiny (sys)} & $\pm$0.0018{\tiny (sys)}\\
\hline
\multicolumn{6}{|c|} { $\Zgg$, muon channel }\\
\hline
MadGraph & 0.37$\pm$ & 0.12$\pm$ & 0.12$\pm$ & 0.12$\pm$ & 0.011$\pm$\\
 & 0.0080{\tiny (stat)} & 0.0046{\tiny (stat)} & 0.0046{\tiny (stat)} & 0.0044{\tiny (stat)} & 0.0014{\tiny (stat)}\\
 & $\pm$0.044{\tiny (sys)} & $\pm$0.014{\tiny (sys)} & $\pm$0.014{\tiny (sys)} & $\pm$0.014{\tiny (sys)} & $\pm$0.0013{\tiny (sys)}\\
\hline
\end{tabular}
\caption
[The SM contributions from the $\Wgg$ and $\Zgg$ to the analysis categories]
{The predicted number of $\lgX$ events in $\apprluminosity\pbinv$
from SM $\Wgg$ and $\Zgg$ production satisfying the analysis cuts from
MadGraph. The uncertainty includes both the statistical uncertainties
and this systematic uncertainty.}
\label{mg_waa_zaa.table}
 \end{center}
\end{table}

The final state of two leptons and two photons, $\llgg$, is
generated through the intermediate photon and $\Z$ states, with
radiation off of any of the charged lines in the diagrams. We denote
this as `$Z\gamma\gamma$' for convenience, but the two amplitudes
modify the mass spectra and angular distributions, and so are both
important.

This process is one of the SM mechanisms that could produce the
$\mmggjj$ event~\cite{toback_thesis} (although with an extra two
jets), and is of interest in the dilepton-diphoton searches as a
background. As one can see below the SM cross sections are small,
typically one femtobarn or less.

The predicted numbers of detected $\eg$ and $\mug$ events in
$\apprluminosity\pbinv$ from SM $\Zgg$ production satisfying the
analysis cuts from MadGraph are given in
Table~\ref{mg_waa_zaa.table}. The agreement on the generator level for
$\Wgg$ between the two monte carlos, CompHep and MadGraph is good
(Figure~\ref{wggzgg_madgraph_vs_comphep.figure}), and we will use
MadGraph $\Wgg$ to get predicted rates. More details, including
kinematic distributions, can be found in Ref.~\cite{CDF6057}.

\section{The Sum of Contributions for SM $\W\gamma$,
  $Z\gamma$, $\W\gg$, $Z\gg$ Processes}

Table~\ref{mc.table} gives the sum of the expected contributions to
the $\eg$ and $\mug$ channels from SM $\W\gamma$, $Z\gamma$, $\W\gg$,
and $Z\gg$ processes.  We have multiplied the average LO predictions
for each channel by the K-factors listed in
Equations~\ref{kfactor_wgwgg.equation}
and~\ref{kfactor_zgzgg.equation}. As we require the event to be
triggered either by the lepton or by the photon trigger, this
combination of triggers is fully efficient.

\begin{table}[!h]
 \begin{center}
\begin{tabular} {|p{1.0cm}||p{2.7cm}|p{2.7cm}|p{2.7cm}|p{2.7cm}|p{2.7cm}|}
\hline

\rotatebox{75}{Categories} &
\rotatebox{75}{All Other } &
\rotatebox{75}{1$\ell$, 1$\gamma$, $\Delta\phi_{\lg}\lt$150} &
\rotatebox{75}{$\lgmet$ } &
\rotatebox{75}{Multi-Lepton } &
\rotatebox{75}{Multi-Photon } \\
\hline
$\eg$ & 38.13$\pm$0.87\xtiny(stat) & 10.15$\pm$0.33\xtiny(stat) & 15.00$\pm$0.76\xtiny(stat) & 12.75$\pm$0.22\xtiny(stat) & 0.17$\pm$0.047\xtiny(stat)\\
& $\pm$4.58\xtiny(sys) & $\pm$1.49\xtiny(sys) & $\pm$1.82\xtiny(sys) & $\pm$1.54\xtiny(sys) & $\pm$0.055\xtiny(sys)\\
\hline
$\mug$ & 27.19$\pm$0.66\xtiny(stat) & 5.72$\pm$0.28\xtiny(stat) & 13.51$\pm$0.57\xtiny(stat) & 7.93$\pm$0.16\xtiny(stat) & 0.039$\pm$0.037\xtiny(stat)\\
& $\pm$3.15\xtiny(sys) & $\pm$1.29\xtiny(sys) & $\pm$1.88\xtiny(sys) & $\pm$0.88\xtiny(sys) & $\pm$0.0064\xtiny(sys)\\
\hline
\hline
$\lg$ & 65.32$\pm$1.09\xtiny(stat) & 15.87$\pm$0.44\xtiny(stat) & 28.51$\pm$0.95\xtiny(stat) & 20.68$\pm$0.28\xtiny(stat) & 0.21$\pm$0.060\xtiny(stat)\\
& $\pm$7.74\xtiny(sys) & $\pm$2.78\xtiny(sys) & $\pm$3.70\xtiny(sys) & $\pm$2.42\xtiny(sys) & $\pm$0.061\xtiny(sys)\\
\hline
\end{tabular}
\caption
[The sum of the expected contributions to the $\lgX$ from
$\W\gamma$, $Z\gamma$, $\W\gg$ and $Z\gg$]
{The sum of the expected contributions to the $\eg$
and $\mug$ channels from SM $\W\gamma$, $Z\gamma$, $\W\gg$, and
$Z\gg$ processes. The average of the two LO predictions for each
channel has been multiplied by the K-factors listed in
Equations~\ref{kfactor_wgwgg.equation} and ~\ref{kfactor_zgzgg.equation}.
}
\label{mc.table}
 \end{center}
\end{table}

\section{$\Wg$ and $\Zg$ Followed by $\Wtaunu$ or $Z\goes \tptm$ 
and $\tau\rightarrow e\nu\nu$ or $\mu\nu\nu$}
\label{tau_contribution}

The last SM direct contribution (as opposed to misidentification) we
consider is $\Wg$ and $\Zg$ production followed by the boson leptonic
decay in the $\tau$ channel(``$\tau\gamma$ background''). The tau can
then decay into an electron or muon. These events are not fakes, in
the sense that the electron or muon is real, although not a direct
product of the vector boson decay.

Table~\ref{tau.table} gives a summary of the contributions to the
$\eg$ and $\mug$ channels from $\tau\gamma$ events ($\Wg$ and $\Zg$
decaying to taus). Shown are numbers of expected tau events from
different processes ($\Wg$, $\Zg$) making a signature of $e\gamma$,
$\mu\gamma$ and $l\gamma$ for the different categories defined in the
analysis. The final lepton is either an electron or muon.

\begin{table}[!h]
 \begin{center}
\begin{tabular} {|p{1.0cm}||p{2.7cm}|p{2.7cm}|p{2.7cm}|p{2.7cm}|p{2.7cm}|}
\hline

\rotatebox{75}{Categories} &
\rotatebox{75}{All Other } &
\rotatebox{75}{1$\ell$, 1$\gamma$, $\Delta\phi_{\lg}\lt$150} &
\rotatebox{75}{$\lgmet$ } &
\rotatebox{75}{Multi-Lepton } &
\rotatebox{75}{Multi-Photon } \\
\hline
$\eg$ & 1.13$\pm$0.20\xtiny(stat) & 0.42$\pm$0.12\xtiny(stat) & 0.71$\pm$0.16\xtiny(stat) & 0$\pm$0.041\xtiny(stat) & 0$\pm$0.041\xtiny(stat)\\
& $\pm$0.14\xtiny(sys) & $\pm$0.051\xtiny(sys) & $\pm$0.085\xtiny(sys) & $\pm$0\xtiny(sys) & $\pm$0\xtiny(sys)\\
\hline
$\mug$ & 0.66$\pm$0.12\xtiny(stat) & 0.38$\pm$0.11\xtiny(stat) & 0.26$\pm$0.072\xtiny(stat) & 0.013$\pm$0.042\xtiny(stat) & 0$\pm$0.041\xtiny(stat)\\
& $\pm$0.079\xtiny(sys) & $\pm$0.045\xtiny(sys) & $\pm$0.031\xtiny(sys) & $\pm$0.0016\xtiny(sys) & $\pm$0\xtiny(sys)\\
\hline
$\lg$ & 1.79$\pm$0.24\xtiny(stat) & 0.80$\pm$0.16\xtiny(stat) & 0.97$\pm$0.18\xtiny(stat) & 0.013$\pm$0.059\xtiny(stat) & 0$\pm$0.059\xtiny(stat)\\
& $\pm$0.21\xtiny(sys) & $\pm$0.096\xtiny(sys) & $\pm$0.12\xtiny(sys) & $\pm$0.0016\xtiny(sys) & $\pm$0\xtiny(sys)\\
\hline
\end{tabular}
\caption
[SM contributions from $\tau\gamma$ events]
{SM contributions from $\tau\gamma$ events ($\Wg$ and $\Zg$ decaying to taus).}
\label{tau.table}
 \end{center}
\end{table}
Background from $\W\goes\tau\nu$, where $\tau\goes\rho\nu$
($\approx$1/4 of the $\tau$ branching fraction), then
$\rho\goes\pi\pi^0$, which could mimic a single track + photon
signature that looks like $\lgmet$ is a part of $\tau\gamma$
background estimated.

\clearpage
\chapter{Backgrounds: Fakes}
\label{fakes}

In addition to the expectations from real SM processes that produce
real $\lg$ events described in Chapter~\ref{predictions},
there are backgrounds due to misidentified leptons and photons, and
also incorrectly calculated $\met$. We generically call these
misidentifications `fakes'. In this chapter we first treat backgrounds
from fake photons, then from fake leptons, including backgrounds to
the W samples due to events with a fake lepton and false $\met$. 

\section{Fake Photons}

We consider two sources of fake photons: QCD jets in which a neutral
hadron or photon from hadron decay mimics a direct photon, and
electron bremsstrahlung, in which an energetic photon is radiated off
of an electron which is then much lower energy and curls away from the
photon.

\subsection{Fake Photons from Jets}
\label{jet_fakes}

High $\Pt$ photons are copiously created from hadron decays in jets
initiated by a scattered quark or gluon. In particular, mesons such as
the $\pi^0$ or $\eta$ decay to photons which may satisfy the photon
selection criteria.

The number of lepton-plus-misidentified-jet events in the $\lgmet$ and
$\llg$ samples is determined by measuring the jet $\Et$ spectrum in
$\ell\met+$jet and $\ell\ell+$jet samples, respectively, and then
multiplying by the probability of a jet being misidentified as a
photon, $P^{jet}_{\gamma}(\Et)$. The uncertainty on the number of such
events is calculated by again using the measured jet spectrum and the
upper and lower bounds on the $\Et$-dependent misidentification
rate. An overview of the fake rate method is given in
Ref.~\cite{hayward_thesis}.

\begin{figure}[!h]
  \begin{center}
    \mbox{
\subfigure
{\epsfig{file=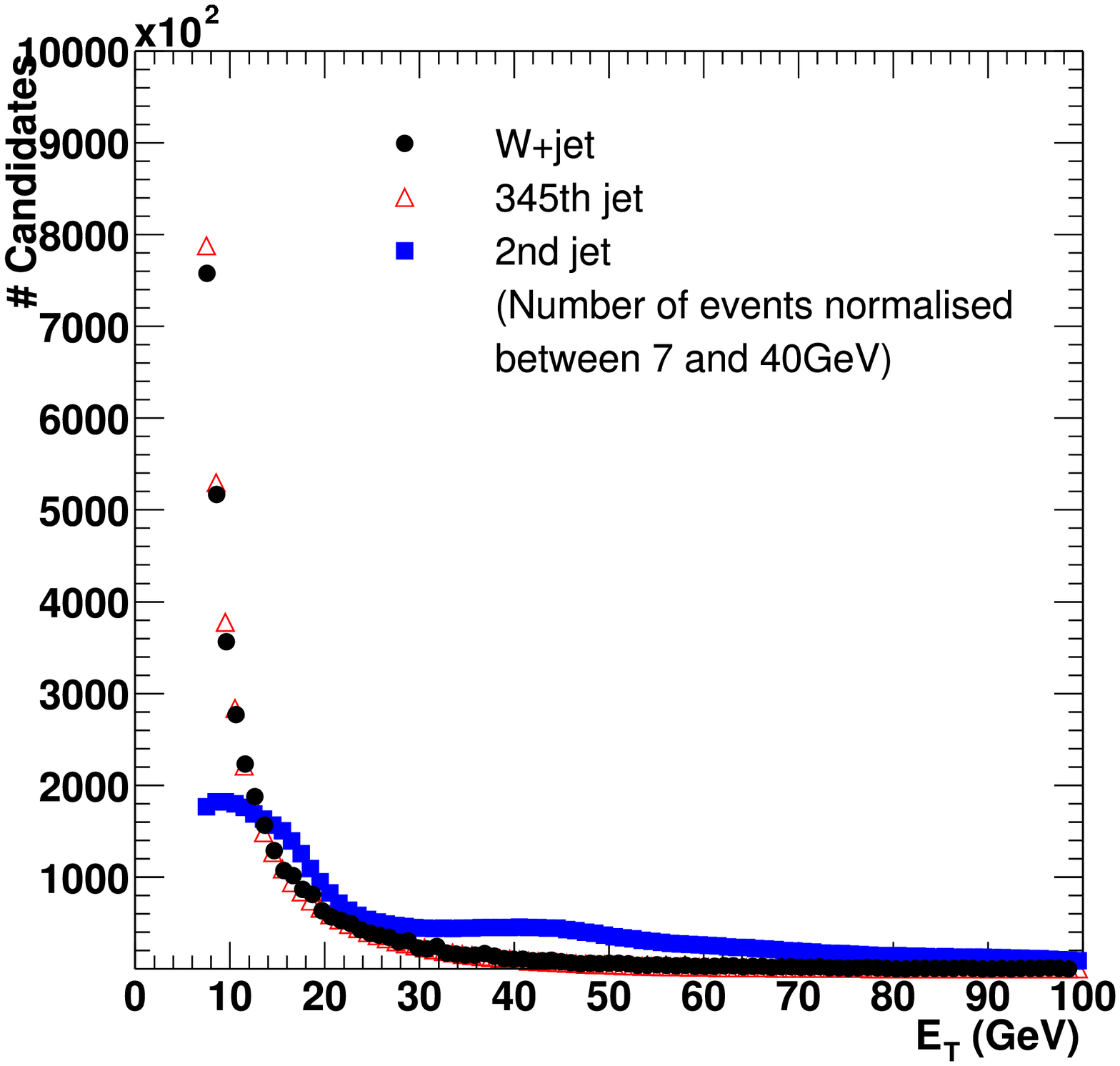,width=2.6in}} \quad
\subfigure
{\epsfig{file=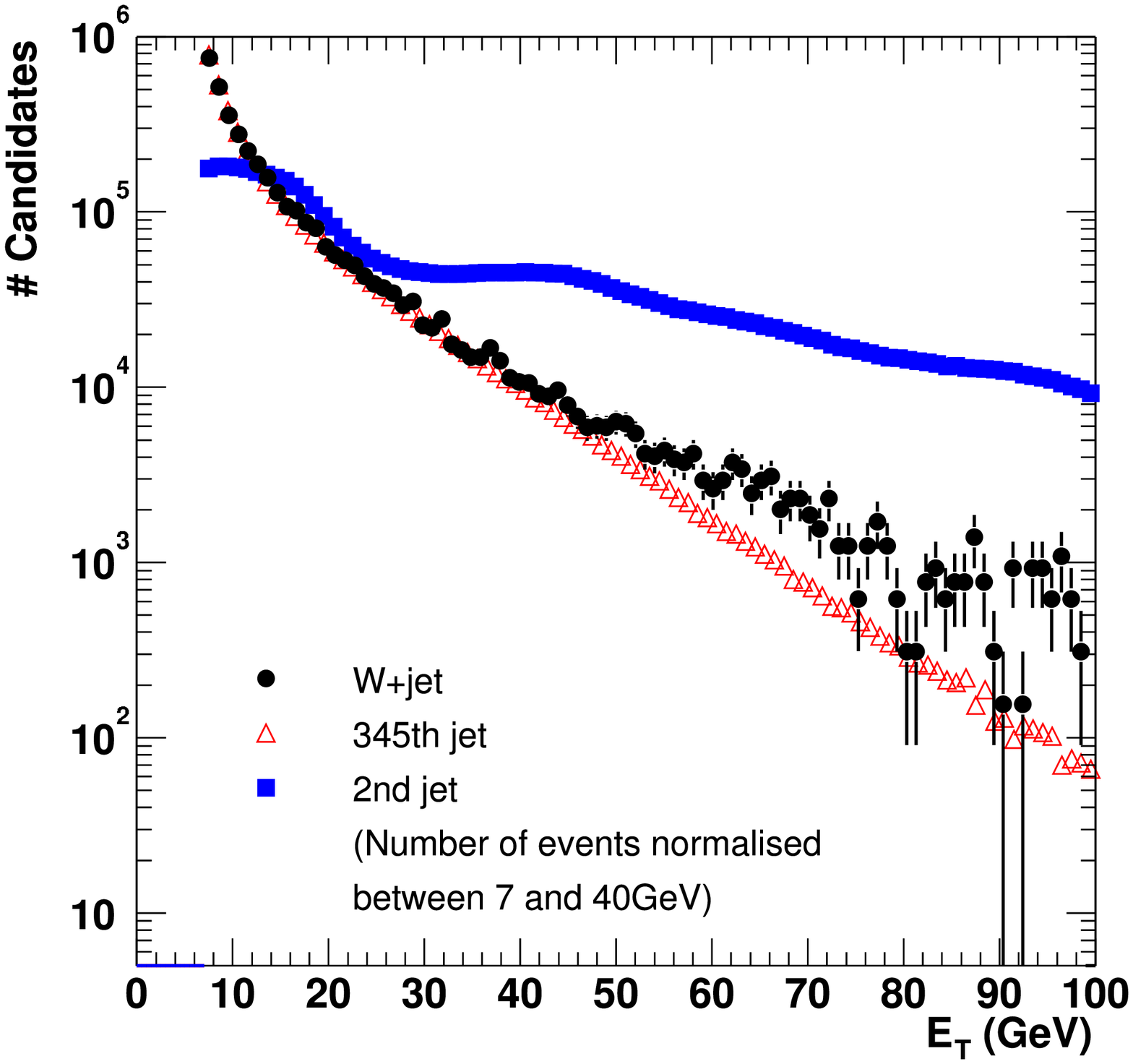,width=2.6in}} \quad
      }
\vskip -0.2in
\caption 
[Distributions for jets in W and jet samples: dN/d$\Et^{jet}$ vs $\Et^{jet}$]
{ The dN/d$\Et^{jet}$ vs $\Et^{jet}$ distribution for jets in W sample
(black points) and in jet samples. 2nd high-$\Et$jet in the jet
samples is shown in blue squares and the 345th jet in red open
triangles. }
\label{jetfakes_wgzg.figure} \end{center}
\end{figure}

Any photon that is due the decay of a meson
($\pi^0\rrr\gg$, $\eta\rrr\gg$) is classified as FAKE and any photon
that is created in the hard scattering process or radiated off a quark
is classified as a TRUE photon. The strategy is first to measure the
RAW fake rate. The fraction of jets which are matched to a photon
candidate

\begin{equation}
P_{RAW}(\Et^{jet})
 = \frac{N_{\gamma-candidate}}{N_{jet}}
 = \frac{N_{\gamma}^{TRUE}+N_{\gamma}^{FAKE}}{N_{jet}}
 \label{fakerate.equation}
\end{equation}

We estimate TRUE fake as

\begin{equation}
P_{TRUE}(\Et^{jet})
 = P_{RAW}(\Et^{jet})\times F_{QCD}, 
~~F_{QCD}
 = \frac{N_{\gamma}^{FAKE}}{N_{\gamma}^{TRUE}+N_{\gamma}^{FAKE}}
 \label{fakerate.equation}
\end{equation}

To distinguish $\gamma$ from $\pi^0$ or other hadrons, the following
variables have been used in the jet fake rate studies:

\begin{itemize}
\setlength{\itemsep}{-0.07in}
\item 
{CES $\chi^2$} \\
-- $\pi^0\rightarrow\gg$ typically have higher $\chi^2$ than prompt photons
\item 
Isolation in a cone in $\eta-\varphi$ space of radius R=0.4 around the
$\gamma$ candidate \\
-- the background is usually produced as a part of a jet $\Rightarrow$
$\Et$ is higher than for $\gamma$
\item
Hit rate in the CPR (Central Preradiator Detector, see Section~\ref{detector_calorimetry}) \\
-- CPR is between solenoid and calorimeter. $\gamma$ converts in the
coil and therefore we measure charge. Photons and fakes have different
conversion probabilities.
\end{itemize}

\begin{figure}[!h]
  \begin{center}
    \mbox{
\subfigure
[$F_{QCD}$]
{\epsfig{file=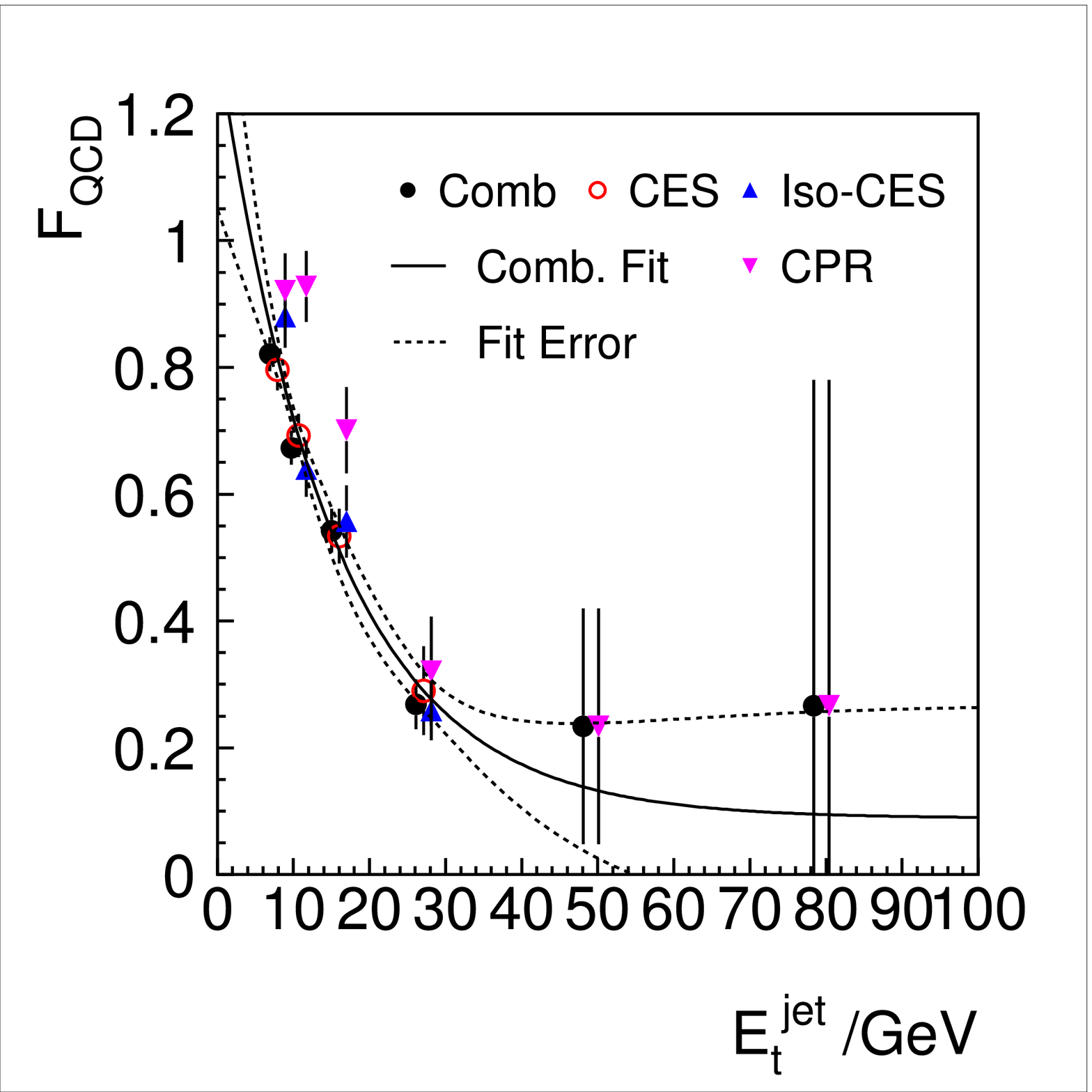,width=2.5in}} \quad
\subfigure
[$P_{TRUE}^{jet\rightarrow\gamma}$]
{\epsfig{file=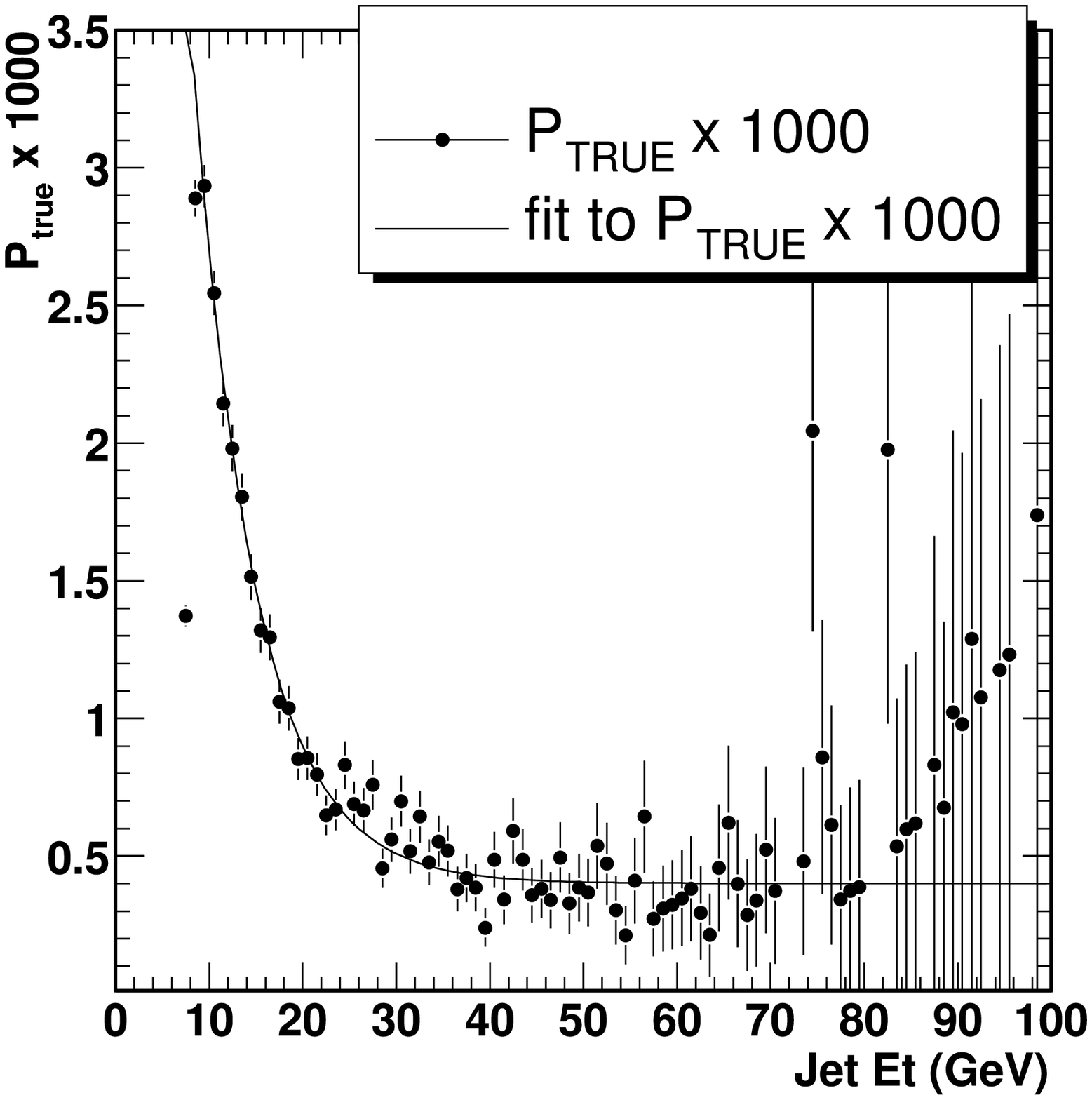,width=2.5in}} \quad
      }
    \caption 
[$F_{QCD}$ and a probability for a jet to fake a photon $P_{TRUE}^{jet\rightarrow\gamma}$]
{(a) $F_{QCD}$; (b) a probability for a jet to fake a photon.}
\label{fqcd_ptrue.figure} \end{center}
\end{figure}

Technical details on the studies of jets faking photons in CDF II are
available in Ref.~\cite{jetfakes}. The most recent numbers are
available in Ref.~\cite{hayward_thesis} and resulting distribution for
$F_{QCD}$ and $P_{TRUE}$ are shown in
Figure~\ref{fqcd_ptrue.figure}. We follow this study closely in our
estimates. We use jets from the inclusive high-$\Pt$ muon and
high-$\Et$ electron lepton samples (Section~\ref{datasets}) to
evaluate the background from jets faking photons.

The above studies show that the (fake) photon carries about 94\% of
the jet $E_T$, and the resolution on the photon $E_T$ is about 5\%. We
consequently scale and smear the jet $E_T$ by these numbers to get the
$E_T$ of the fake photon. We then select jets with $|\eta|\lt 1.1$ and
$E_T>$ 25 GeV after the scaling and smearing.  The jets are then
weighted with by the jet fake rate~\cite{jetfakes}. 

\begin{figure}[!h]
  \begin{center}
    \mbox{
\subfigure
[$\lgmet$ (Electron Channel)]
{\epsfig{file=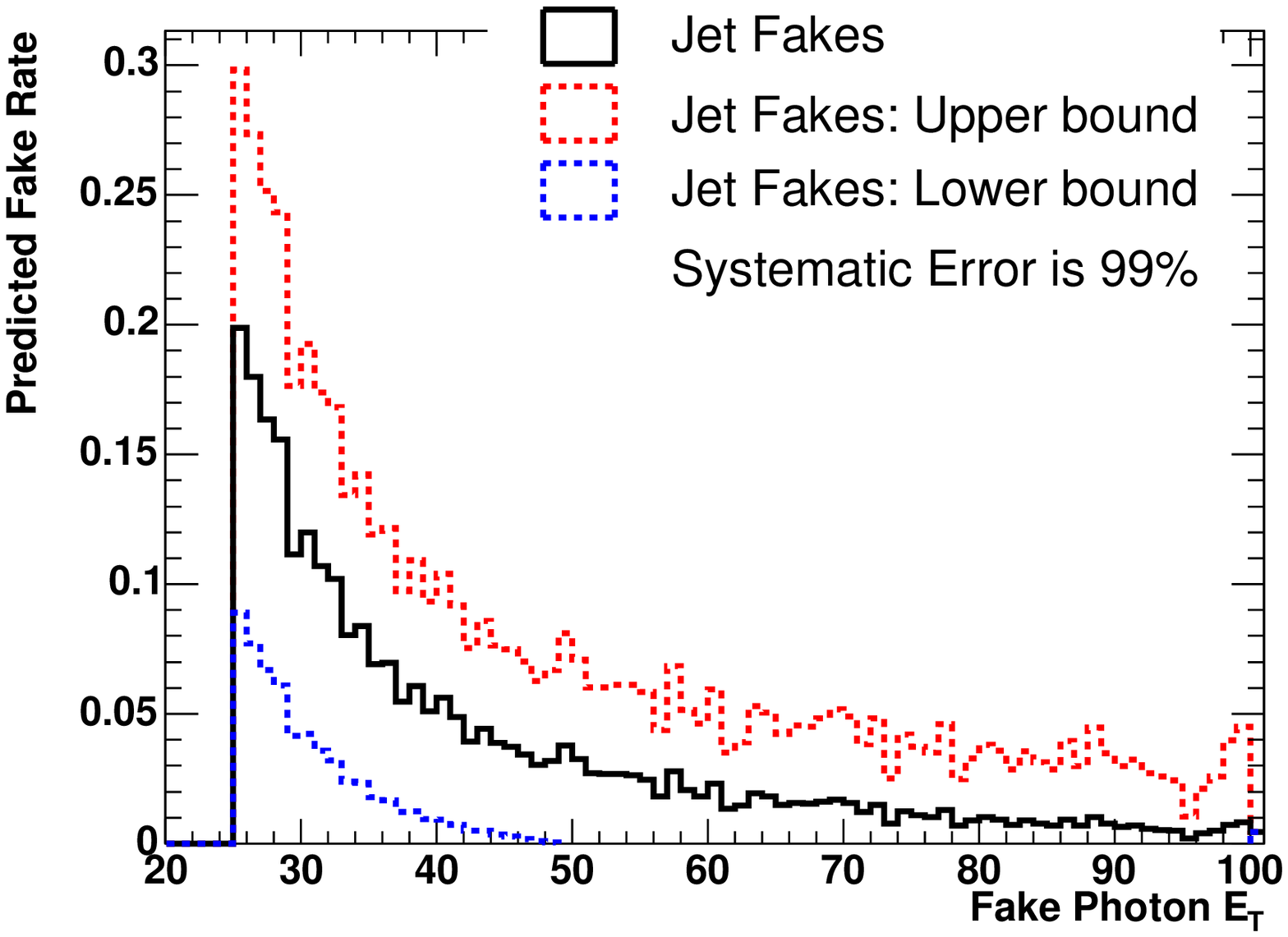,width=2.5in}} \quad
\subfigure
[$\llg$ (Electron Channel)]
{\epsfig{file=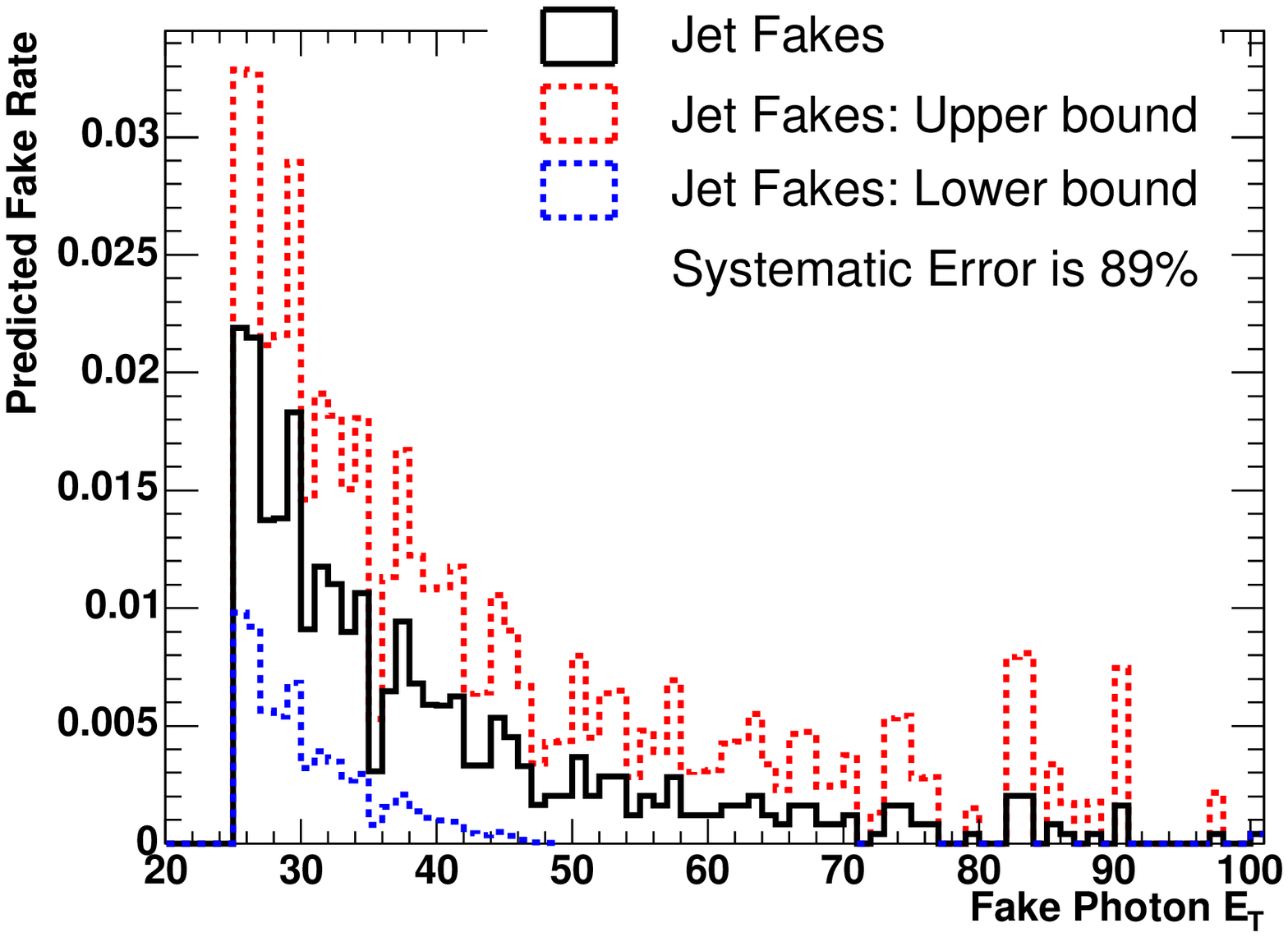,width=2.5in}} \quad
      }
    \mbox{
\subfigure
[$\lgmet$ (Muon Channel)]
{\epsfig{file=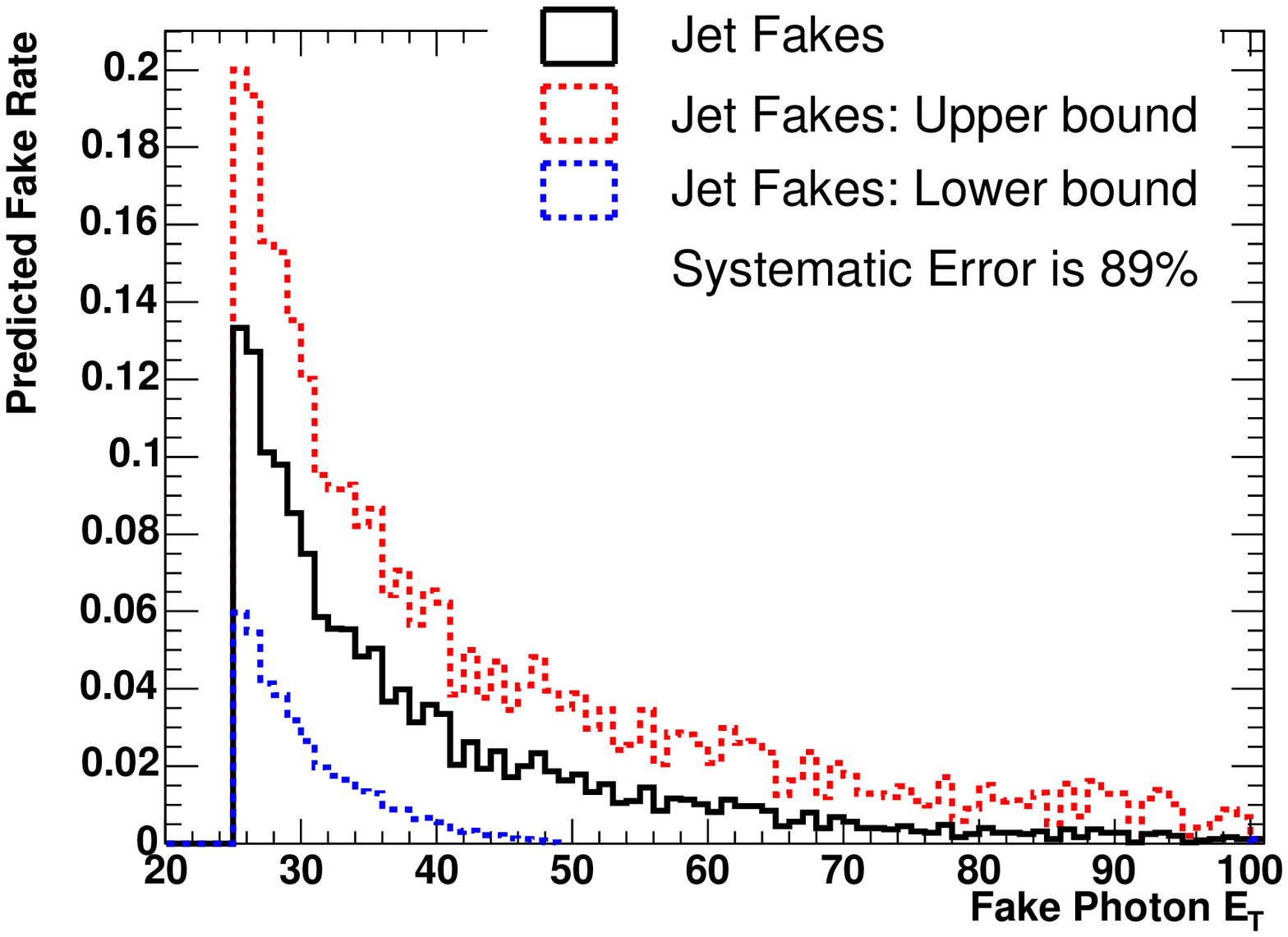,width=2.5in}} \quad
\subfigure
[$\llg$ (Muon Channel)]
{\epsfig{file=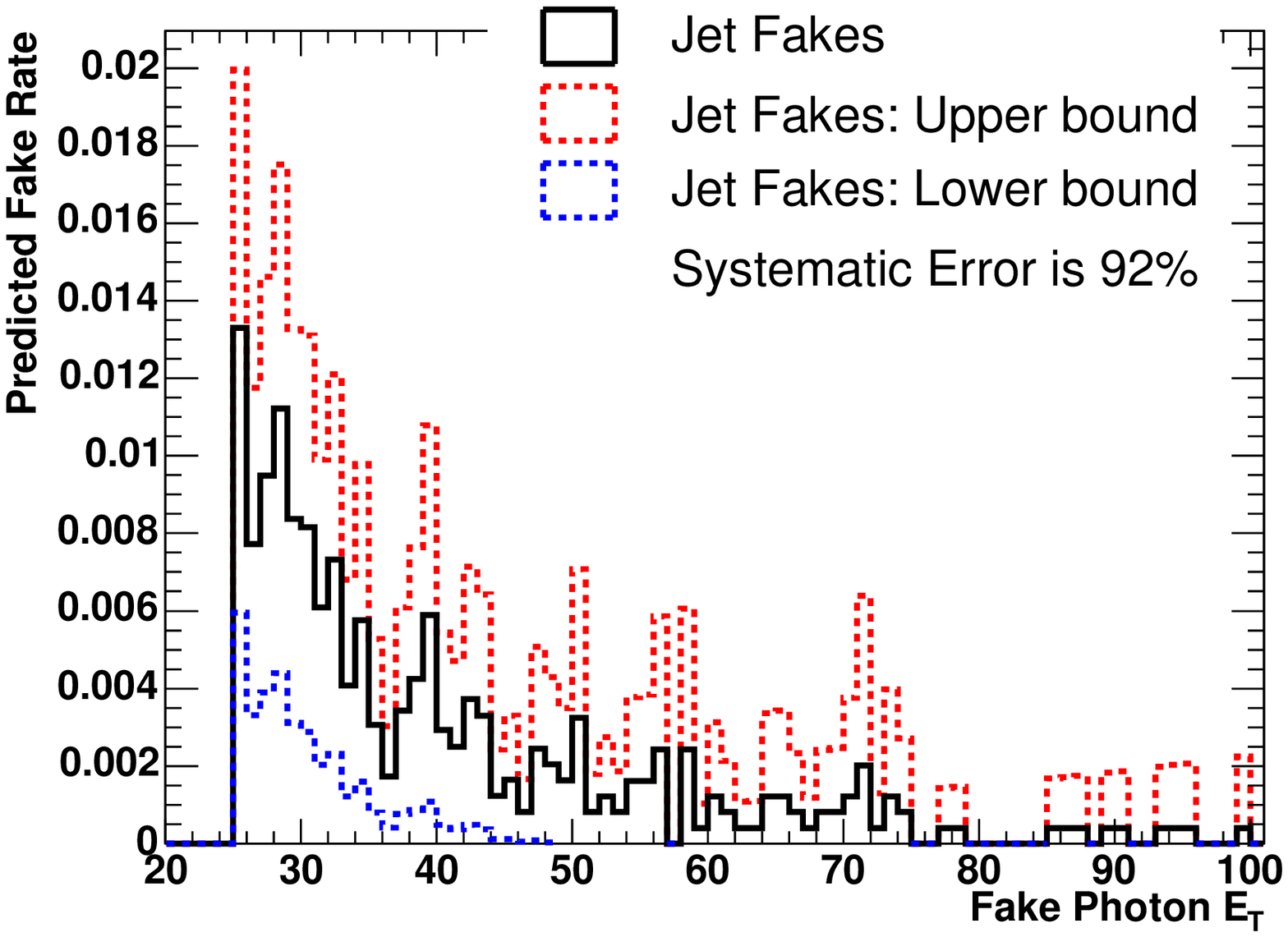,width=2.5in}} \quad
      }
\caption 
[The $E_T$ spectrum of $j\rrr\gamma$ events expected in $\lgmet$ and $\llg$ signatures]
{The $E_T$ spectrum of events expected in $\apprluminosity\pbinv$ with
a fake photon from a jet versus the $E_T$ of the fake photon. The four
plots show the number of events per 2 GeV bin expected in the $\lgmet$ and
$\llg$ signatures in both the electron and muon channels. The upper and
lower error bands from Ref.~\cite{hayward_thesis} are shown.}
\label{jetfakes_wgzg.figure} 
\end{center}
\end{figure}

\begin{table}[!h]
\begin{tabular} {|p{1.0cm}||p{2.5cm}|p{2.5cm}|p{2.5cm}|p{2.5cm}|p{3.8cm}|}
\hline

\rotatebox{75}{Categories} &
\rotatebox{75}{All Other } &
\rotatebox{75}{1 $l$, 1 $\gamma$, $\Delta \phi < $150 } &
\rotatebox{75}{$\lgmet$ } &
\rotatebox{75}{Multi-Lepton } &
\rotatebox{75}{Multi-Photon } \\

\hline
$\eg$                                         & 
$\nofejgallother$$\pm$$\dsysnofejgallother$\xtiny(tot)       &
$\nofejgzunlike$ $\pm$$\dsysnofejgzunlike$\xtiny(tot)        &
$\nofejglgmet$   $\pm$$\dsysnofejglgmet$\xtiny(tot)          &
$\nofejgmultil$  $\pm$$\dsysnofejgmultil$\xtiny(tot)         &
$\nofejgmultig$  $\pm$$\dsysnofejgmultig$\xtiny(tot)         \\
\hline
$\mug$                                               &
$\nofmujgallother$$\pm$$\dsysnofmujgallother$\xtiny(tot)       &
$\nofmujgzunlike$ $\pm$$\dsysnofmujgzunlike$\xtiny(tot)        &
$\nofmujglgmet$   $\pm$$\dsysnofmujglgmet$\xtiny(tot)          &
$\nofmujgmultil$  $\pm$$\dsysnofmujgmultil$\xtiny(tot)         &
$\nofmujgmultig$  $\pm$$\dsysnofmujgmultig$\xtiny(tot)         \\
\hline
\hline
$\lg$                                             &
$\nofljgallother$$\pm$$\dsysnofljgallother$\xtiny(tot)       &
$\nofljgzunlike$ $\pm$$\dsysnofljgzunlike$\xtiny(tot)        &
$\nofljglgmet$   $\pm$$\dsysnofljglgmet$\xtiny(tot)          &
$\nofljgmultil$  $\pm$$\dsysnofljgmultil$\xtiny(tot)         &
$\nofljgmultig$  $\pm$$\dsysnofljgmultig$\xtiny(tot)         \\
\hline
\end{tabular}
\caption
[The predicted backgrounds from $j\rrr\gamma$ in the analysis subcategories]
{The predicted backgrounds from jets faking photons
in the analysis subcategories. We estimate systematic uncertainty on
the predicted number to be 100\%. Statistical errors are
negligible. The uncertainties in the electron and muon samples are
assumed to be 100\% correlated.}
\label{jfakep.table}
\end{table}

The number of fake events versus the $E_T$ of the fake photon are
shown in Figure~\ref{jetfakes_wgzg.figure} for the $\lgmet$ and $\llg$
samples in both the electron and muon channels. We count lepton-`fake
photon` candidates in the same way as we do for real photons to
calculate background for each subcategory.  Table~\ref{jfakep.table}
summarizes the number of events in the $\eg$ and $\mug$ samples for
the sub-categories used in the analysis.

\subsection{Fake Photons from Electron Bremsstrahlung}
\label{egamma_fakes}

We can measure the probability that a high-$\Et$ electron `brems' an
energetic photon in the material before the COT tracking volume in
such a way that it fakes a photon, by using the $\Z$ as a source of
`tagged' electrons. We look for a back-to-back $\epem$ pair close to
the $\Z$ mass ($\Delta\phi_{ee} > 150\deg$ and $81~\GeV<
M_{ee}<101~\GeV$) - this is the sub-category labeled `Z-like'. For the
$\eg$ sample the requirements are an exactly 1 electron and 1 photon,
with $\Delta\phi_{\eg} > 150\deg$ and $81~\GeV<
m_{e\gamma}<101~\GeV$. From the number of these events and the number
of $\Zee$ events we can measure the probability per `leg' of a $\Z$
that an electron is misidentified as a photon.  We require both
electrons in $\Zee$ to be central and pass tight cuts, so that the
kinematic requirements will be the same as for the $\eg$ sample
(i.e. two tight central electromagnetic objects).

The events in the 2nd row (labeled as $\Zee$) of the
Table~\ref{efakep.table} have the same signature as $\lg$ events (see
Figure~\ref{flowchart_electron.figure}), but instead of a photon we
require a tight central electron(Table~\ref{electron.table}).

For instance, to estimate contribution from electron faking photon to
$\lgmet$ category we count events with ee+$\met$ (both electrons are
tight central, Table~\ref{electron.table}).  For Multi-Lepton + Photon
we count events with three electrons, of which two electrons should be
tight central, and the third one can be tight or loose central
(Table~\ref{electron_cuts.table}), or phoenix tight
(Table~\ref{plugelectron_cuts.table}). Finally, for Multi-Photon +
Lepton category we use $ee\gamma$ events, both electrons are Tight
Central.

Using the numbers in Table~\ref{efakep.table} we calculate the
background from electrons misidentified as photons.  For example, for
$\egmet$ subcategory the estimated number of electron-fake-photon
events(``$e~\rightarrow~\gamma~fakes$'') is calculated as follows:

\begin{equation}
N^{e~\rightarrow~\gamma~fakes}_{\egmet} = 
N^{\Zee}_{\eemet}\times
\frac{N^{e\gamma}_{\Z-like} - N^{diboson}_{\Z-like} - N^{jets}_{\Z-like}}
{N^{\Zee}_{\Z-like}}
 \label{efakep.equation}
\end{equation}

\begin{itemize}
\setlength{\itemsep}{-0.05in}
\item $N^{\Zee}_{\eemet}~\equiv$  $\eemet$ events in the $\Zee$ sample

\item $N^{e\gamma}_{\Z-like}~~~~~~~~\equiv$ $\eg$ $\Z$-like events

\item $N^{diboson}_{\Z-like}~~~~~~~~\equiv$ $\eg$ $\Z$-like events expected from diboson events($\Wg$, $\Zg$)

\item $N^{jets}_{\Z-like}~~~~~~~~\equiv$ $e$+{\it jet faking photon} $\Z$-like events expected from misidentified jets

\item $N^{\Zee}_{\Z-like}~\equiv$ $\Zee$ events (81 GeV $< M_\Z < $ 101 GeV, $\Delta\phi_{ee} >$ 150)

\end{itemize}

We take all numbers from data, with the exception that we get number
of $\Z-like$ events expected from diboson events, $\Wg$ and $\Zg$,
from MC (we take into account contribution from $\Zgg$ and $\Wgg$,
although it's $\approx$ 1\% of that from $\Zg$ and $\Wg$).

Finally, we estimate the number of electron-fake-photon events in
$\lgmet$ to be

\begin{equation}
N^{e~\rightarrow~\gamma~fakes}_{\egmet} = 
\nofzeemet*\frac{\nofegzlike\pm\dnofegzlike - \mcnofegzlike\pm\mcdnofegzlike - \nofejgzlike\pm\dsysnofejgzlike}{\nofzeetight}
=\efpnofeglgmet\pm\efpdnofeglgmet
 \label{efakep_lgmet.equation}
\end{equation}

\begin{table}
\small
\begin{tabular} {|p{1.8cm}||p{2.3cm}||p{1.9cm}|p{1.9cm}|p{1.9cm}|p{1.9cm}|p{2.9cm}|}
\hline

\rotatebox{75}{Categories} &
\rotatebox{75}{$\Z$-like } &
\rotatebox{75}{All Other } &
\rotatebox{75}{1 $l$, 1 $\gamma$, $\Delta \phi < $150 } &
\rotatebox{75}{$\lgmet$ } &
\rotatebox{75}{Multi-Lepton } &
\rotatebox{75}{Multi-Photon } \\
\hline
$\Zee$         &
$\nofzeetight$       &
$\nofzeeallother$    &
$\nofzeezunlike$     &
$\nofzeemet$         &
$\nofzeemultil$      &
$\nofzeemultig$      \\
\hline
Diboson        &
$\mcnofegzlike     \pm  \mcdnofegzlike$          &
$\mcnofegallother  \pm  \mcdnofegallother$       &
$\mcnofegzunlike   \pm  \mcdnofegzunlike$        &
$\mcnofeglgmet     \pm  \mcdnofeglgmet$          &
$\mcnofegmultil    \pm  \mcdnofegmultil$         &
$\mcnofegmultig    \pm  \mcdnofegmultig$         \\
\hline
Jet fakes      &
$\nofejgzlike      \pm  \dsysnofejgzlike$          &
$\nofejgallother   \pm  \dsysnofejgallother$       &
$\nofejgzunlike    \pm  \dsysnofejgzunlike$        &
$\nofejglgmet      \pm  \dsysnofejglgmet$          &
$\nofejgmultil     \pm  \dsysnofejgmultil$         &
$\nofejgmultig     \pm  \dsysnofejgmultig$         \\
\hline
\hline
$e\goes\gamma$ fakes          &
$\efpnofegzlike    \pm  \efpdnofegzlike$          &
$\efpnofegallother \pm  \efpdnofegallother$       &
$\efpnofegzunlike  \pm  \efpdnofegzunlike$        &
$\efpnofeglgmet    \pm  \efpdnofeglgmet$          &
$\efpnofegmultil   \pm  \efpdnofegmultil$         &
$\efpnofegmultig   \pm  \efpdnofegmultig$         \\
\hline
\end{tabular}
	
\caption
[The predicted backgrounds from $e\rrr\gamma$ in the analysis subcategories]
{Bottom row: The calculated number of events in each analysis
subcategory for electrons faking photons by catastrophic
bremsstrahlung. The upper three rows are the input numbers used in the
calculation, which is given by Equation~\ref{efakep.equation}. Only
statistical errors are quoted. Systematic errors estimated by varying
Z mass window are found to be negligible.}
\label{efakep.table}
\end{table}

Table~\ref{efakep.table} summarizes the calculated number of events in
each analysis subcategory for electrons faking photons by catastrophic
bremsstrahlung. The upper three rows are the input numbers used in the
calculation, which is given by Equation~\ref{efakep.equation}. The last
row gives the estimated number of events detected with fake photons
from electron bremsstrahlung.

\section{QCD ('Non-W/Z') Backgrounds}
\label{nonwz}
To measure the QCD backgrounds from fake leptons and or fake $\met$,
we form a `non-W/Z' sample we expect to have very little real lepton
content~\cite{Sacha_Kopp_thesis} by selecting on loose leptons,
rejecting events from the W or Z.

To estimate Non-W/Z background we use four samples:

\begin{itemize}
\setlength{\itemsep}{-0.09in}
\item {\it Non-W/Z sample}: 1 $\ell$ + jet(s), no W or Z candidates
\item {\it Signal Sample}: $\lgmet$ or $ll\gamma$
\item {\it Golden-Lepton}: tight e's, CC $\Zee$; tight $\mu$'s $\Zmumu$
\item {\it Jet Faking Photon}: $\ljmet$ or $\llj$
\end{itemize}

In these samples we define three track isolation regions:

\begin{itemize}
\setlength{\itemsep}{-0.09in}
\item {\it Track-Isolated}: 0 GeV $\lt$ Track Isolation $\lt$ 2 GeV
\item {\it Non-Track-Isolated}: Track Isolation $\gt$4 GeV
\item {\it Intermediate}: 2 GeV $\lt$ Track Isolation $\lt$4 GeV
\end{itemize}

The procedure we use is to:

\begin{itemize}
\setlength{\itemsep}{-0.09in}
\item
Estimate fraction of {\it golden} leptons with bad track isolation
\item
Estimate fraction of {\it non-W/Z} leptons with good track isolation
\item
Estimate {\it QCD(Jet faking lepton and $\met$)} background
\item
Vary track isolation regions to get systematics
\end{itemize}

We describe the selection and the procedure in detail below. We define
'Non-W/Z' sample in Section~\ref{nonwz_sample}. We describe the basic
method in Section~\ref{TrackIsoMethod}, and then modify it to avoid
double counting and to include systematics in
Section~\ref{ModifiedTrackIsoMethod}. We estimate QCD (non-W/Z)
background for $\Wenu$ and $\Wmunu$ in Section~\ref{nonwz_for_w} to
make sure we obtain results consistent with Ref.~\cite{CDF_WZ}.

\subsection{Non-W/Z Sample}
\label{nonwz_sample}

To select an event for the Non-W/Z sample we require the event to have
no W or Z candidates. Therefore, we require the event to have exactly
one tight lepton and no additional loose leptons or $\met>10~\GeV$. We
also require the event to have at least one hadronic jet. For the
electron Non-W/Z sample the jet is required to have EM Fraction $\lt$
0.8, $N_{tracks~in~the~jet}\gt 2$, $E_T^{jet}\gt 20$ GeV. For the muon
Non-W/Z sample the jet is required to have 0.1 $<$ EM Fraction $<$
0.9, $N_{tracks~in~the~jet}\gt 2$, $E_T^{jet}\gt 20$ GeV.

To reject $\zmumu$ events in which one muon is not identified, events
with a second track with $\Pt>10$ GeV and associated EM and HAD
calorimeter energies less than 3 and 9 GeV, respectively, are rejected
(we used the same requirements to select $\Wmunu$ control sample in
Section~\ref{wmunu_control_sample}).

\begin{figure}[!h]
  \begin{center}
    \mbox{
\subfigure[{Track Isolation}]
{\epsfig{file=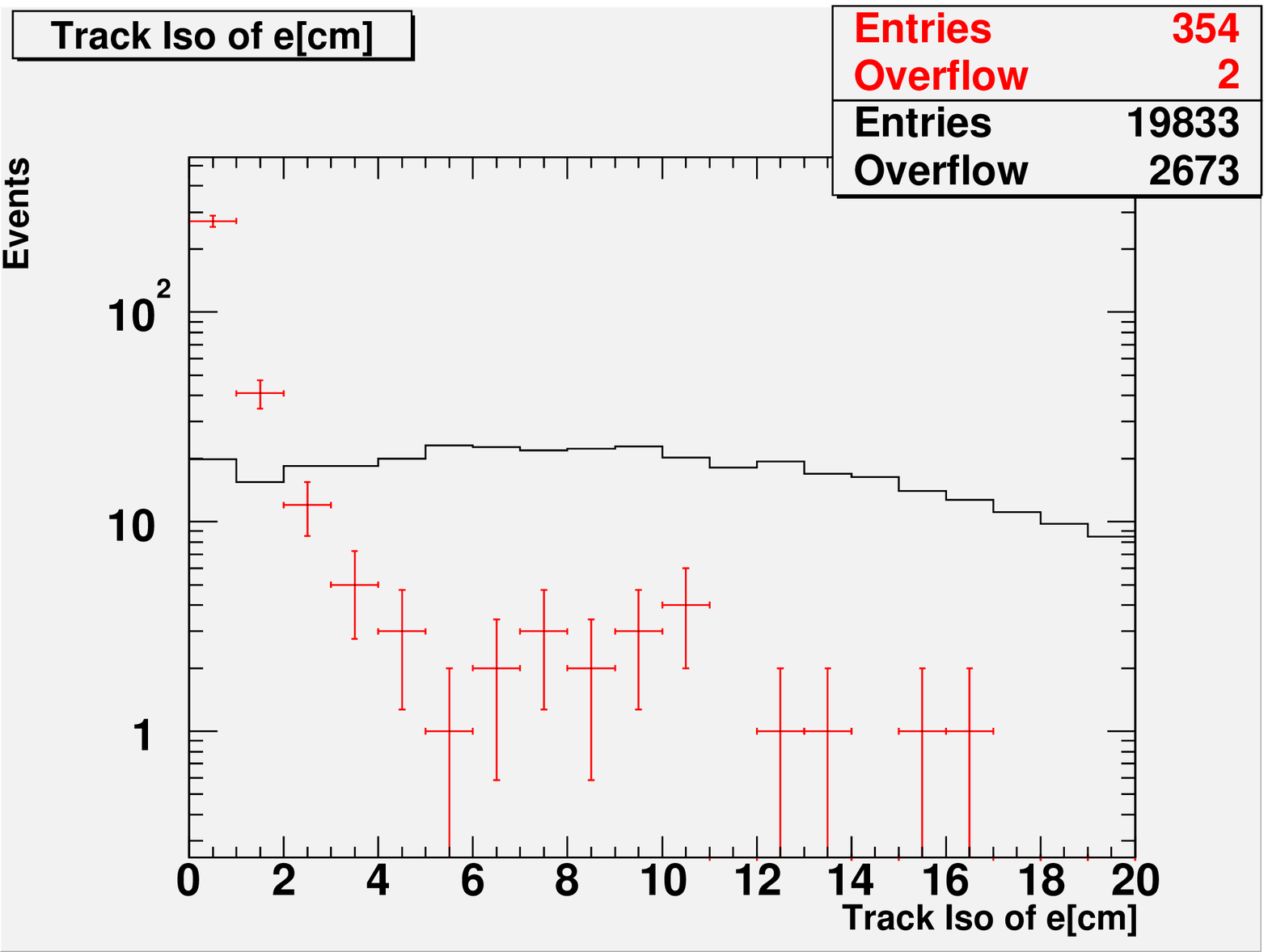, width=1.5in}} \quad
\subfigure[{Iso $E_T$}]
{\epsfig{file=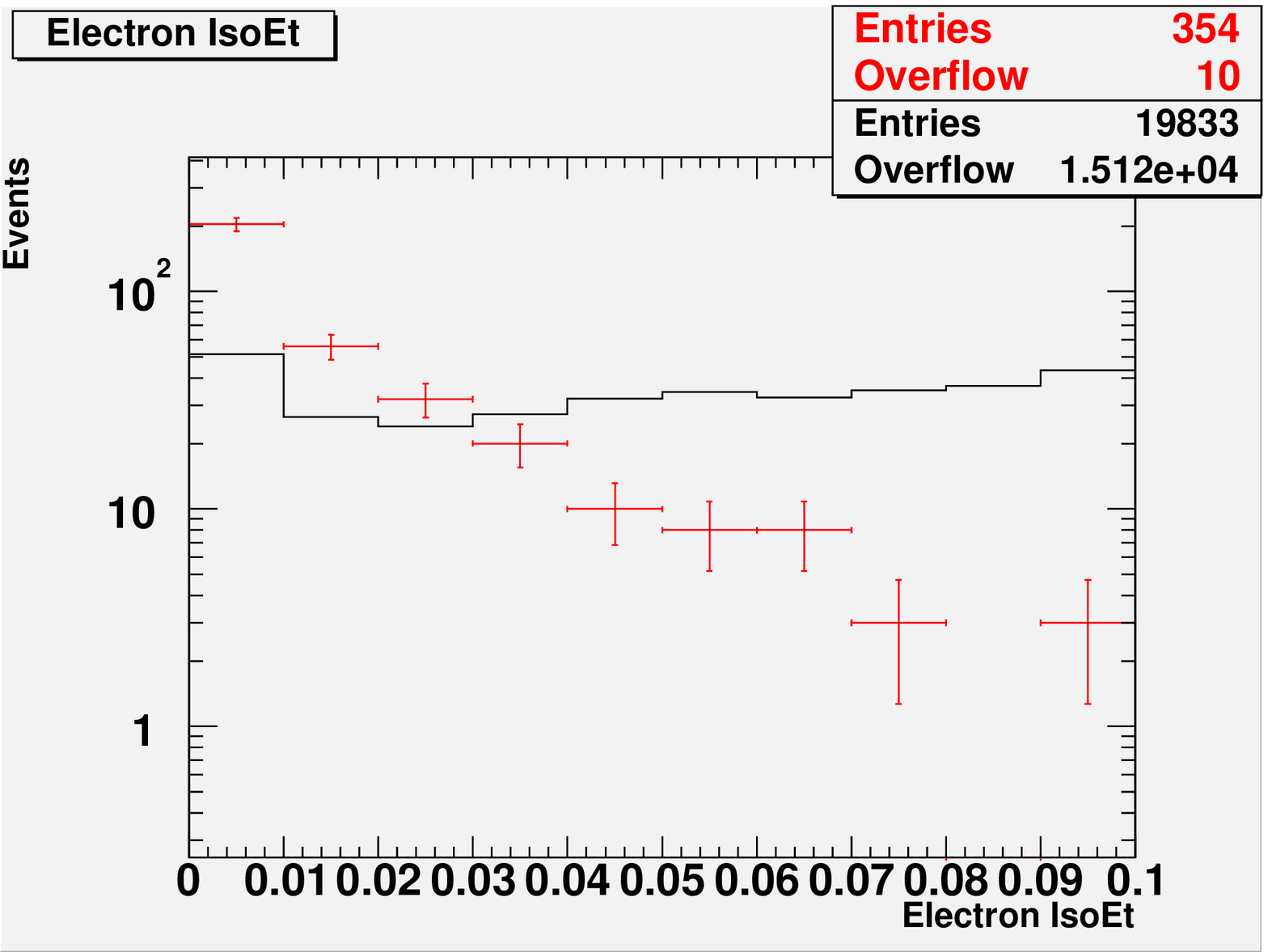, width=1.5in}} \quad
\subfigure[{Electron $E_T$}]
{\epsfig{file=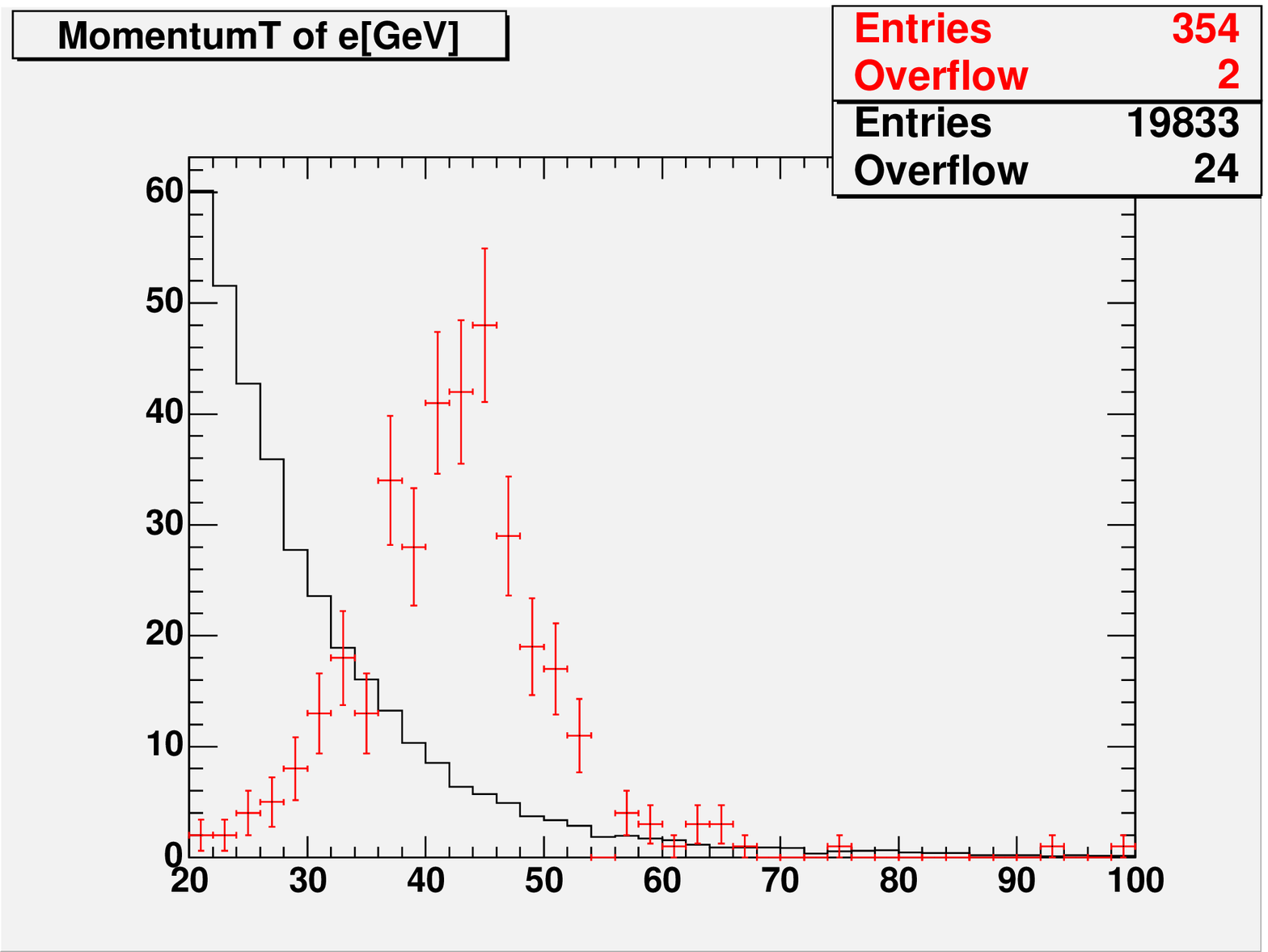, width=1.5in}} \quad
\subfigure[{Conversion Flag}]
{\epsfig{file=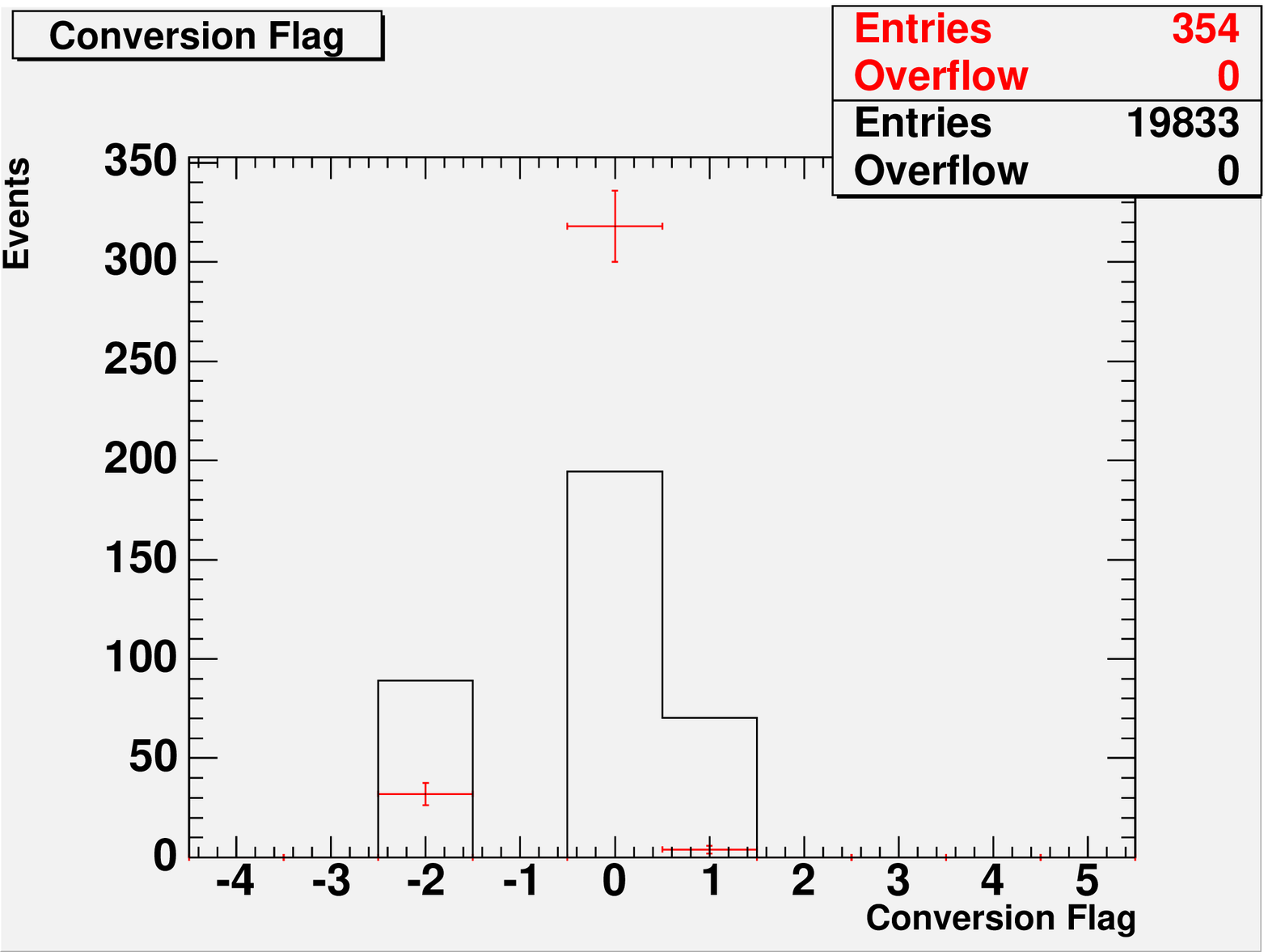, width=1.5in}} \quad
}
    \mbox{
\subfigure[{Track Isolation}]
{\epsfig{file=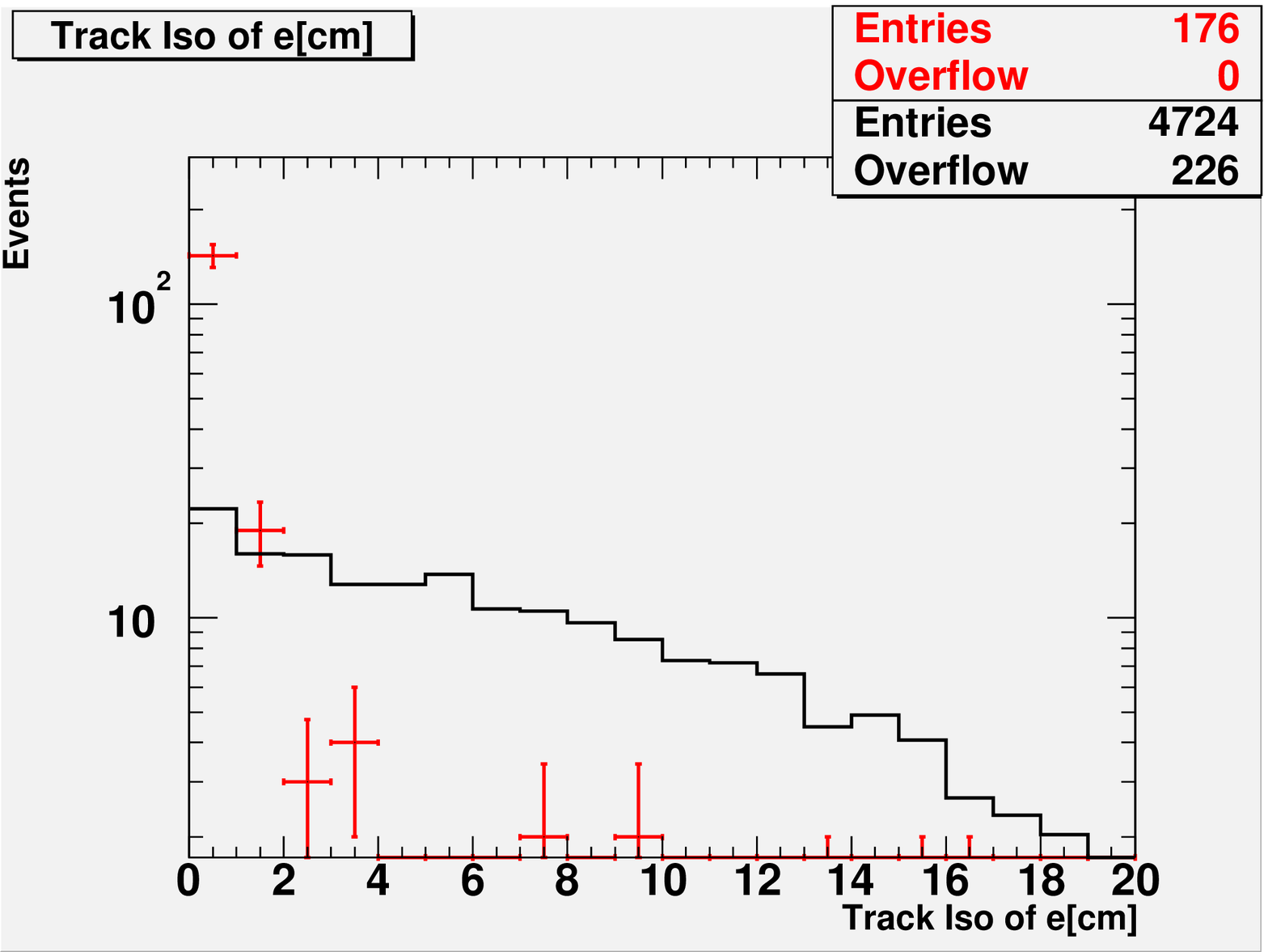, width=1.5in}} \quad
\subfigure[{Iso $E_T$}]
{\epsfig{file=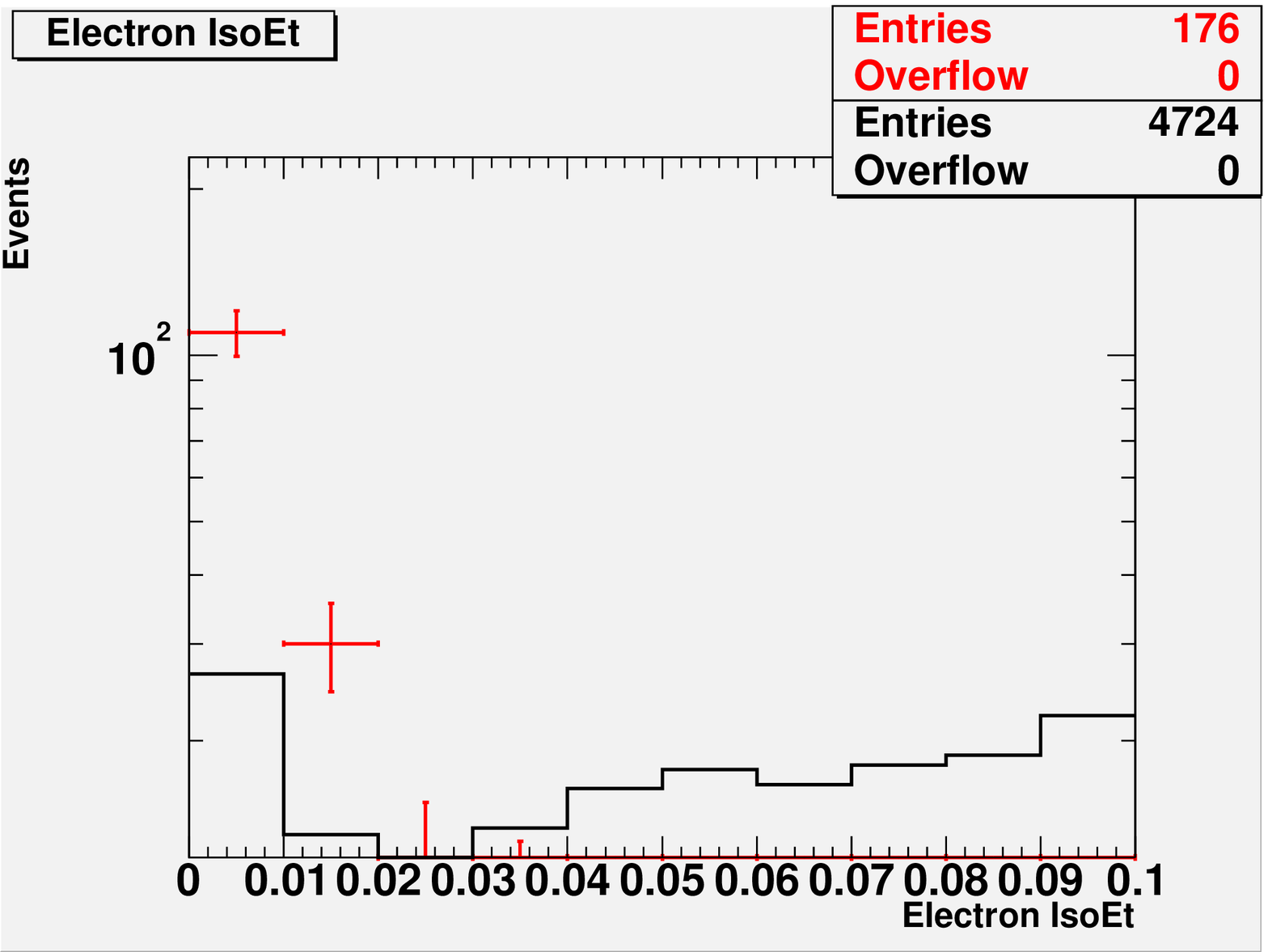, width=1.5in}} \quad
\subfigure[{Electron $E_T$}]
{\epsfig{file=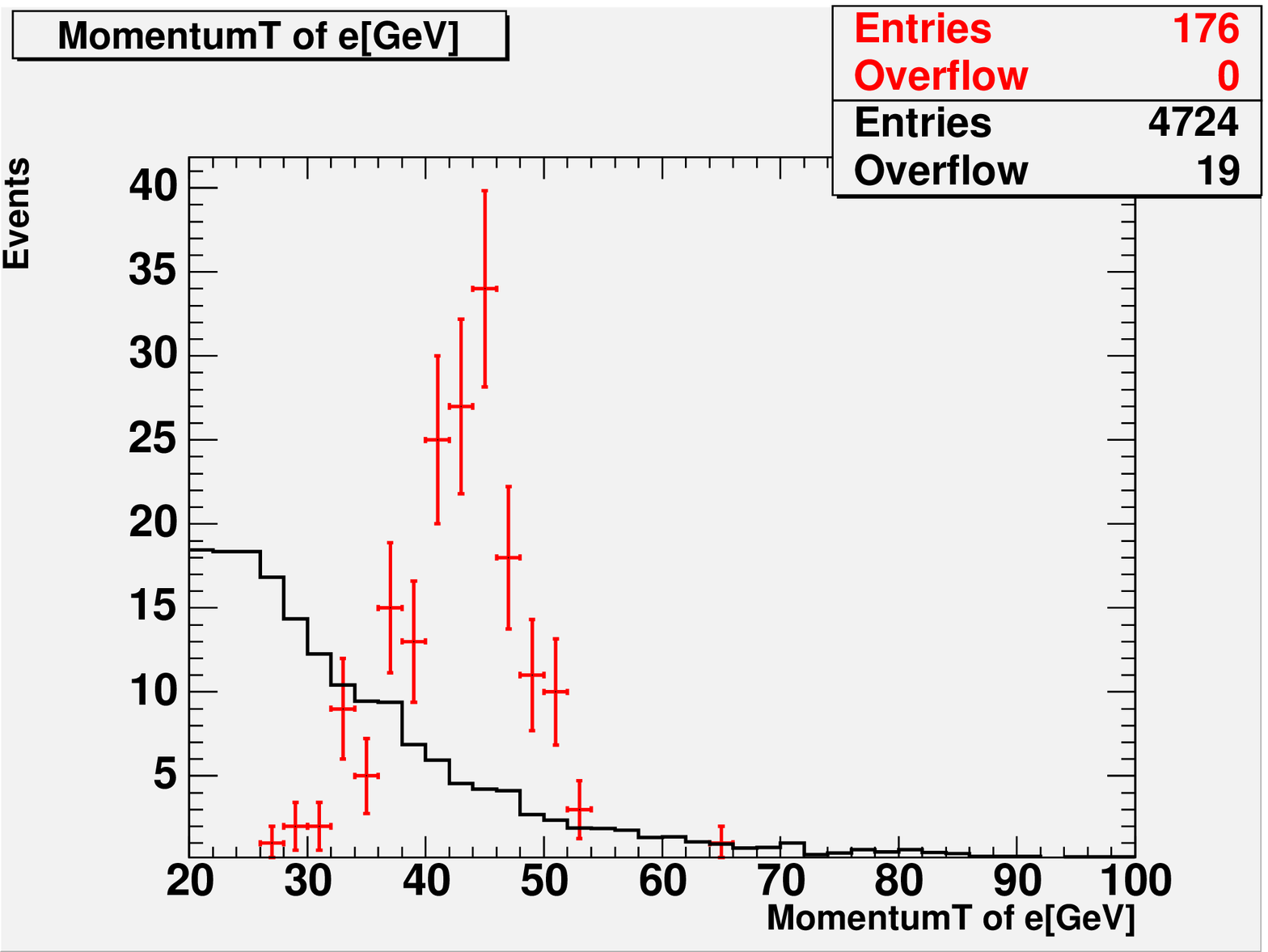, width=1.5in}} \quad
\subfigure[{Conversion Flag}]
{\epsfig{file=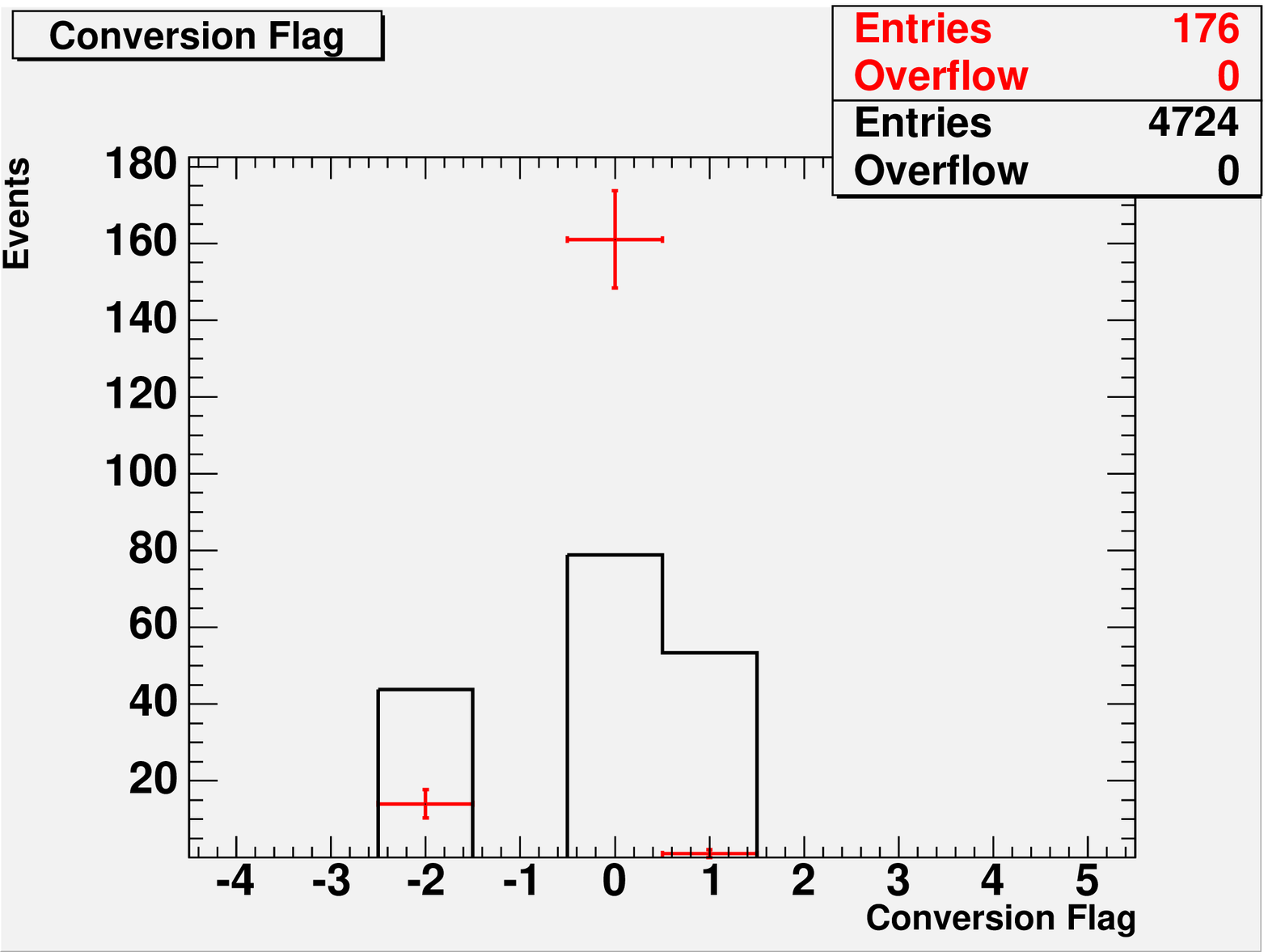, width=1.5in}} \quad
}
    \caption 
[Comparison of distributions for the electrons from Non-WZ and $\Zee$ samples]
{A comparison of distributions from fake `electrons'
from the Non-WZ sample(black histogram) and from electrons from tight
Z's (red dots). The top set of plots has no (calorimeter) 
IsoEt cut applied; the IsoEt cut is applied in the  bottom plots.
The plot labeled `Conversion' shows the value of the conversion flag,
where the meaning is: 0 - not a conversion, 1 - conversion
electron, -2 - trident.  }
    \label{nonwz_electrons.figure}
  \end{center}
\end{figure}

\begin{figure}[!h]
  \begin{center}
    \mbox{
\subfigure[{Track Isolation}]
{\epsfig{file=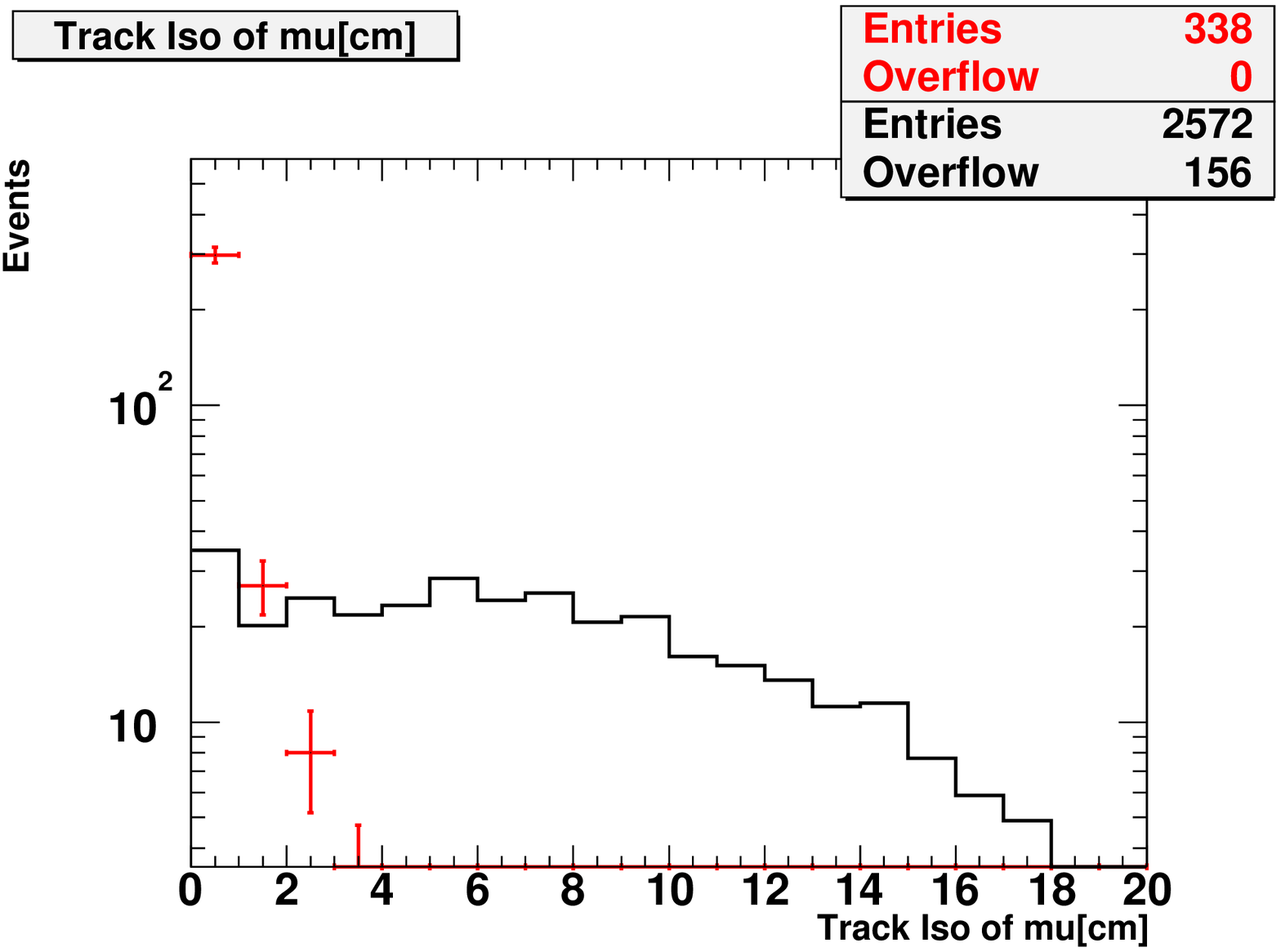, width=2.1in}} \quad
\subfigure[{Iso $E_T$}]
{\epsfig{file=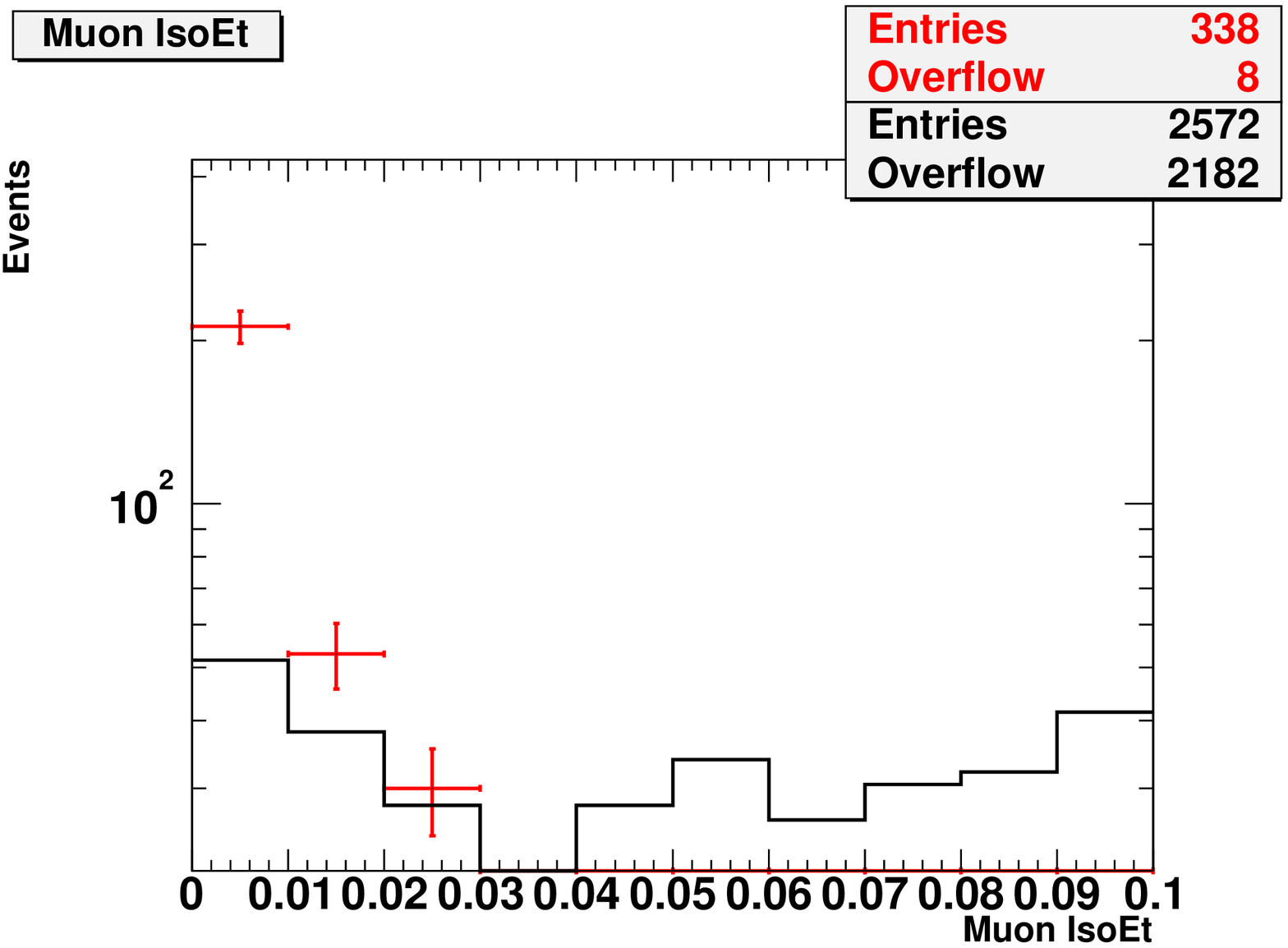, width=2.1in}} \quad
\subfigure[{Muon $P_T$}]
{\epsfig{file=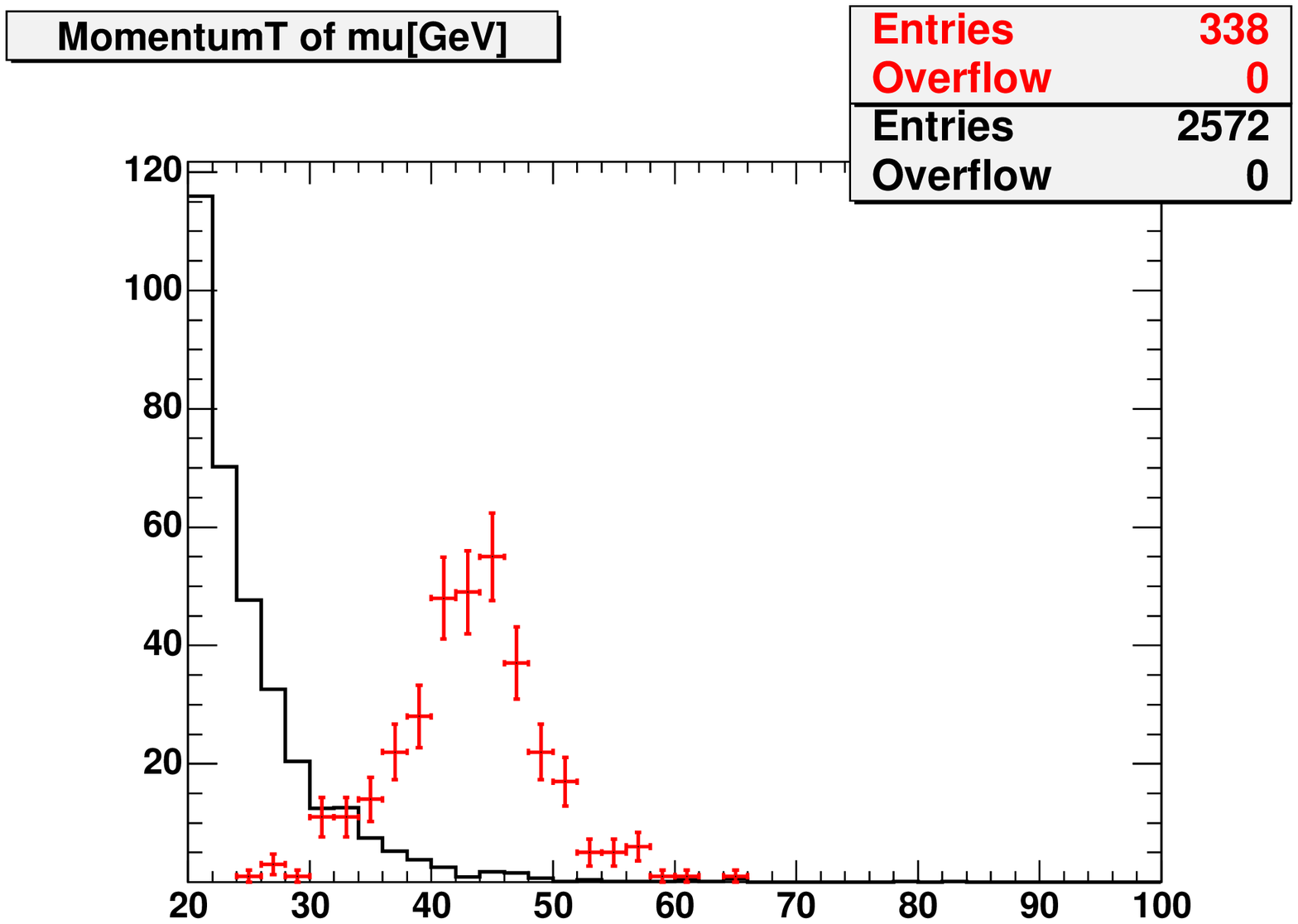, width=2.1in}} \quad
}
    \mbox{
\subfigure[{Track Isolation}]
{\epsfig{file=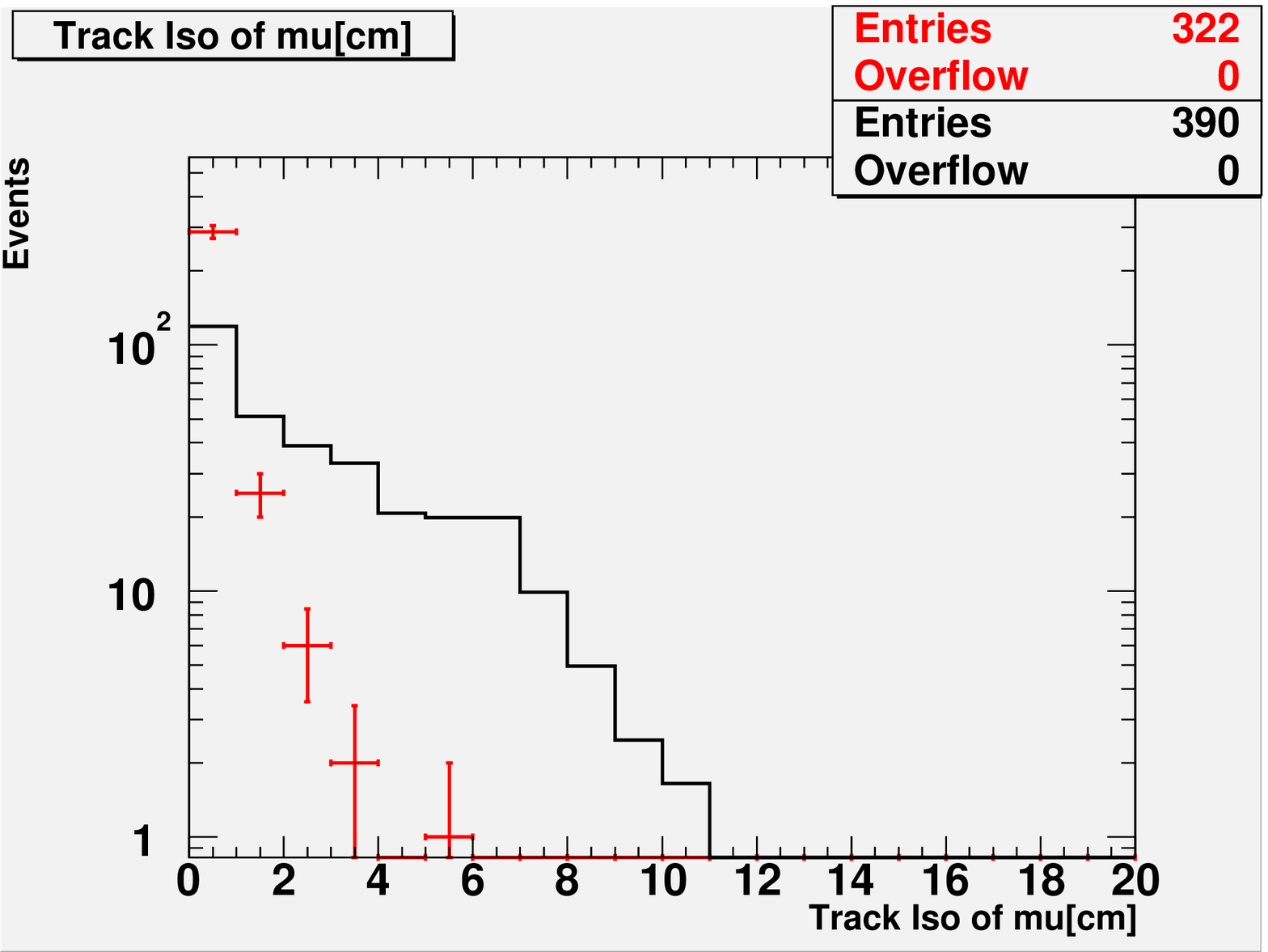, width=2.1in}} \quad
\subfigure[{Iso $E_T$}]
{\epsfig{file=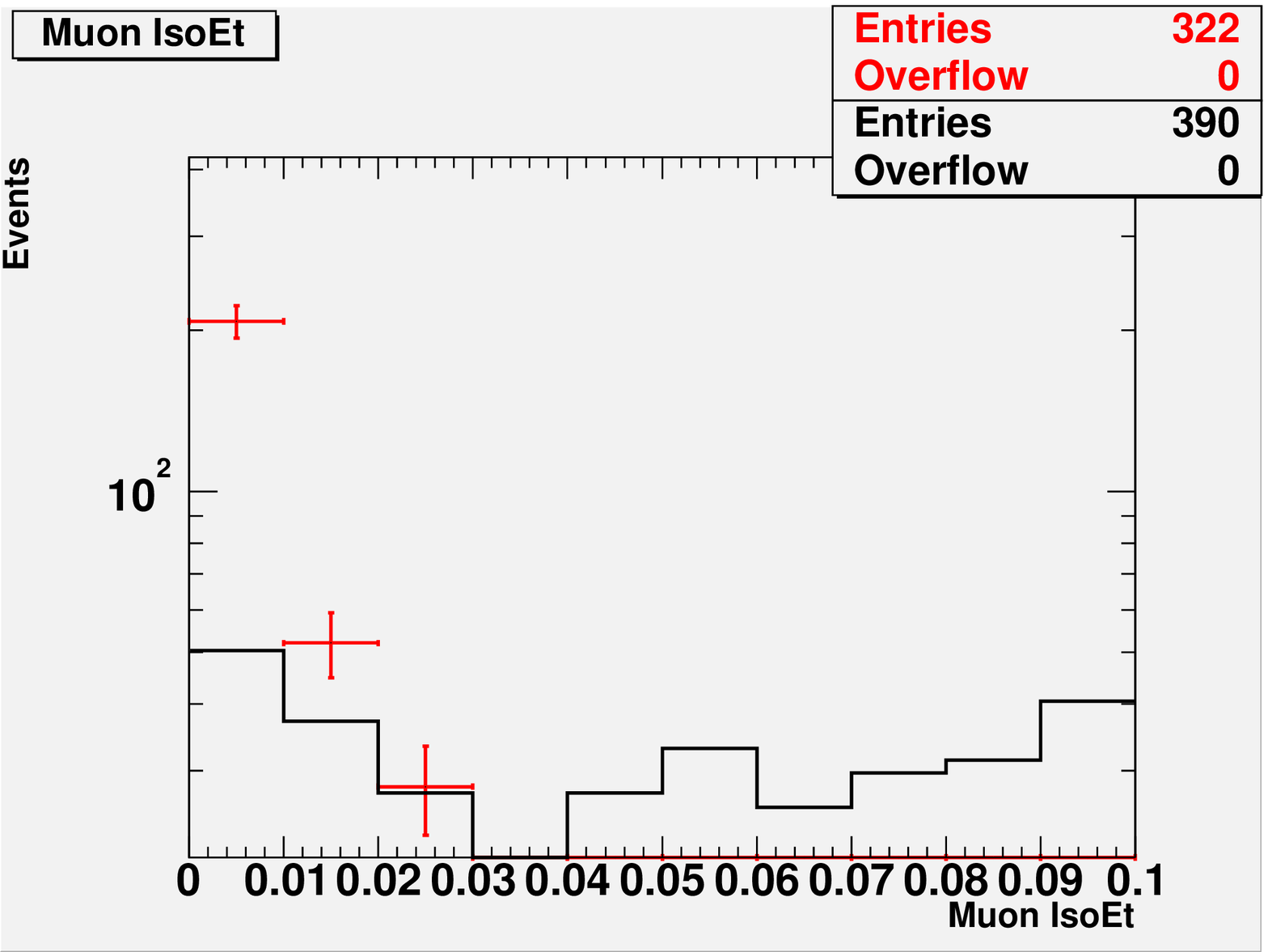, width=2.1in}} \quad
\subfigure[{Muon $P_T$}]
{\epsfig{file=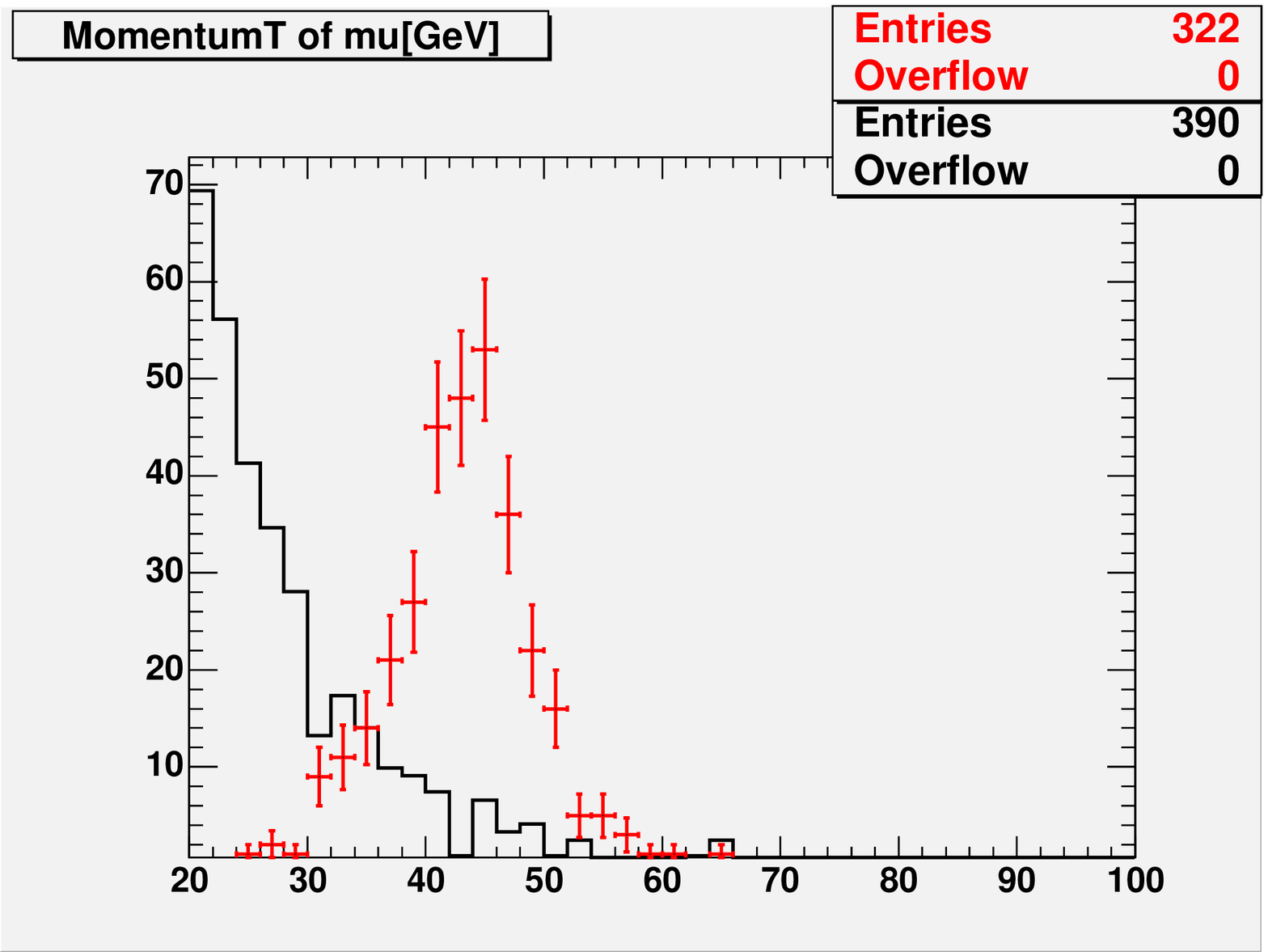, width=2.1in}} \quad
}
    \caption 
[Comparison of distributions for the muons from Non-WZ and $\Zee$ samples]
{A comparison of distributions from fake `muons'
from the Non-WZ sample(black histogram) and from muons from tight
Z's (red dots). The top set of plots has no (calorimeter) 
IsoEt cut applied; the IsoEt cut is applied in the  bottom plots.}
    \label{nonwz_muons.figure}
  \end{center}
\end{figure}

\subsection{Track Isolation Method}
\label{TrackIsoMethod}
To calculate the backgrounds from fake W's and fake Z's in which the
lepton comes from a jet, we use the track isolation of the lepton and
the samples of good electrons from Z's and QCD background from 
the non-WZ sample described above. The procedure is as follows:

\begin{itemize}
\setlength{\itemsep}{-0.07in}
\item
     Assume W, Z, $\Wg$, $\Zg$ all have the same underlying event
     structure, including jets (good assumption to first order in the
     SM, see for example Figure~\ref{trackiso_mumug_details.figure}).
\begin{figure}[!h]
  \begin{center}
    \mbox{
\subfigure
[Number of Tracks in a cone 0.4]
{\epsfig{file=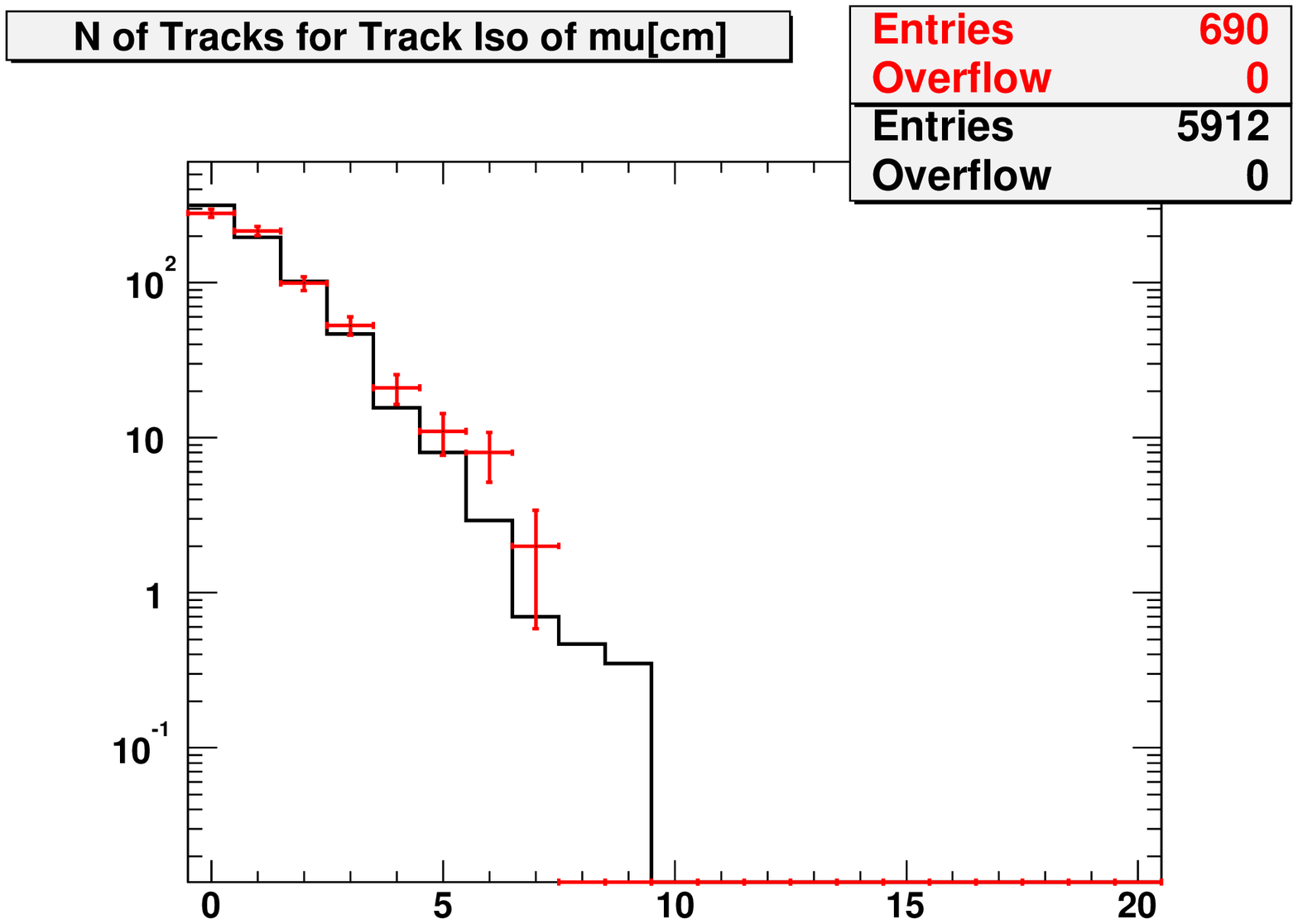,width=3.0in}} \quad
\subfigure
[Average Track $\Pt$ in a cone 0.4]
{\epsfig{file=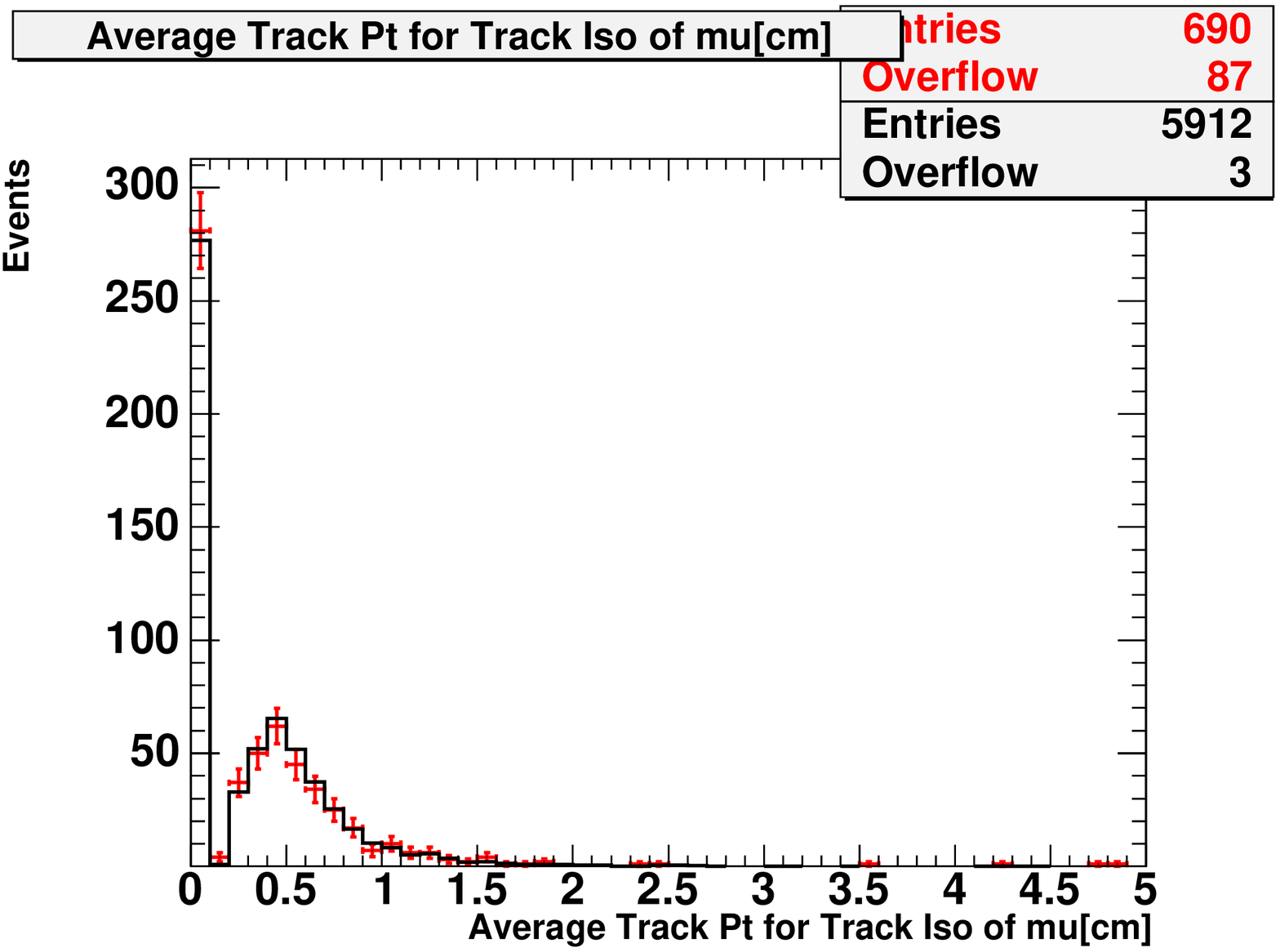,width=3.0in}} \quad
      }
\caption 
[Check of the same underlying event structure assumption]
{Check of the same underlying event structure assumption: muons from
$\mu\mu j$ (red points) vs ``golden'' Z (black histogram).}
\label{trackiso_mumug_details.figure}
  \end{center}
\end{figure}

\item
Define 3 samples:
   \begin{itemize}
     \setlength{\itemsep}{-0.07in}
     \item $\lgmet$ or $ll\gamma$ (signal region)
     \item golden-lepton (tight central-central electrons $\Zee$, tight muons $\Zmumu$)
     \item Non-W/Z QCD background
   \end{itemize}

\item 
     {\bf[0]} $N_{TOT}$: {\bf Number of events in a signal sample}\\
Count $N_{TOT}$ events of $\lgmet$($ll\gamma$)

\item 
     {\bf[1]} $f_S$: {\bf fraction of golden leptons with bad track isolation}\\
From the golden-lepton sample find the fraction $f_S$ of golden
leptons that have bad track isolation (tiso$\gt$4).

\item
     {\bf[2]} $N_S^{tiso\gt 4}$: {\bf number of signal in a $tiso\gt 4$ region}\\
$f_S \times N_{TOT}$ represents the number of real electrons we will
lose by subtracting off electrons with tiso$\gt$4 in the
$\lgmet$($ll\gamma$) sample. This has an uncertainty ($\delta f_S
\times N_{TOT}$) $\Rightarrow$ \\
$N_S^{tiso\gt 4} = f_S \times N_{TOT} \pm \delta f_S \times N_{TOT}$

\item
     {\bf[3]} $R_B$: {\bf ratio of background with good track isolation to the bad track isolation}\\
From the non-WZ sample find the ratio $R_B$ of background with good
track tiso ($tiso\lt 2$) to those with bad track tiso ($tiso\gt
4$). We use this to estimate how many of the tiso$\lt$2 candidates are
really QCD background.\\
$R_B = {{Non-WZ^{tiso\lt 2}}\over{Non-WZ^{tiso\gt 4}}}$

\item
     {\bf[4]} $N_B^{tiso\gt 4}$: {\bf the number of background with $tiso\gt 4$}\\
From the $\lgmet$($ll\gamma$) sample, the number of background with
tiso$\gt$4 should be the number of candidates with tiso$\gt$4 minus
the expected number of real electrons with tiso$\gt$4: \\
$N_B^{tiso\gt 4} = N_{data}^{tiso\gt 4} - N_S^{tiso\gt 4}$

\item
     {\bf[5]} $N_B^{tiso \lt 2}$: {\bf the number of background with $tiso\lt 2$}\\
The number of background with $tiso\lt 2$ is \\
$N_B^{tiso \lt 2}=(R_B)(N_B^{tiso \gt 4}) \pm \delta N_B^{tiso \lt 2}$

\item
     {\bf[6]} $N_B^{2\lt tiso\lt 4}$: {\bf the number of background with $2\lt tiso\lt 4$}\\
From number of background in $tiso\gt 4$ $N_B^{tiso\gt 4}$ and number
of events in Non-WZ sample with $2\lt tiso\lt 4$ and $tiso\gt 4$ we
estimate number of background with $2\lt tiso\lt 4$. The number of
background with $2\lt tiso\lt 4$ is\\
$N_B^{2\lt tiso\lt 4}=(N_B^{tiso\gt 4})\times{{Non-WZ^{2\lt tiso\lt 4}}\over{Non-WZ^{tiso\gt 4}}}$

\item
     {\bf[7]} $N_{QCD}$: {\bf QCD background} \\
Resulting Number of QCD background \\
        $N_{QCD} = N_B^{tiso\lt 2} + N_B^{2 \lt tiso\lt 4} + N_B^{tiso\gt 4}$

\end{itemize}

\subsection{Non-W/Z Results}

We have used track isolation to make estimates of the non-W backgrounds in
the $\Wlnu$ control sample (Section~\ref{nonwz_for_w}).

Table~\ref{qcd.table} summarizes the numbers of
track-isolated and non-track-isolated events in the good electron,
non-WZ, and signal samples. In addition, we add numbers of events from
W+jet and Z+jet categories, used to estimate jet faking photon
background.

The good electrons are heavily track-isolated ($0 \lt tiso \lt 2$);
the QCD background `electrons' are predominantly track-non-isolated
($4 \lt tiso$).

\begin{table}[!h]
\centering
\begin{tabular}{|l|l|l|l|l|}
\hline\hline
            & $0~\GeV\lt tiso \lt 2~\GeV$ & $2~\GeV \lt tiso \lt 4~\GeV$ & $tiso>4~\GeV$ & total \\
\hline\hline
\multicolumn{5}{|c|} {
Electrons:}\\
\hline\hline
$Z_{tight}^{CC}$ & 11204 & 610 & 534 & 12348\\
Non-WZ           & 1179 & 870 & 3709 & 5758\\
$\egmet$        & 21 & 2 & 2 & 25\\
$\eeg$          & 35 & 1 & 2 & 38\\
$\ejmet$        & 1.98655 & 0.164143 & 0.62847 & 2.77916\\
$  eej$          & 0.46559 & 0.0241221 & 0.0371426 & 0.526854\\
\hline\hline
\multicolumn{5}{|c|} {
Muons:}\\
\hline\hline
$Z_{tight}$      & 5686 & 190 & 36 & 5912\\
Non-WZ           & 173 & 78 & 93 & 344\\
$\mugmet$       & 16 & 1 & 0 & 17\\
$\mumug$        & 22 & 2 & 0 & 24\\
$\mujmet$       & 1.41881 & 0.0601055 & 0.0618934 & 1.54081\\
$\mumuj$        & 0.264149 & 0.00932185 & 0.0318951 & 0.305366\\
\hline\hline
\end{tabular}

    \caption 
[The numbers of events in different track isolation regions for $\Zll$, Non-WZ, $\lgmet$, $\llg$, $\ljmet$, $\llj$]
{The numbers of events that are track-isolated ($0 \lt
tiso \lt 2$) versus non-track-isolated ($tiso\gt 4$) for good
electrons, QCD background `electrons', and the signal samples. The
good electrons are heavily track-isolated; the
QCD background `electrons' are predominantly track-non-isolated.
             }
\label{qcd.table}
\end{table}

Table~\ref{qcd.steps} shows step-by-step calculation of predicted QCD
(Non-WZ) backgrounds.

\begin{sidewaystable}
 \begin{center}
  \begin{tabular}{|l|l|l|l|l|l|l|l|l|l|}
\hline
  Sample &
  $f_S^{tiso>4}$  & $N_S^{tiso>4}$ & $R_B$          & $N_B^{tiso>4}$ & $N_B^{tiso<2}$& $N_B^{2<tiso<4}$ & $N_B^{tot}$\\ 
\hline
$\egmet$ & 
0.043$\pm$0.0023 & 
1.1$\pm$0.057 & 
0.32$\pm$0.014 & 
0.92$\pm$0.057 & 
0.29$\pm$0.031 & 
0.22$\pm$0.024 & 
1.4$\pm$0.11 \\ 
$\eeg$ & 
0.043$\pm$0.0023 & 
1.6$\pm$0.086 & 
0.32$\pm$0.014 & 
0.36$\pm$0.086 & 
0.11$\pm$0.032 & 
0.084$\pm$0.024 & 
0.55$\pm$0.14 \\ 
$\mugmet$ & 
0.0061$\pm$0.0011 & 
0.1$\pm$0.019 & 
1.9$\pm$0.33 & 
0$\pm$0.019 & 
0$\pm$0.035 & 
0$\pm$0.016 & 
0$\pm$0.069 \\ 
$\mumug$ & 
0.0061$\pm$0.0011 & 
0.15$\pm$0.026 & 
1.9$\pm$0.33 & 
0$\pm$0.026 & 
0$\pm$0.049 & 
0$\pm$0.022 & 
0$\pm$0.097 \\ 
\hline
\end{tabular}

  \caption
[Predicted QCD backgrounds]
{Predicted QCD backgrounds: step-by-step calculation.}
\vskip 0.3in
\label{qcd.steps}

  \begin{tabular}{|l|l|l|l|l|l|l|l|l|l|}
\hline
  Sample &
  $f_S^{tiso>4}$  & $N_S^{tiso>4}$ & $R_B$          & $N_B^{tiso>4}$ & $N_B^{tiso<2}$& $N_B^{2<tiso<4}$ & $N_B^{tot}$\\ 
\hline
$\egmet$ & 
0.043$\pm$0.0023 & 
0.96$\pm$0.05 & 
0.32$\pm$0.014 & 
0.41$\pm$0.05 & 
0.13$\pm$0.022 & 
0.096$\pm$0.017 & 
0.64$\pm$0.089 \\ 
$\eeg$ & 
0.043$\pm$0.0023 & 
1.6$\pm$0.085 & 
0.32$\pm$0.014 & 
0.34$\pm$0.085 & 
0.11$\pm$0.032 & 
0.08$\pm$0.024 & 
0.53$\pm$0.14 \\ 
$\mugmet$ & 
0.0061$\pm$0.0011 & 
0.094$\pm$0.017 & 
1.9$\pm$0.33 & 
0$\pm$0.017 & 
0$\pm$0.031 & 
0$\pm$0.014 & 
0$\pm$0.063 \\ 
$\mumug$ & 
0.0061$\pm$0.0011 & 
0.14$\pm$0.026 & 
1.9$\pm$0.33 & 
0$\pm$0.026 & 
0$\pm$0.048 & 
0$\pm$0.022 & 
0$\pm$0.096 \\ 
\hline
\end{tabular}

  \caption
[Predicted QCD backgrounds: with jet fakes subtracted]
{Predicted QCD backgrounds: with jet fakes subtracted. Step-by-step calculation.}
\vskip 0.3in
\label{jqcd.steps}

  \begin{tabular}{|l|l|l|l|l|l|l|l|l|l|}
\hline
  Sample &
  $f_S^{tiso>4}$  & $N_S^{tiso>4}$ & $R_B$          & $N_B^{tiso>4}$ & $N_B^{tiso<2}$& $N_B^{2<tiso<4}$ & $N_B^{tot}$\\ 
\hline
$\egmet$ & 
0.043$\pm$0.0023 & 
0.84$\pm$0.044 & 
0.32$\pm$0.014 & 
0$\pm$0.044 & 
0$\pm$0.014 & 
0$\pm$0.01 & 
0$\pm$0.068 \\ 
$\eeg$ & 
0.043$\pm$0.0023 & 
1.6$\pm$0.084 & 
0.32$\pm$0.014 & 
0.33$\pm$0.084 & 
0.1$\pm$0.031 & 
0.077$\pm$0.023 & 
0.51$\pm$0.14 \\ 
$\mugmet$ & 
0.0061$\pm$0.0011 & 
0.085$\pm$0.015 & 
1.9$\pm$0.33 & 
0$\pm$0.015 & 
0$\pm$0.028 & 
0$\pm$0.013 & 
0$\pm$0.056 \\ 
$\mumug$ & 
0.0061$\pm$0.0011 & 
0.14$\pm$0.026 & 
1.9$\pm$0.33 & 
0$\pm$0.026 & 
0$\pm$0.048 & 
0$\pm$0.021 & 
0$\pm$0.095 \\ 
\hline
\end{tabular}

  \caption
[Predicted QCD backgrounds: with jet fakes double-subtracted]
{Predicted QCD backgrounds: with jet fakes double-subtracted. Step-by-step calculation.}
\vskip 0.3in
\label{jjqcd.steps}
 \end{center}
\end{sidewaystable}

\subsection{Modified Track Isolation Method}
\label{ModifiedTrackIsoMethod}

From the jet fake rate we obtain (fake gamma + real W) and (fake gamma
+ fake W) backgrounds. From Track Isolation Method we obtain (real
gamma + fake W) and possibly (fake gamma + fake W).

To summarize, there are three sources of faking $\Wg$ with either fake gamma and/or fake W:
\begin{enumerate}
\setlength{\itemsep}{-0.07in}
  \item fake gamma + real W (from jet fake rate)
  \item fake gamma + fake W (also part of jet fake rate and possibly part of non-W/Z background)
	  \item real gamma + fake W (from non-W/Z background)
\end{enumerate}

If fake W is a part of non-W/Z background it should contribute to
track-non-isolated regions. Real W's from W+jet should have the same
track isolation distribution as W+$\gamma$ (we have performed the
checks for all the samples we use, see for example
Figure~\ref{trackiso_mumug_details.figure}).

To avoid double counting, we modify Track Isolation Method as follows:
we subtract W+jet faking photon contribution from $\lgmet$ trackiso
regions, we subtract Z+jet faking photon contribution from $\llg$
trackiso regions. Then we repeat procedure, documented in
Section~\ref{TrackIsoMethod}, with these modified $\lgmet$ and $\llg$
numbers (Table~\ref{jqcd.steps}).

To take into account systematic error of 100\% for jet faking photon
rate (Table~\ref{jqcd.steps}) we repeat these procedure, subtracting
W+jet and Z+jet contribution multiplied by 2. Therefore, to avoid
double-counting we use background estimates from ``Predicted QCD
Backgrounds with Jet Fakes Subtracted'' (Table~\ref{jqcd.steps}).

\begin{table}[!h]
\begin{center}
\begin{tabular}{|l||l|l|l|l|l|l|}
\hline
Region, GeV & $\egmet$ & $\eeg$ & $\mugmet$ & $\mumug$ \\
\hline
0-2; 2-$\infty$ &
1.4$\pm$0.11 &
0$\pm$0.17 &
0.58$\pm$0.12 &
2.1$\pm$0.23 \\
\hline
0-2; 4-$\infty$ &
0.64$\pm$0.09 &
0.53$\pm$0.14 &
0$\pm$0.06 &
0$\pm$0.10 \\
\hline
0-2; 5-$\infty$ &
1.2$\pm$0.09 &
1.2$\pm$0.14 &
0$\pm$0.07 &
0$\pm$0.1 \\
\hline
0-2.5; 4.5-$\infty$ &
0.93$\pm$0.09 &
0.9$\pm$0.14 &
0$\pm$0.07 &
0$\pm$0.1 \\
\hline
0-3; 3-$\infty$ &
0.03$\pm$0.08 &
0$\pm$0.14 &
0$\pm$0.08 &
1.6$\pm$0.2 \\
\hline
0-3; 5-$\infty$ &
1.2$\pm$0.09 &
1.2$\pm$0.14 &
0$\pm$0.07 &
0$\pm$0.1 \\
\hline
Final &
0.7 &
0.6 &
0.3 &
1.0 \\
 &
$\pm$0.1{\tiny (stat)} &
$\pm$0.1{\tiny (stat)} &
$\pm$0.1{\tiny (stat)} &
$\pm$0.1{\tiny (stat)} \\
 &
$\pm$0.7{\tiny (sys)} &
$\pm$0.6{\tiny (sys)} &
$\pm$0.3{\tiny (sys)} &
$\pm$1.0{\tiny (sys)} \\
\hline
\end{tabular}
\caption
[QCD background summary table for $\lgmet$ and $\llg$]
{QCD studies table: estimating the QCD background faking a $\lgmet$
or $\llg$ event. The method uses track isolation and two regions, one at
low track-iso and one at high. Leptons from the $\Zll$ sample are
used to estimate the number of leptons in the high-track-iso region;
`leptons' from the non-WZ sample are used to estimate the number of
leptons in the low-track-iso region. The first column of the table
gives the ranges in GeV for the low and high track-iso regions,
respectively. The last line gives the final estimates.}
\end{center}
\label{qcd_all_loginov}
\end{table}

Finally, Non-W/Z backgrounds for $\lgmet$ and $\llg$ signatures are
summarized in Table~\ref{qcd_all_loginov}. Systematic errors are
obtained by varying track isolation regions. Further checks for Non-WZ
background for $\Wenu$ and $\Wmunu$ are described below in
Section~\ref{nonwz_for_w}.

\subsection{Non-W/Z background for $\Wenu$ and $\Wmunu$}
\label{nonwz_for_w}

We have used track isolation to make estimates of the non-WZ backgrounds in
the $\Wlnu$ control sample and $\llg$ and $\lgmet$ signal samples.
Table~\ref{QCDbackgroundsforWs} shows the estimated `non-W' background
in the W samples for 5 different track-iso regions. This is for a
check- the values are consistent with known QCD backgrounds for
W's~\cite{CDF_WZ}. The method uses track isolation and two regions,
one at low track-iso and one at high. Leptons from the golden Z sample
are used to estimate the number of leptons in the high-track-iso
region; `leptons' from the non-WZ sample are used to estimate the
number of leptons in the low-track-iso region. The first column of the
table gives the ranges in GeV for the low and high track-iso regions,
respectively.

\begin{table}[!h]
\begin{center}
\begin{tabular}{|l|l|l|l|}
\hline
Region, GeV &  $\Wenu$ & $\Wmunu$ \\
\hline
0-2; 4-$\infty$ &
6.5e+02$\pm$6.6e+02 &
1.8e+03$\pm$7.3e+02 \\
0-3; 5-$\infty$ &
1.1e+03$\pm$6.5e+02 &
1.6e+03$\pm$7.2e+02 \\
0-2; 5-$\infty$ &
1.1e+03$\pm$6.5e+02 &
1.6e+03$\pm$7.3e+02 \\
0-2; 2-$\infty$ &
56$\pm$8.3e+02 &
5.7e+02$\pm$7.7e+02 \\
0-3; 3-$\infty$ &
4.5e+02$\pm$7.1e+02 &
1.1e+03$\pm$7e+02  \\
\hline
\end{tabular}
\caption
[QCD background summary table for $\Wlnu$]
{QCD backgrounds for W's
The method uses track isolation and two regions, one at
low track-iso and one at high. Leptons from the golden Z sample are
used to estimate the number of leptons in the high-track-iso region;
`leptons' from the non-WZ sample are used to estimate the number of
leptons in the low-track-iso region. The first column of the table
gives the ranges in GeV for the low and high track-iso regions,
respectively.
}
\label{QCDbackgroundsforWs}
\end{center}
\end{table}

\clearpage
\chapter{Systematic Uncertainties}
\label{systematics}
In this chapter we summarize estimates of the systematic uncertainties
on the SM predicted rates and on the measured event counts.

The errors are categorised as theoretical
(Section~\ref{theoretical_systematics}), luminosity
(Section~\ref{luminosity_systematics}) and experimental
(Section~\ref{experimental_systematics}. The contributing effects for
the SM predictions we have considered are:

\begin{itemize}
\setlength{\itemsep}{-0.11in}
\item  7\% error is on the total theoretical prediction, including
the NLO uncertainties.
\item  Luminosity: 6\% 
\item  Trigger Efficiencies: 2\% for muons and 1\% for electrons
for lepton triggers only. We OR'ing lepton trigger with photon
trigger, and therefore this combination of triggers is fully
efficient.
\item  $|$z\_vert$|$ $<$ 60: 1\% 
\item  Muon ID Efficiencies: 2\%
\item  Electron ID Efficiencies: 1\% 
\item  Photons ID Efficiencies:  4\%
\end{itemize}

The systematic uncertainties on the backgrounds are included in the
background estimates, discussed in Chapter~\ref{predictions} and
Chapter~\ref{fakes}. For the SM predictions the total systematic
uncertainty is 10.2\% for $\Wg$ and $\Zg$ for electrons, and 10.5\%
for $\Wg$ and $\Zg$ for muons (Chapter~\ref{predictions}).

\section{Theoretical Systematic Uncertainties}
\label{theoretical_systematics}

Limitations in the theoretical precision of the calculation, result in
an uncertainty on the cross-section prediction. The effect of these
errors on the cross-section is studied
in~\cite{CDF_WZgamma,hayward_thesis,kirby_thesis} and is summarized in
Table~\ref{mc_systematics.table}

\begin{table}[h]
 \begin{center}
\begin{tabular}{|p{5.0cm}|p{2.0cm}|}
\hline
Source               & \% \\
\hline
Factorization Scale  & 2  \\
PDF                  & 6  \\
K-factor             & 3  \\
\hline
Total                & 7  \\
\hline
\end{tabular}
 \end{center}
\caption{Systematic errors on the $\Zg$, $\Wg$, $\Zgg$ and $\Wgg$ generation}
\label{mc_systematics.table}
\end{table}

\subsection{Factorization Scale}

The factorization scale is the minimum $Q^2$ value
calculated~\cite{hayward_thesis,kirby_thesis} for photon emission in
the ZGAMMA and WGAMMA programs~\cite{diboson_mc}. This value will
affect the maximum $Q^2$ value for post generation Pythia
fragmentation. The default factorization scale was $\hat{s}$, the
square of the collision energy of the event. The cross-section and
acceptance were measured using four other values, 2 $\hat{s}$, 3/2
$\hat{s}$, 2/3 $\hat{s}$and 1/2 $\hat{s}$. The greatest variation in
the cross-section from the default value of Q2 was 2\%.

\subsection{Parton Density Function Choice}

Protons and anti-protons are composite particles. Therefore, any
interactions between them must be described using parton density
functions (PDF). The PDF describes the energy distributions of the
valence quarks, gluons and sea quarks inside the proton/anti-proton.

The PDF chosen for use with ZGAMMA~\cite{diboson_mc} was the CTEQ5L
PDF. In order to determine the systematic error from this choice, the
LO cross-section is compared to the corresponding predictions
calculated from the MRST
72 - 76 PDFs. The MRST cross-sections range between 1.604 and
1.625 $\invpb$ whereas the cross-section using CTEQ5L is 1.72
$\invpb$. The difference between the two was taken to be the
systematic error~\cite{hayward_thesis,kirby_thesis}, of 6\%.

\subsection{K-factor }

The calculated K-factor only takes into account O($\alpha_s$)
corrections. To take into account higher order corrections, the Q
scale in the NLO calculation was varied by factors of 2 and 1/2. A 3\%
variation in the cross-section calculation was observed, and taken to
be a systematic error.

\section{Luminosity Systematic Uncertainties}
\label{luminosity_systematics}
The luminosity error is estimated to be 6\%, which includes a 4.4\%
contribution from the acceptance and operation of the luminosity
monitor and 4.0\% from the theoretical uncertainty on the calculation
of the total $\ppbar$ cross-section~\cite{inelastic_x_section}.

\section{Experimental Systematic Uncertainties}
\label{experimental_systematics}

The sources of experimental systematic
errors~\cite{CDF_WZgamma,hayward_thesis,kirby_thesis} for the $\lgX$
analysis subcategories are summarized in
Table~\ref{systematic_errors.table}. Jet Fake systematic error is
discussed in Section~\ref{jet_fakes}.

\begin{table}[h]
 \begin{center}
\begin{tabular} {|p{5.0cm}|p{2.0cm}|p{1.5cm}|p{1.5cm}|p{1.5cm}|p{1.5cm}|}
\hline
Source & \% & Central & Plug & CMUP & CMX \\ 
\hline
Jet Fake                     & $\approx$100         & x & x & x & x \\
\hline
$Z_0$ cut eff                & 1.0                  & x & x & x & x \\ 
photon cut eff		     & 2.0                  & x & x & x & x \\
energy scale ($\gamma$)	     & 3.0                  & x & x & x & x \\
conversion rate uncertainty  & 1.5                  & x & x & x & x \\
momentum scale ($\mu$)	     & 2.0                  &   &   & x & x \\
acceptance ($e$)	     & 1.0                  & x & x &   &   \\
acceptance ($\mu$)	     & 2.0                  &   &   & x & x \\
central $e$ ID		     & 1.0                  & x &   &   &   \\
central $e$ trigger	     & 1.0                  & x &   &   &   \\
energy scale ($e$)	     & 1.0                  & x & x &   &   \\
plue $e$ ID		     & 2.5                  &   & x &   &   \\
plug trig eff		     & 1.0                  &   & x &   &   \\
plug $e$ vertex eff	     & 1.0                  &   & x &   &   \\
plug $e$ track eff	     & 1.5                  &   & x &   &   \\
cosmic			     & 0.01                 &   &   & x & x \\
Cot track reconstruction     & 0.4                  & x &   & x & x \\
CMUP ID			     & 0.7                  &   &   & x &   \\
CMUP reconstruction	     & 0.6                  &   &   & x &   \\
CMUP trigger		     & 0.7                  &   &   & x &   \\
CMX ID			     & 0.8                  &   &   &   & x \\
CMX reconstruction	     & 0.3                  &   &   &   & x \\
CMX trigger		     & 0.6                  &   &   &   & x \\
\hline
\end{tabular}
 \end{center}
\caption
[Systematic errors summary for $\lgX$]
{Systematic errors summary for $\lg$. 'x' means that channel needs to
take into account its systematic uncertainty. Jet Fake systematic
error is discussed in Section~\ref{jet_fakes}}
\label{systematic_errors.table}
\end{table}

Systematic uncertainty on jet faking photon rate is one of the
dominating errors. The uncertainty is limited by the statistics for
the high-$\Et$ photons, so we expect it to significantly improve with
more data. At that point we'll be dominated by the systematic
uncertainty. Therefore, the biggest contribution will be from the SM
estimates on $\Wg$, $\Zg$, $\Wgg$ and $\Zgg$ production.

\clearpage

\chapter{The $\lgX$ Search}
\label{photonlepton}

This chapter presents the results of the $\lgX$ search for the three
signatures of interest - $\lgmet$, $\llg$ and $\llg$.

Section~\ref{topologies} describes the `analysis subcategories'
established in the Run I analysis~\cite{jeff_prd}, and used again here
so as to be {\it a priori}. Section~\ref{number_observed} presents the
number of events in each analysis subcategory for the $\egX$, $\mugX$,
and $\lgX$ (the sum of $e+\mu$) samples.
Section~\ref{observed_stability} discusses the stability of the
observed numbers during the course of the run. The predicted and
observed totals for the $\lgmet$, $\llg$ and $\lgg$ and comparison of
observed kinematic distributions to the SM predicted shapes is done in
Section~\ref{comparison}.

\section{Defining the Event Categories by Topology}
\label{topologies}

Categories of photon-lepton events were defined {\it a priori} in a
way that characterized the different possibilities for new physics.
For each category, the inclusive event total and basic kinematic
distributions can be compared with standard model expectations. The
decay products of massive particles are typically isolated from other
particles, and possess large transverse momentum and low rapidity.

Therefore, inclusive $\lg$ events are selected by requiring a
central tight photon with $\Etgamma>25$ $\GeV$ and a central $e$ or
$\mu$ with $\Etlepton>25$ $\GeV$. Both signal and control samples are
drawn from this $\lg$ sample
(Figures~\ref{flowchart_lepton.figure},~\ref{flowchart_electron.figure}
and ~\ref{flowchart_muon.figure}).

Considering the control samples first, from the $\lg$ sample we
select back-to-back events with exactly one photon and one lepton
(i.e. $\met<25~\GeV$); this is the dominant contribution to the
$\lg$ sample, and has a large Drell-Yan component. A subset of this
sample is the `Z-like' sample, which provides the calibration for the
probability that an electron radiates and is detected as a photon, as
discussed in Section~\ref{egamma_fakes}. The remaining back-to-back
events are called the Two-Body Events and were not used in this
analysis.

All events which either have more than one lepton or photon, or in
which the lepton and photon are not back-to-back (and hence the event
cannot be a Two-Body event), are classified as `Inclusive Multi-Body
$\lgX$'. These are further subdivided into three categories: $\lgmet$
(Section~\ref{lgmet.section}) (`Multi-Body $\lgmet$ Events'), for
which the $\met$ (Section~\ref{metid}) is greater than 25 GeV , $\llg$
(Section~\ref{llg.section}) and $\lgg$ (Section~\ref{lgg.section})
(`Multi-Photon and Multi-Lepton Events'), and events with exactly one
lepton and exactly one photon, which are not back-to-back. The events
with exactly one lepton and exactly one photon, which are not
back-to-back were not used in the analysis.

\section{The Number of Events Observed}
\label{number_observed}

Figure~\ref{flowchart_electron.figure} shows the results of this
classification for the inclusive electron data sample. We find
$\nofeg$ $\eg$ events, of which $\nofegallother$ are in the Inclusive
Multi-Body category. Of these, $\nofeglgmet$ are classified as
$\egmet$ events and $\nofegmultig$ and $\nofegmultil$ as Multi-Photon
and Multi-Electron events respectively.

Figure~\ref{flowchart_muon.figure} shows the results for the inclusive
muon sample.We find $\nofmug$ $\mug$ events, of which
$\nofmugallother$ are in the Inclusive Multi-Body category. Of these,
$\nofmuglgmet$ are classified as $\mugmet$ events and $\nofmugmultig$
and $\nofmugmultil$ as Multi-Photon and Multi-Muon events.

Figure~\ref{flowchart_lepton.figure} shows the sum of the electron and
muon entries in the analysis subcategories. There are $\noflglgmet$
$l\gamma\met$ events, $\noflgmultig$ Multi-Photon events, and
$\noflgmultil$ Multi-Lepton events. It is these categories, shown in
red in the figures, that are of particular interest due to the Run I
results.

\begin{huge}
\begin{figure*}[h]
\thicklines
\begin{center}
\framebox(155,70)[s]{
{
\begin{minipage}{150pt}
\begin{center}
Electron-Photon Sample \\
1 Electron and 1 Photon \\
$E_T\gt 25$ GeV \\
{\bf\nofeg} Events
\end{center}
\end{minipage}
}
}
\put(-10,0){\vector(0,-1){60}}
\put(-140,0){\vector(0,-1){60}}
\vspace{-1pt}

\framebox(150,90)[0,0]{
{
\begin{minipage}{150pt}
\begin{center}
Exactly 1 Electron \\
Exactly 1 Photon \\
$\Delta\phi_{\eg}\gt 150$ \\
$\met\lt 25$ \\
{\bf\nofegdphi} Events
\end{center}
\end{minipage}
}
}
\put(-20,0){\vector(0,-1){32}}
\put(-130,0){\vector(0,-1){135}}
\makebox(30,60)[l]{}
\framebox(150,90)[0,0]{
{
\begin{minipage}{150pt}
\begin{center}
Inclusive Multi-Body Events \\
(All Other Electron-Photon) \\
{\bf\nofegallother} Events
\end{center}
\end{minipage}
}
}
\put(-20,0){\vector(0,-1){135}}
\put(-130,0){\vector(0,-1){135}}
\put(-75,0){\vector(0,-1){32}}
\end{center}

\vspace{-118pt}

\makebox(110,60)[l]{}
\framebox(150,90)[0,0]{
{
\begin{minipage}{150pt}
\begin{center}
Z-Like Electron-Photon \\
81 Gev $\lt M_{\eg}\lt$ 101 Gev\\
(Background Calibration) \\
{\bf\nofegzlike} Events
\end{center}
\end{minipage}
}
}
\makebox(20,60)[l]{}
\framebox(100,90)[0,0]{
{
\begin{minipage}{150pt}
\begin{center}
Exactly 1 Electron \\
Exactly 1 Photon \\
$\Delta\phi_{\eg}\lt 150$ \\
$\met\lt 25$ GeV \\
{\bf\nofegzunlike} Events
\end{center}
\end{minipage}
}
}

\begin{center}
\framebox(115,70)[0,0]{
{
\begin{minipage}{115pt}
\begin{center}
Two-Body Events \\
{\bf \nofegtwobody} Events
\end{center}
\end{minipage}
}
}
\framebox(115,70)[0,0]{
{
\color{red}
\begin{minipage}{115pt}
\begin{center}
Multi-Body $\egmet$ Events\\
$\met\gt 25$ GeV \\
{\bf \nofeglgmet} Events
\end{center}
\end{minipage}
}
}
\framebox(115,70)[0,0]{
{
\color{red}
\begin{minipage}{115pt}
\begin{center}
Multi-Photon and \\
Multi-Electron Events \\
{\bf\nofegmultig} and {\bf\nofegmultil} Events, resp.
\end{center}
\end{minipage}
}
}
\end{center}
\caption
[The subsets of electron-photon sample]
{Electron-photon sample: the subsets of inclusive $\eg$ events analyzed.}
\label{flowchart_electron.figure}
\end{figure*}

\begin{figure*}[h]
\thicklines
\begin{center}
\framebox(150,70)[s]{
{
\begin{minipage}{150pt}
\begin{center}
Muon-Photon Sample \\
1 Muon and 1 Photon \\
$E_T\gt 25$ GeV \\
{\bf\nofmug} Events
\end{center}
\end{minipage}
}
}
\put(-10,0){\vector(0,-1){60}}
\put(-140,0){\vector(0,-1){60}}
\vspace{-1pt}

\framebox(150,90)[0,0]{
{
\begin{minipage}{150pt}
\begin{center}
Exactly 1 Muon \\
Exactly 1 Photon \\
$\Delta\phi_{\mug}\gt 150$ \\
$\met\lt 25$ \\
{\bf\nofmugdphi} Events
\end{center}
\end{minipage}
}
}
\put(-20,0){\vector(0,-1){32}}
\put(-130,0){\vector(0,-1){135}}
\makebox(30,60)[l]{}
\framebox(150,90)[0,0]{
{
\begin{minipage}{150pt}
\begin{center}
Inclusive Multi-Body Events \\
(All Other Muon-Photon) \\
{\bf\nofmugallother} Events
\end{center}
\end{minipage}
}
}
\put(-20,0){\vector(0,-1){135}}
\put(-130,0){\vector(0,-1){135}}
\put(-75,0){\vector(0,-1){32}}
\end{center}

\vspace{-118pt}

\makebox(110,60)[l]{}
\framebox(150,90)[0,0]{
{
\begin{minipage}{150pt}
\begin{center}
Z-Like Muon-Photon \\
81 Gev $\lt M_{\mug}\lt$ 101 Gev\\
(Background Calibration) \\
{\bf\nofmugzlike} Events
\end{center}
\end{minipage}
}
}
\makebox(20,60)[l]{}
\framebox(100,90)[0,0]{
{
\begin{minipage}{150pt}
\begin{center}
Exactly 1 Muon \\
Exactly 1 Photon \\
$\Delta\phi_{\mug}\lt 150$ \\
$\met\lt 25$ GeV \\
{\bf\nofmugzunlike} Events
\end{center}
\end{minipage}
}
}

\begin{center}
\framebox(115,70)[0,0]{
{
\begin{minipage}{115pt}
\begin{center}
Two-Body Events \\
{\bf \nofmugtwobody} Events
\end{center}
\end{minipage}
}
}
\framebox(115,70)[0,0]{
{
\color{red}
\begin{minipage}{115pt}
\begin{center}
Multi-Body $\mugmet$ Events\\
$\met\gt 25$ GeV \\
{\bf \nofmuglgmet} Events
\end{center}
\end{minipage}
}
}
\framebox(115,70)[0,0]{
{
\color{red}
\begin{minipage}{115pt}
\begin{center}
Multi-Photon and \\
Multi-Muon Events \\
{\bf\nofmugmultig} and {\bf\nofmugmultil} Events, resp.
\end{center}
\end{minipage}
}
}
\end{center}
\caption
[The subsets of muon-photon sample]
{Muon-photon sample: the subsets of inclusive $\mug$ events analyzed.}
\label{flowchart_muon.figure}
\end{figure*}

\begin{figure*}[h]
\thicklines
\begin{center}
\framebox(155,70)[s]{
{
\begin{minipage}{150pt}
\begin{center}
Lepton-Photon Sample \\
1 Lepton and 1 Photon \\
$E_T~>~25$ GeV \\
{\bf\noflg} Events
\end{center}
\end{minipage}
}
}
\put(-10,0){\vector(0,-1){60}}
\put(-140,0){\vector(0,-1){60}}
\vspace{-1pt}

\framebox(150,90)[0,0]{
{
\begin{minipage}{150pt}
\begin{center}
Exactly 1 Lepton \\
Exactly 1 Photon \\
$\Delta\phi_{l\gamma}\gt 150$ \\
$\met\lt 25$ \\
{\bf\noflgdphi} Events
\end{center}
\end{minipage}
}
}
\put(-20,0){\vector(0,-1){32}}
\put(-130,0){\vector(0,-1){135}}
\makebox(30,60)[l]{}
\framebox(150,90)[0,0]{
{
\begin{minipage}{150pt}
\begin{center}
Inclusive Multi-Body Events \\
(All Other Lepton-Photon) \\
{\bf\noflgallother} Events
\end{center}
\end{minipage}
}
}
\put(-20,0){\vector(0,-1){135}}
\put(-130,0){\vector(0,-1){135}}
\put(-75,0){\vector(0,-1){32}}
\end{center}

\vspace{-118pt}

\makebox(110,60)[l]{}
\framebox(150,90)[0,0]{
{
\begin{minipage}{150pt}
\begin{center}
Z-Like Lepton-Photon \\
81 Gev $\lt M_{\ell\gamma}\lt$ 101 Gev\\
(Background Calibration) \\
{\bf\noflgzlike} Events
\end{center}
\end{minipage}
}
}
\makebox(20,60)[l]{}
\framebox(100,90)[0,0]{
{
\begin{minipage}{150pt}
\begin{center}
Exactly 1 Lepton \\
Exactly 1 Photon \\
$\Delta\phi_{l\gamma}\lt 150$ \\
$\met\lt 25$ GeV \\
{\bf\noflgzunlike} Events
\end{center}
\end{minipage}
}
}

\begin{center}
\framebox(115,70)[0,0]{
{
\begin{minipage}{115pt}
\begin{center}
Two-Body Events \\
{\bf\noflgtwobody} Events
\end{center}
\end{minipage}
}
}
\framebox(115,70)[0,0]{
{
\color{red}
\begin{minipage}{115pt}
\begin{center}
Multi-Body $\lgmet$ Events\\
$\met\gt 25$ GeV \\
{\bf\noflglgmet} Events
\end{center}
\end{minipage}
}
}
\framebox(115,70)[0,0]{
{
\color{red}
\begin{minipage}{115pt}
\begin{center}
Multi-Photon and \\
Multi-Lepton Events \\
{\bf\noflgmultig} and {\bf\noflgmultil} Events, resp. 
\end{center}
\end{minipage}
}
}
\end{center}
\caption
[The subsets of lepton-photon sample]
{Lepton-photon sample: the subsets of inclusive $\lg$ events analyzed.}
\label{flowchart_lepton.figure}
\end{figure*}

\end{huge}
\clearpage
\section{Stability of the Event Rates versus Run Number}
\label{observed_stability}
\begin{figure}[!h]
  \begin{center}
    \mbox{
\subfigure[$\ell\ell\gamma$]{\epsfig{file=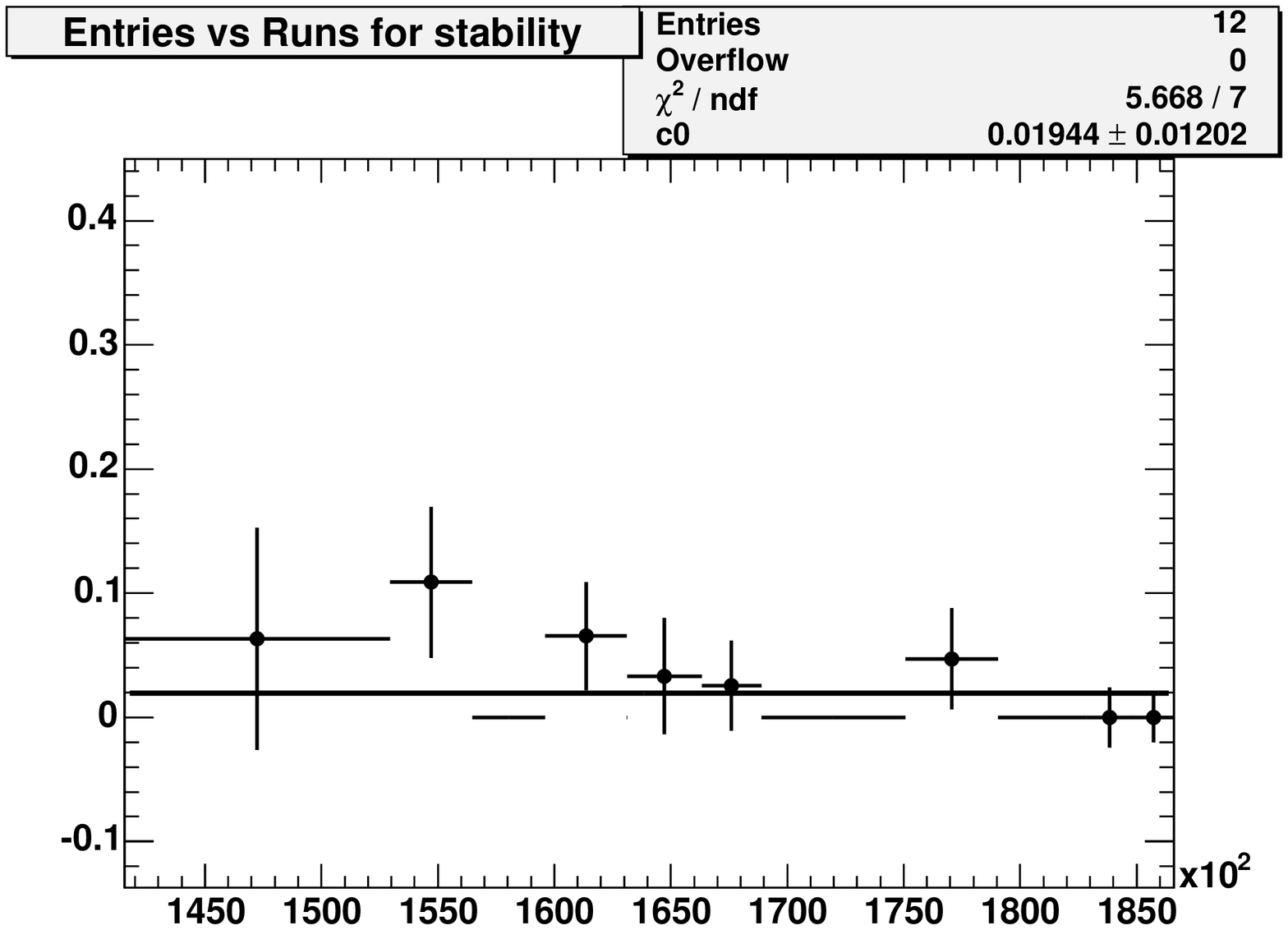,width=2.5in}} \quad
\subfigure[$\ell\gamma\met$]{\epsfig{file=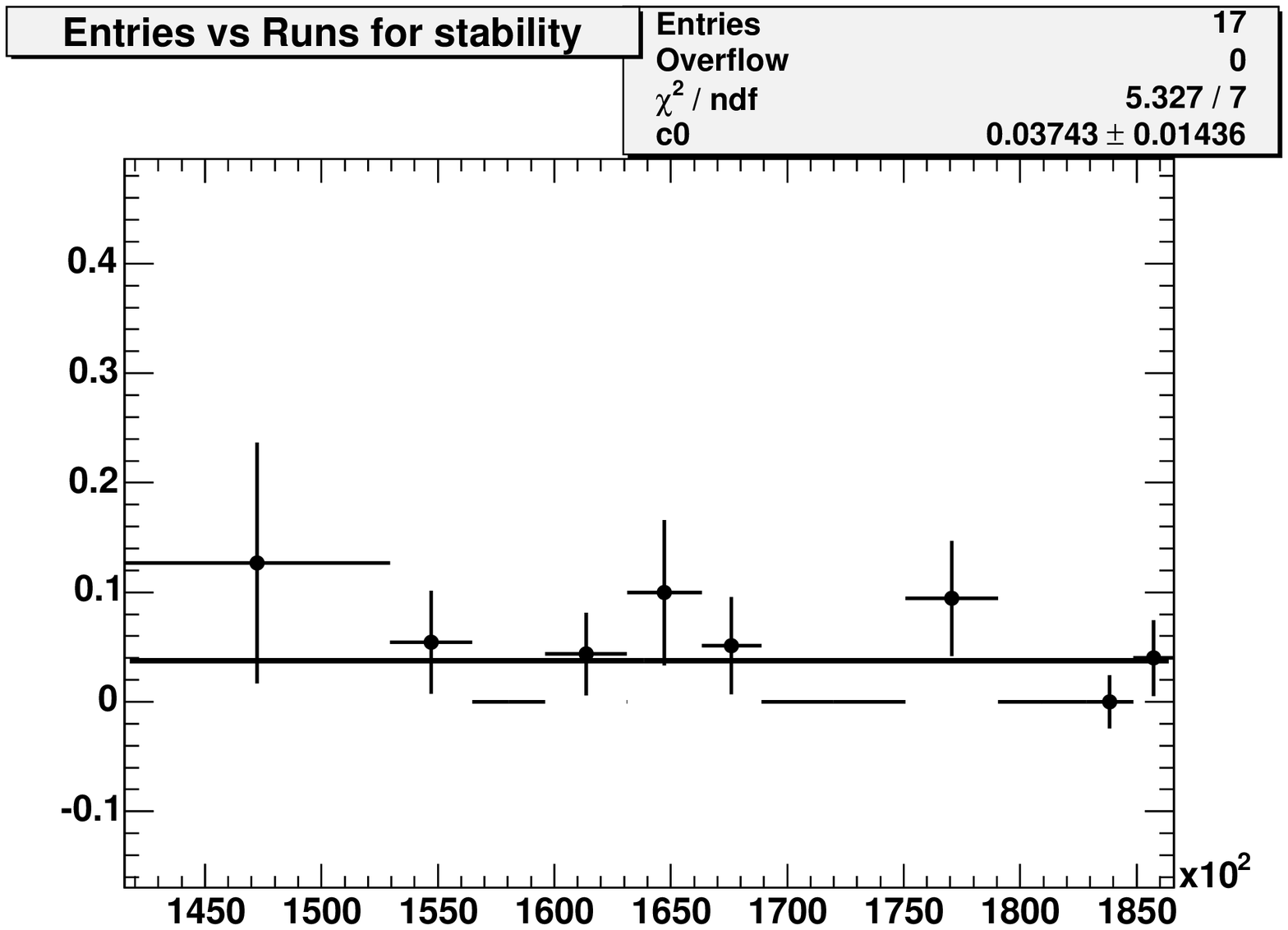,width=2.5in}} \quad
      }
    \caption 
[Stability plots for $\mumug$ and $\mugmet$]
{ Stability plots for the muon channels of the rate in
events per $\invpb$ for the 8 run segments (see
Section~\ref{datasets}) for: a) $\ell\ell\gamma$, and b) $\ell\gamma\met$.  
}
    \label{zmumuawmunuastability.figure}
  \end{center}
\end{figure}

\begin{figure}[!h]
  \begin{center}
    \mbox{
\subfigure[$\ell\ell\gamma$]{\epsfig{file=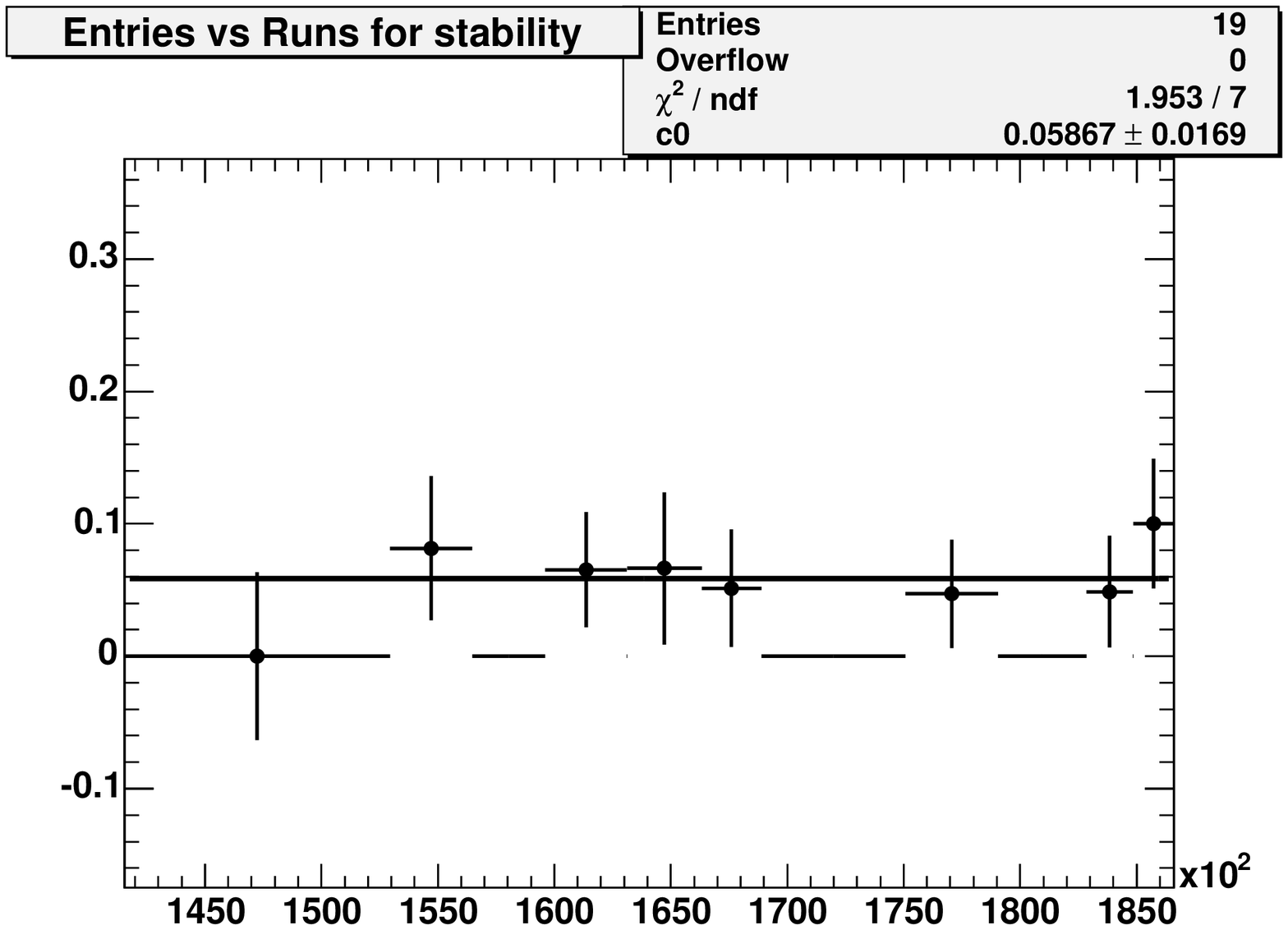,width=2.5in}} \quad
\subfigure[$\ell\gamma\met$]{\epsfig{file=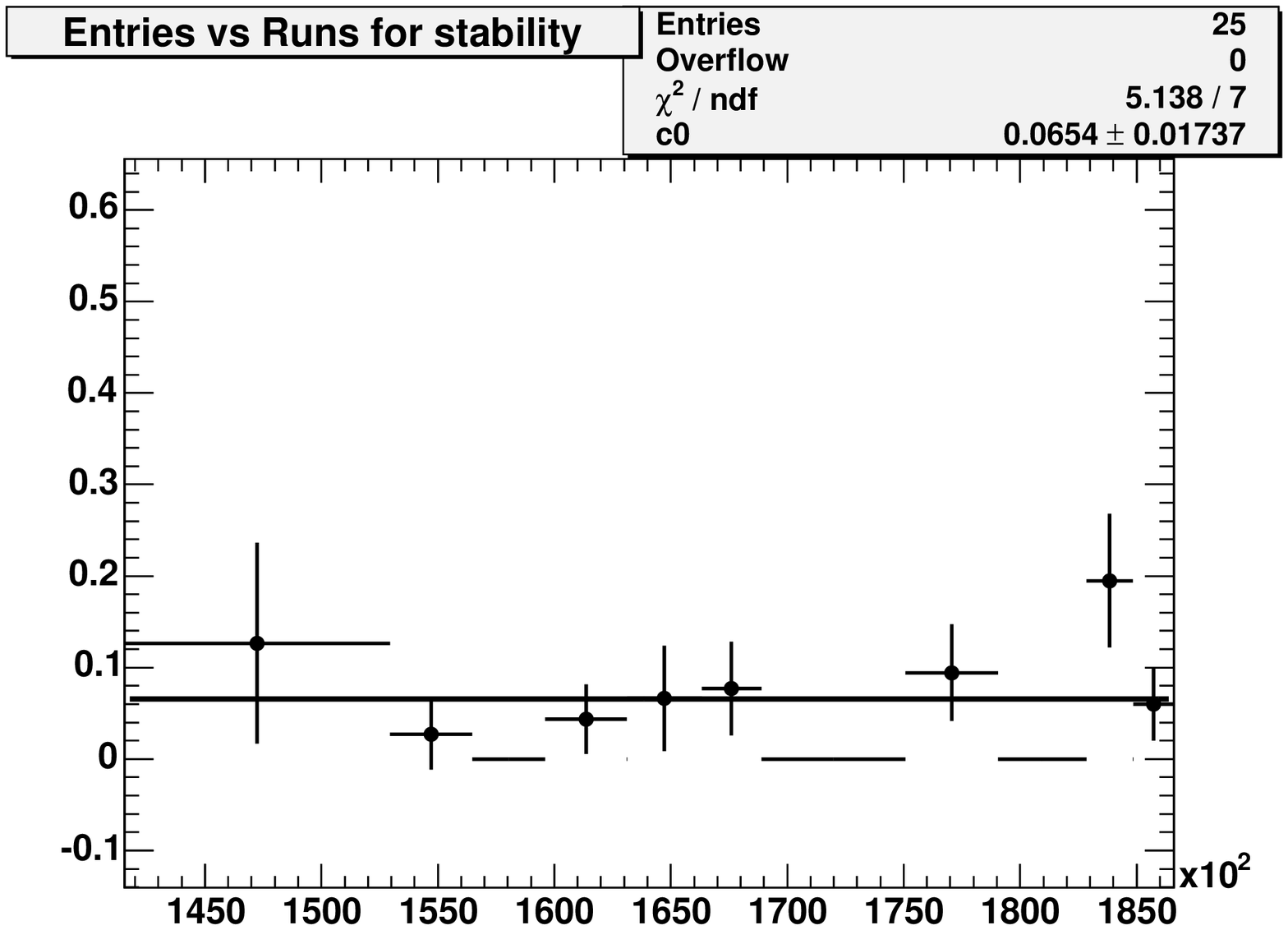,width=2.5in}} \quad
      }
    \caption 
[Stability plots for $\eeg$ and $\egmet$]
{ Stability plots for the electron channels of the rate in
events per $\invpb$ for the 8 run segments (see
Section~\ref{datasets}) for: a) $\ell\ell\gamma$, and b)
$\ell\gamma\met$.  }
    \label{zeeawenuastability.figure}
  \end{center}
\end{figure}

This sub-section looks at the rate for the analysis subcategories as a
function of run number. We use the same eight luminosity bins used to
check the stability of the control samples, described in
Section~\ref{datasets}. We see no obvious problems, although the
statistics are low. Figure~\ref{zmumuawmunuastability.figure} shows
the rates in events/$\invpb$ for the $\llg$ and $\lgmet$ signal
subcategories in each run segment in the muon channel;
Figure~\ref{zeeawenuastability.figure} does the same for the electron
channel.


\section{Results}
\label{comparison}

In this section we summarize the predicted and observed totals for the
$\lgmet$, $\llg$ and $\lgg$ searches. We compare observed kinematic
distributions to the SM predicted shapes.

\section{$\lgmet$ Search}
\label{lgmet.section}

The predicted and observed totals for the $\lgmet$ search are shown in
Table~\ref{lgmet_summary.table}. We observe $\noflglgmet$ $\lgmet$
events compared to the expectation of $\smnoflglgmet \pm
\totdsysnoflglgmet$ events. 

\begin{table}[!h]
\begin{center}
\begin{tabular}{lccc}

\multicolumn{4}{c} {
}\\
\hline
\multicolumn{4}{c} {
}\\
\multicolumn{4}{c} {
{\bf Lepton+Photon+$\bf\met$ Events}}\\
\hline
{\bf SM Source} & {\bf   $e\gamma\met$} & {\bf   $\mu\gamma\met$} & {\bf   $(e+\mu)\gamma\met$} \\
\hline
$W^{\pm}\gamma$                         & 
13.70$\pm$1.89    &
 8.84$\pm$1.35    &
22.54$\pm$2.80    \\
$\Zgstar + \gamma$                              & 
1.16$\pm$0.40     &
4.49$\pm$0.64    &
5.65$\pm$1.03     \\
$W^{\pm}\gamma\gamma, \Zgstar \plus \gamma\gamma$   & 
0.14$\pm$0.02   &
0.18$\pm$0.02   &
0.32$\pm$0.03  \\
$W^{\pm}\gamma,\Zgstar\plus\gamma\goes\tau\gamma$     & 
0.71$\pm$0.18  &
0.26$\pm$0.08  &
0.97$\pm$0.22  \\
\hline
$W^{\pm}$+Jet faking $\gamma$                     & 
2.8$\pm$2.8       &
1.6$\pm$1.6       &
4.4$\pm$4.4       \\
$\Zgstar \goes \epem,e\goes\gamma$           & 
2.45$\pm$0.33    &
-                                       &
2.45$\pm$0.33                           \\
Jets faking $\ell+\met$                         & 
0.7$\pm$0.7   &
0.3$\pm$0.3   &
1.0$\pm$0.8   \\
\hline
{\bf Total SM }   & & & \\ 
{\bf Prediction} & 
{\bf 21.7$\pm$3.4}     & 
{\bf 15.7$\pm$2.2}     & 
{\bf 37.3$\pm$5.4}     \\

\hline
{\bf Observed }   & & & \\ 
{\bf in Data} & 
{\bf   25}                          & 
{\bf   17}                          & 
{\bf   42}                          \\
\hline
\end{tabular}
\caption
[Summary table for the $\lgmet$ search]
{A comparison of the numbers of events predicted by the
standard model(SM) and the observations for the $\lgmet$ search.  The
SM predictions for the search are dominated by $\Wg$ production,
respectively~\cite{madgraph,baur,comphep}. Other contributions come
from $\Zg$ production, from the tri-boson processes $\Wgg$ and $\Zgg$,
leptonic $\tau$ decays, and misidentified leptons, photons, or
$\met$.}
\label{lgmet_summary.table} 
\end{center}
\end{table}

\begin{figure}[!h]
\begin{center}
\includegraphics*[angle =90,width=0.8\textwidth]{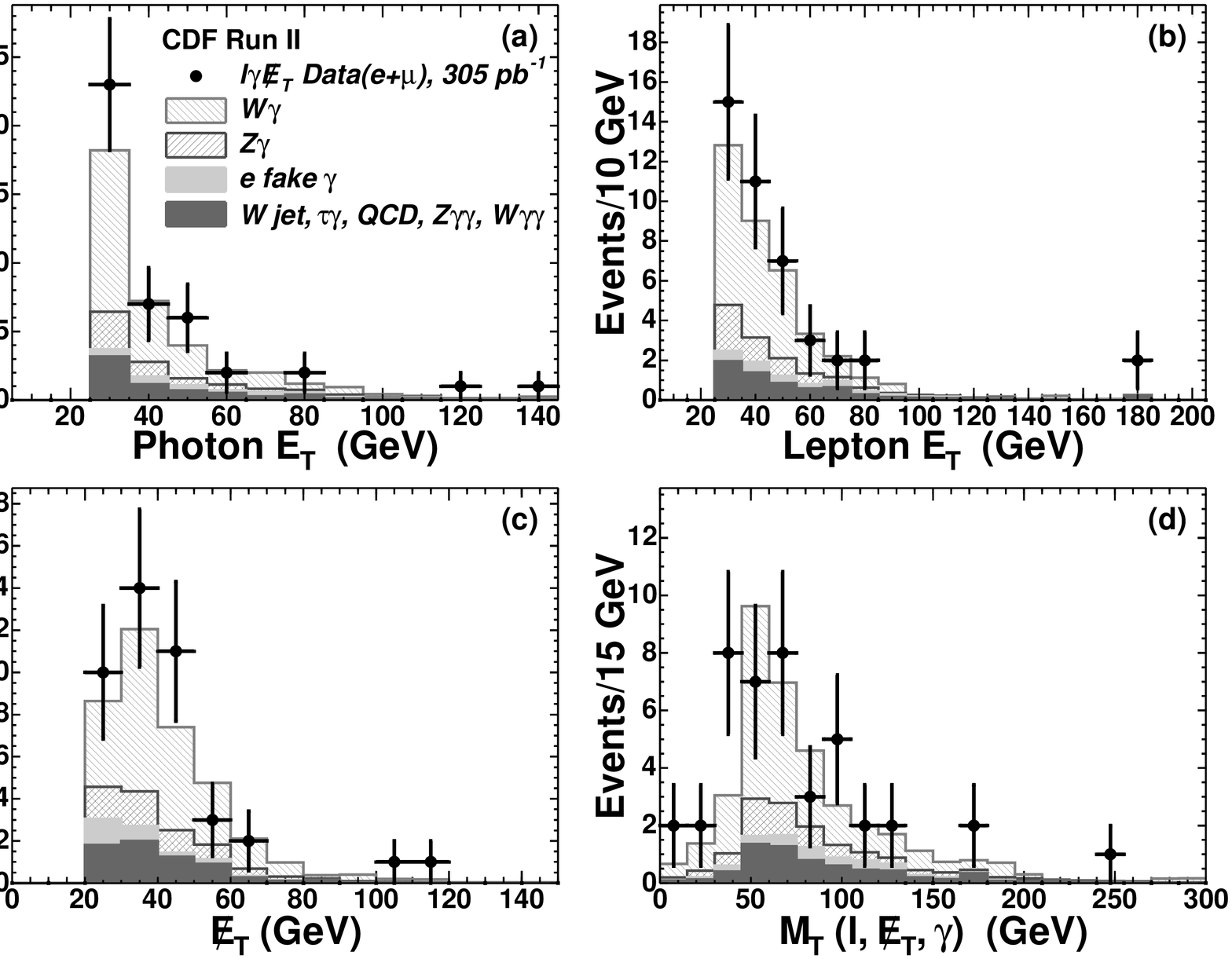}
\end{center}
\caption
[Distributions for the events in the $\lgmet$ sample]
{Distributions for the events in the $\lgmet$ sample (points)
  in a) the $\Et$ of the photon; b) the $\Et$ of the lepton; c) the
  missing transverse energy, $\met$; and d) the transverse mass of the
  $\lgmet$ system.  The histograms show the expected SM contributions,
  including estimated backgrounds from misidentified photons and
  leptons.}
\label{lgmet.figure}
\end{figure}

There is no significant excess in the $\lgmet$
signature. Figure~\ref{lgmet.figure} shows the observed distributions
summed over the $\egmet$ and $\mugmet$ events in a) the $\Et$ of the
photon; b) the $\Et$ of the lepton; c) the missing transverse energy,
$\met$; and d) the transverse mass of the $\lgmet$ system, where
$\rm{M_T} = [{(\rm E_T^\ell+E_T^\gamma + \met)^2}$ - $(\lepvec +
\phovec + \metvec)^2]^{1/2}$. 

\begin{figure}[!h]
\begin{center}
\includegraphics*[angle =90,width=0.8\textwidth]{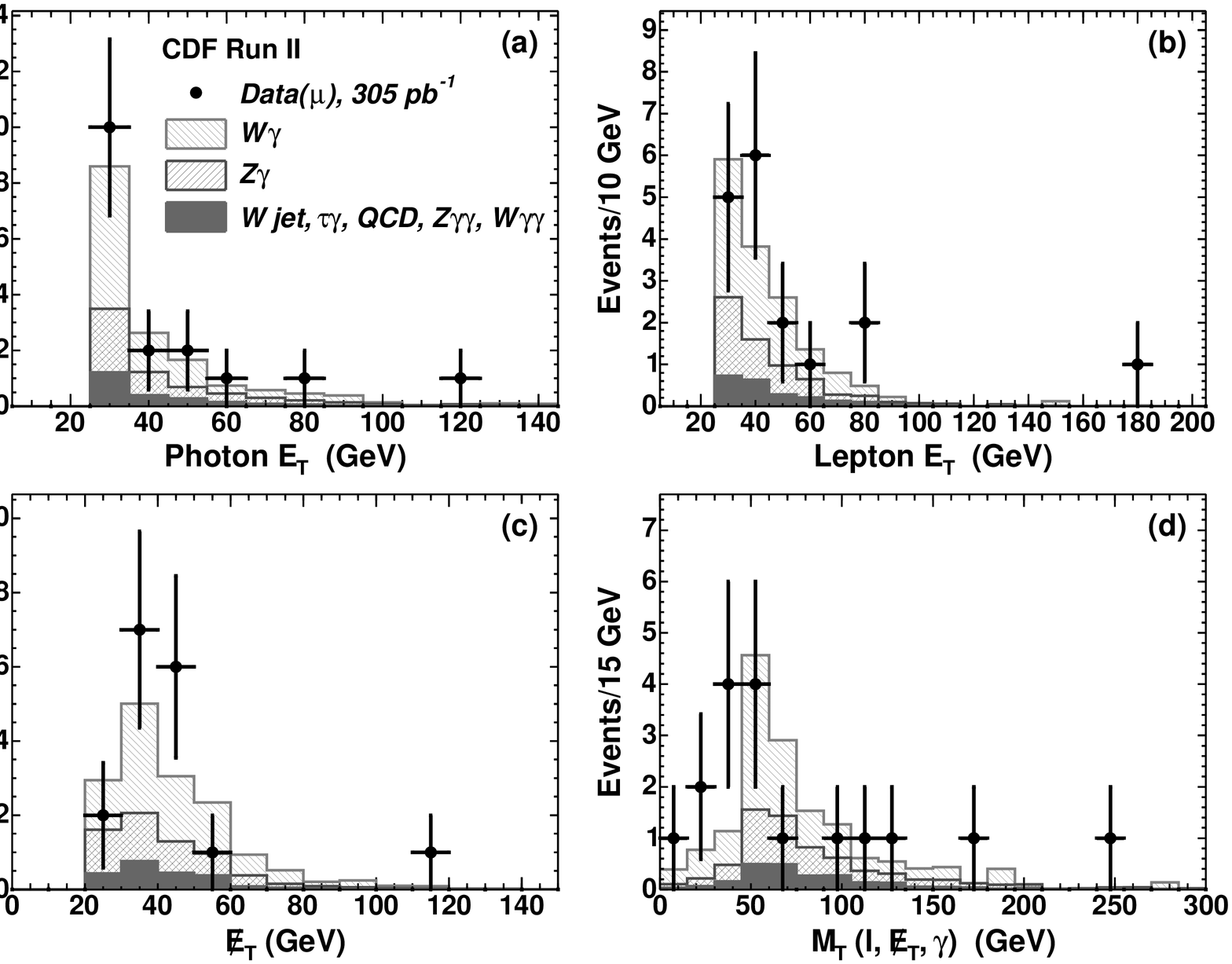}
\end{center}
\caption
[Distributions for the events in the $\mugmet$ sample]
{Distributions for the events in the $\mugmet$ sample (points)
  in a) the $\Et$ of the photon; b) the $\Et$ of the muon; c) the
  missing transverse energy, $\met$; and d) the transverse mass of the
  $\lgmet$ system.  The histograms show the expected SM
  contributions, including estimated backgrounds from misidentified
  photons and leptons.}
\label{mugmet.figure}
\end{figure}

\begin{figure}[!h]
\begin{center}
\includegraphics*[angle =90,width=0.8\textwidth]{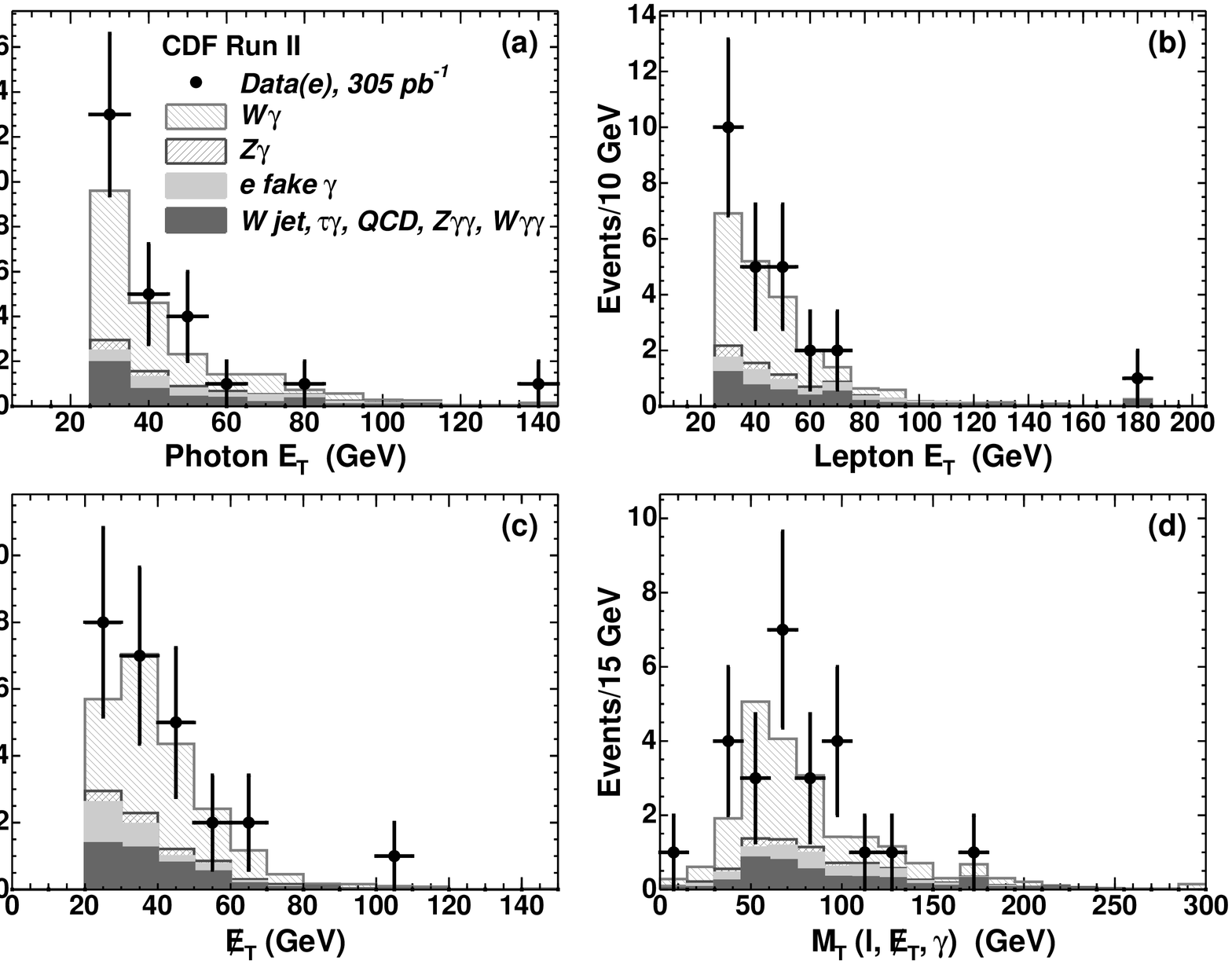}
\end{center}
\caption
[Distributions for the events in the $\egmet$ sample]
{Distributions for the events in the $\egmet$ sample (points)
  in a) the $\Et$ of the photon; b) the $\Et$ of the electron; c) the
  missing transverse energy, $\met$; and d) the transverse mass of the
  $\lgmet$ system.  The histograms show the expected SM
  contributions, including estimated backgrounds from misidentified
  photons and leptons.}
\label{egmet.figure}
\end{figure}

The predicted and observed kinematic distributions for $\mugmet$ are
compared in Figure~\ref{mugmet.figure}. The distributions for $\egmet$
signature are compared in Figure~\ref{egmet.figure}.

The additional plots of the identification variables for $\egmet$ and
$\mugmet$ are available in Section~\ref{lgmet_plots}.

\clearpage
\newpage
\section{$\llg$ Search}
\label{llg.section}

The predicted and observed totals for the $\llg$ search are shown in
Table~\ref{llg_summary.table}. We observe $\noflgmultil$ $\llg$ events
compared to the expectation of $\smnoflgmultil \pm
\totdsysnoflgmultil$ events. 

\begin{table}[!h]
\begin{center}
\begin{tabular}{lccc}
\hline
\multicolumn{4}{c} {
}\\
\multicolumn{4}{c}{
{\bf Multi-Lepton + Photon Events}}\\ 
\hline
{\bf SM Source} & {\bf  $ee\gamma$} & {\bf  $\mu\mu\gamma$} & {\bf  $ll\gamma$}  \\
\hline
$\Zgstar + \gamma$            		          & 
12.50$\pm$1.53    &
 7.81$\pm$0.88    &
20.31$\pm$2.40    \\
$\Zgstar + \gamma\gamma$ 	     	& 
0.24$\pm$0.03  &
0.12$\pm$0.02 &
0.36$\pm$0.04  \\
\hline
$\Zgstar+$Jet faking $\gamma$   & 
0.3$\pm$0.3     &
0.2$\pm$0.2     &
0.5$\pm$0.5     \\
$\Zgstar \goes \epem,e\goes\gamma$           & 
0.23$\pm$0.09    &
-                                       &
0.23$\pm$0.09                           \\
Jets faking $\ell+\met$                         & 
0.6$\pm$0.6  &
1.0$\pm$1.0  &
1.6$\pm$1.2  \\
\hline
{\bf Total SM }          & & & \\ 
{\bf Prediction} & 
{\bf 13.9$\pm$1.7}    & 
{\bf  9.1$\pm$1.4}    & 
{\bf 23.0$\pm$2.7}   \\

\hline
{\bf Observed }   & & & \\ 
{\bf in Data} & 
{\bf   19}                          & 
{\bf   12}                          &                 
{\bf   31}                          \\
\hline
\end{tabular}
\caption
[Summary table for the $\llg$ search]
{A comparison of the numbers of events predicted by the
standard model(SM) and the observations for the $\llg$ search. The SM
predictions for the search are dominated by $\Zg$
production~\cite{madgraph,baur,comphep}. Other contributions come from
the tri-boson process $\Zgg$, and misidentified leptons or photons.}
\label{llg_summary.table} 
\end{center}
\end{table}

\begin{figure}[!h]
\begin{center}
\includegraphics*[angle =90,width=0.8\textwidth]{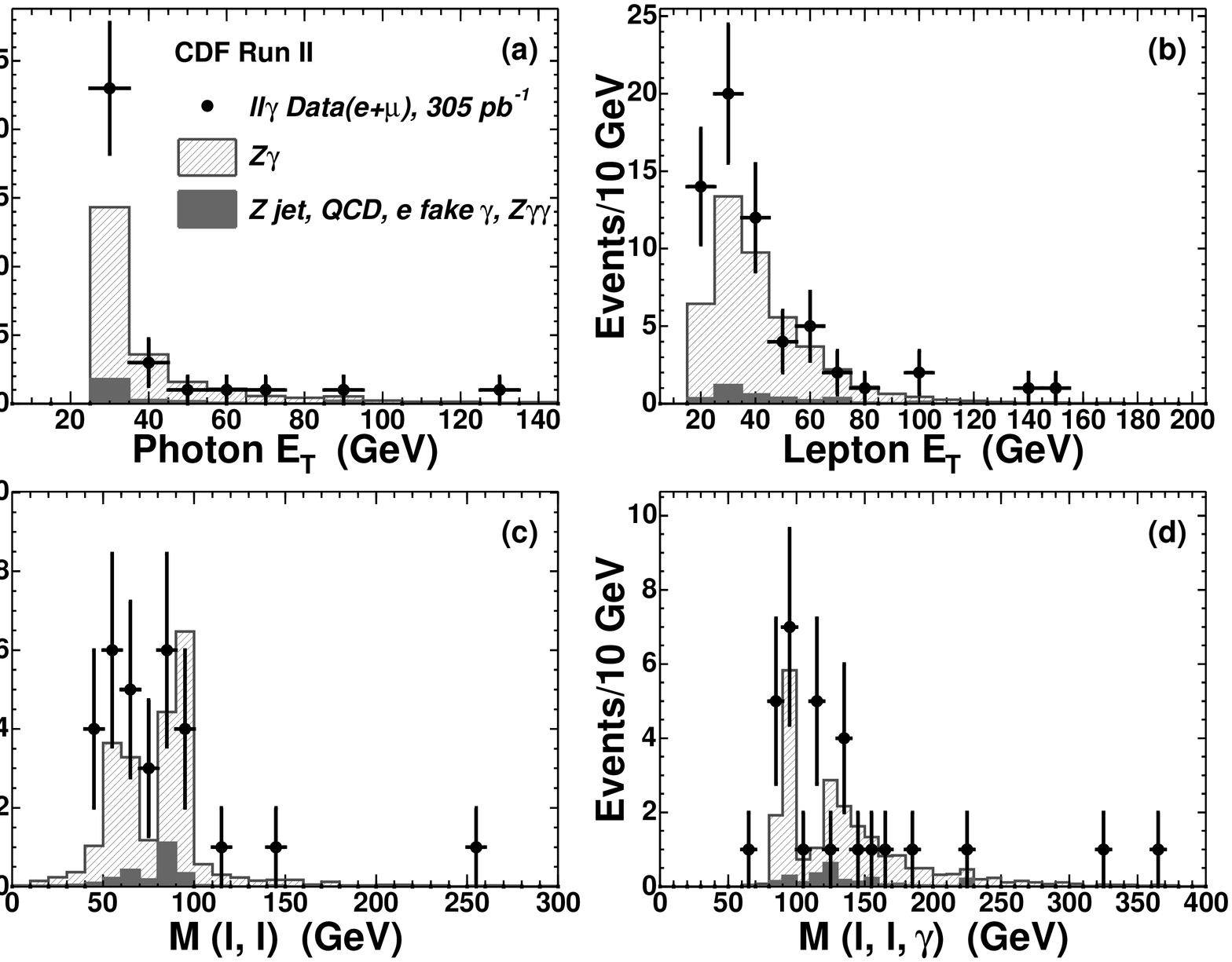}
\end{center}
\caption
[Distributions for the events in the $\llg$ sample]
{Distributions for the events in the $\llg$ sample (points)
in a) the $\Et$ of the photon; b) the $\Et$ of the leptons (two
entries per event); c) the 2-body mass of the dilepton system; and d)
the 3-body mass $m_{\llg}$. The histograms show the expected SM
contributions.}
\label{llg.figure}
\end{figure}

The $\llg$ search criteria select $\noflgmultil$ events
($\nofegmultil$ $\eeg$ and $\nofmugmultil$ $\mumug$) of the $\noflg$
$\lg$ events. No $e\mu\gamma$ events are
observed. Figure~\ref{llg.figure} shows the observed distributions in
a) the $\Et$ of the photon; b) the $\Et$ of the leptons; c) the 2-body
mass of the dilepton system; and d) the 3-body mass $m_{\llg}$. For
the $\Zg$ process occurring via initial state radiation, the dilepton
invariant mass $m_{\ll}$ distribution is peaked around the
$Z^0$-pole. For the final state radiation, the three body invariant
mass $m_{\llg}$ distribution is peaked about the $Z^0$-pole.

\begin{figure}[!h]
\begin{center}
\includegraphics*[angle =90,width=0.8\textwidth]{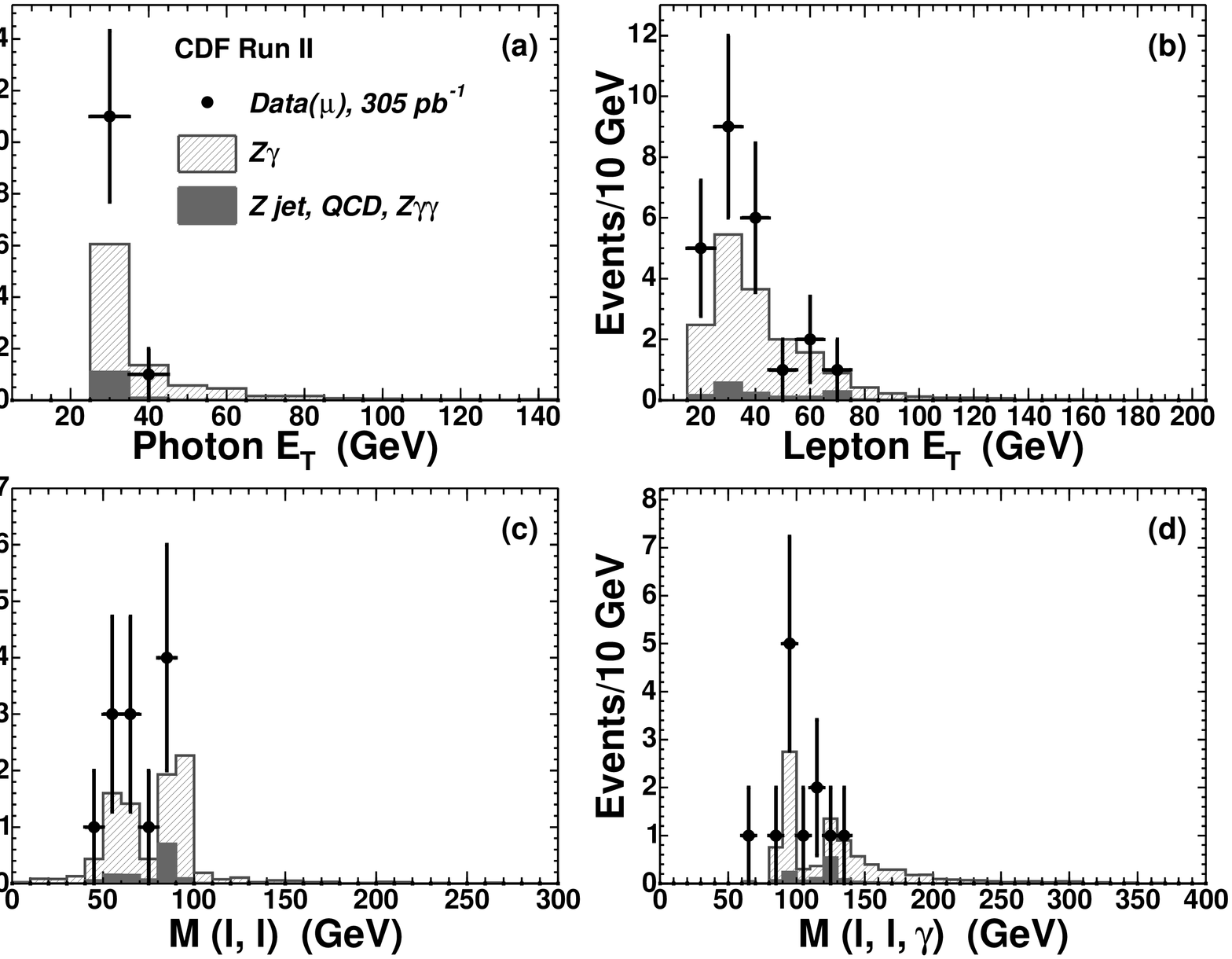}
\end{center}
\caption
[Distributions for the events in the $\mumug$ sample]
{Distributions for the events in the $\mumug$ sample (points)
in a) the $\Et$ of the photon; b) the $\Et$ of the muons (two
entries per event); c) the 2-body mass of the dimuon system; and d)
the 3-body mass $m_{\mumug}$. The histograms show the expected SM
contributions.}
\label{mumug.figure}
\end{figure}

\begin{figure}[!h]
\begin{center}
\includegraphics*[angle =90,width=0.8\textwidth]{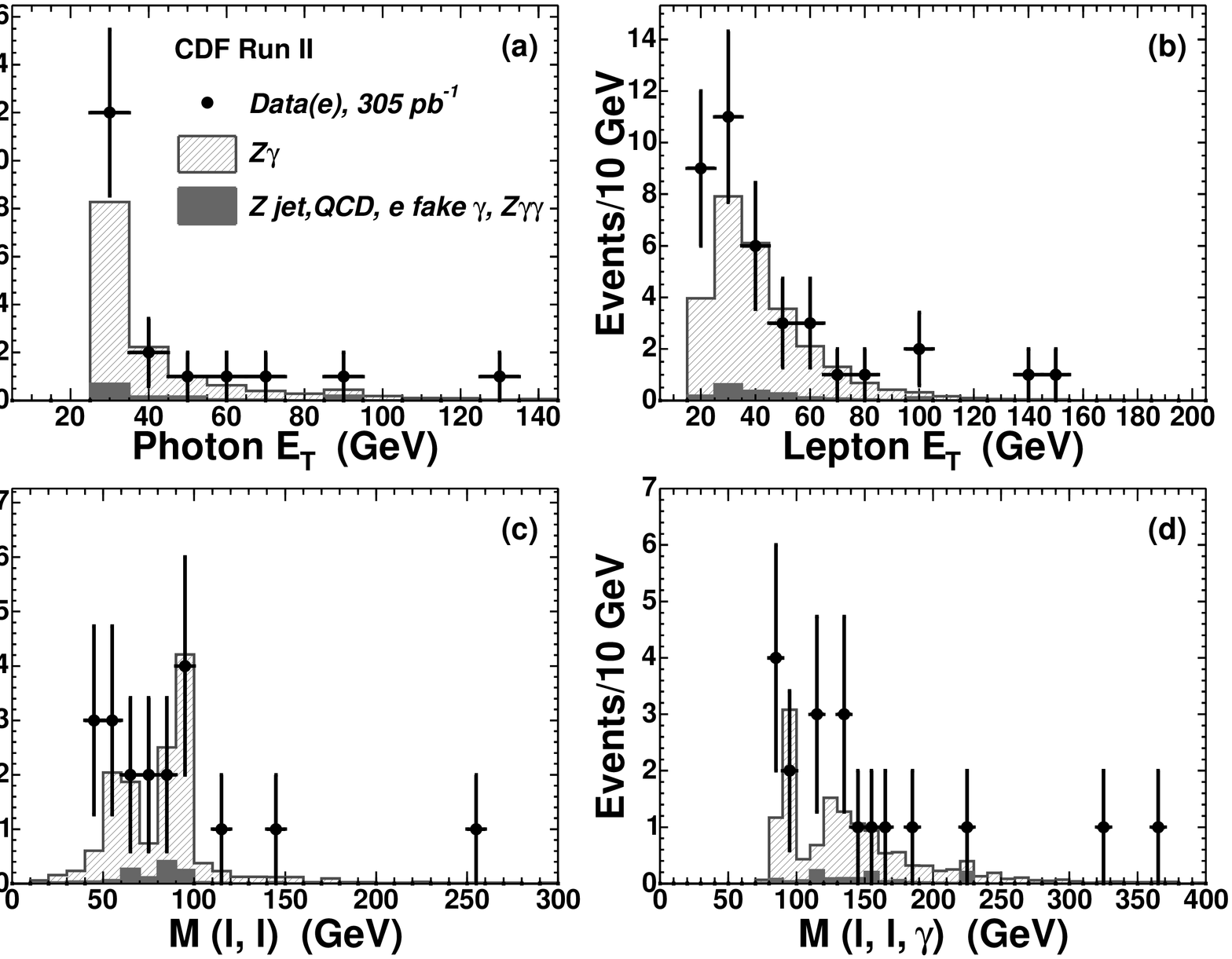}
\end{center}
\caption
[Distributions for the events in the $\eeg$ sample]
{Distributions for the events in the $\eeg$ sample (points)
in a) the $\Et$ of the photon; b) the $\Et$ of the electrons (two
entries per event); c) the 2-body mass of the dielectron system; and d)
the 3-body mass $m_{\eeg}$. The histograms show the expected SM
contributions.}
\label{eeg.figure}
\end{figure}

The predicted and observed kinematic distributions for $\mumug$ are
compared in Figure~\ref{mumug.figure}. The distributions for $\eeg$
signature are compared in Figure~\ref{eeg.figure}. The dominant
contribution for the $\eeg$ and $\mumug$ signatures is from the SM
$\Zg$ production.

\begin{figure}[!b]
\begin{center}
\includegraphics*[width=0.35\textwidth, angle=90,clip=]{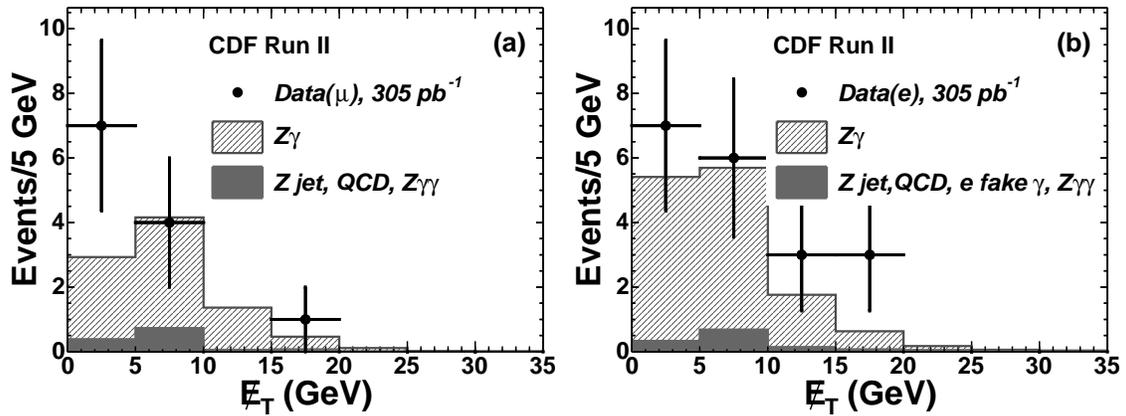}
\end{center}
\caption
[Distributions in $\met$ for the $\mumug$ and $\eeg$ events]
{Distributions in missing transverse energy $\met$
  observed in the inclusive search for a) $\mumug$ events and
  b) $\eeg$ events. The histograms show the expected SM
  contributions.}
\label{llg_met.figure}
\end{figure}

We do not expect events with large $\met$ in the $\llg$ sample, based
on the SM backgrounds; the Run I $\eeggmet$ event was of special
interest in the context of supersymmetry~\cite{susy,allanach-2002-89}
due to the large value of $\met$ (55 $\pm$ 7
$\GeV$). Figure~\ref{llg_met.figure} shows the distributions in $\met$
for the $\mumug$ and $\eeg$ subsamples of the $\llg$ sample. No events
are observed with $\met> 25$ $\GeV$.

The additional plots of the identification variables for $\eeg$ and
$\mumug$ are available in Section~\ref{llg_plots}.

\clearpage
\newpage
\section{$\lgg$ Search}
\label{lgg.section}

The predicted and observed totals for the $\lgg$ search are shown in
Table~\ref{lgg_summary.table}. We do not observe any $\lgg$ candidate
events compared to the expectation of $\smnoflgmultig \pm
\totdsysnoflgmultig$ events. 

\begin{table}[!h]
\begin{center}
\begin{tabular}{lccc}
\hline
\multicolumn{4}{c}{
} \\
\multicolumn{4}{c}{
\bf Multi-Photon + Lepton Predicted Events} \\
\hline
{\bf SM Source} & $e\gg$ & $\mu\gg$ & $\l\gg$ \\
\hline
$W^{\pm}\gamma\gamma$                     & 
$\waanofegmultig  \pm \waadnofegmultig$   &
$\waanofmugmultig \pm \waadnofmugmultig$  &
$\waanoflgmultig  \pm \waadnoflgmultig$   \\
$\Z\gamma\gamma$                        & 
$\zaanofegmultig  \pm \zaadnofegmultig$   &
$\zaanofmugmultig \pm \zaadnofmugmultig$  &
$\zaanoflgmultig  \pm \zaadnoflgmultig$   \\
$\Z\gamma$, $e\goes\gamma$                & 
$\efpnofegmultig  \pm \efpdnofegmultig$   &
-                                         &
$\efpnofegmultig  \pm \efpdnofegmultig$   \\
\hline
{\bf Total SM  }             & & & \\
{\bf Prediction }            & 
$\bf\smnofegmultig  \pm \totdsysnofegmultig $    &
$\bf\smnofmugmultig \pm \totdsysnofmugmultig$   &
$\bf\smnoflgmultig  \pm \totdsysnoflgmultig $    \\
\hline
{\bf Observed}                        & & & \\
{\bf in Data }                        & 
$\nofegmultig$                           & 
$\nofmugmultig$                          & 
$\noflgmultig$                           \\
\hline
\end{tabular}
%
%
\caption
[Summary table for the $\lgg$ search]
{The predicted number of multi-body events with additional photons in
$\apprluminosity\pbinv$ from Standard Model sources, and the numbers
of events observed. The $Z\gamma\gamma$ and $W\gamma\gamma$
predictions are from MadGraph.  The ``$\Z\gamma$,
$e\rightarrow\gamma$'' prediction is taken from the data. }
\label{lgg_summary.table}
\end{center}
\end{table}

\chapter*{Conclusions}
\label{conclusions}
\addcontentsline{toc}{section}{Conclusions}

In conclusion, we have searched for physics beyond the Standard Model
in a channel in which new phenomena due to supersymmetry or extra
dimensions, for example, could well appear. In particular, we have
repeated the Run I CDF search for inclusive lepton + photon
production, a final state in which both a very rare event appeared and
also in which there seemed to be an excess over SM predictions. The
new analysis, the subject of this thesis, was done specifically with
the same kinematic requirements as the Run I search, but with a
significantly larger data sample and a higher collision energy.

We conclude that the excess in the $\lgmet$ signature in Run I of 2.7
sigma was at least largely a statistical fluctuation. If the Run I
ratio of the number of $\lgmet$ events observed to the number
expected, 16/7.6, had held up, the ``2.7 $\sigma$ excess'' in this
channel observed in Run I would have resulted in an observation of
78$\pm$11 events when $\smnoflglgmet \pm \totdsysnoflglgmet$ are
expected in the Run II repeat of the analysis, versus the
$\noflglgmet$ events observed.

We find that the numbers of events in the $\lgmet$ and $\llg$
subsamples of the $\lgX$ sample agree with the SM predictions. We find
no events like the $\eeggmet$ candidate event of Run I, and, even more
generally, observe no $\llg$ events with anomalous large $\met$ or
with multiple photons. We have no explanation for the Run I $\eeggmet$
event, and nothing we have measured leads us to believe that is
background. 

However, we still find that in the $\lgmet$ signature in the Run II
data the number of observed events is higher than predicted, although
the excess is now slight. It is possible that the leading-order +
K-factor theoretical calculations of the diboson $\Wg$ and $\Zg$
channels, which contain many diagrams including initial state
radiation, are not precise enough for the precision we have now
reached. 

Alternatively, or perhaps in addition, this analysis has observed a
small number of events on the `tails' of the kinematic distributions,
in regions we expect few SM events. These events contribute to the
observation of more events than expected in the $\lgmet$ signature,
much as the $\eeggmet$ event contributed to the excess in the Run I
search.  Whether these are very rare backgrounds or something new will
require yet more data.

The Fermilab plan is to have a factor of 10-20 more data than
presented here by the end of Run II of the Tevatron. The increased
statistics will require an improved understanding of backgrounds as
well as better SM predictions. In particular the estimate of the rate
for a jet to be misidentified as a photon is limited now by the
statistics for high-$\Et$ photons; we expect the estimate will
significantly improve with more data.

In summary, while it would have been very exciting to find physics
beyond the Standard Model, we found no more $\eeggmet$ events in a
much larger sample than in Run I, and the Run I excess in $\lgmet$
became less significant rather than more. However, we have
conclusively settled a question that generated much interest in the
theoretical community. The channels we have investigated will remain
interesting, and the techniques we have developed and the knowledge
gained will be useful for similar searches at the LHC.

\chapter*{Acknowledgments}
\label{acknowledgments}
\addcontentsline{toc}{section}{Acknowledgments}

I thank my thesis co-supervisors Henry Frisch (UC) and Andrey
Rostovtsev (ITEP) for their encouragement and support. I learned many
things from my co-supervisors, and most important of which is to enjoy
what I do and keep my eyes wide open.

To keep my body as active as my mind really helped me to survive years
of my graduate study. I learned two basic things: ``no pain - no
gain'' and ``what doesn't break me makes me stronger''. These two
simple things still keep me going.

I would like to thank Eduard Boos and Lev Dudko for their support and
development of CompHep, and Tim Steltzer, Steve Mrenna, and Fabio
Maltoni for their similar responsiveness for MadGraph and
Pythia. Alexander Belyaev, and Alexander Sherstnev generated the
CompHep $\Wgg$ and $\Zgg$ datasets, respectively, for the search. Uli
Baur provided wisdom on the effective Born approximation parameters
for the LO parameters. Steve Levy, Peter Onyisi, Alexander Paramonov,
Carla Pilcher, and Collin Wolfe provided invaluable support for the
UCNtuple and datasets.

I want to thank GodParents of the $\lgX$ PRL publication Bob Blair,
Ron Moore and Antonio Sidoti, for all the time and effort they put
into the paper in order to improve it and see it published in time for
my thesis defense. I also want to thank my thesis opponents, Leonid
Gladilin and Vladimir Gavrilov, for their questions and comments. I
wish to thank Paul Tipton for careful proof reading my thesis.

I would like to thank my friends in Russia, who helped me to have
necessary (many!) papers to have the pre-defense process started while
I was in the US - Dmitry Liventsev, Roman Vishnitsky, Alexander and
Elena Zykovy.

I want to thank many people at ITEP who helped me to go through all
the formalities for my thesis defense - Tagir Aushev, Elena
Filimonova, Valentina Korchagina, Elena Minervina, Mikhail Trusov,
Valery Vasil'ev and Tatiana Tokareva. I would like to thank those who
spent a lot of time helping me to start doing research at ITEP -
Alexei Drutskoy, Alexander Fedotov, Dmitry Ozerov, Vyacheslav Zaharov,
Alexander Zhokin.

I am grateful to Aspasia Sotir-Plutis and Veronica ("Vicki")
McClain-Stone, who helped me with different formalities at the
University of Chicago, which allowed me to concentrate on performing
the search.

I also would like to thank my CDF collaborators Anadi Canepa, Kathy
Copic, Ray Culbertson, Max Goncharov, Helen Hayward, Beate Heinemann,
Heather Gerberich, Al Goshaw, Mike Kirby, Bruce Knuteson, Konstantin
Kotelnikov, Slava Krutelyov, Giulia Manca, Pasha Murat, Jane Nachtman,
Alexei Safonov, Tara Shears, Irina Shreyber, Reda Tafirout, Stan
Thompson, Jason Tsui, Soushi Tsuno, Song Ming Wang, and Un-Ki Yang for
their many contributions.

\appendix
\chapter{Appendices}
\label{a.appendix}

\section{List Of Lepton-Photon Events}
\label{list_of_lgx}
\begin{table}[h]
\small
 \begin{center}
\input{local_tables/en_muonphotonmet.table}
 \end{center}
\caption
[List of muon + photon + $\met$ events]
{List of muon + photon + $\met$ events. $\Et$, $\Pt$, $\met$ and
$m_{\lg}$ are in $\GeV$. Column ``T'' shows if an event has been
triggered by the high-$\Et$ muon trigger. ``+'' in front of $\Etgamma$
value means that an event has been triggered by the high-$\Et$ photon
trigger.}
\label{muonphotonmet.table}
\end{table}

\begin{table}[h]
\small
 \begin{center}
\input{local_tables/en_muonphotonmulti.table}
 \end{center}
\caption
[List of multi-muon + photon events]
{List of multi-muon + photon events. $\Et$, $\Pt$, $\met$ and $m_{\lg}$
are in $\GeV$. Column ``T'' shows if an event has been
triggered by the high-$\Et$ muon trigger. ``+'' in front of $\Etgamma$ value
means that an event has been triggered by the high-$\Et$ photon trigger.}
\label{muonphotonmulti.table}
\end{table}

\begin{table}[!h]
\small
 \begin{center}
\input{local_tables/en_electronphotonmet.table}
\caption
[List of electron + photon + $\met$ events]
{List of electron + photon + $\met$ events. $\Et$, $\Pt$, $\met$ and
$m_{\lg}$ are in $\GeV$. TCEM stands for Tight CEM electron. Column
``T'' shows if an event has been triggered by the high-$\Et$ electron
trigger. ``+'' in front of $\Etgamma$ value means that an event has
been triggered by the high-$\Et$ photon trigger.}  
\end{center}
\label{electronphotonmet.table}
\end{table}

\clearpage
\newpage
\begin{small}
\begin{longtable} {@{\extracolsep{0.3cm}}l|@{\extracolsep{0.3cm}}r@{\extracolsep{0.3cm}}r@{\extracolsep{0.3cm}}r|@{\extracolsep{0.3cm}}l@{\extracolsep{0.3cm}}r@{\extracolsep{0.3cm}}r@{\extracolsep{0.3cm}}r@{\extracolsep{0.3cm}}l|@{\extracolsep{0.3cm}}r@{\extracolsep{0.3cm}}r|@{\extracolsep{0.3cm}}r@{\extracolsep{0.3cm}}r}
\endfirsthead
\multicolumn{13}{c}%
{{\bfseries \tablename\ \thetable{} -- continued}} \\
\hline 
\endhead
\hline \multicolumn{13}{|l|}{{Continued on the next page}} \\ \hline
\endfoot
\hline
\endlastfoot
\hline
\input{local_tables/en_electronphotonmulti.table}
\caption
[List of multi-electron + photon events]
{List of multi-electron + photon events. $\Et$, $\Pt$, $\met$ and
$m_{\lg}$ are in $\GeV$. TCEM stands for Tight CEM Electron. LCEM
stands for Loose CEM Electron. HCEM stands for Tight100 CEM
Electron. EAST stands for Phoenix East Plug Electron, WEST stands for
Phoenix West Plug Electron. Column ``T'' shows if an event has been
triggered by the high-$\Et$ electron trigger. ``+'' in front of
$\Etgamma$ value means that an event has been triggered by the
high-$\Et$ photon trigger. } 
\label{electronphotonmulti.table}
\end{longtable}
\end{small}

\clearpage
\newpage
\section{Additional $\lgmet$ Plots}
\label{lgmet_plots}
In this section we present additional plots of the identification
variables for $\mugmet$ and $\egmet$ signatures.
\subsection{Additional $\mugmet$ Plots}
\begin{figure}[!h]
  \begin{center}
\vskip-0.2in
    \mbox{
\subfigure
[$\Delta R$($\mu\gamma$)]
{\epsfig{file=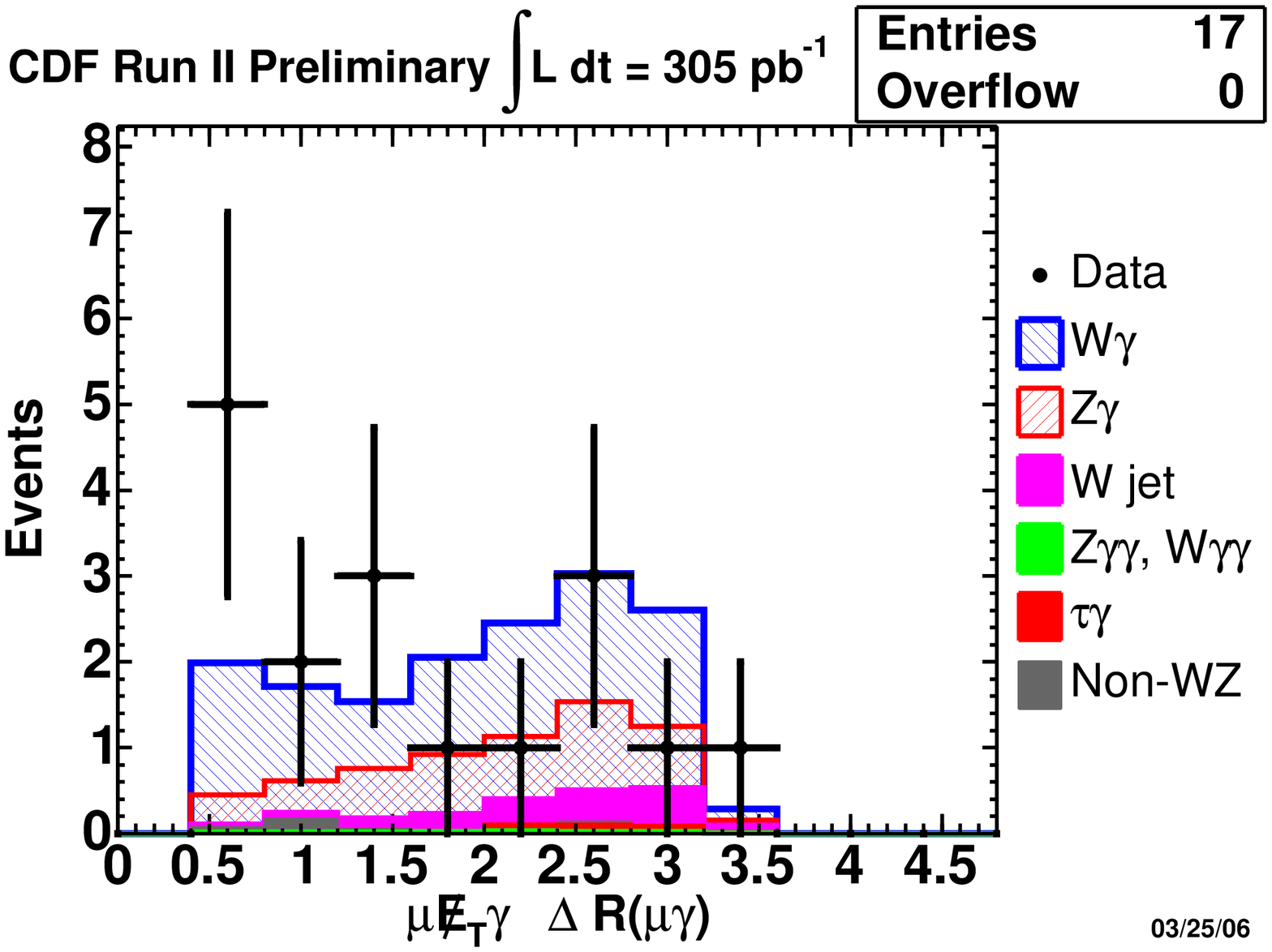,width=3.3in}} \quad
\subfigure
[$H_T$ for $\mugmet$ Candidate Events]
{\epsfig{file=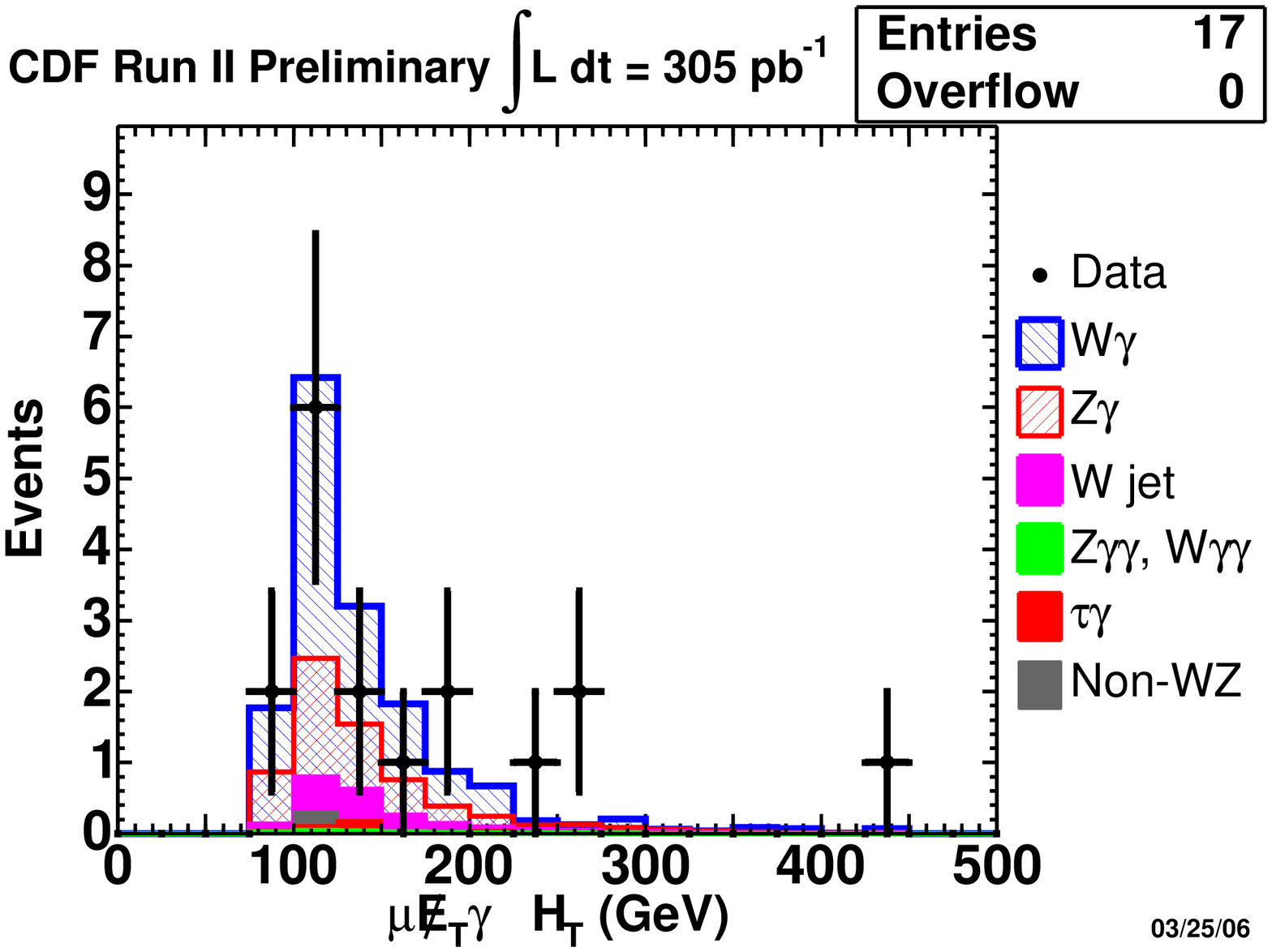,width=3.3in}} \quad
      }
\vskip-0.2in
    \mbox{
\subfigure
[Muon Detector Type]
{\epsfig{file=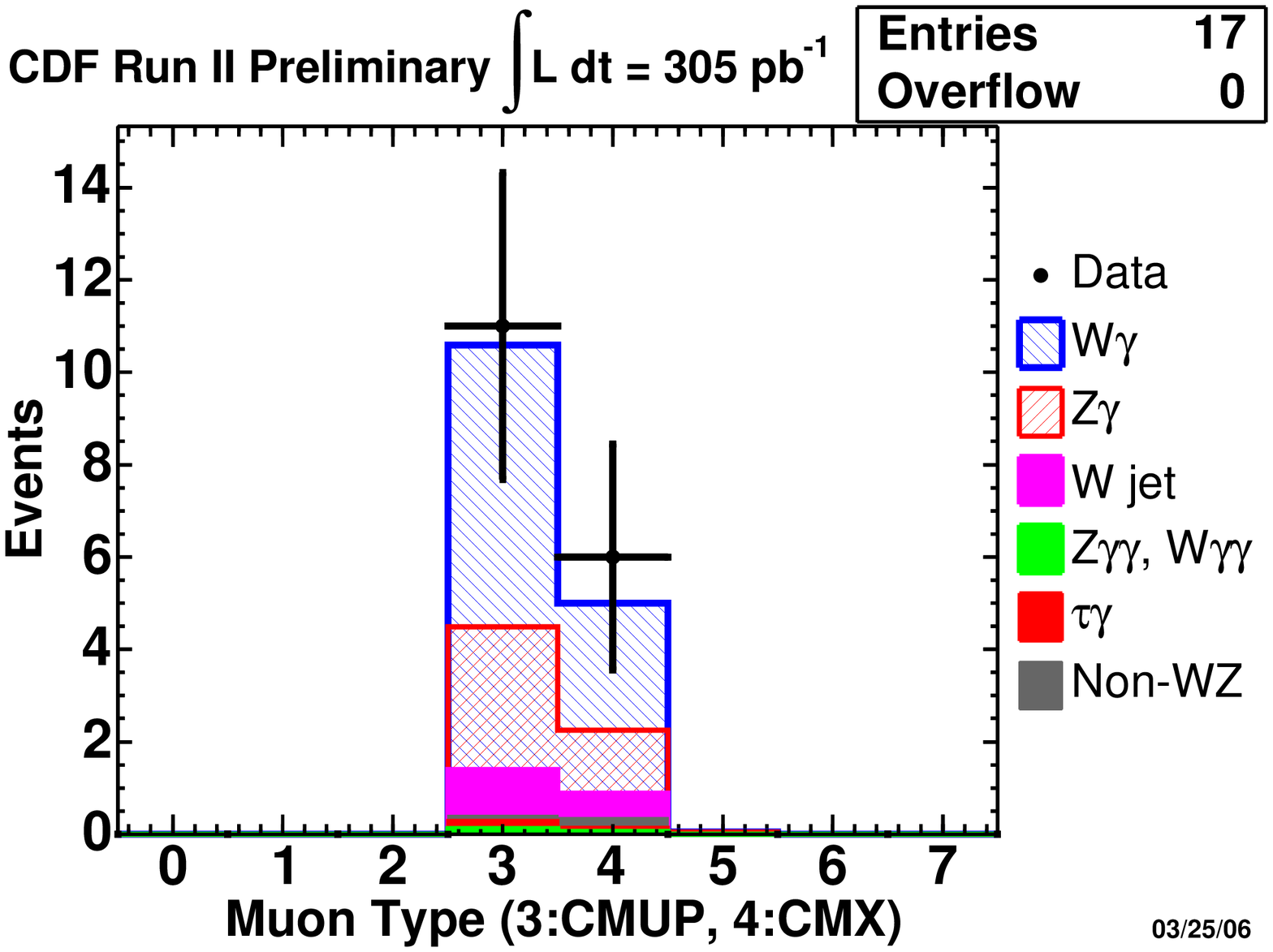,width=3.3in}} \quad
\subfigure
[Muon $\eta$]
{\epsfig{file=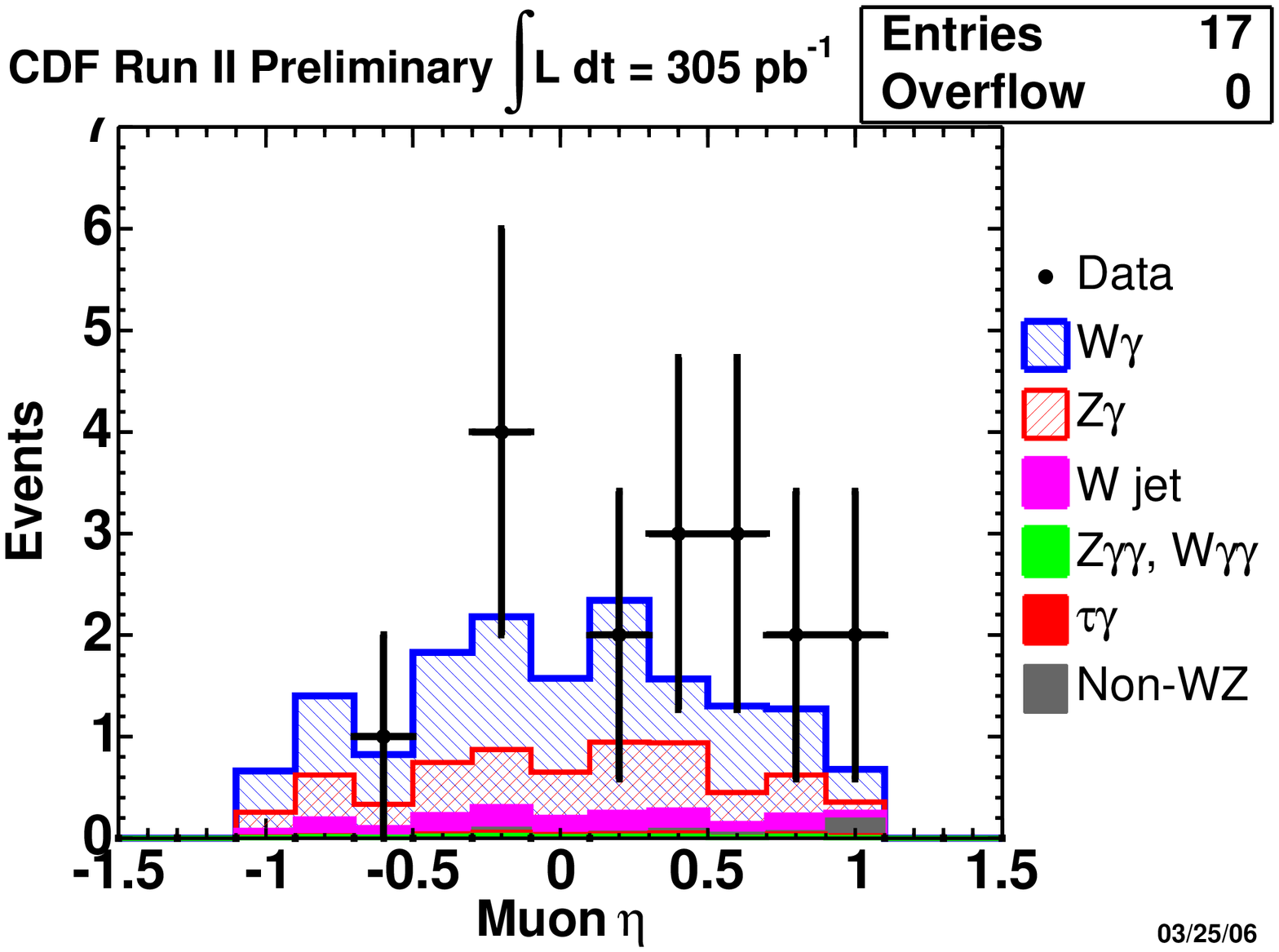,width=3.3in}} \quad
      }
    \caption
[Muon + photon + $\met$ distributions]
{Muon + photon + $\met$ distributions: $\Delta R$($\mu\gamma$), $H_T$,
Detector Type($\mu$), $\eta(\mu)$.}
    \label{lgmet_muonchannel_1.figure}
  \end{center}
\end{figure}

\begin{figure}[!t]
  \begin{center}
\vskip-0.2in
    \mbox{
\subfigure
[Muon Chi2]
{\epsfig{file=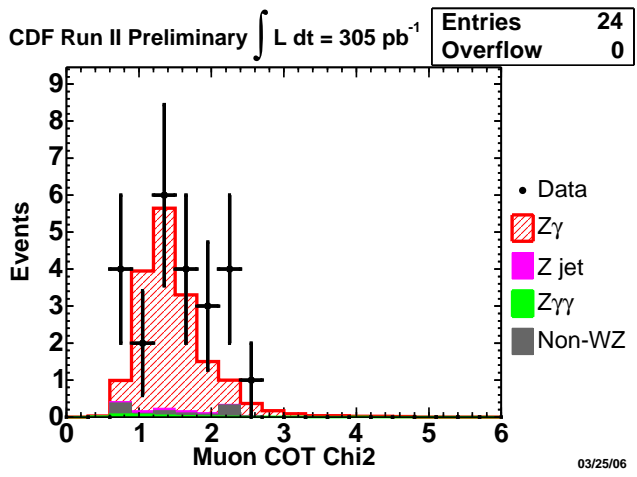,width=3.3in}} \quad
\subfigure
[Muon Track $d_0$]
{\epsfig{file=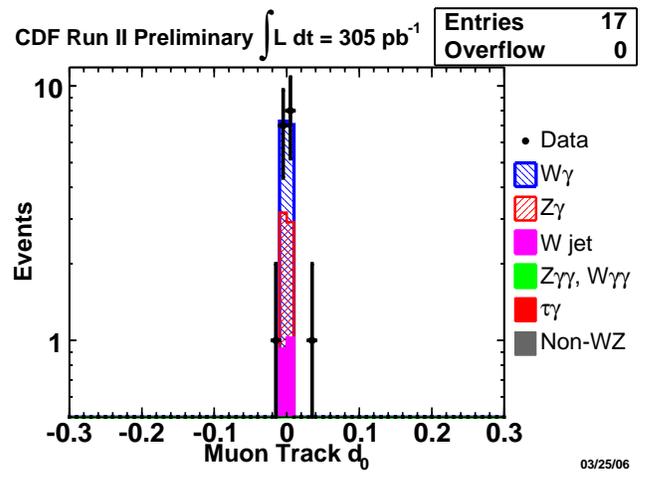,width=3.3in}} \quad
      }
    \caption
[Muon + photon + $\met$ distributions: continued]
{Muon + photon + $\met$ distributions: $\chi^2(\mu)$, $d_0(\mu)$.}
    \label{lgmet_muonchannel_2.figure}
  \end{center}
\end{figure}

\clearpage
\subsection{Additional $\egmet$ Plots}
\begin{figure}[!h]
  \begin{center}
\vskip-0.2in
    \mbox{
\subfigure
[$\Delta R$($e\gamma$)]
{\epsfig{file=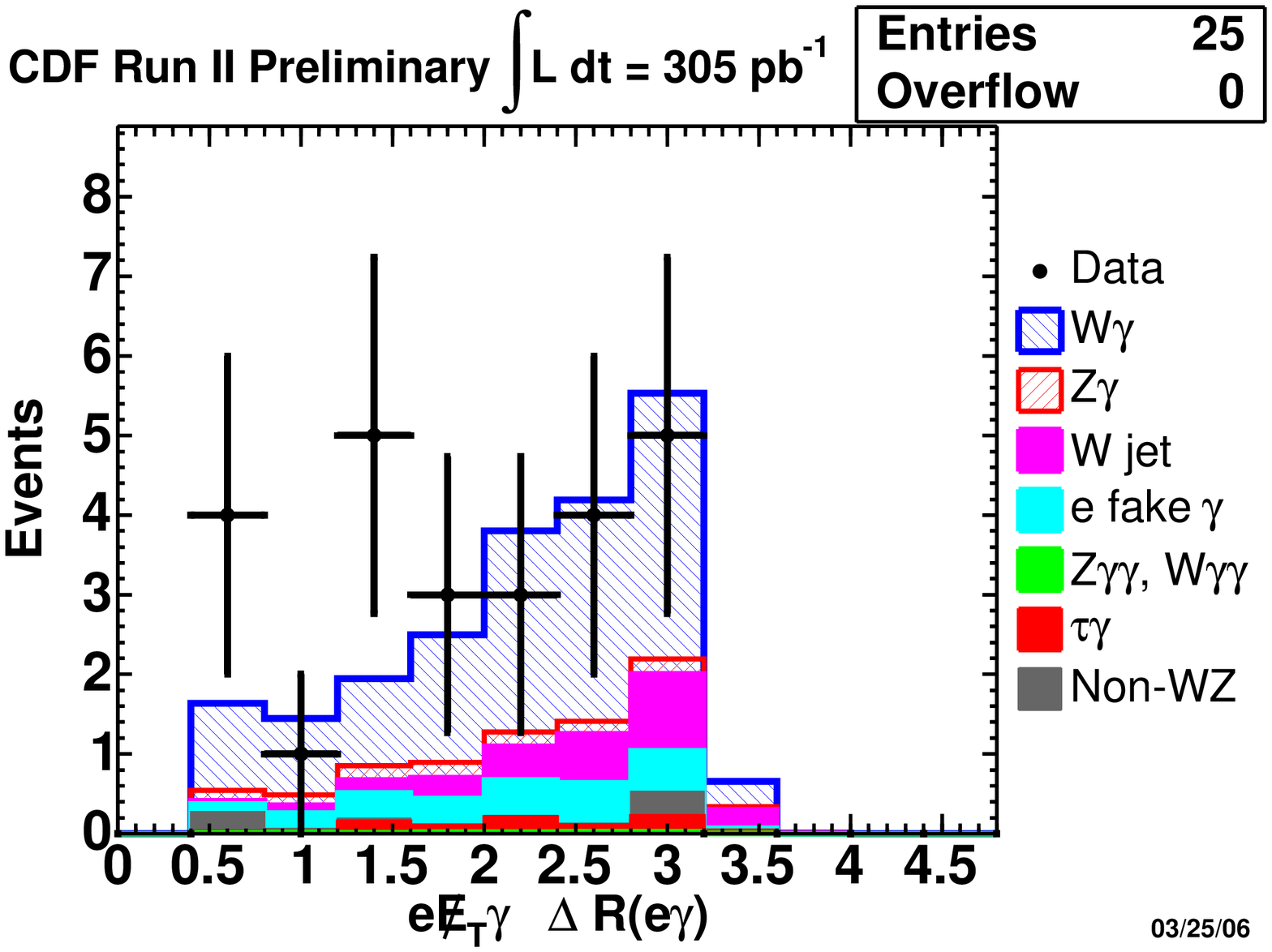,width=3.2in}} \quad
\subfigure
[$H_T$ for $\egmet$ Candidate Events]
{\epsfig{file=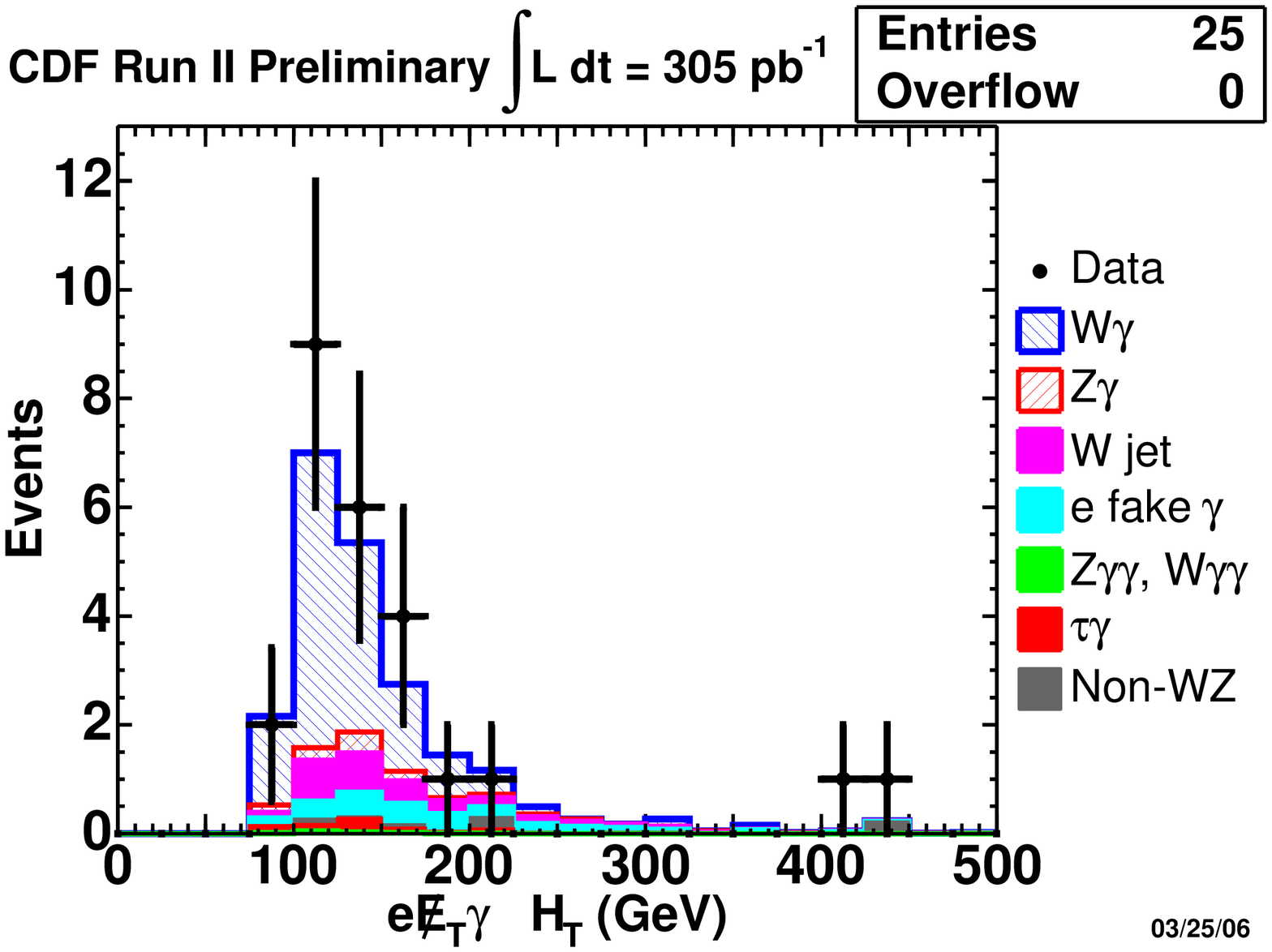,width=3.2in}} \quad
      }
\vskip-0.2in
    \mbox{
\subfigure
[Electron Fiduciality]
{\epsfig{file=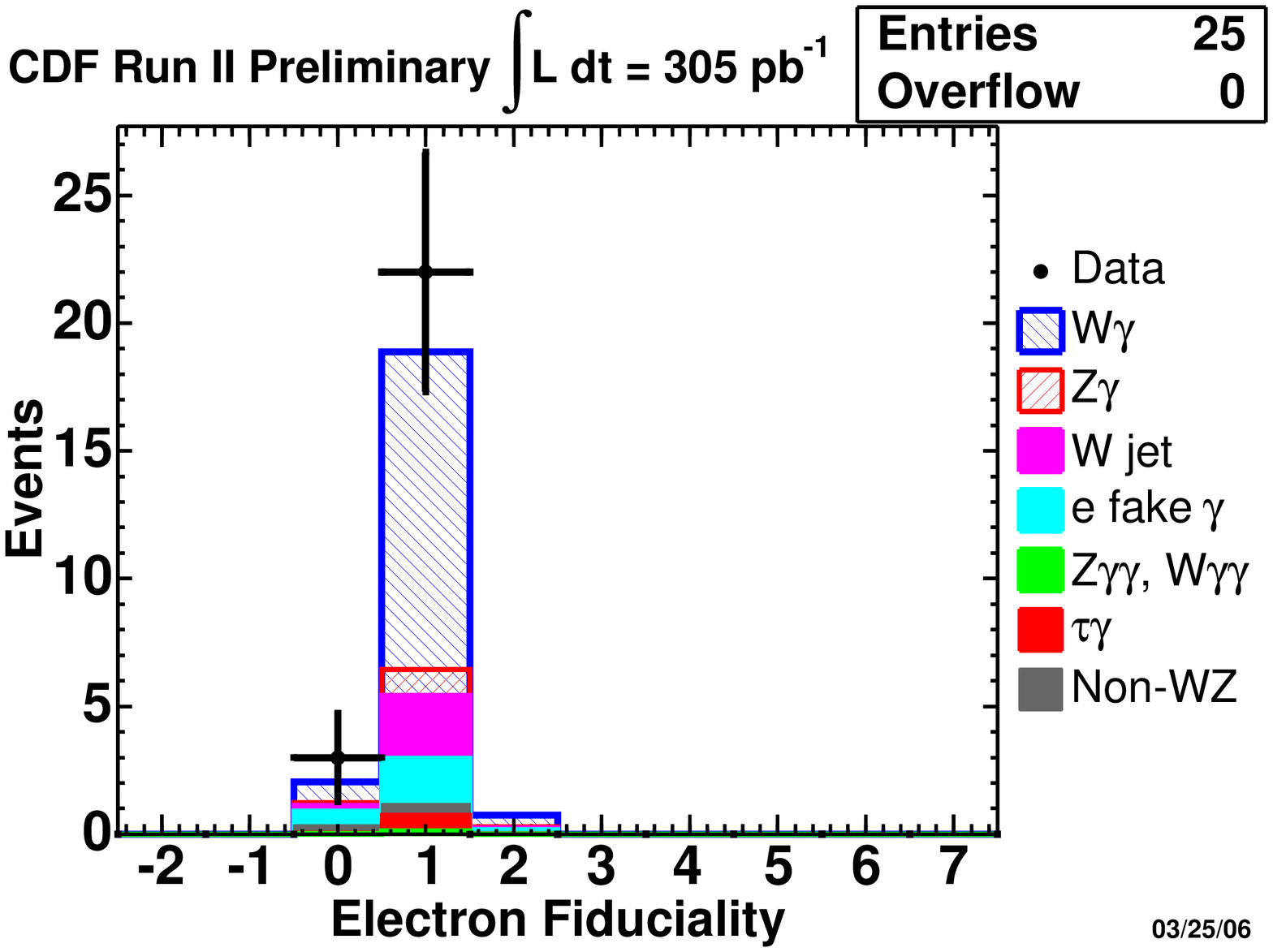,width=3.2in}} \quad
\subfigure
[Electron $\eta$]
{\epsfig{file=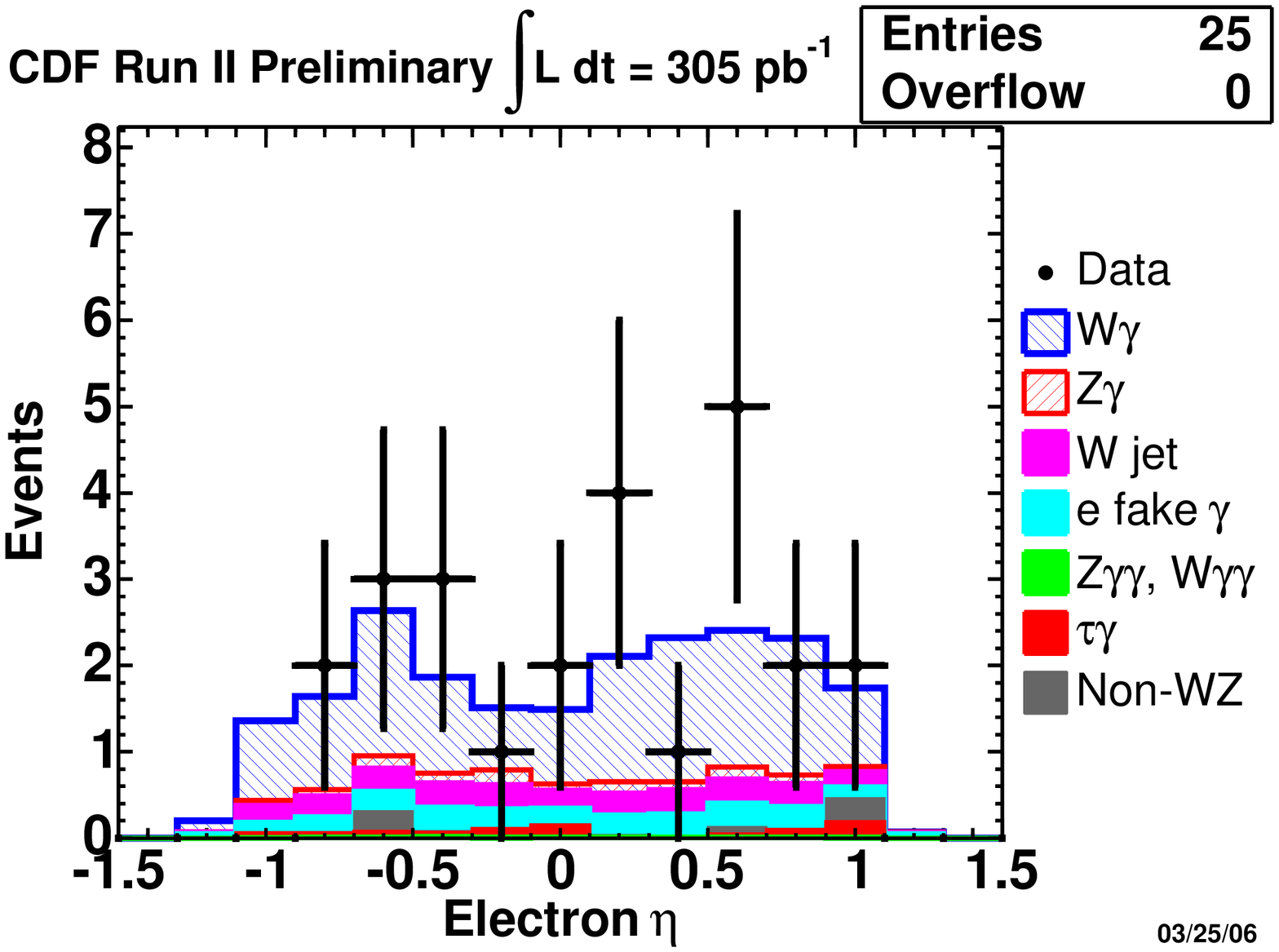,width=3.2in}} \quad
      }
\caption
[Electron + photon + $\met$ distributions]
{Electron + photon + $\met$ distributions: $\Delta R$($e\gamma$),
$H_T$, fiduciality (see Chapter~\ref{electronid}), electron
$\eta$. There are 3 electron candidates non-fiducial in central or
plug, which is in the agreement with the expectation. }
\label{lgmet_electronchannel_1.figure}
  \end{center}
\end{figure}
\begin{figure}[!h]
  \begin{center}
\vskip-0.2in
    \mbox{
\subfigure
[Electron CES X]
{\epsfig{file=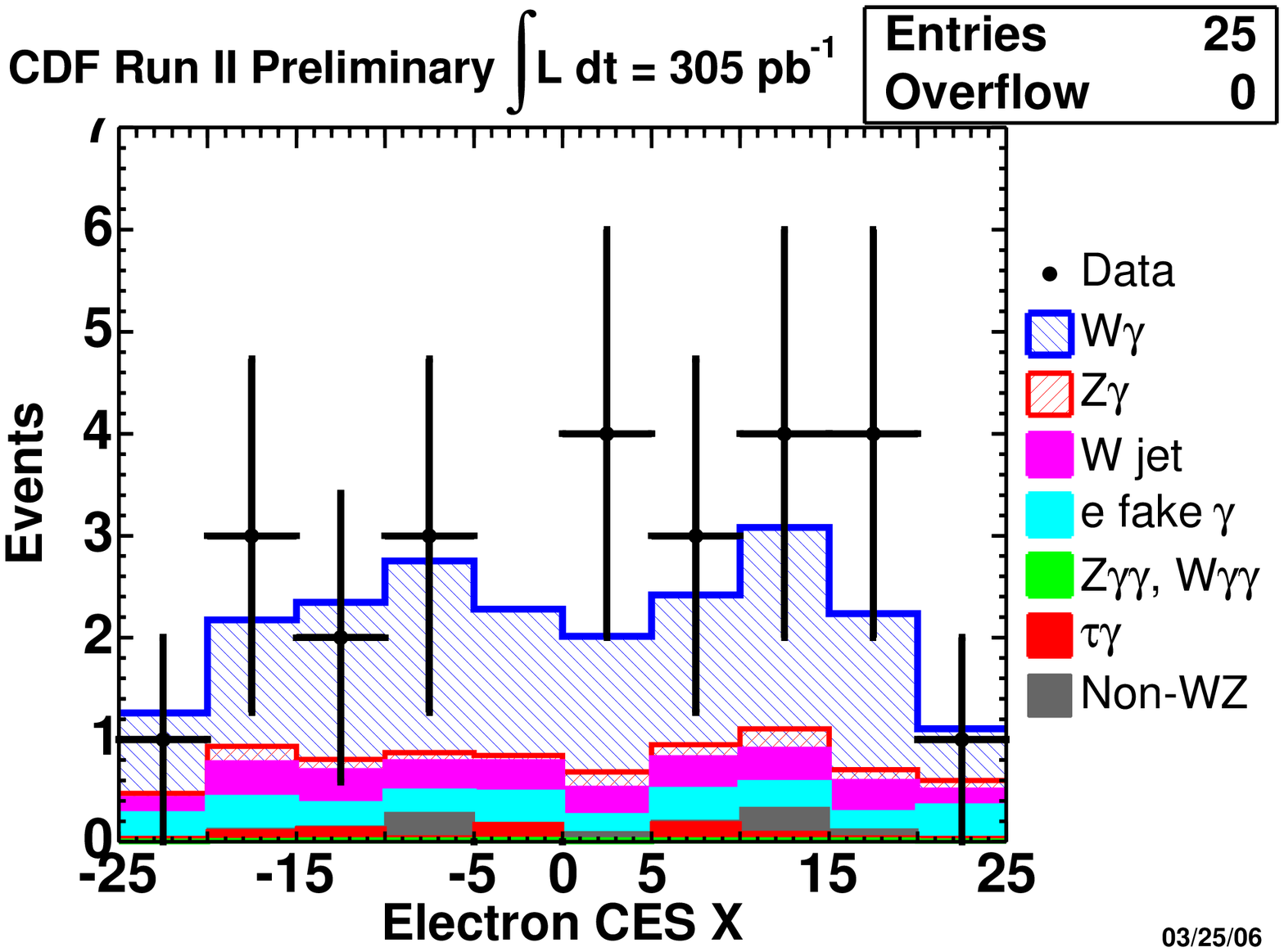,width=3.2in}} \quad
\subfigure
[Electron CES Z]
{\epsfig{file=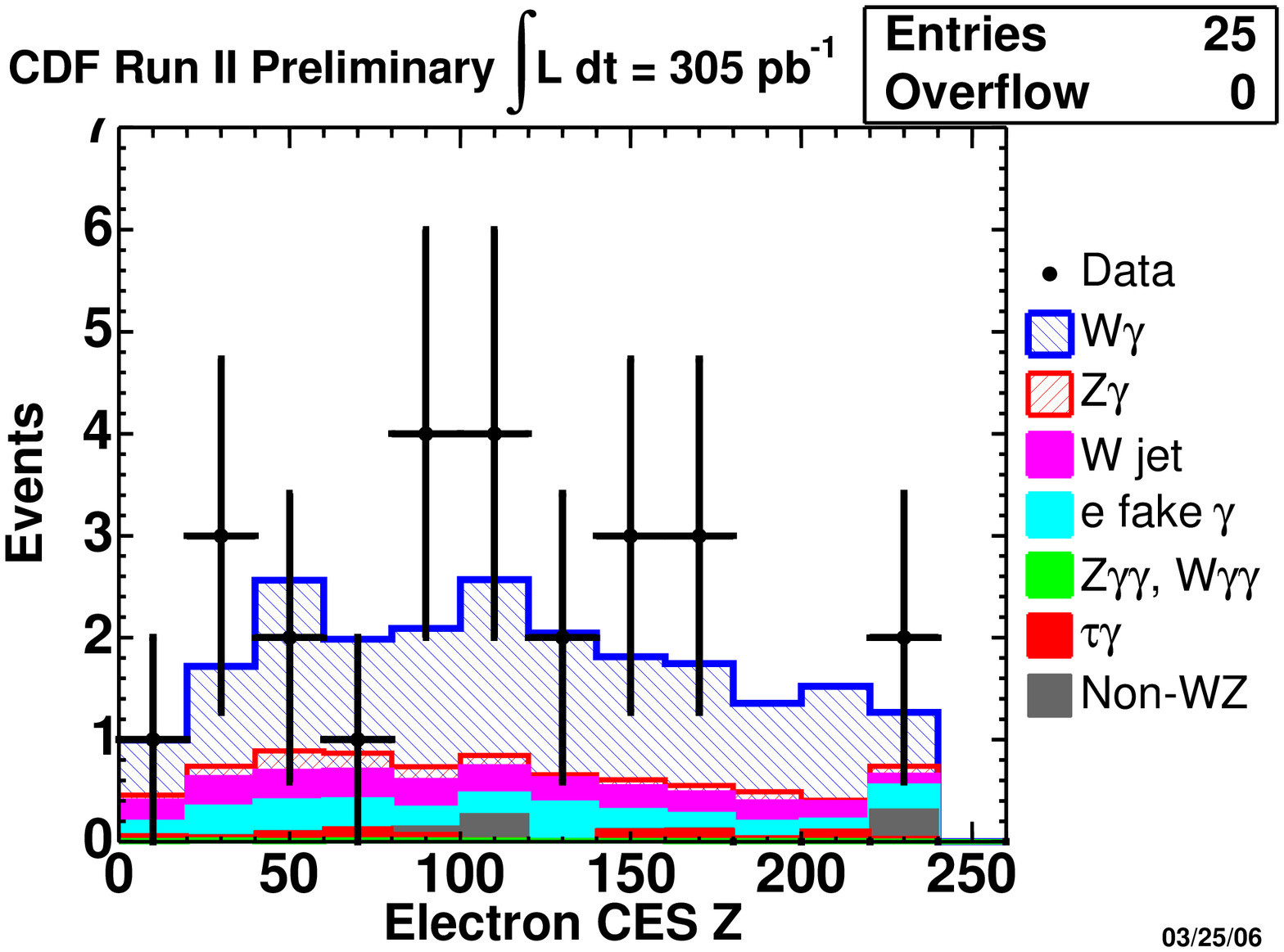,width=3.2in	}} \quad
      }
\vskip-0.2in
    \mbox{
\subfigure
[$\Delta cot(\theta)$]
{\epsfig{file=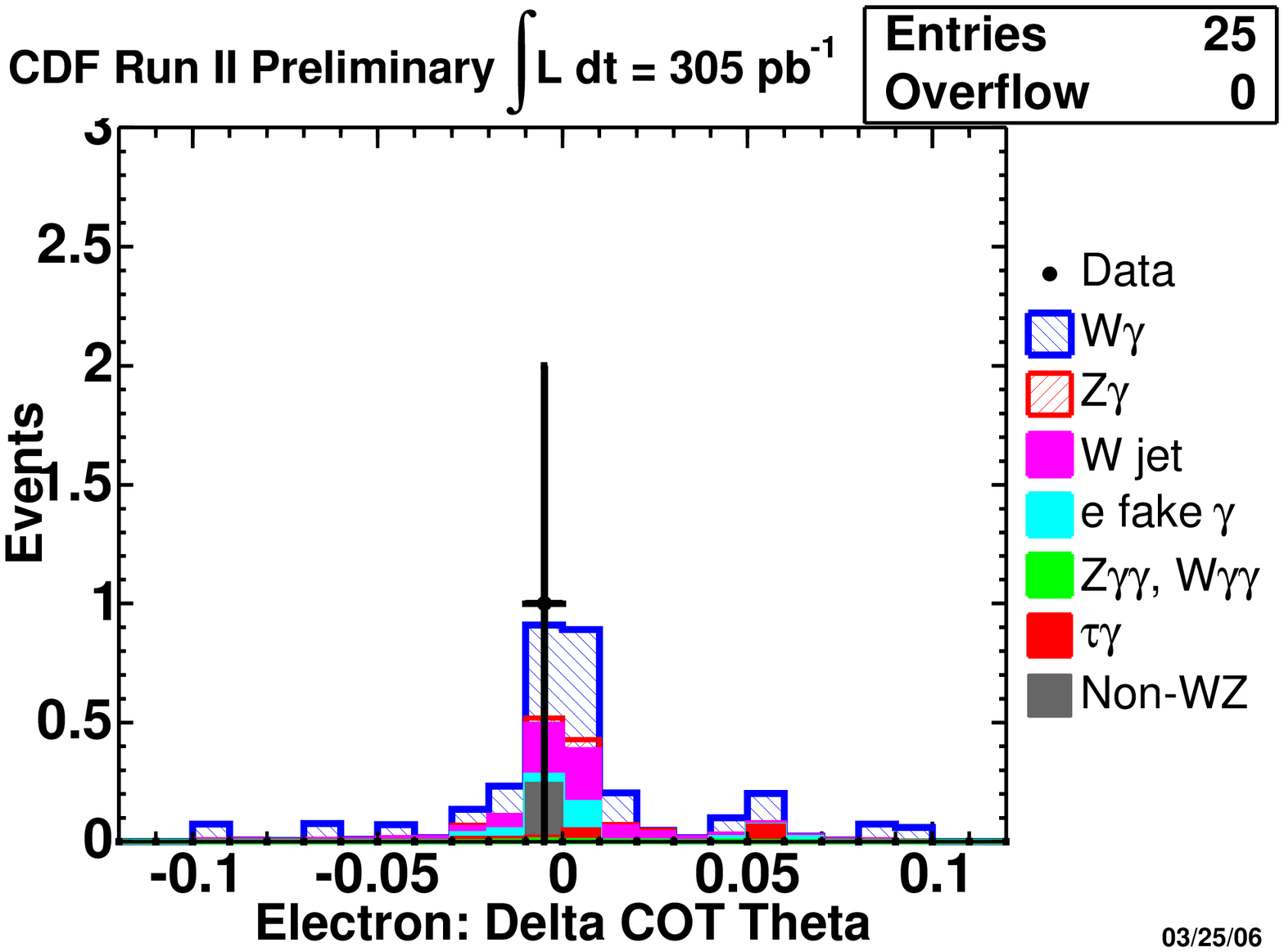,width=3.2in}} \quad
\subfigure
[$\Delta xy$]
{\epsfig{file=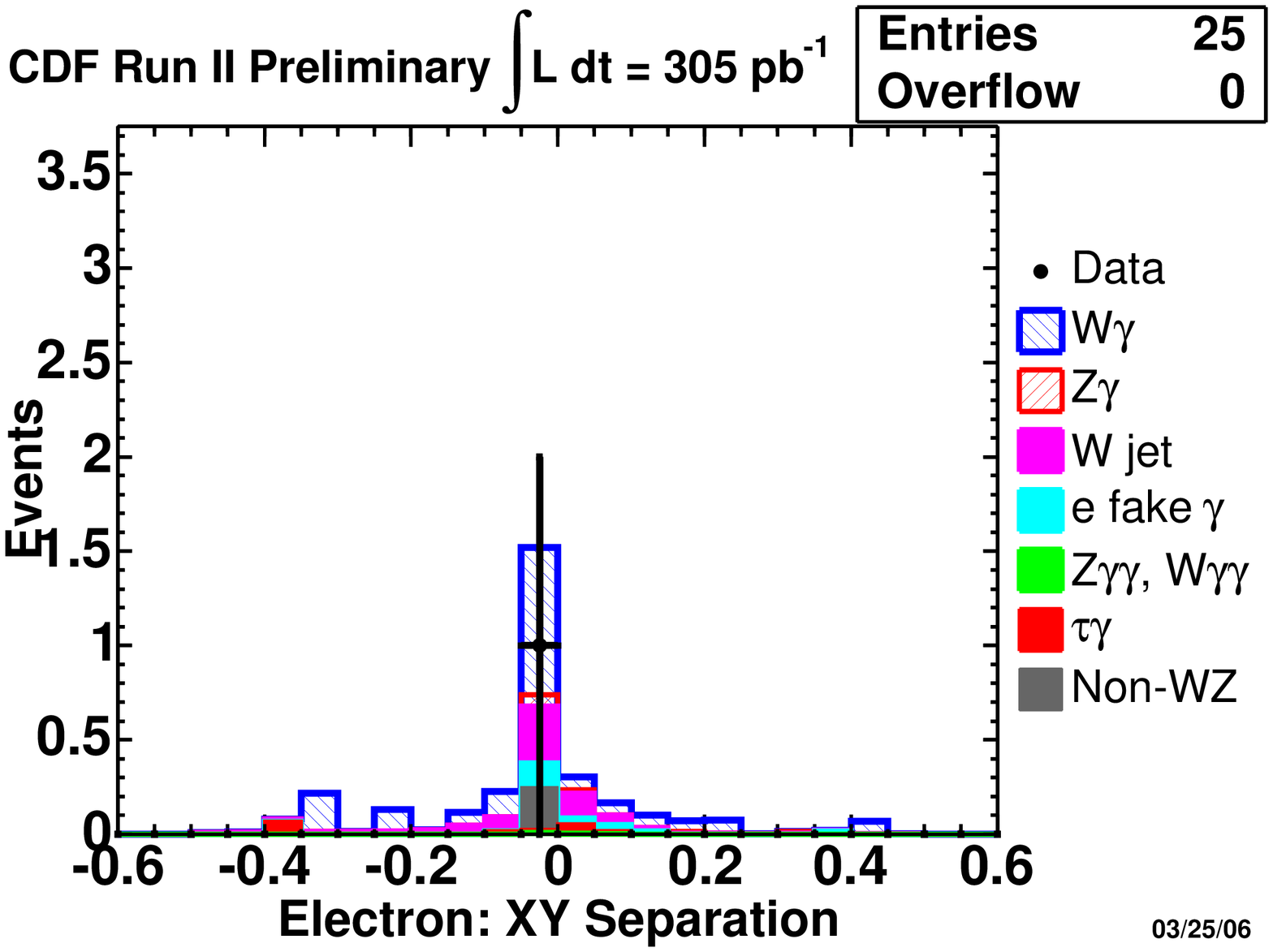,width=3.2in}} \quad
      }
    \caption
[Electron + photon + $\met$ distributions: continued]
{Electron + photon + $\met$ distributions: CES X, CES Z,
$\Delta cot(\theta)$ and $\Delta xy$ (see
Section~\ref{conversions}). There is only one trident in the $\egmet$
sample.}
    \label{lgmet_electronchannel_2.figure}
  \end{center}
\end{figure}

\clearpage
\newpage
\section{Additional $\llg$ Plots}
\label{llg_plots}
In this section we present additional plots of the identification
variables for $\mumug$ and $\eeg$ signatures.
\subsection{Additional $\mumug$ Plots}
\begin{figure}[!h]
  \begin{center}
\vskip-0.2in
    \mbox{
\subfigure
[$\Delta R$($\mu\gamma$), for both muons]
{\epsfig{file=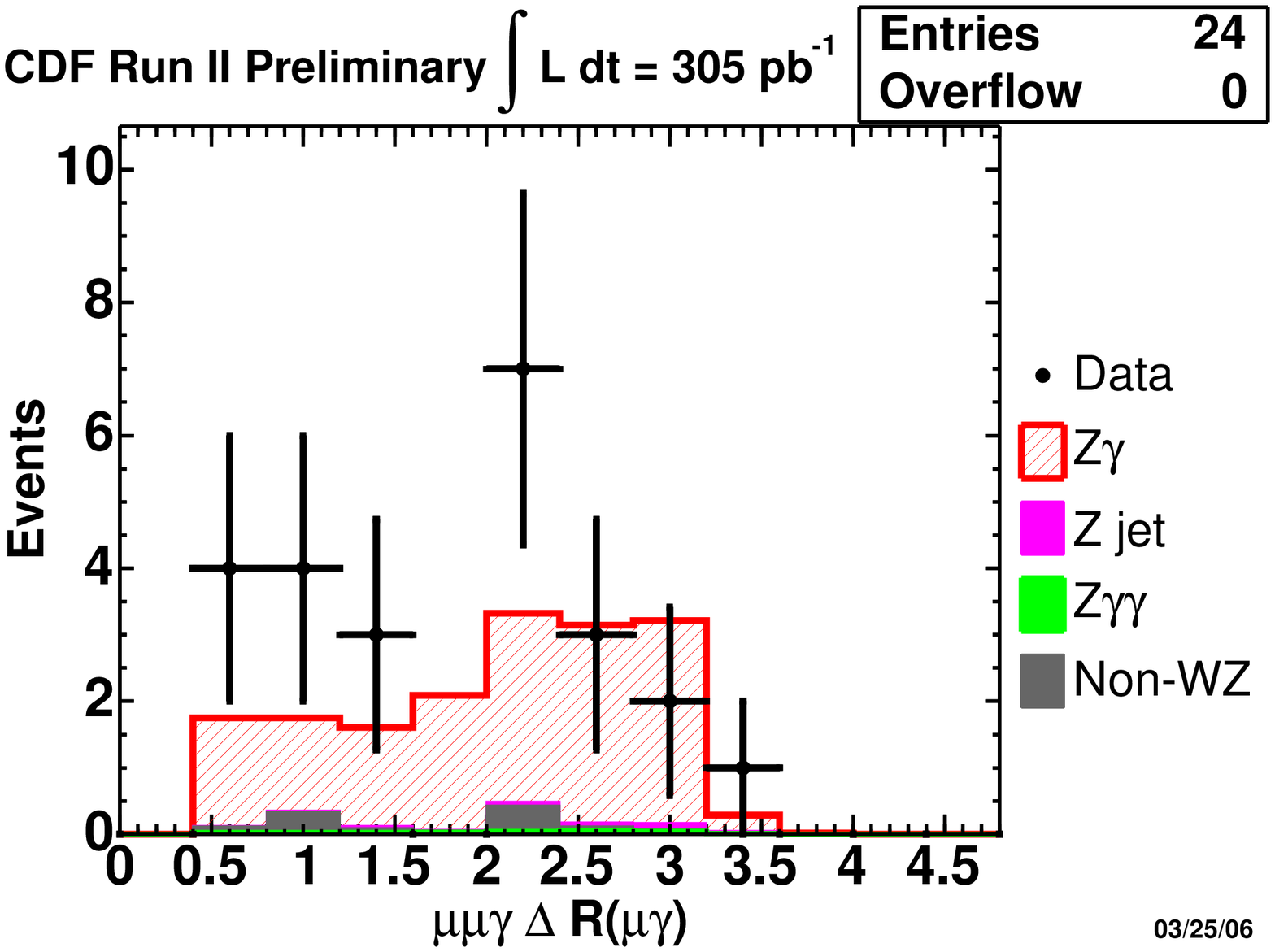,width=3.1in}} \quad
\subfigure
[$H_T$ for $\mumug$ Candidate Events]
{\epsfig{file=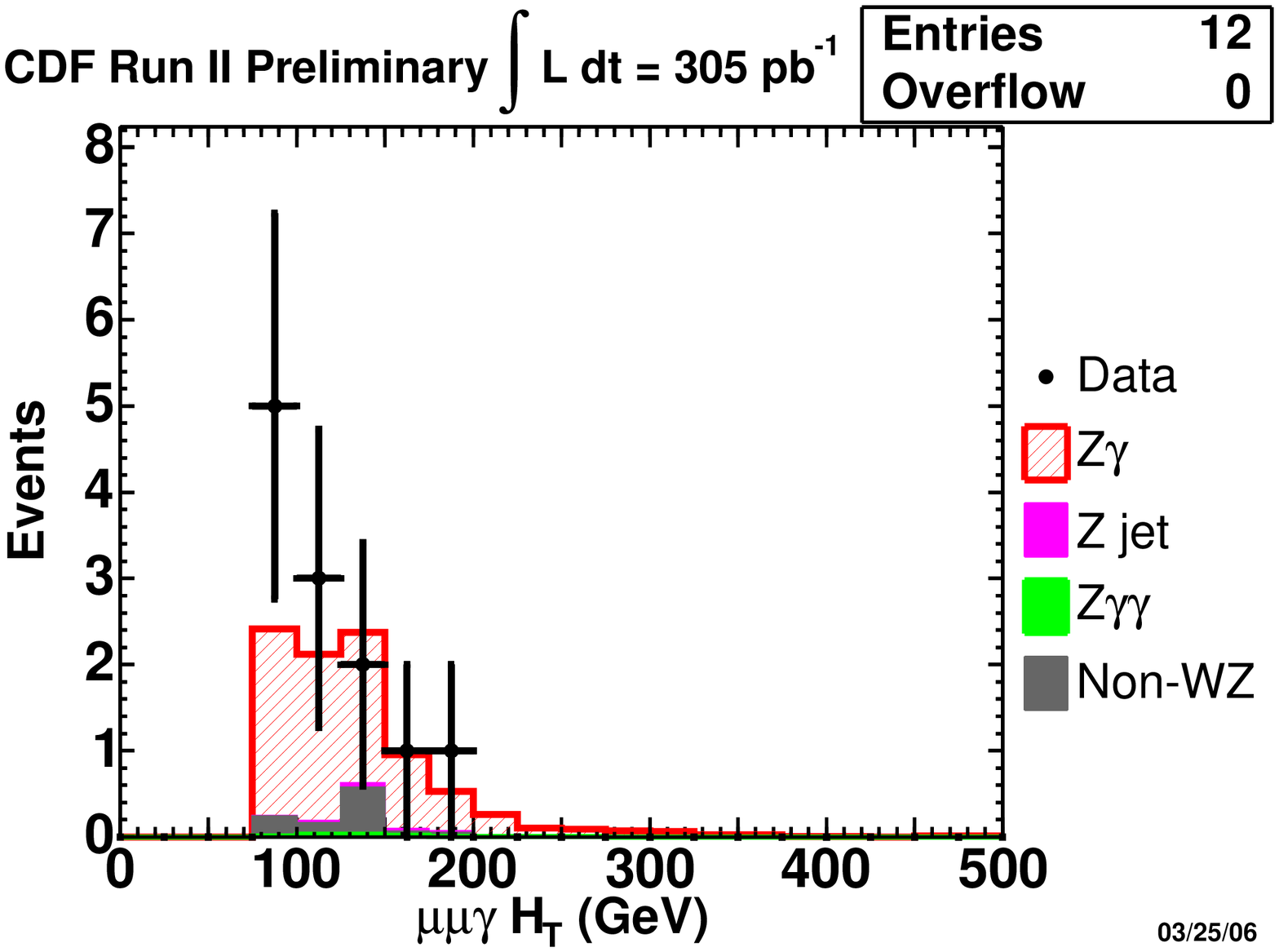,width=3.1in}} \quad
	      }
\vskip-0.2in
    \mbox{
\subfigure
[Muon $\eta$]
{\epsfig{file=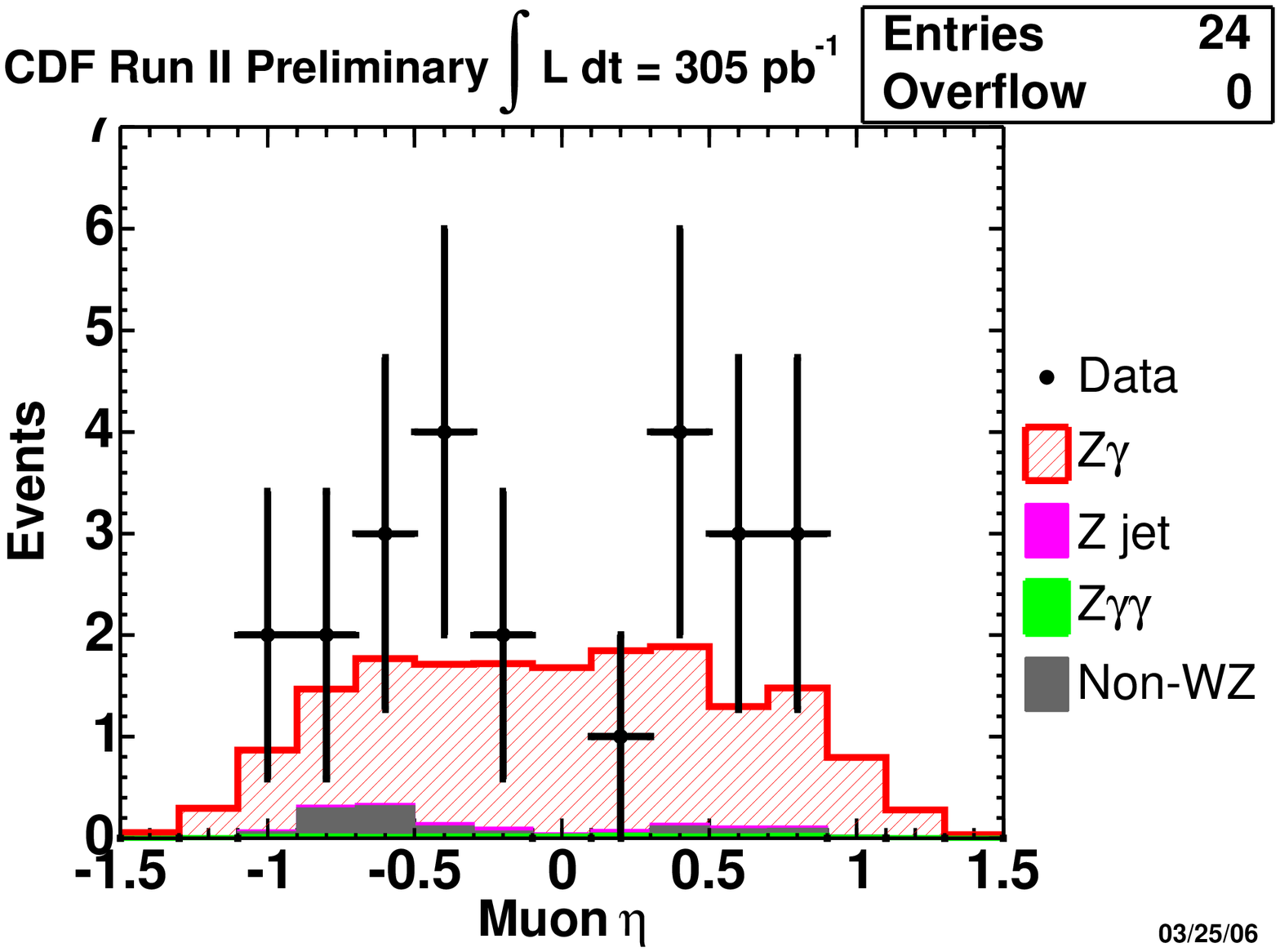,width=3.1in}} \quad
\subfigure
[Muon Track $d_0$]
{\epsfig{file=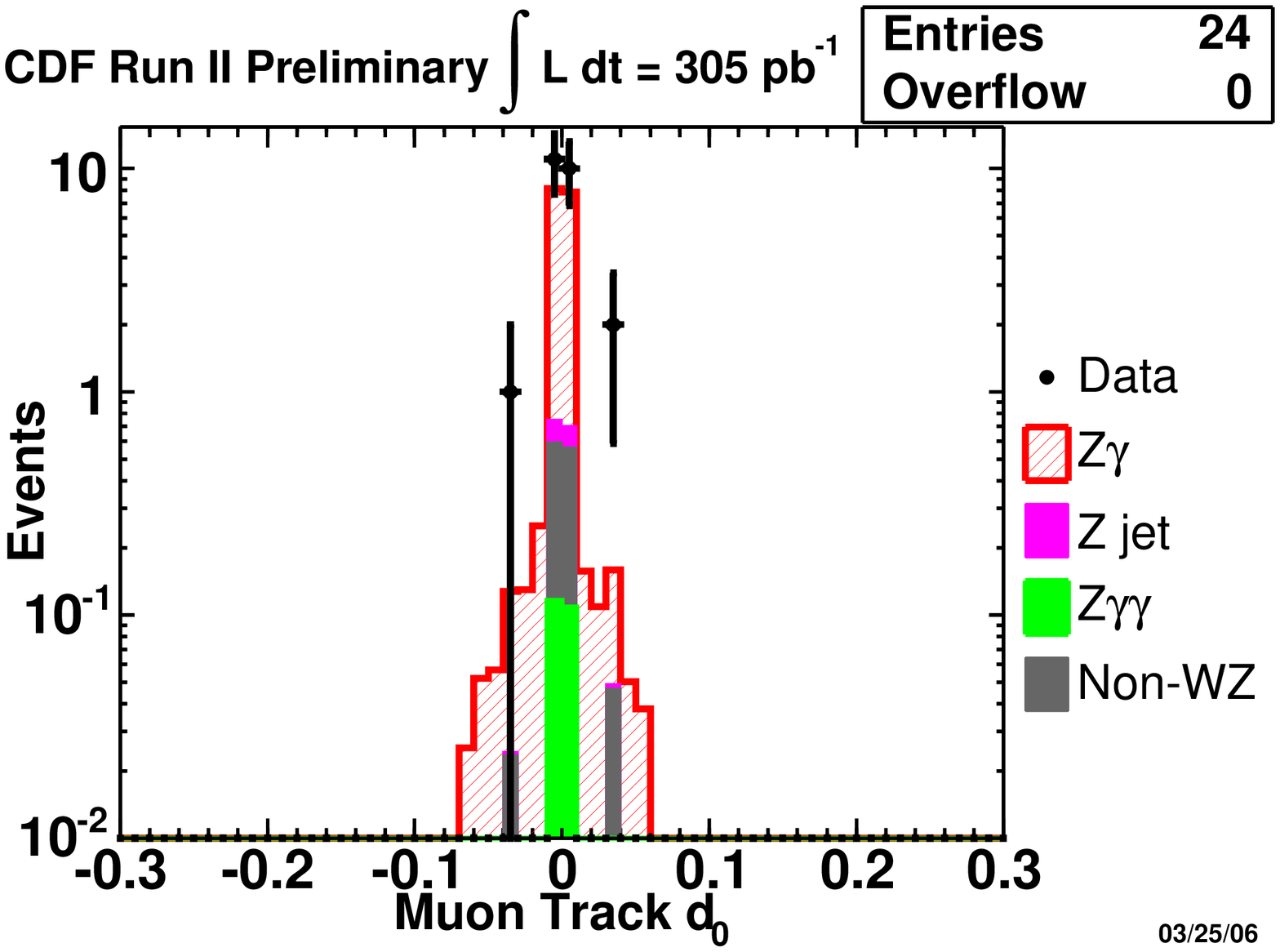,width=3.1in}} \quad
      }
    \caption 
[Multi-muon + photon distributions]
{Multi-muon + photon distributions: $\Delta R$($\mu\gamma$), $H_T$, $\eta(\mu)$, $d_0(\mu)$}
    \label{llg_muonchannel.figure}
  \end{center}
\end{figure}

\begin{figure}[!h]
  \begin{center}
\vskip-0.2in
    \mbox{
\subfigure
[Muon Chi2]
{\epsfig{file=golden_plots/zg_mu_chi2_13.eps,width=3.3in}} \quad
\subfigure
[Muon N Si Hits]
{\epsfig{file=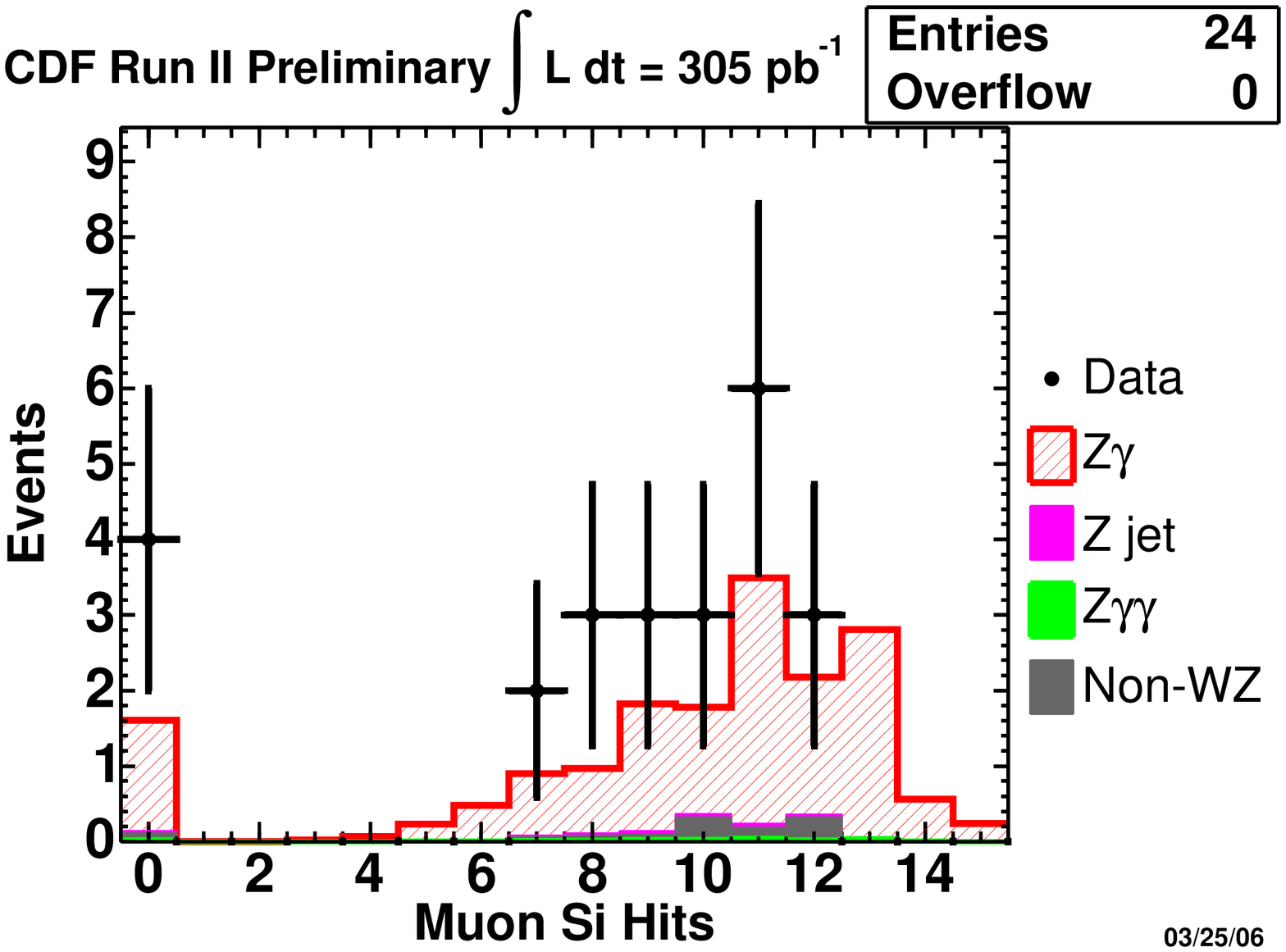,width=3.3in}} \quad
      }
\vskip-0.2in
    \mbox{
\subfigure
[Muon N Axial Segments]
{\epsfig{file=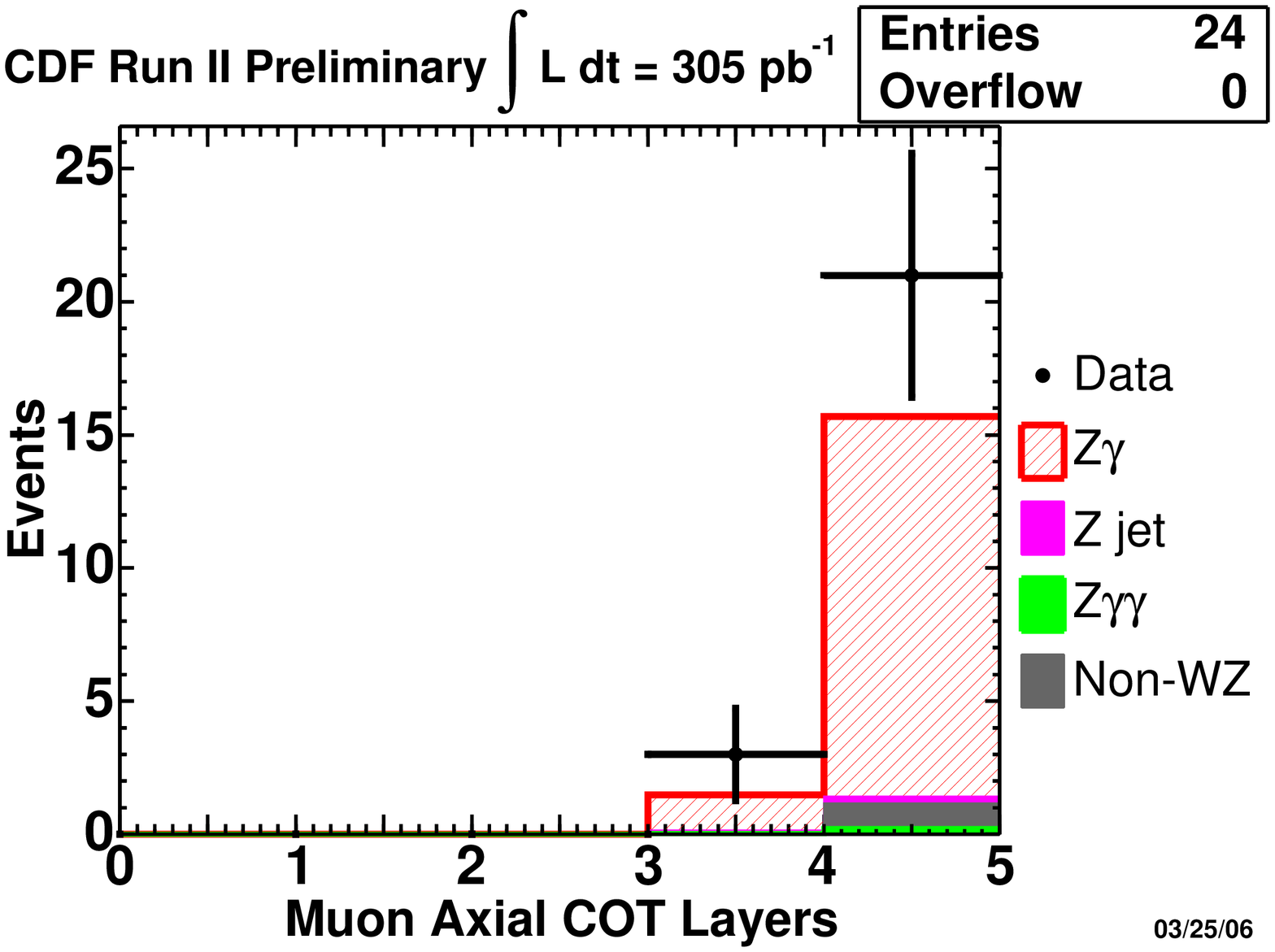,width=3.3in}} \quad
\subfigure
[Muon N Stereo Segments]
{\epsfig{file=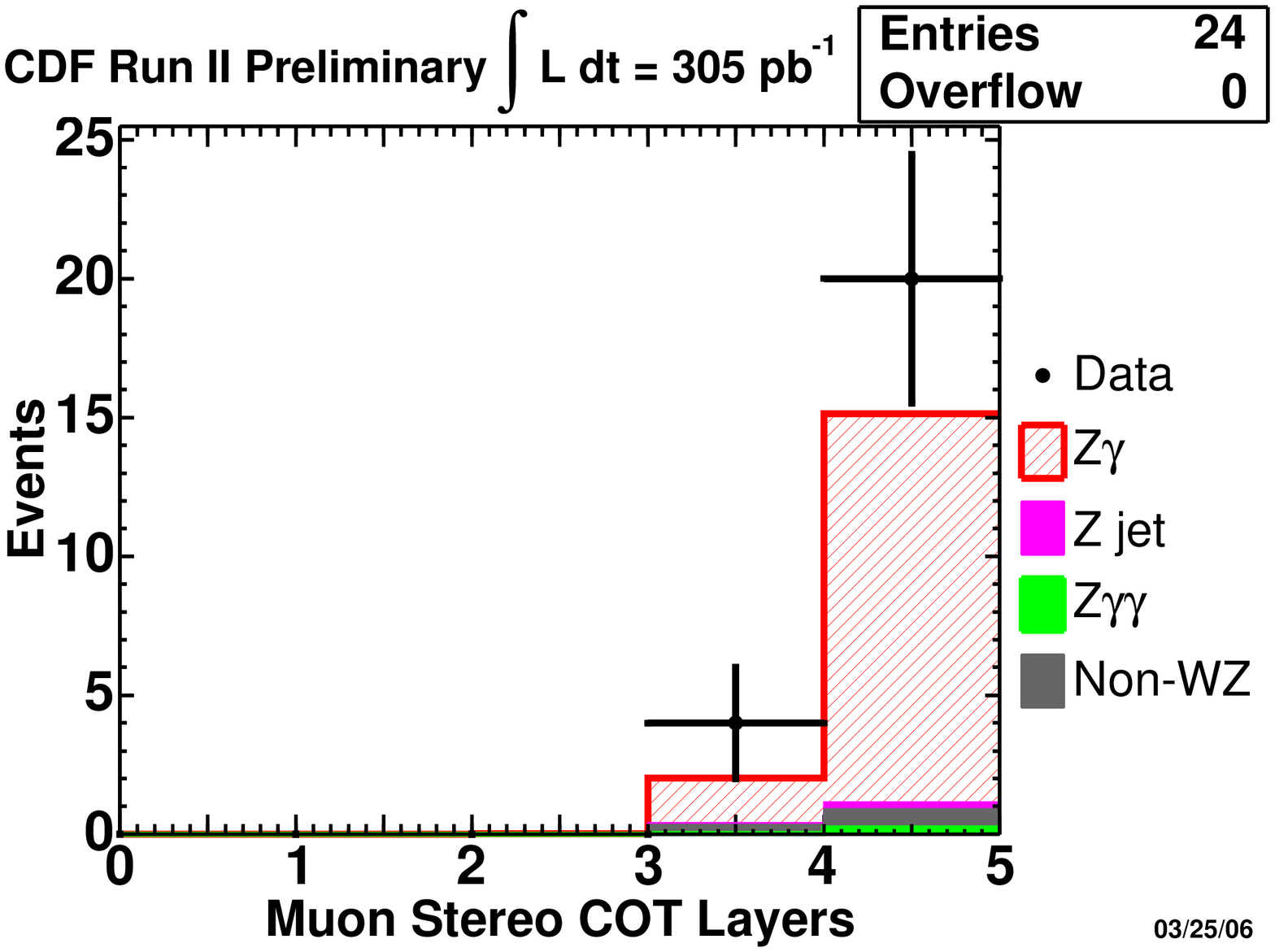,width=3.3in}} \quad
      }
    \caption
[Multi-muon + photon distributions: continued]
{Multi-muon + photon distributions: $\chi^2(\mu)$, Number of si hits, Numbers of Axial and Stereo Segments in COT}
    \label{llg_muonchannel_2.figure}
  \end{center}
\end{figure}

\begin{figure}[!h]
  \begin{center}
\vskip-0.2in
    \mbox{
\subfigure
[Muon Detector Type]
{\epsfig{file=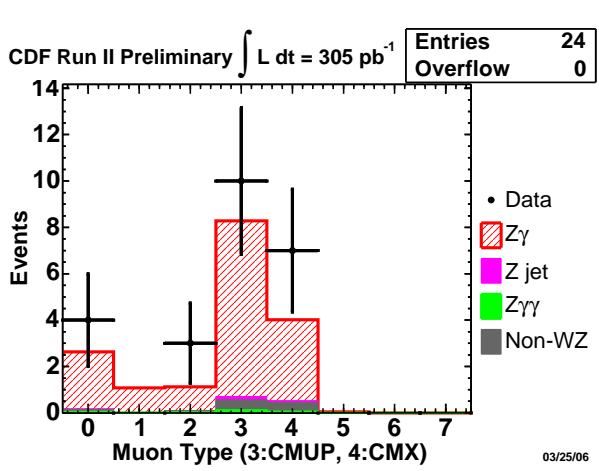,width=3.1in}} \quad
\subfigure
[Muon $\dx$ CMU]
{\epsfig{file=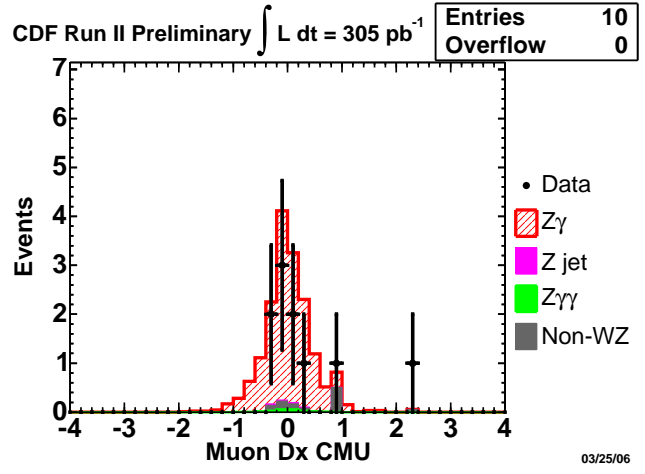,width=3.3in}} \quad
      }
\vskip-0.2in
    \mbox{
\subfigure
[Muon $\dx$ CMP]
{\epsfig{file=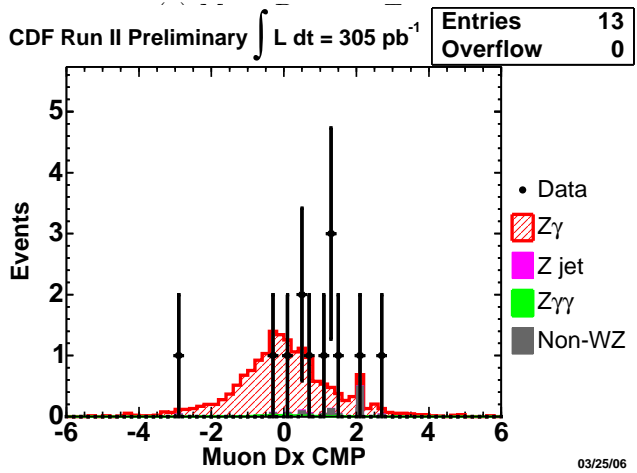,width=3.3in}} \quad
\subfigure
[Muon $\dx$ CMX]
{\epsfig{file=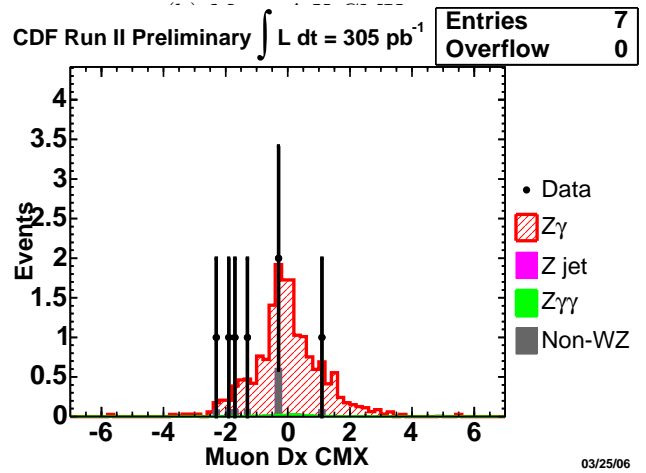,width=3.3in}} \quad
}
    \caption
[Multi-muon + photon distributions: continued]
{Multi-muon + photon distributions: detector type ($\mu$), $\dx(CMU)$, $\dx(CMP)$, $\dx(CMX)$}
    \label{llg_muonchannel_3.figure}
  \end{center}
\end{figure}

\begin{figure}[!h]
  \begin{center}
\vskip-0.2in
    \mbox{
\subfigure
[Muon Iso $\Et$]
{\epsfig{file=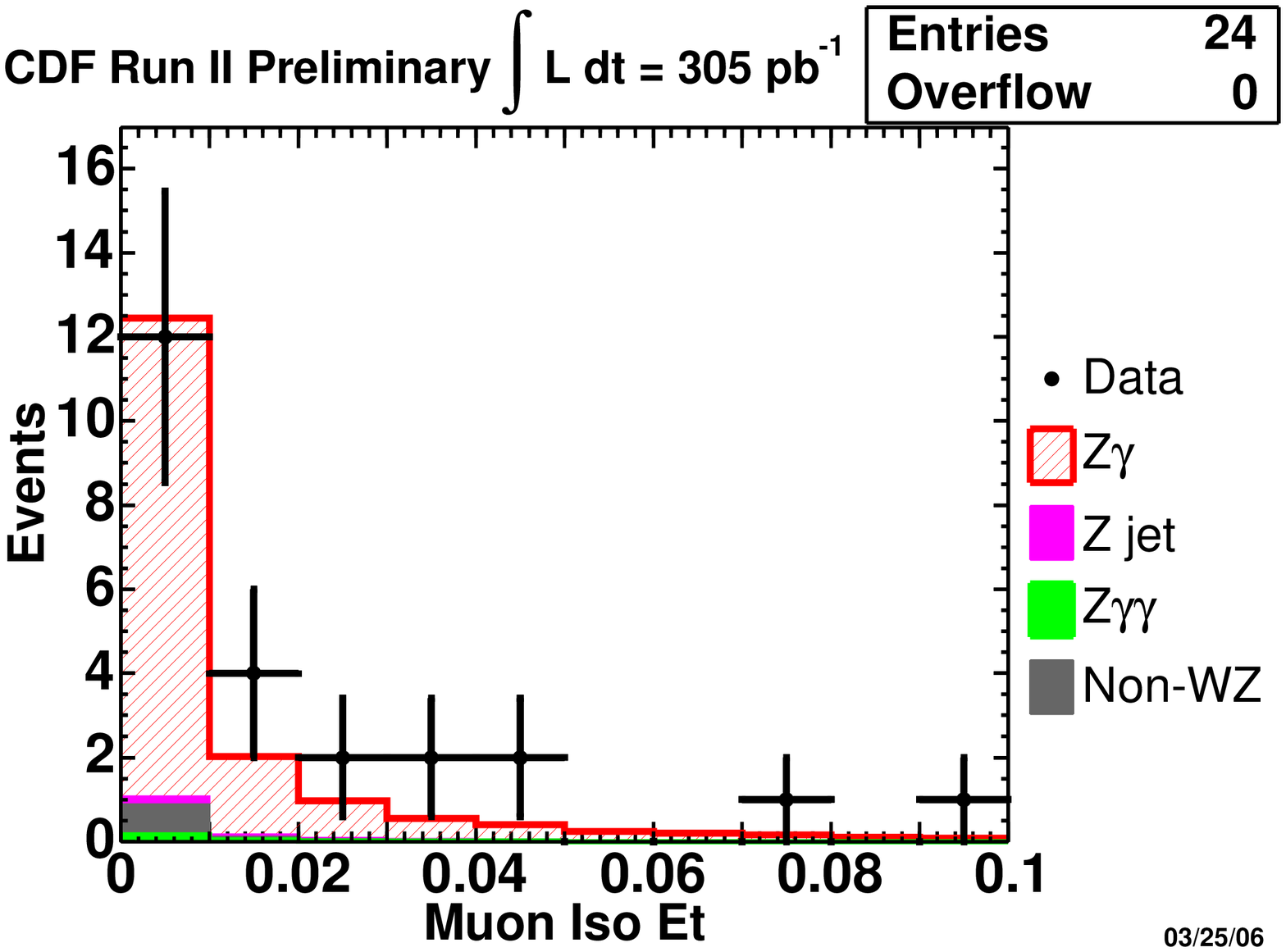,width=3.3in}} \quad
\subfigure
[Muon Iso $\Pt$]
{\epsfig{file=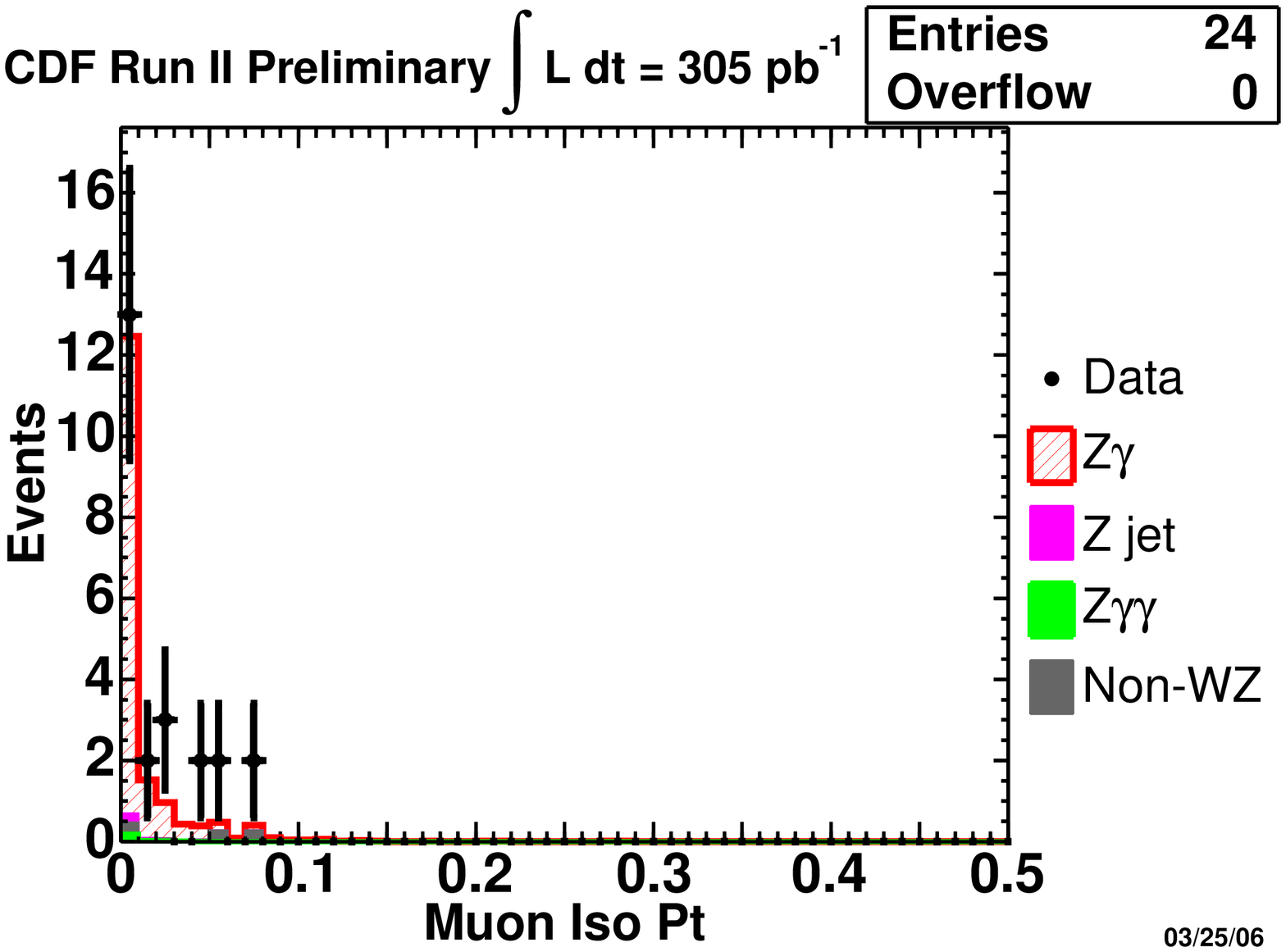,width=3.3in}} \quad
      }
\vskip-0.2in
    \mbox{
\subfigure
[Muon Acollinearity (3D Angle($\mumu$))]
{\epsfig{file=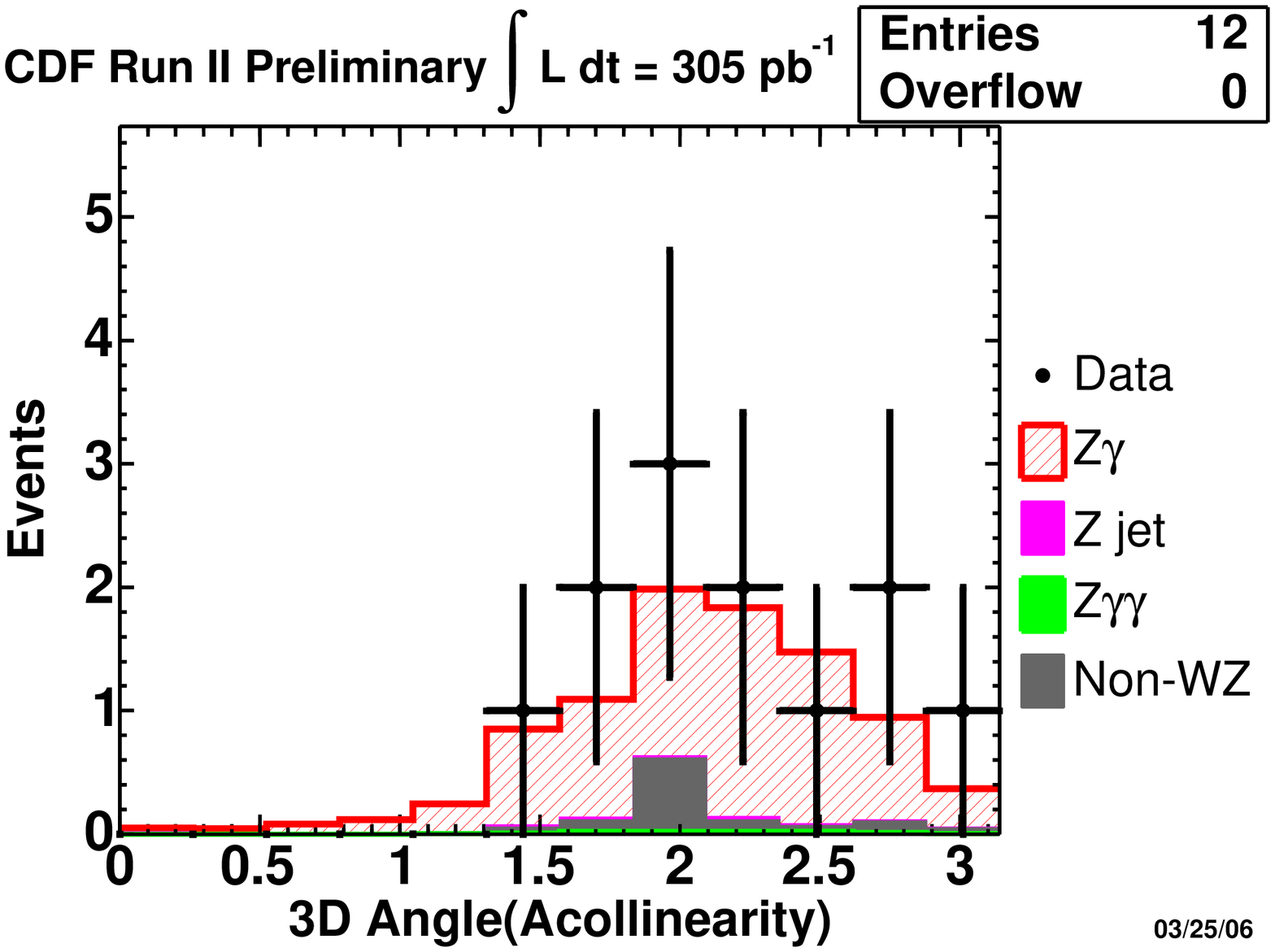,width=3.1in}} \quad
      }
    \caption
[Multi-muon + photon distributions: continued]
{Multi-muon + photon distributions: muon relative calorimeter isolation (Iso $\Et$), relative track isolation (Iso $\Pt$), 3D angle($\mumu$)}
    \label{llg_muonchannel_4.figure}
  \end{center}
\end{figure}

\clearpage
\subsection{Additional $\eeg$ Plots}
\begin{figure}[!h]
  \begin{center}
\vskip-0.2in
    \mbox{
\subfigure
[$\Delta R$($e\gamma$), for both electrons]
{\epsfig{file=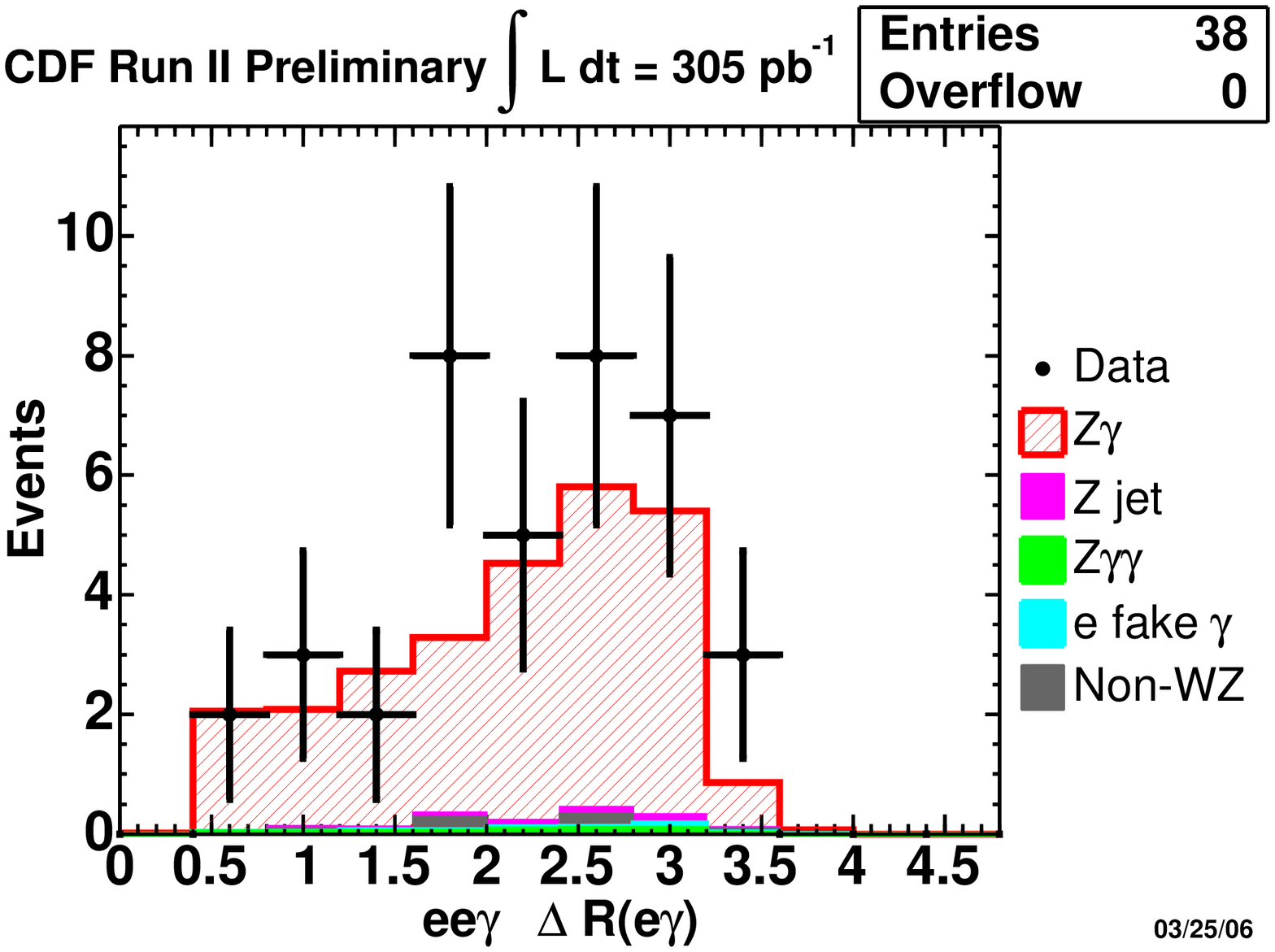,width=3.1in}} \quad
\subfigure
[$H_T$ for $\eeg$ Candidate Events]
{\epsfig{file=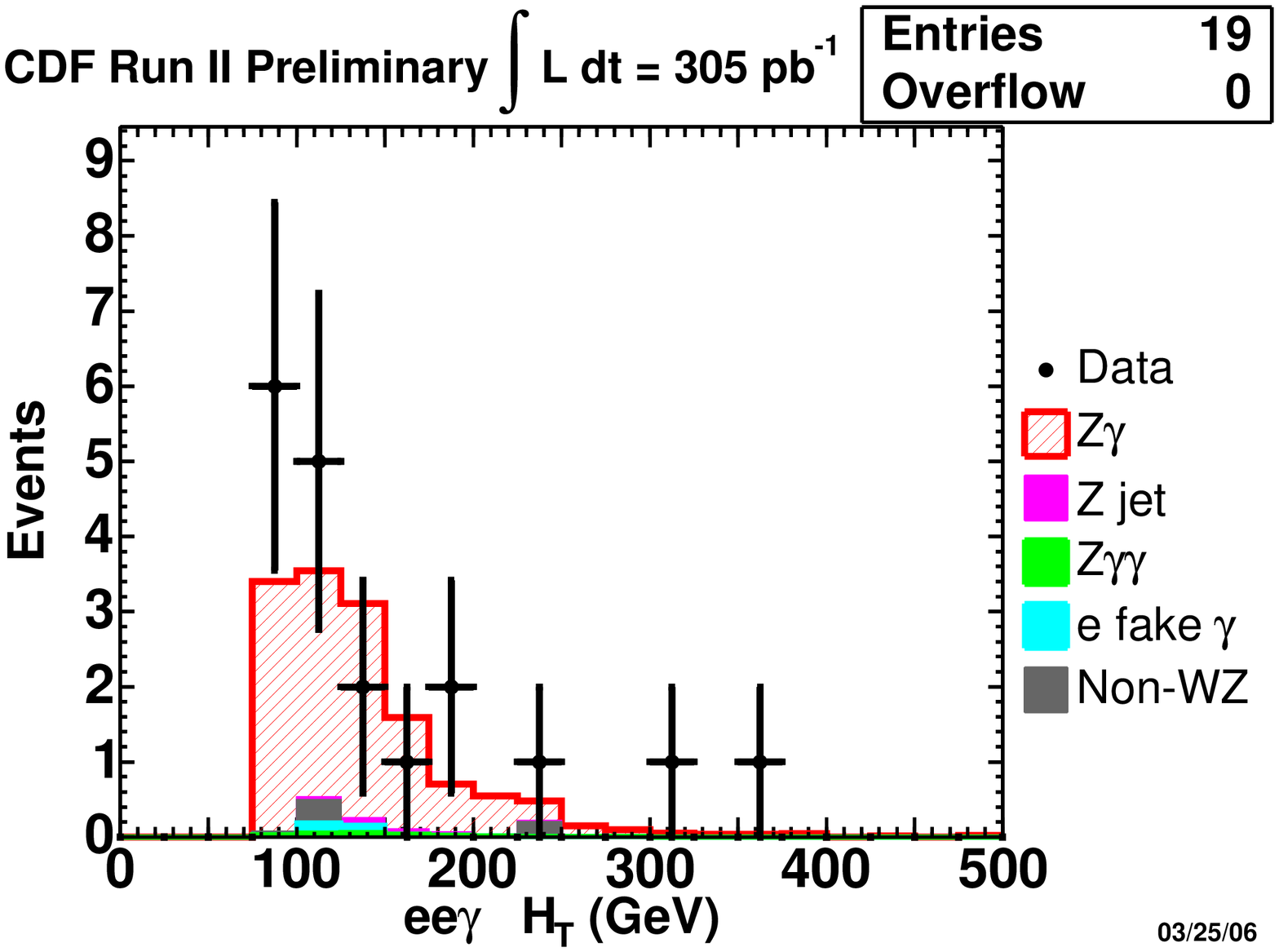,width=3.1in}} \quad
      }
\vskip-0.2in
    \mbox{
\subfigure
[Electron Fiduciality]
{\epsfig{file=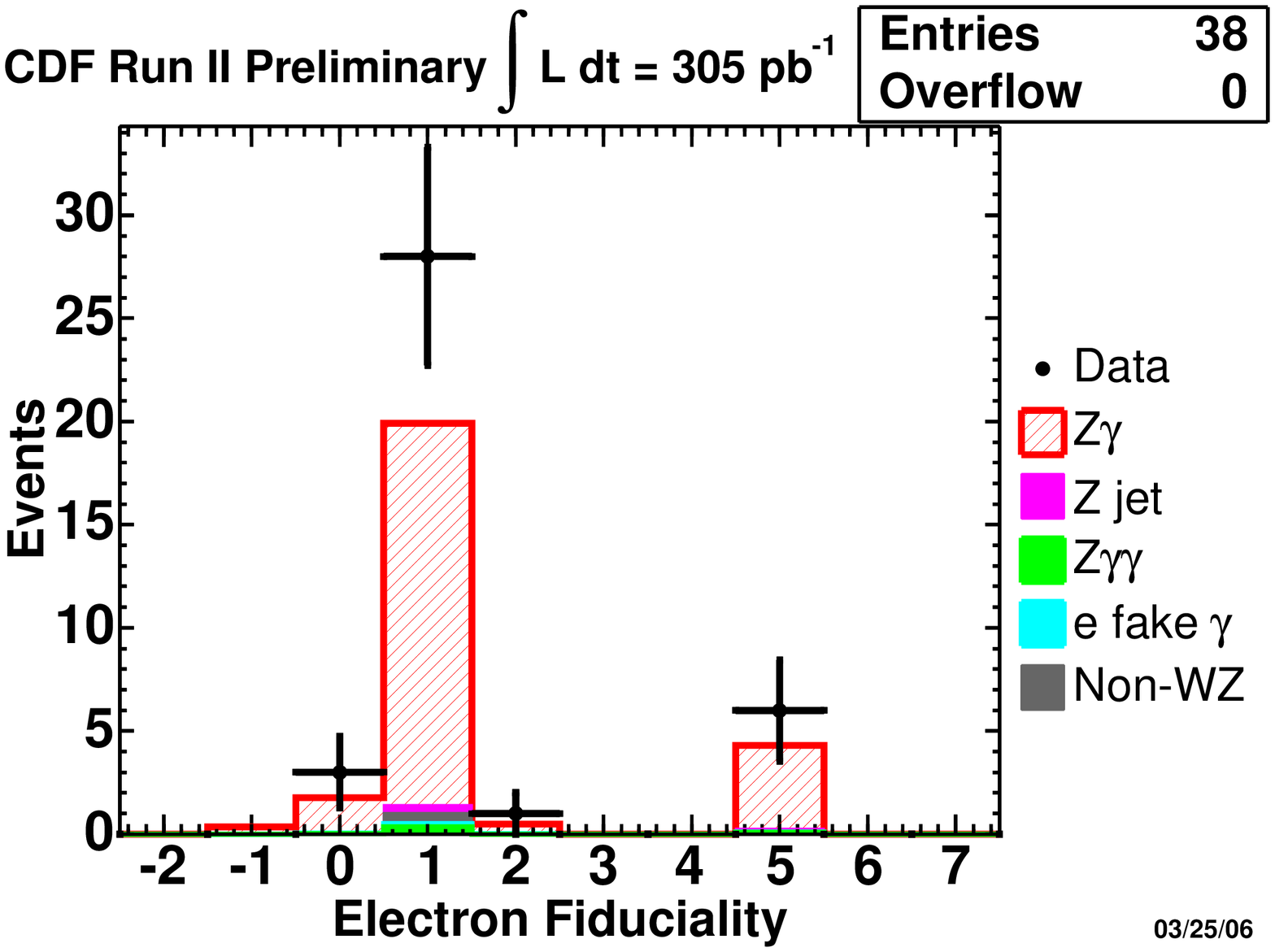,width=3.1in}} \quad
\subfigure
[Electron $\eta$]
{\epsfig{file=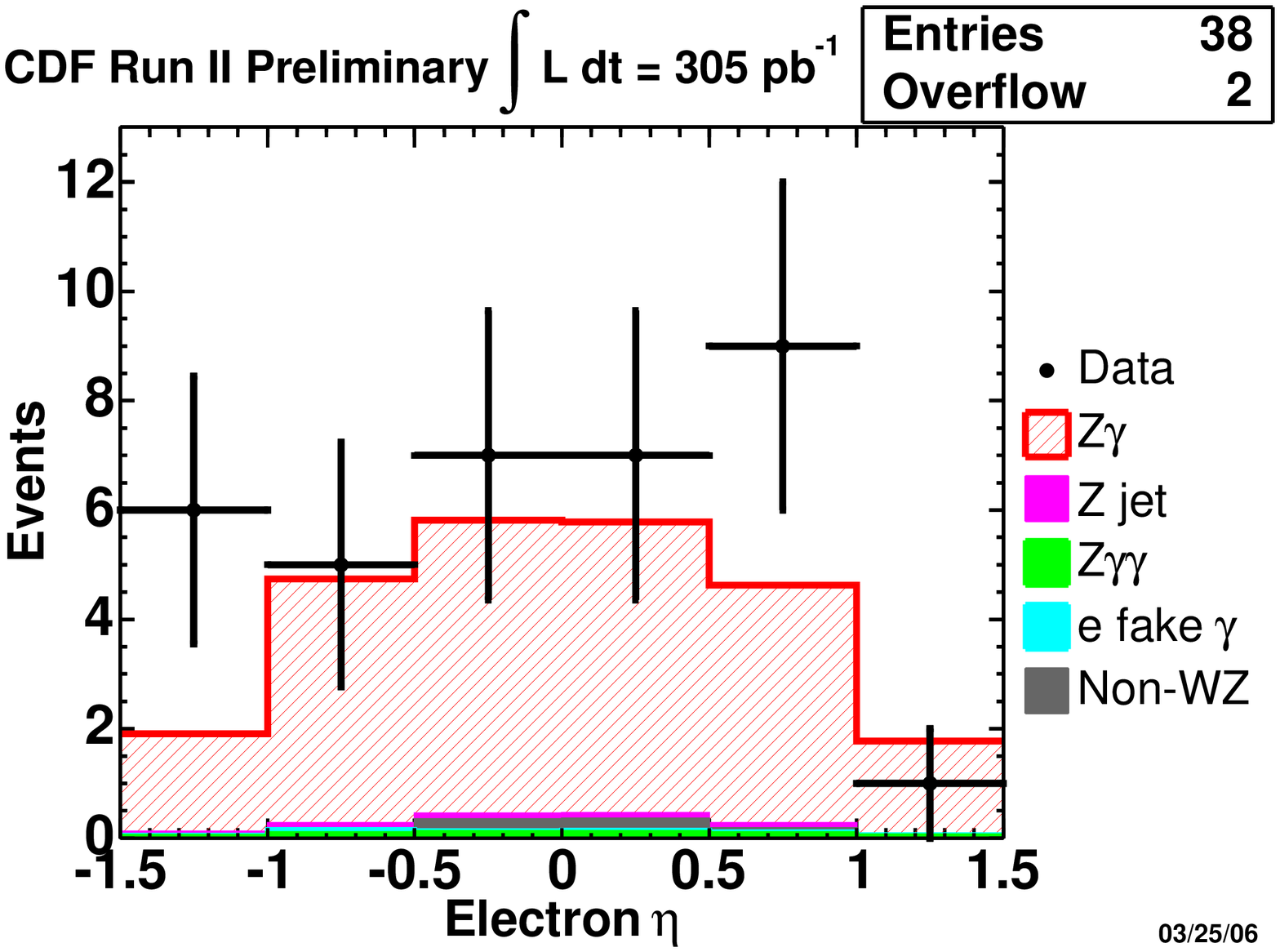,width=3.1in}} \quad
      }
    \caption 
[Multi-electron + photon distributions]
{Multi-electron + photon distributions: $\Delta R$($e\gamma$), $H_T$, electron fiduciality (see Chapter~\ref{electronid}), $\eta(e)$.}
    \label{llg_electronchannel_1.figure}
  \end{center}
\end{figure}
\begin{figure}[!h]
  \begin{center}
\vskip-0.2in
    \mbox{
\subfigure
[Electron CES X]
{\epsfig{file=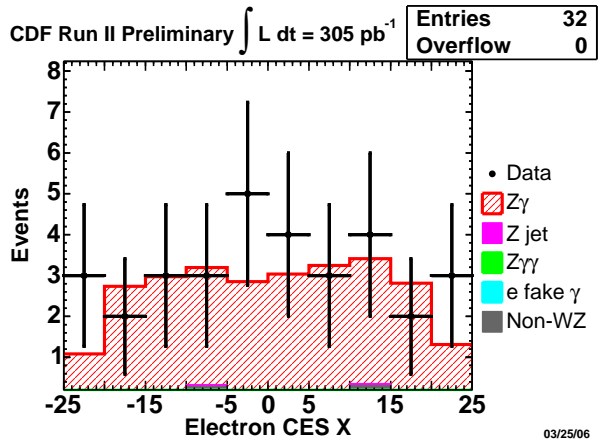,width=3.1in}} \quad
\subfigure
[Electron CES Z]
{\epsfig{file=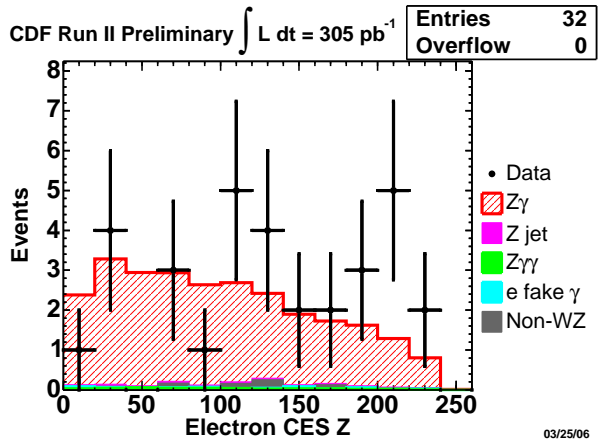,width=3.1in}} \quad
      }
\vskip-0.2in
    \mbox{
\subfigure
[$\Delta cot(\theta)$]
{\epsfig{file=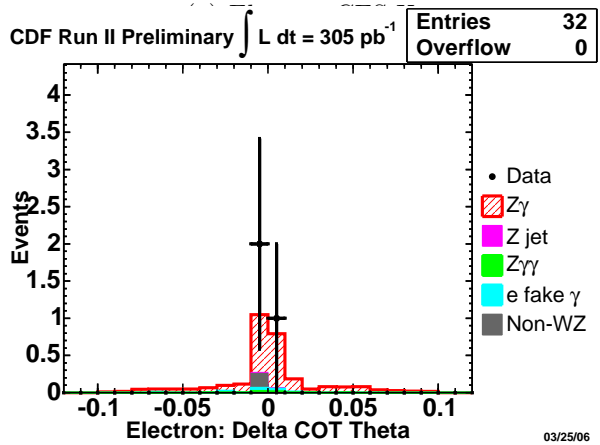,width=3.1in}} \quad
\subfigure
[$\Delta xy$]
{\epsfig{file=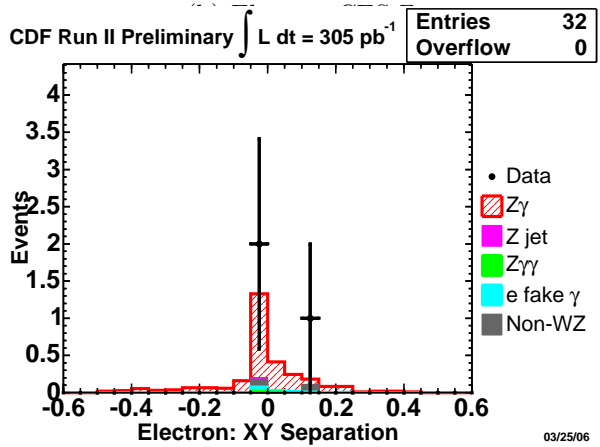,width=3.1in}} \quad
      }
    \caption 
[Multi-electron + photon distributions: continued]
{Multi-electron + photon distributions: CES X, CES Z, $\Delta
cot(\theta)$ and $\Delta xy$ (see Section~\ref{conversions}). There is
one conversion electron and two tridents in the $\eeg$ sample. }
\label{llg_electronchannel_2.figure}
  \end{center}
\end{figure}

\clearpage
\newpage
\section{Stability Plots for $\Zj$ and $\Wj$}
\label{stability_plots_zj_wj}

\begin{figure}[!h]
  \begin{center}
    \mbox{
\subfigure
[$\Zj$ $\mu$ channel]
{\epsfig{file=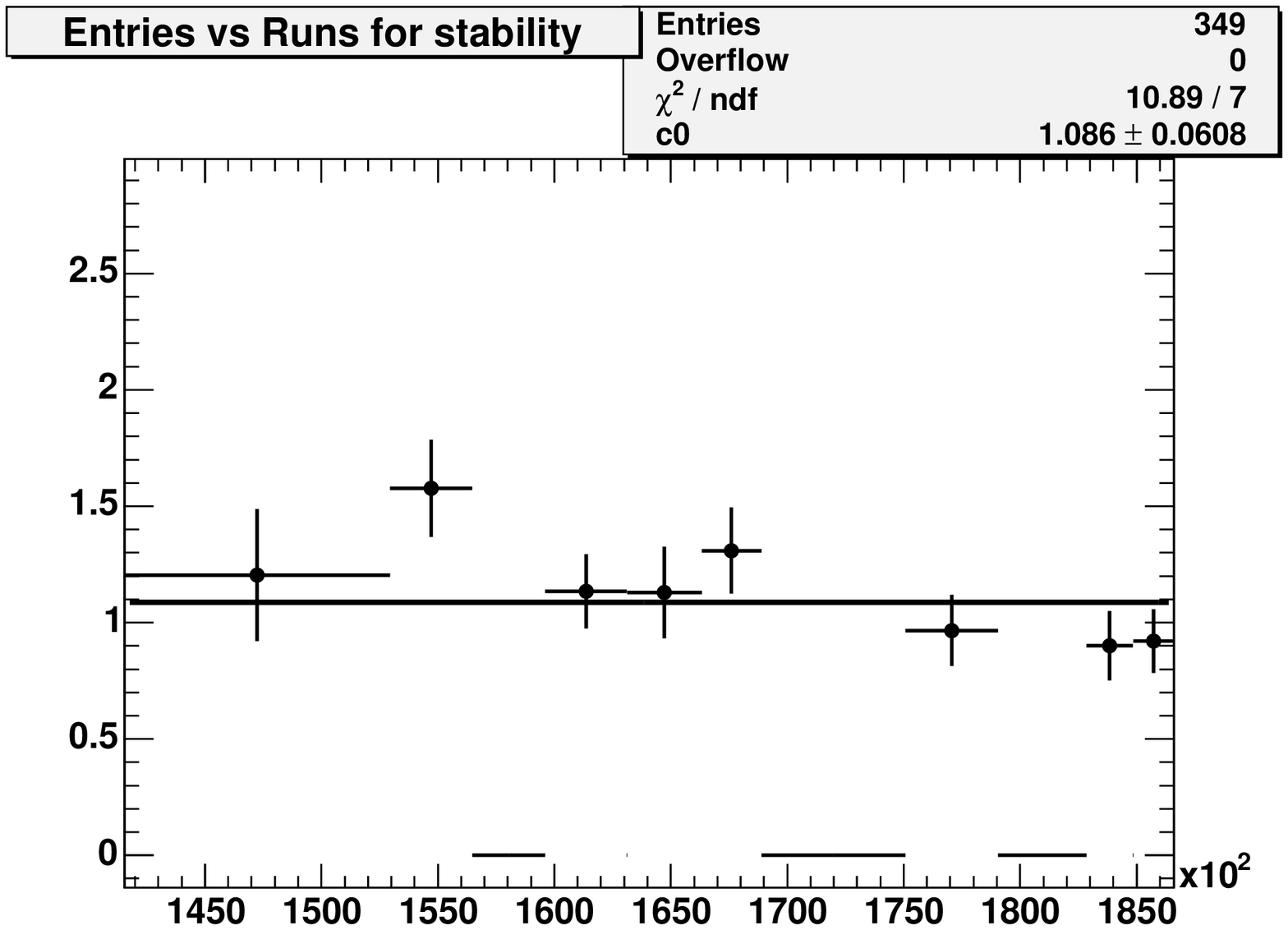,width=3.1in}} \quad
\subfigure
[$\Wj$, $\mu$ channel]
{\epsfig{file=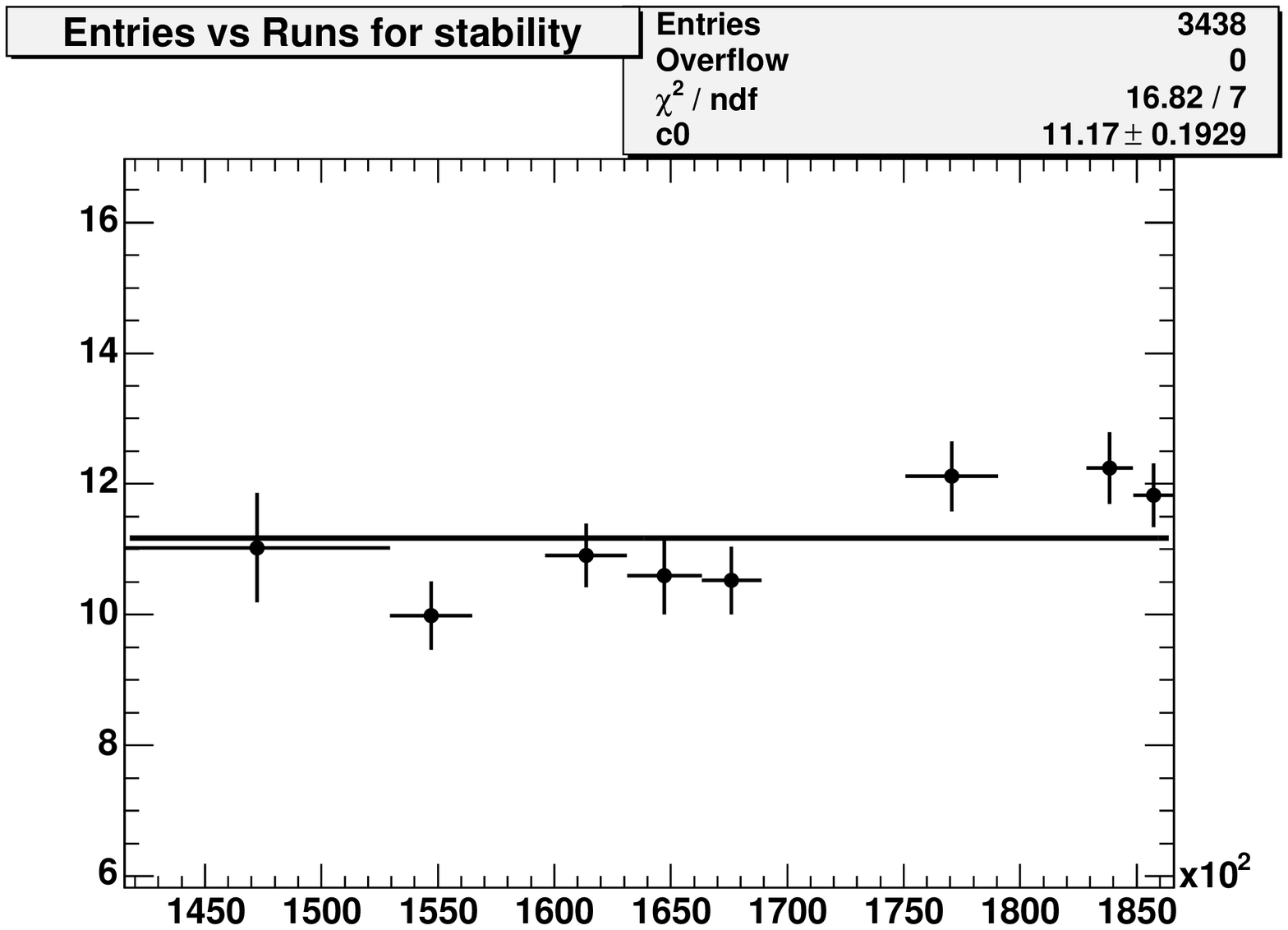,width=3.1in}} \quad
      }
    \mbox{
\subfigure
[$\Zj$ $e$ channel]
{\epsfig{file=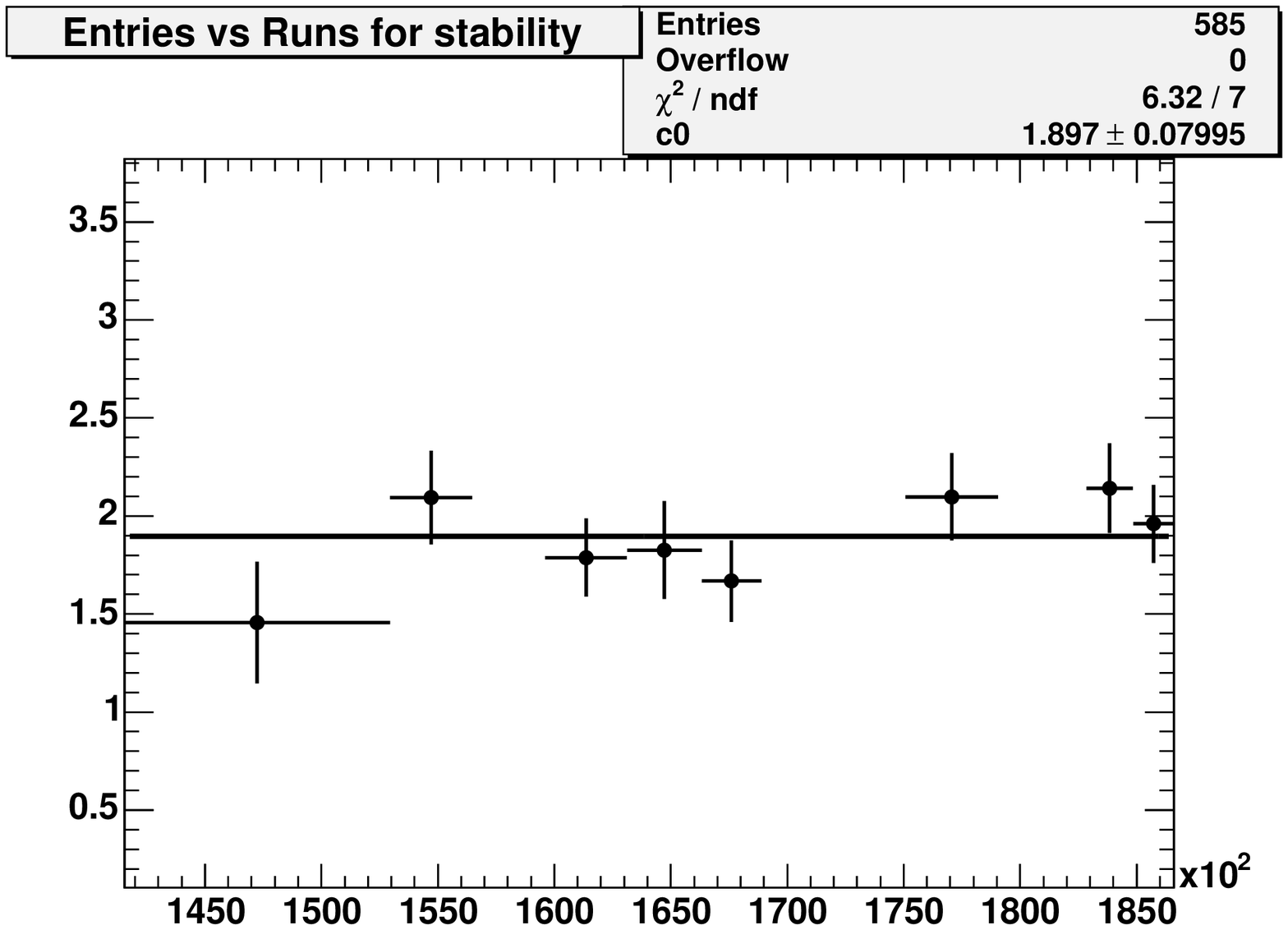,width=3.1in}} \quad
\subfigure
[$\Wj$, $e$ channel]
{\epsfig{file=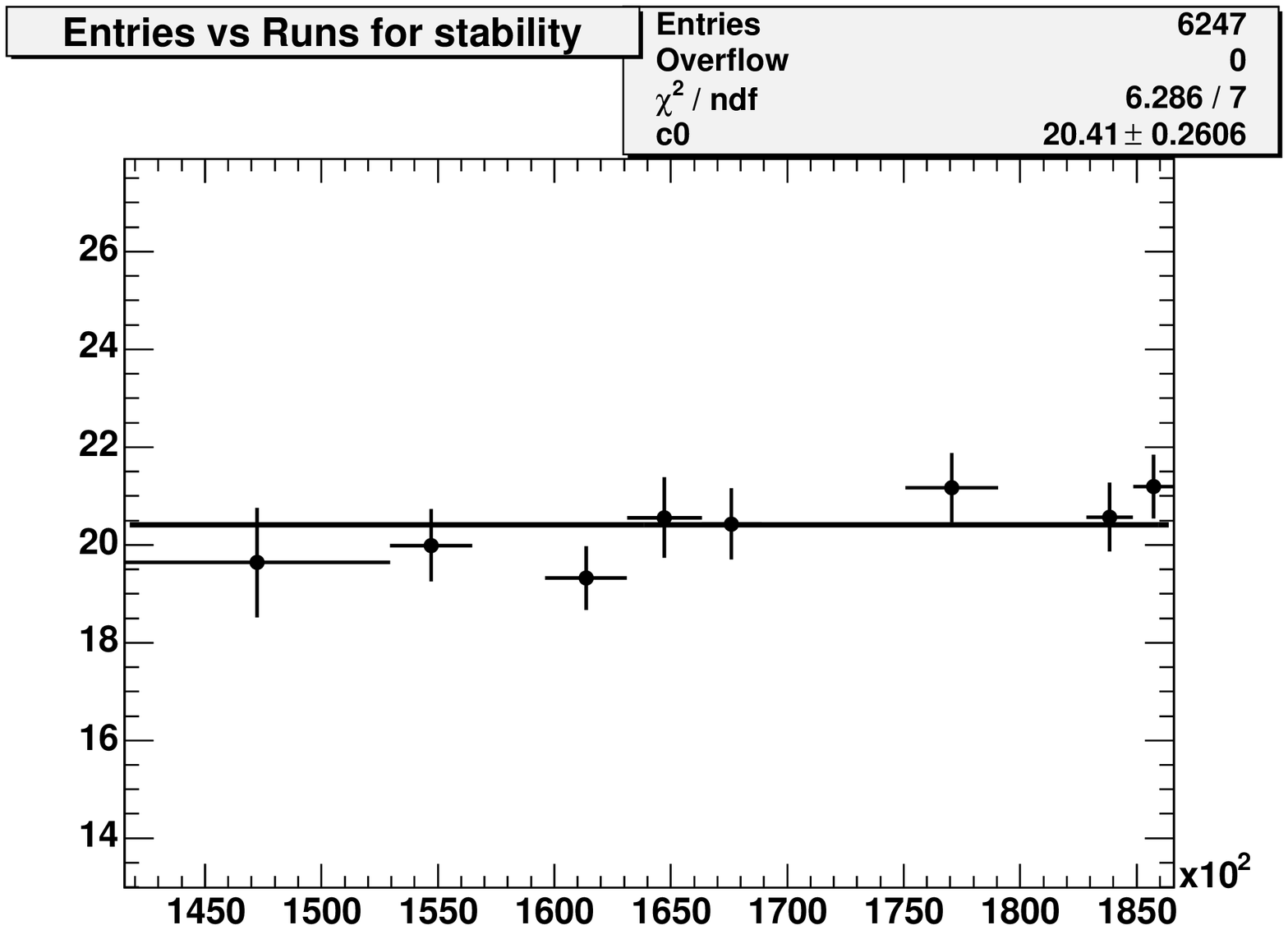,width=3.1in}} \quad
      }
\caption 
[Stability plots for $\Zj$ and $\Wj$]
{Stability plots for $\Zj$ (a), $\Wj$ (b) for muons; $\Zj$ (c), $\Wj$
(d) for electrons. The bins are those of
Table~\ref{lumi_stability.table}. The uncertainties are statistical
only. The luminosity systematic error of 6\%
(Chapter~\ref{systematics}) is not included}
\label{zjwj_stability.figure}
  \end{center}
\end{figure}

To check electron and muon identification in events with an extra
object (such as a photon in the signal channel) we plot the rate for
$\Zj$ and $\Wj$ in the 8 bins of luminosity
(Table~\ref{lumi_stability.table}) in
Figure~\ref{zjwj_stability.figure}.  

\section{CMX vs CMUP muons: Comparison of Isolation Variables}
\label{cmx_vs_cmup}

\begin{figure}[!h]
  \begin{center}
    \mbox{
\subfigure
[$\Zmumu$: CMX vs {\color{red} CMUP} muons. IsoEt]
{\epsfig{file=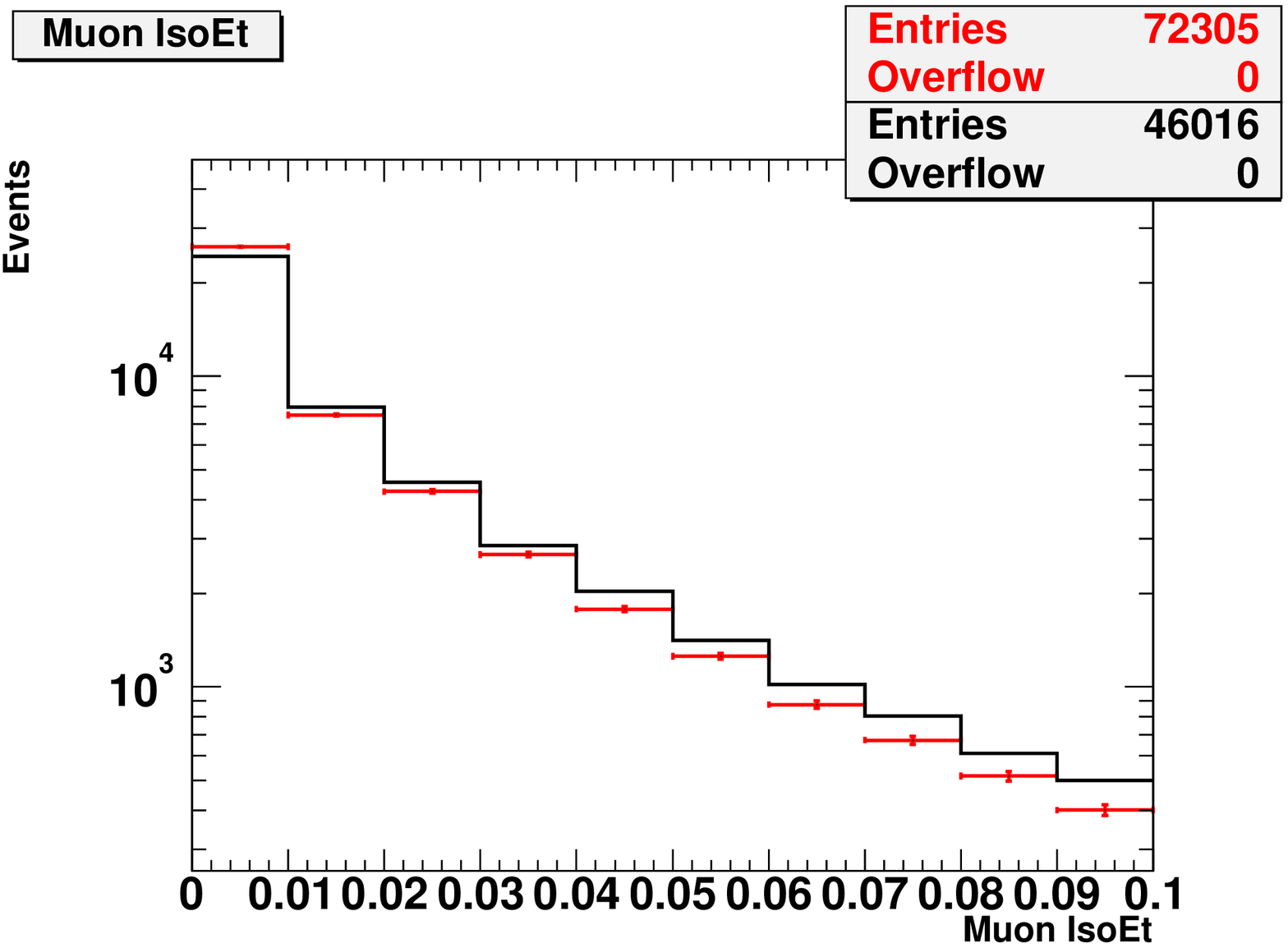,width=2.0in}} \quad
\subfigure
[$\Zmumu$: CMX vs {\color{red} CMUP} muons. IsoPt]
{\epsfig{file=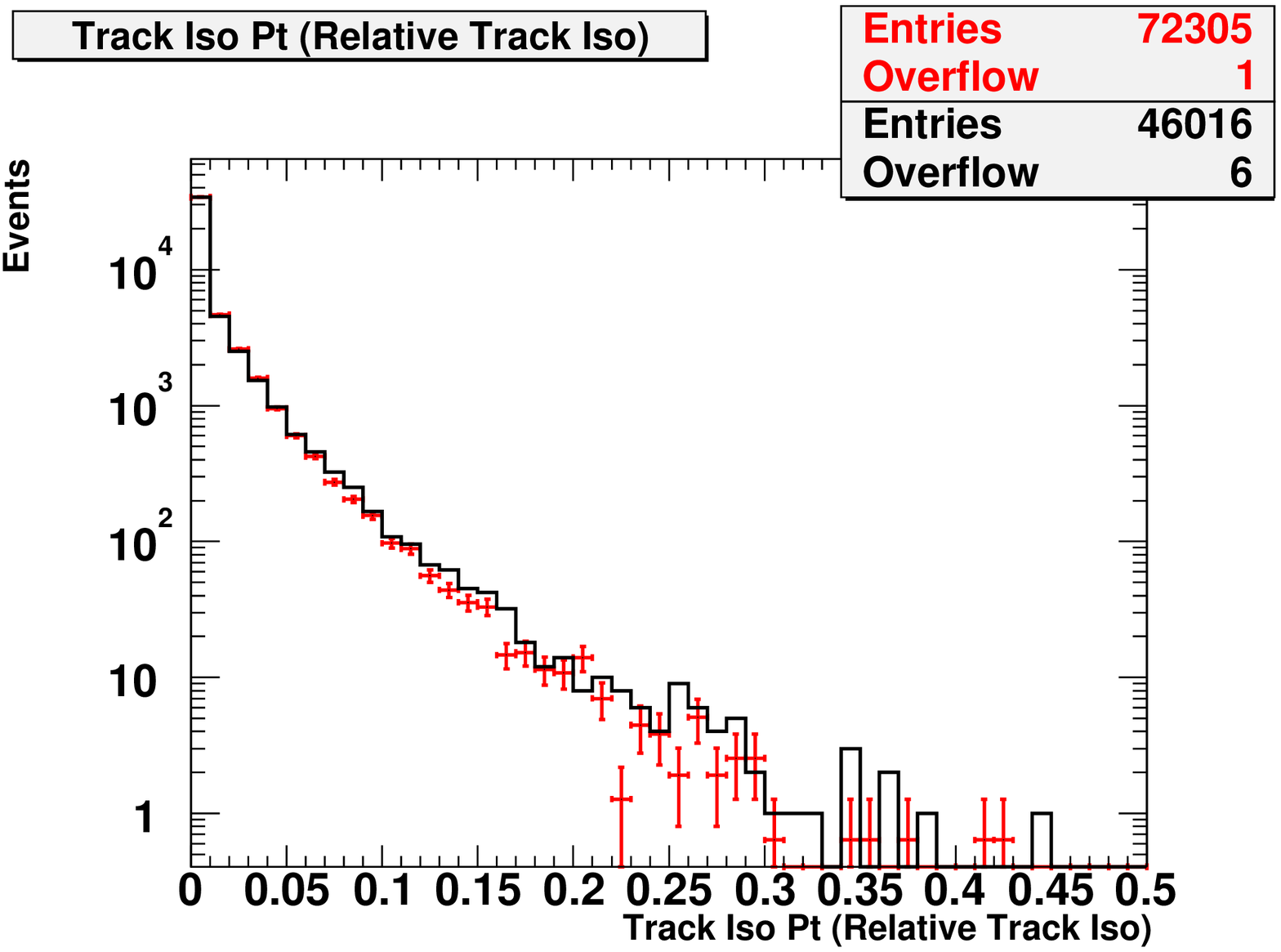,width=2.0in}} \quad
\subfigure
[$\Zmumu$: CMX vs {\color{red} CMUP} muons. Track Iso]
{\epsfig{file=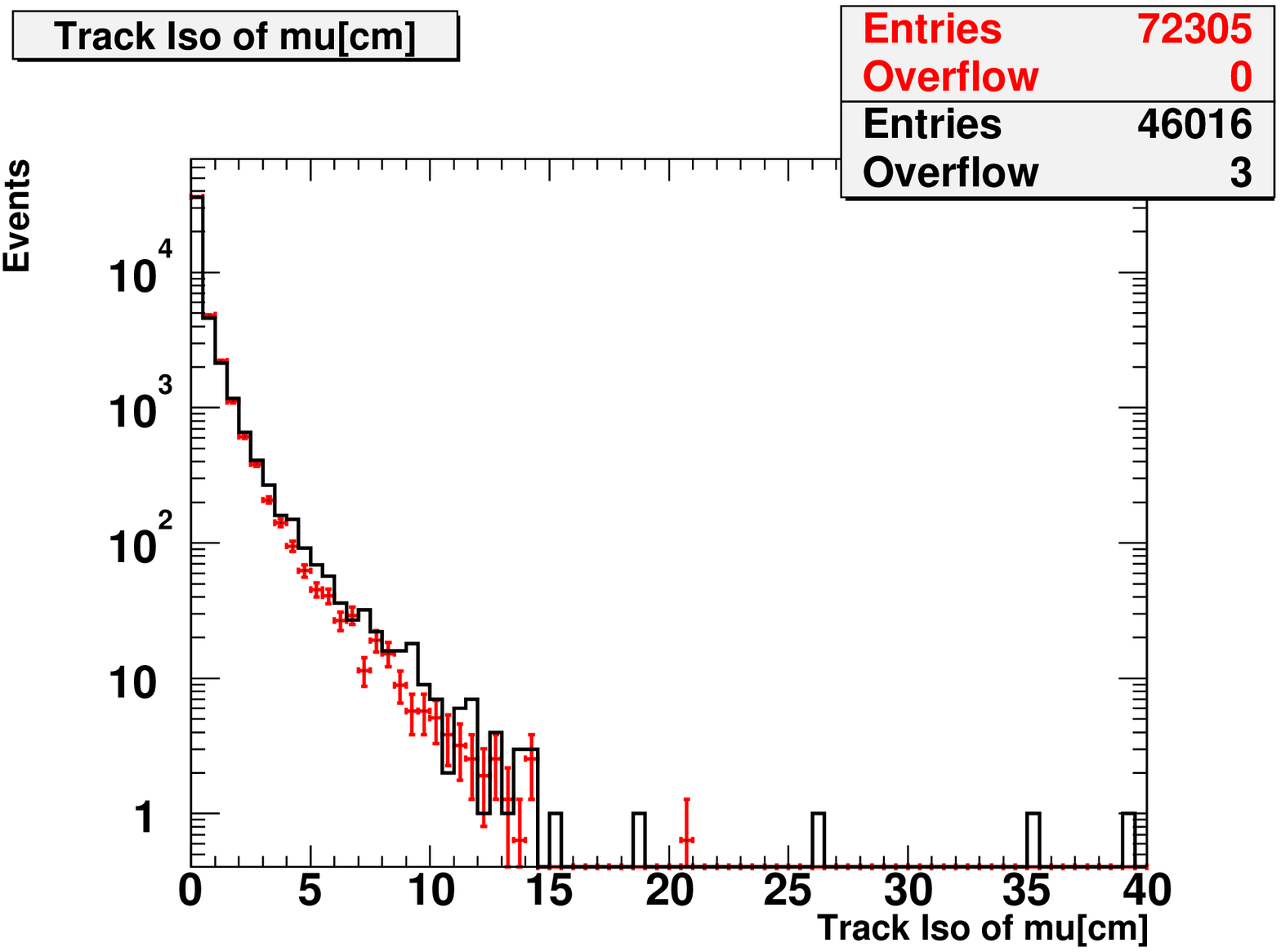,width=2.0in}} \quad
      }
    \mbox{
\subfigure
[$\Wmunu$: CMX vs {\color{red} CMUP} muons. IsoEt]
{\epsfig{file=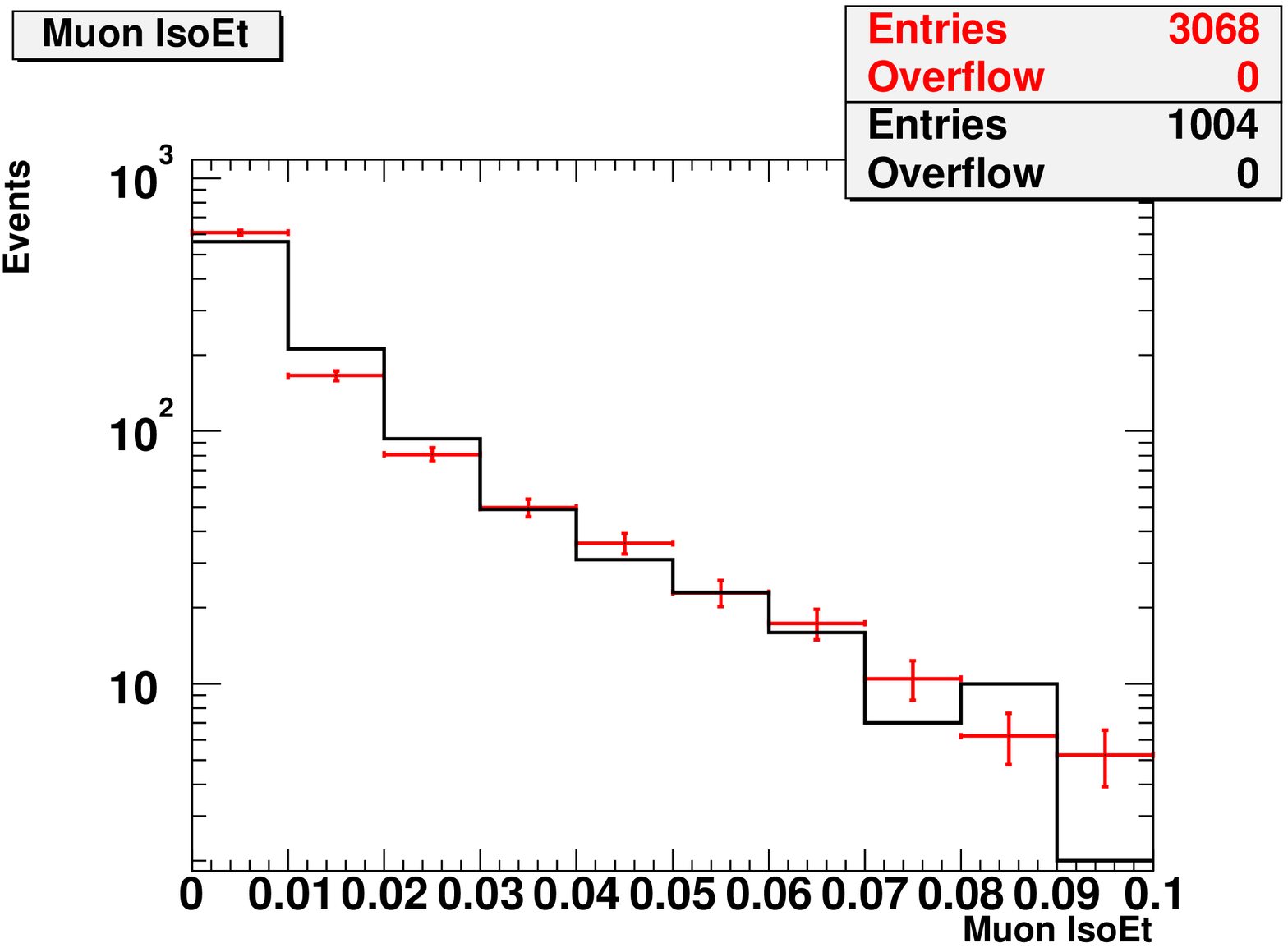,width=2.0in}} \quad
\subfigure
[$\Wmunu$: CMX vs {\color{red} CMUP} muons. IsoPt]
{\epsfig{file=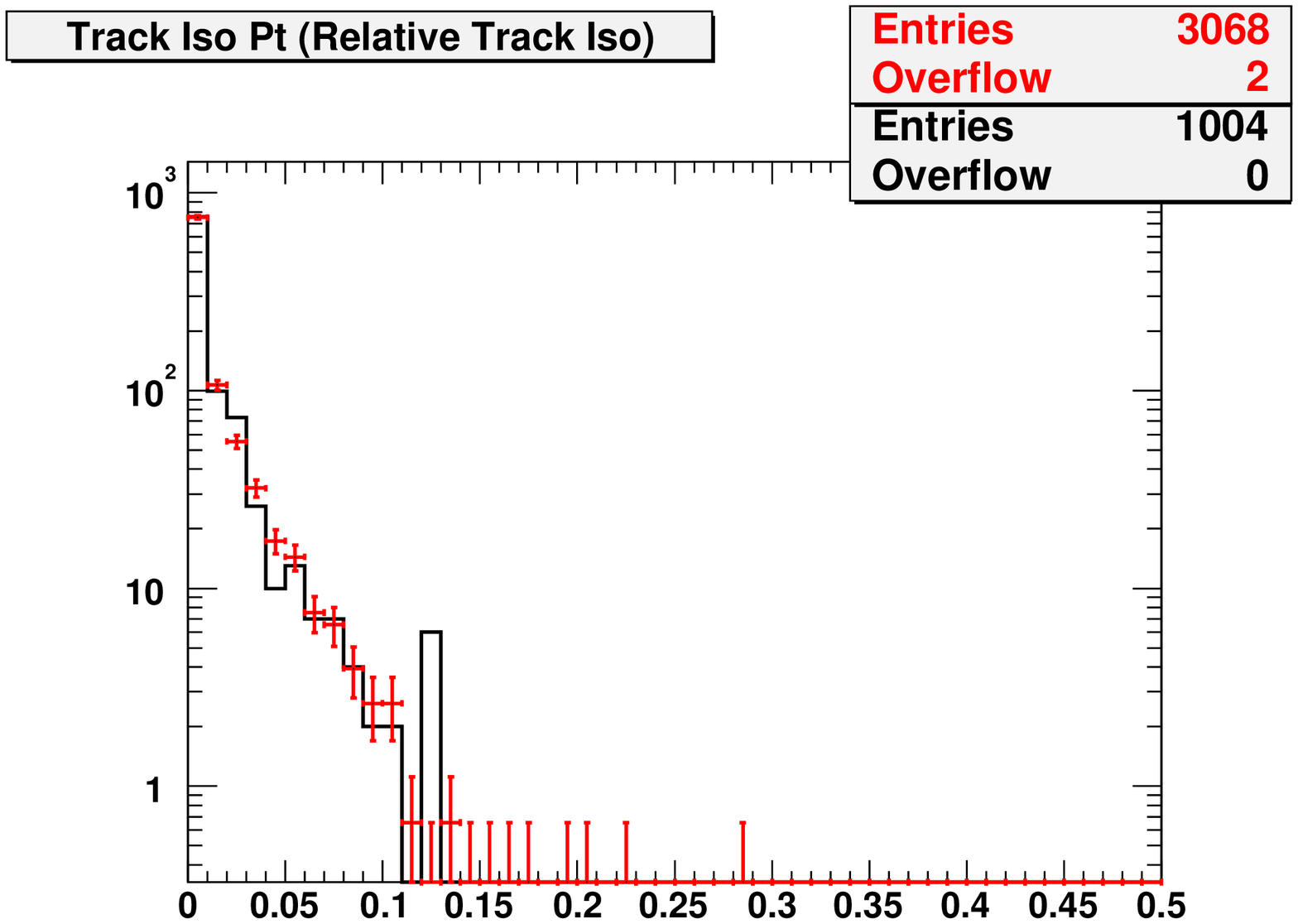,width=2.0in}} \quad
\subfigure
[$\Wmunu$: CMX vs {\color{red} CMUP} muons. Track Iso]
{\epsfig{file=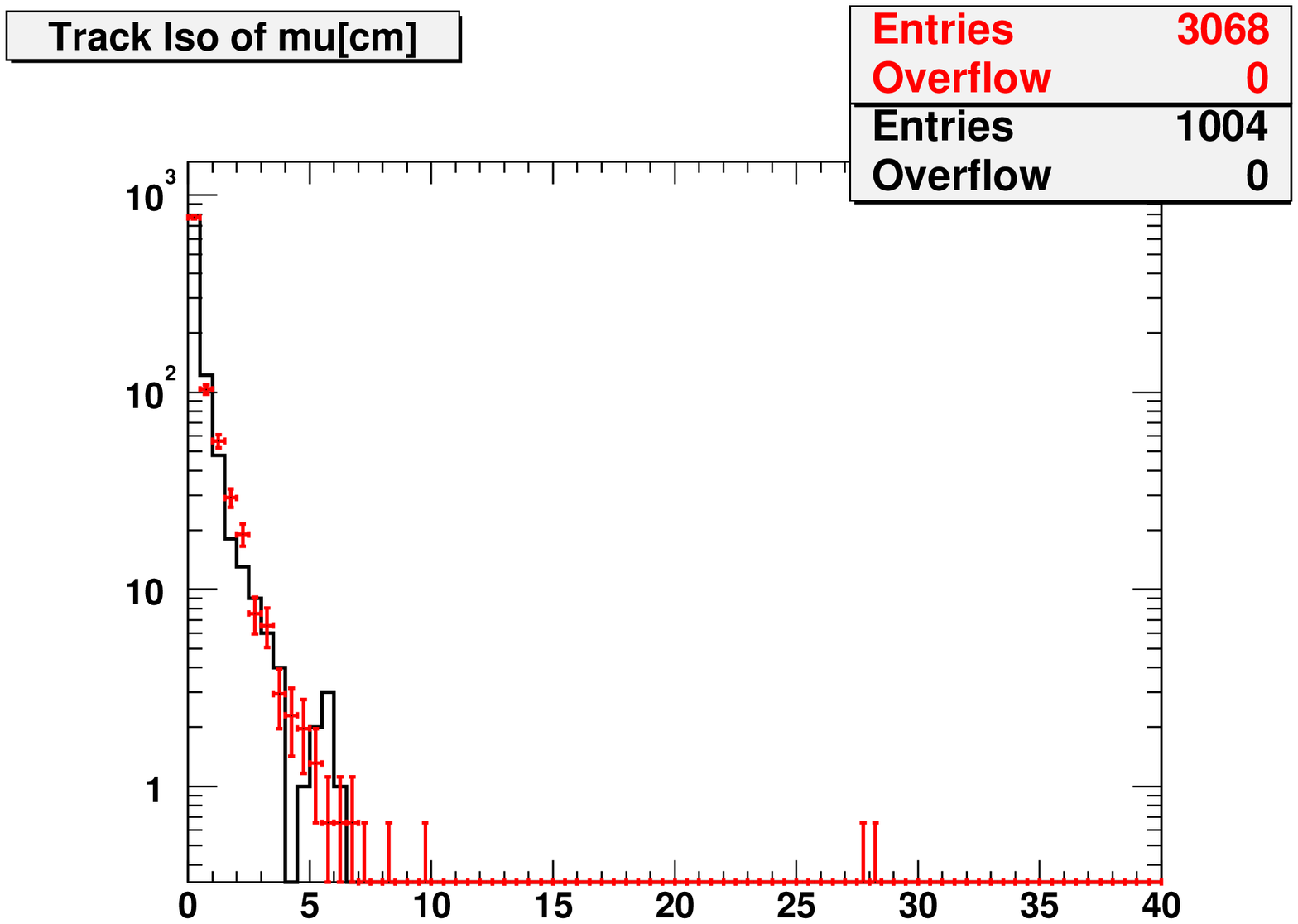,width=2.0in}} \quad
      }
\caption
[CMX vs CMUP muons: comparison of isolation variables] 
{The distributions for CMX (black histogram) and {\color{red} CMUP(red
points)} muons of calorimeter isolation, relative track isolation (see
text), absolute track isolation, for $\Zmumu$ events (top row) and
$\Wmunu$ events (bottom row). For the $\Zmumu$ plots both muons are
required to be either CMUP or CMX.}
\label{cmx_vs_cmup.figure}
  \end{center}
\end{figure}

We have checked that muons that go into the CMX system have the same
isolation properties as CMUP muons. Figure~\ref{cmx_vs_cmup.figure}
shows the distributions in calorimeter isolation, relative track
isolation (total $\Pt$ of tracks in a cone in $\eta-\varphi$ space of
radius $R=0.4$ around the muon track divided by $\Pt^{muon}$) and
absolute track isolation for $\Zmumu$ and $\Wmunu$ events.
\clearpage

\clearpage
\newpage
\section{Fake Electrons from Photon Conversions}
\label{conversions}

There are three dominant sources of fake electrons: a) photons from
$\pizero$, $\eta$ and other mesons, that convert into asymmetric
$\epem$ pairs in the material before the COT volume, b) charged
hadrons in jets that either interact in the electromagnetic volume of
the calorimeter or overlap with a $\pizero$ or secondary photon in the
jet, and c) electrons from the decay of heavy flavor ($b$, $c$, and
maybe even $s$). We estimated these backgrounds in Section~\ref{nonwz}
by studying the total $\Pt$ of tracks in a cone in $\eta-\varphi$
space of radius $R=0.4$ around the lepton track.

We consider fake electrons from photon conversions below. Electrons
coming from photon conversion are identified by conversion algorithm,
which looks for couple of opposite sign tracks with $|\Delta xy|<$0.2
cm and $|\Delta cot(\theta)|<$0.04.

For each electron a conversion flag is tested. We define if the
electron is flagged as coming from a conversion ($\gamma\rrr e^+e^-$)
or from trident events where a conversion is caused by a
bremsstrahlung photon ($e\rrr e\gamma, \gamma\rrr e^+e^-$). We study
same-sign events in $\Zee$ sample and then check how many of them
contain electrons tagged as conversions or tridents.

\begin{figure}[!h]
  \begin{center}
    \mbox{
\subfigure[Invariant Mass]{\epsfig{file=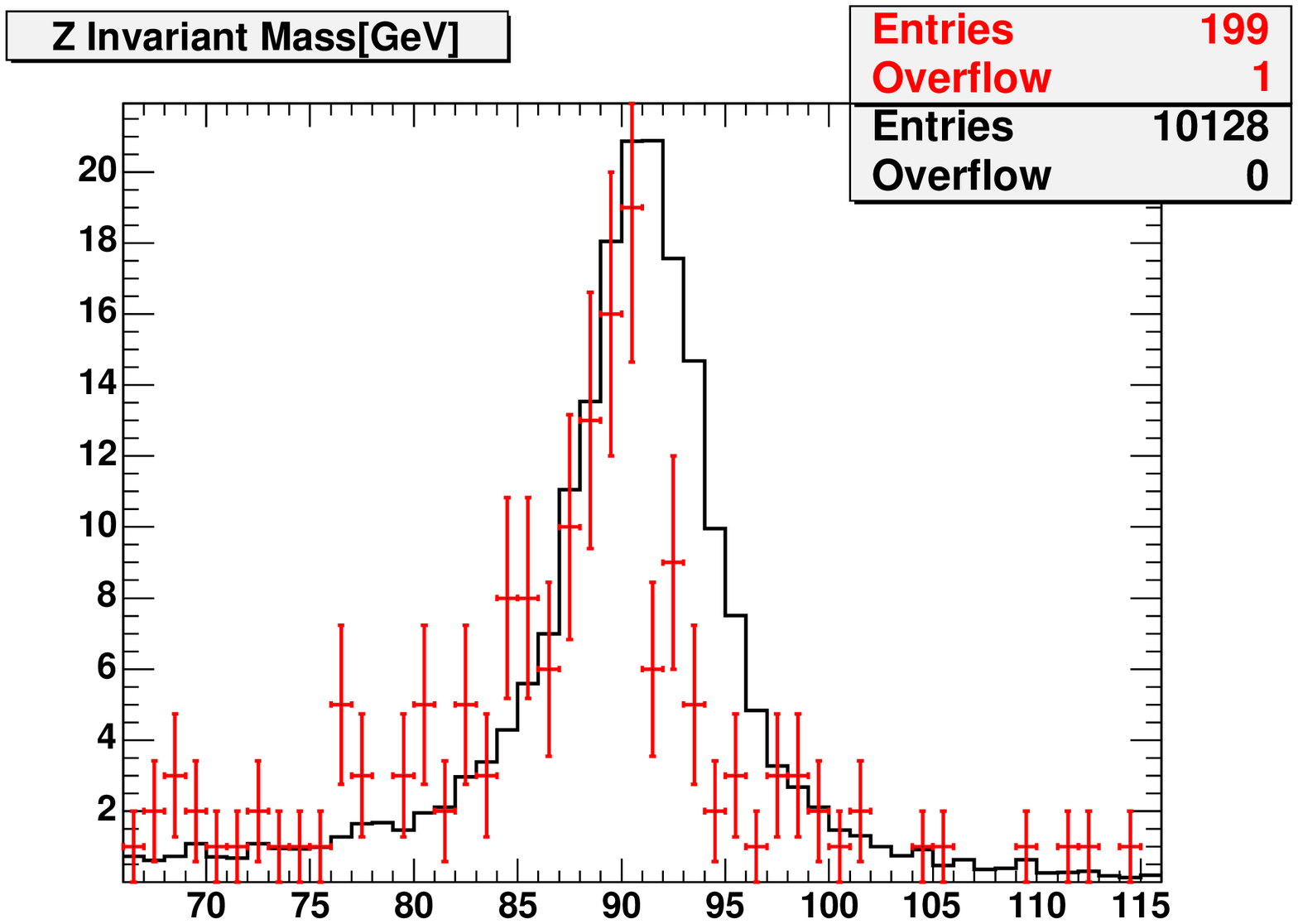,width=2.0in}} \quad
\subfigure[$P_T$ of $\Zee$]{\epsfig{file=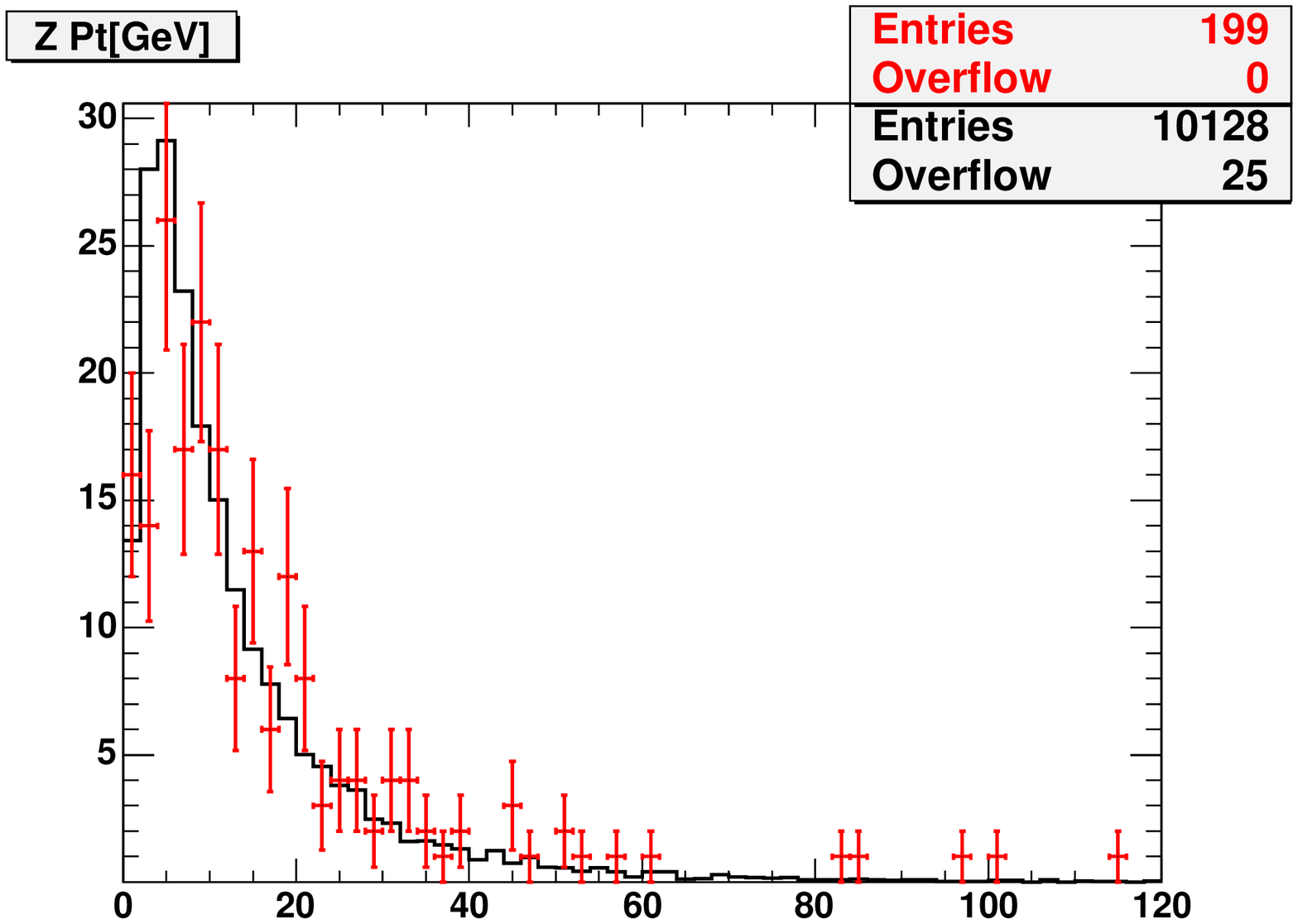,width=2.0in}} \quad
\subfigure[$\Delta\phi$ of $\Zee$]{\epsfig{file=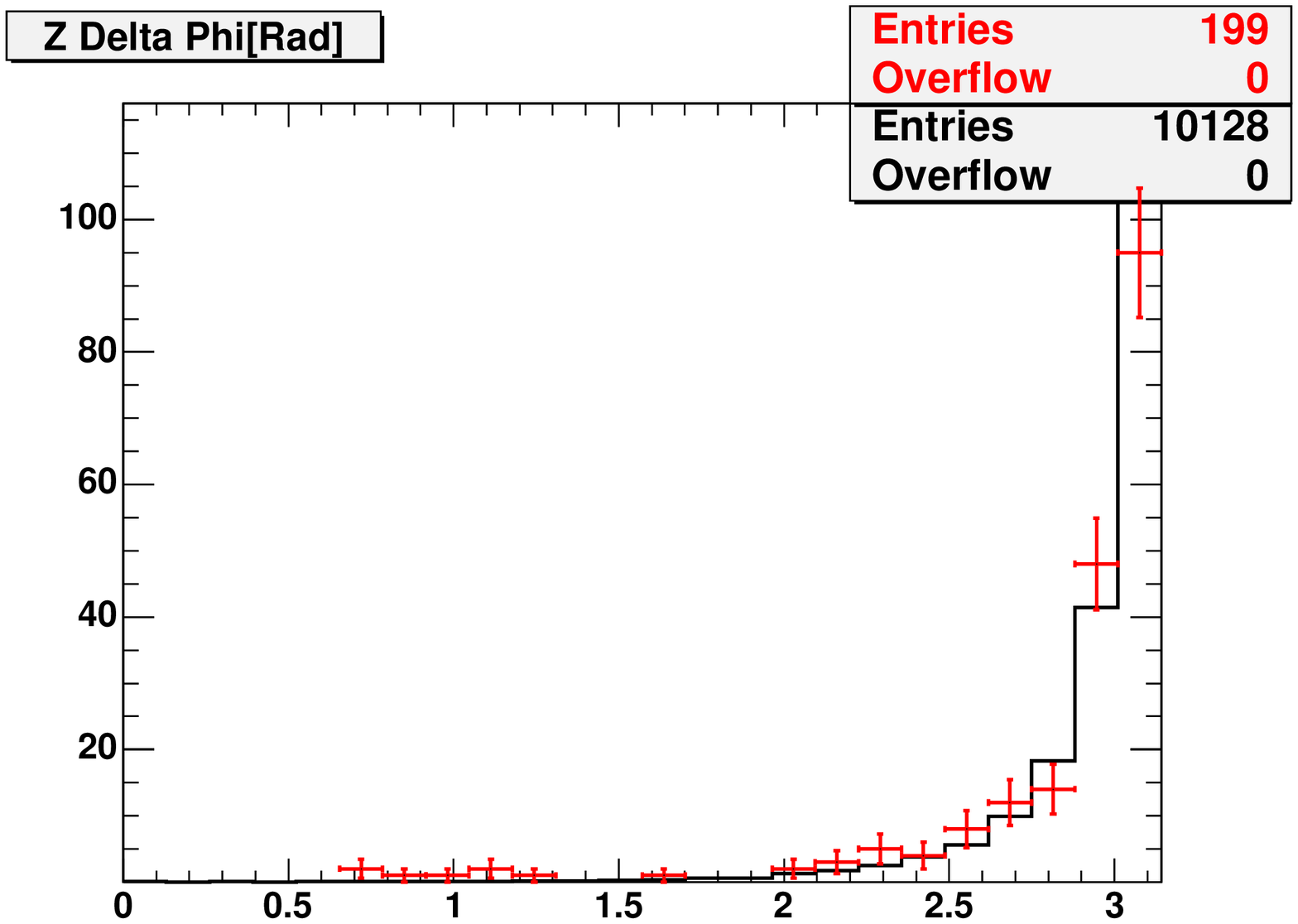,width=2.0in}} \quad
      }
    \mbox{
\subfigure[Invariant Mass (log)]{\epsfig{file=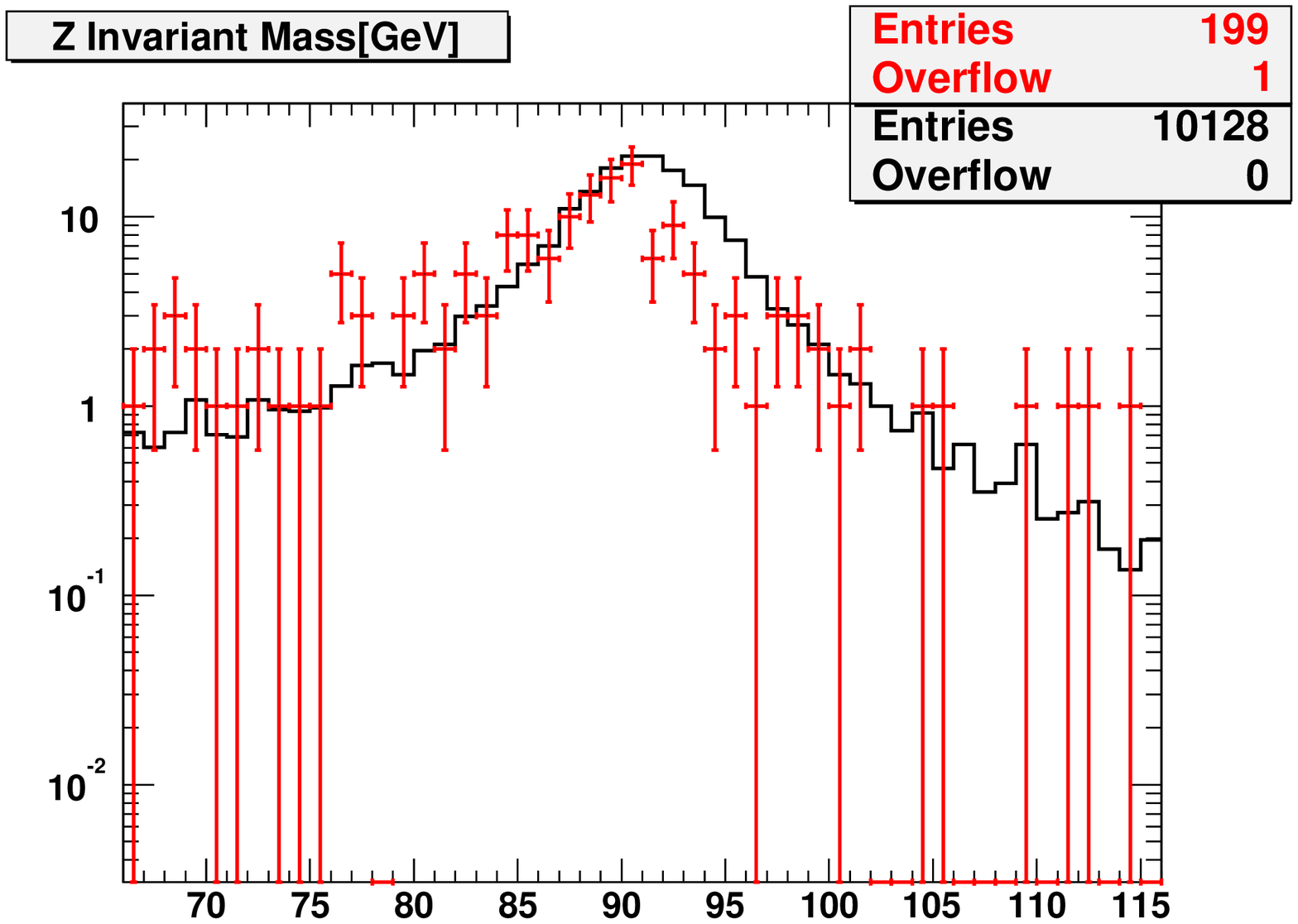,width=2.0in}} \quad
\subfigure[$P_T$ of $\Zee$ (log)]{\epsfig{file=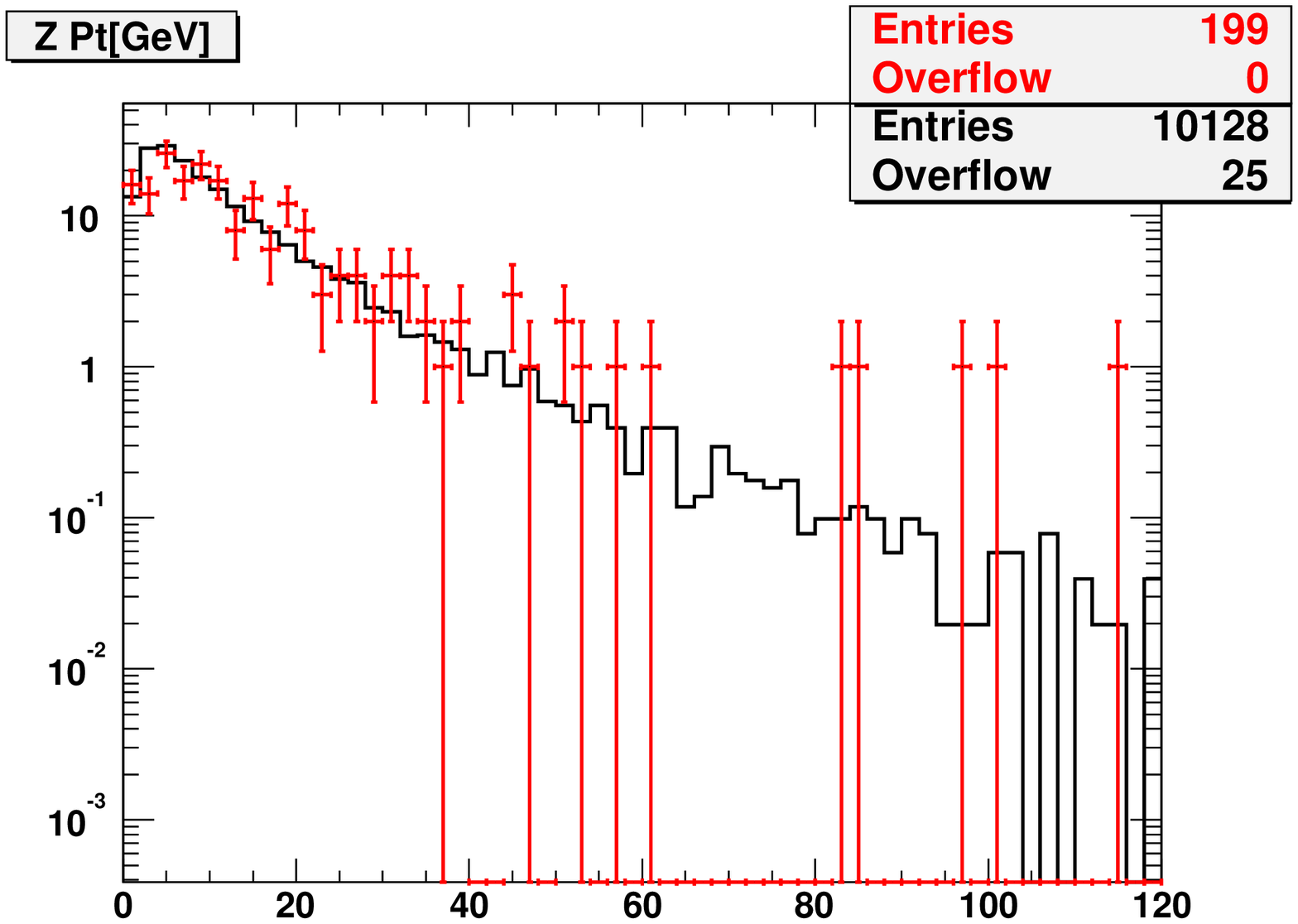,width=2.0in}} \quad
\subfigure[$\Delta\phi$ of $\Zee$ (log)]{\epsfig{file=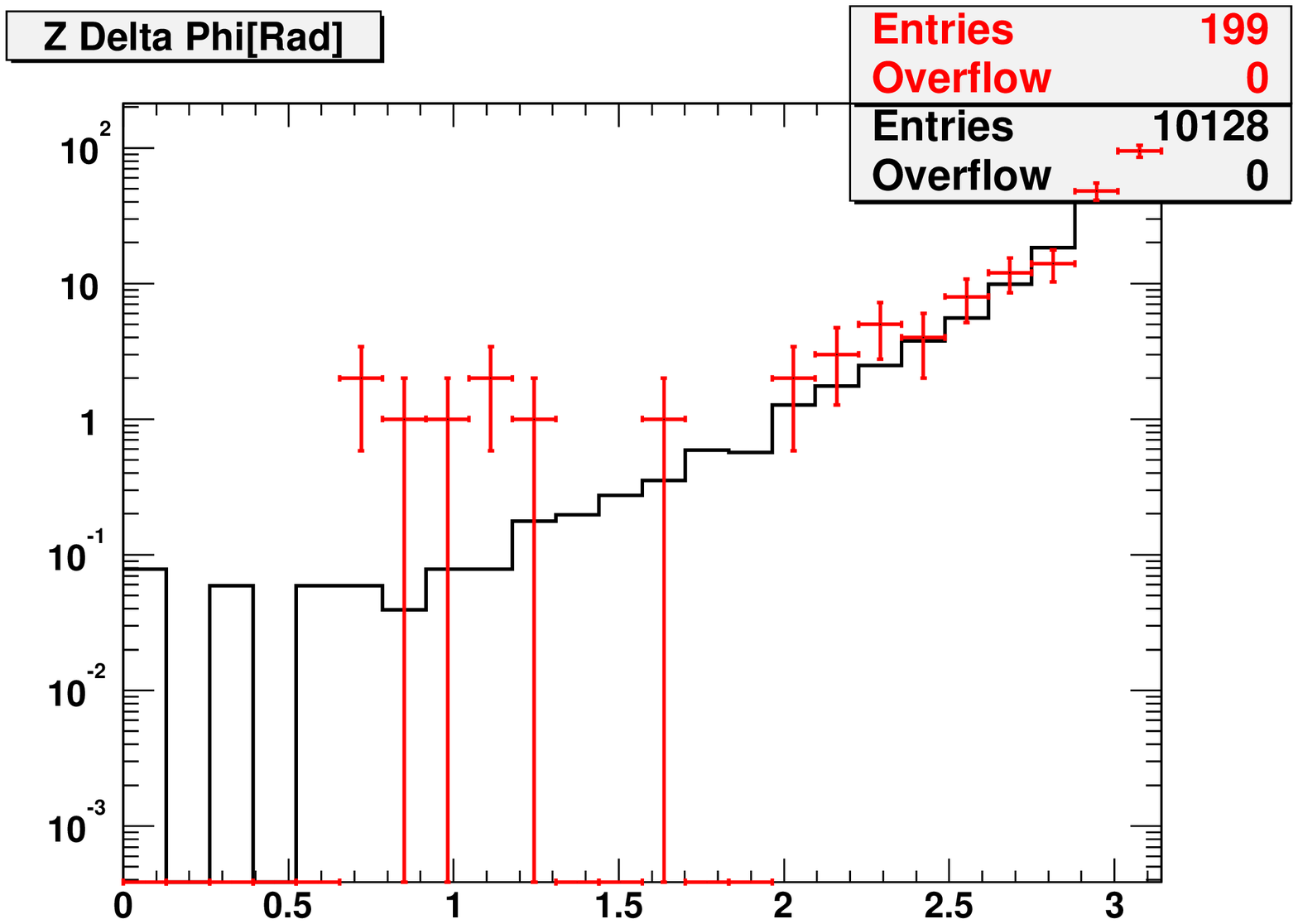,width=2.0in}} \quad
      }
    \caption 
[The distributions for same-sign and opposite-sign $e^+e^-$ pairs for data]
{ The distributions for same-sign (red points) and opposite-sign
(black histogram) $e^+e^-$ pairs in invariant mass, $P_T$, and $\Delta
\phi$; each distribution is shown twice, in linear plots(a, b, c), and
in log plots(d, e, f).}
    \label{zeess.figure}
  \end{center}
\end{figure}

\begin{figure}[!h]
  \begin{center}
    \mbox{
{\epsfig{file=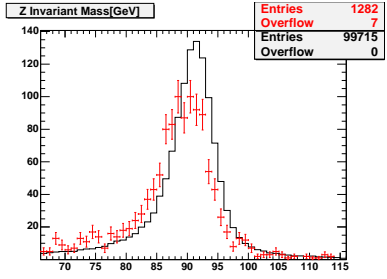,width=2.0in,clip=}} \quad
      }
    \caption 
[The distributions for same-sign and opposite-sign $e^+e^-$ pairs for $\Zee$ MC]
{ The distributions for same-sign (red) and opposite-sign (black)
$e^+e^-$ pairs in invariant mass for $\Zee$ MC sample.
The observed behavior is similar to what we have in data.
             }
   \label{ztop2i_zee_z_mass.figure}
  \end{center}
\end{figure}

Figure~\ref{zeess.figure} shows the distributions for same-sign and
opposite-sign $e^+e^-$ pairs. The invariant mass of the same-sign
electrons is shifted with respect to the invariant mass of $\Zee$
(Figure~\ref{zeess.figure}, a). We observe similar behavior (shift in
invariant mass distribution) in the $\Zee$ MC as in data
(Figure~\ref{ztop2i_zee_z_mass.figure}).

To develop an understanding of the conversion cut for the candidate
events, we summarize the same-sign events in $\Zee$ sample. Out of
$\nofzeesamesign$ events in this sample only $\nofzeessneither$ are
not tagged as either a conversion electron or a trident. We find that
we have $\nofzmumusamesign$ same sign muon events, which is comparable
to $\nofzeessneither$ non-conversion/trident same-sign electron
events.

\begin{table}[!h]
\begin{center}
\begin{tabular}{|l|r|}
\hline
$\Zee$ same sign                      & \nofzeesamesign      \\
\hline
$\Zee$ SS, one conversion             & \nofzeessconversion  \\
\hline
$\Zee$ SS, one trident                & \nofzeesstrident     \\
\hline
$\Zee$ SS, two conversions (CC)       & \nofzeessconversiond \\
\hline
$\Zee$ SS, two tridents (TT)          & \nofzeesstridentd    \\
\hline
$\Zee$ SS  trident + conversion (TC)  & \nofzeessweird       \\
\hline
$\Zee$ SS neither conversion/trident  & \nofzeessneither     \\
\hline
\end{tabular}
    \caption 
[Sources of same-sign electrons in the $\Zee$ data sample]
{ A breakdown of the source of same-sign electrons in the
	        $\Zee$ sample. Most same-sign events are tagged by the
	        conversion filter as conversions or tridents.}
    \label{zsamesign.table}
\end{center}
\end{table}

\clearpage
\newpage
\section{Checking the $\mumug$ and $\mugmet$ for Additional Backgrounds}
\label{mumug_checks}

We have used a number of techniques, described below, to look for
additional backgrounds in the $\mumug$ and $\mugmet$ samples. 

\subsubsection{Same-Sign Leptons to Estimate Jets Faking one or More Leptons}

We used results from Ref.~\cite{CDF7560} on the numbers of same-sign
(SS) and other-sign (OS) muon pairs in the dimuon sample. The expected
background from SS muon pairs to $\mumug$ is calculated as follows.

First we obtain the ratio of W+1 jet events to OS events, x, expected
in the dimuon sample: x = SS/OS $\times$ 1.51$\pm$0.05. For the
$\mu\mu$ sample ratio of SS/OS is of order of 0.05\% and therefore
x=0.1\%. In the $\mumug$ sample we have 12 OS events, and therefore
the expected background from W+1 jet is negligible, 0.1\% $\times$ 12
= 0.012$\pm$0.001 events.

\subsubsection{Decays in flight of Low Momentum $K^\pm$ Faking a High-Momentum Muon}

\begin{figure}[!h]
\begin{center}
   \begin{minipage}[b]{2.8in}
   \begin{center}
      \parbox{2.8in}{\epsfxsize=\hsize\epsffile{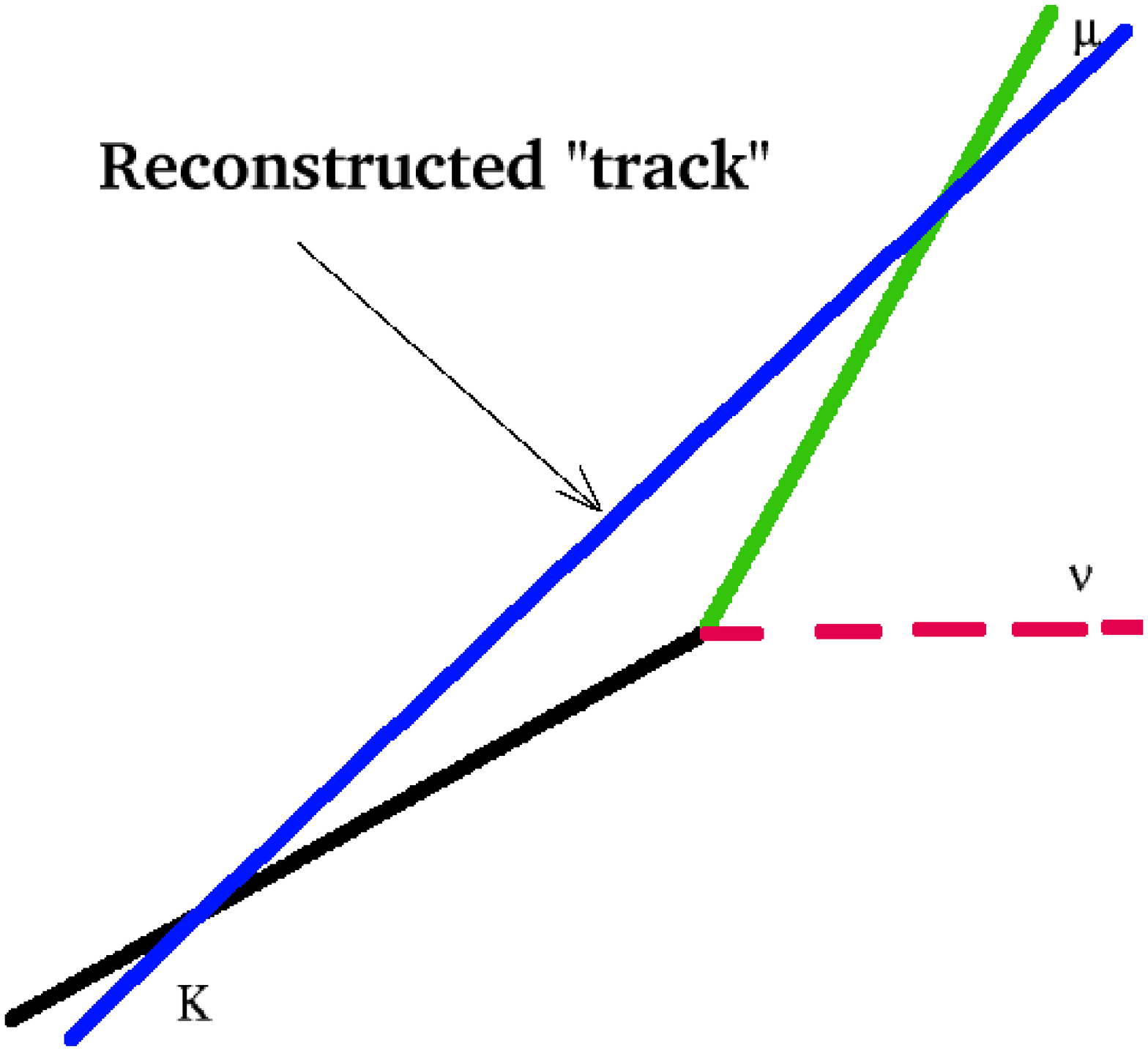}}
      \caption[Decays-In-Flight: schematic figure]
              {Decays-In-Flight: schematic figure, $K\rrr\mu\nu$. Two track segments from $K$ and $\mu$ misreconstructed as one track.}
      \label{dif_scatch.figure}
   \end{center}
\vskip2.0in
   \end{minipage}
\hspace{0.2in}
   \begin{minipage}[b]{3.6in}
   \begin{center}
      \parbox{3.6in}{\epsfxsize=\hsize\epsffile{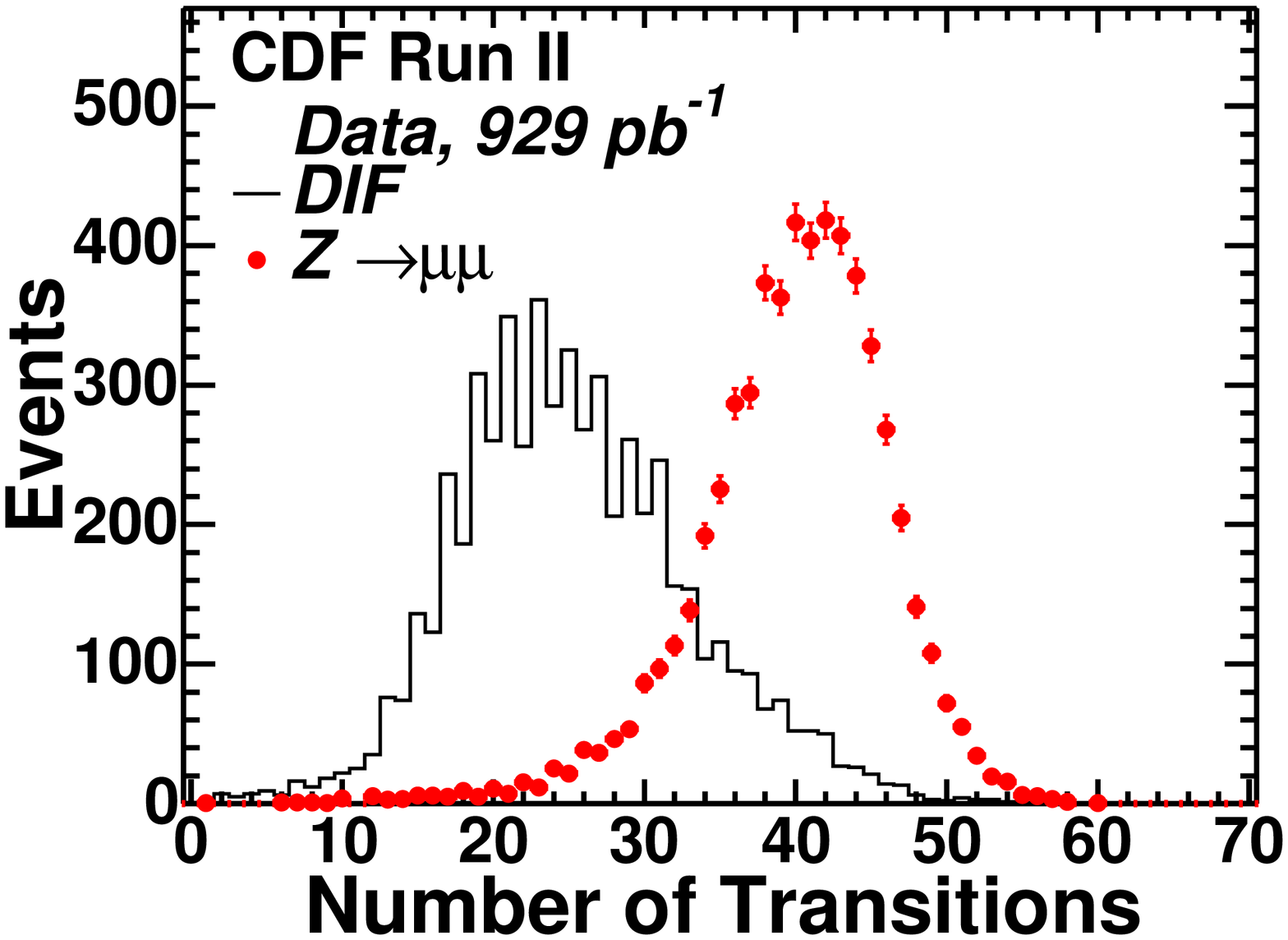}}
\caption 
[$\mu$'s from DIF sample vs. $\mu$'s from $\Zmumu$]
{$\mu$'s from DIF sample(histogram) vs. $\mu$'s from $\Zmumu$ (dots).}
\label{ztight_mu_transitions_good_mu_transitions}
   \end{center}
   \end{minipage}
\end{center}
\end{figure}

A low-momentum hadron, not in an energetic jet, can decay to a muon
forming a ``kink'' between the hadron and muon trajectories
(Figure~\ref{dif_scatch.figure}). In this case a high-momentum track
may be reconstructed from the initial track segment due to the hadron
and the secondary track segment from the muon. A kaon that decays
before the COT volume results in a muon whose momentum is correctly
measured; a kaon that decays after the COT is itself correctly
measured. These contributions are included in the total background
estimate (see Section~\ref{nonwz}).

The contribution from this background is estimated by identifying
tracks consistent with a ``kink'' in the COT. We count the number of
times that, proceeding radially along a COT track, a ``hit'' in the
n+1 layer of sense-wires is on the other side of the fitted track from
the hit in the nth layer. Real tracks will have hits distributed on
both sides of the fit, and will therefore have many ``transitions''. A
mis-measured track from a 5-$\GeV$ $K^+$ (for example), on the other
hand, will consist of two intersecting low-momentum arcs fit by a high
momentum track, and will have a small number of
transitions~\cite{paramonov}.

\begin{figure}[!h]
  \begin{center}
    \mbox{
\hspace{-0.2in}
\subfigure
[$\mumug$: \newline $\mu$ tracks with SVX hits]
{\epsfig{file=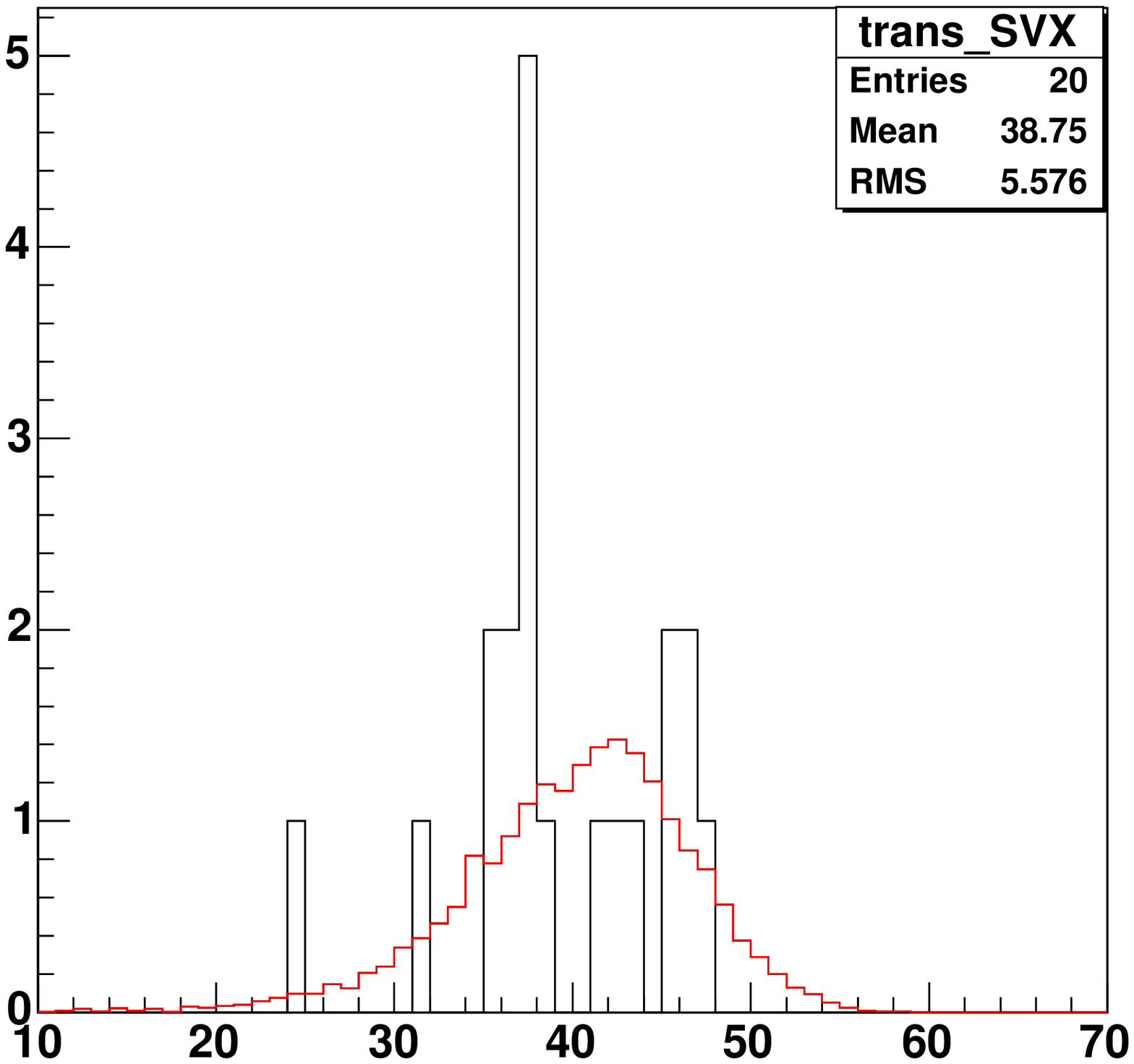,width=1.9in}} \quad
\hspace{-0.4in}
\subfigure
[$\mumug$: \newline $\mu$ tracks without SVX hits]
{\epsfig{file=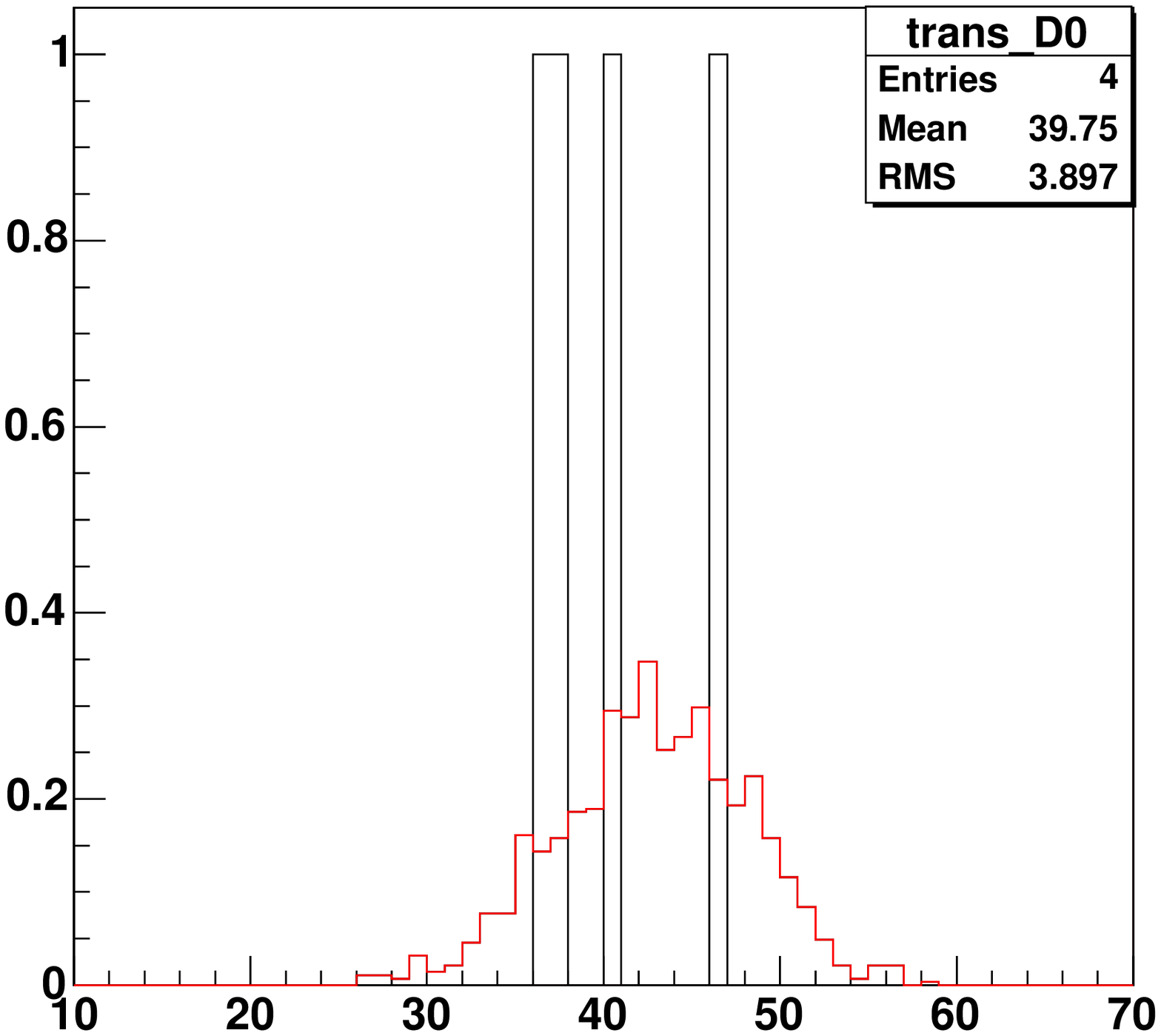,width=1.9in}} \quad
\hspace{-0.4in}
\subfigure
[$\mugmet$: \newline $\mu$ tracks with SVX hits]
{\epsfig{file=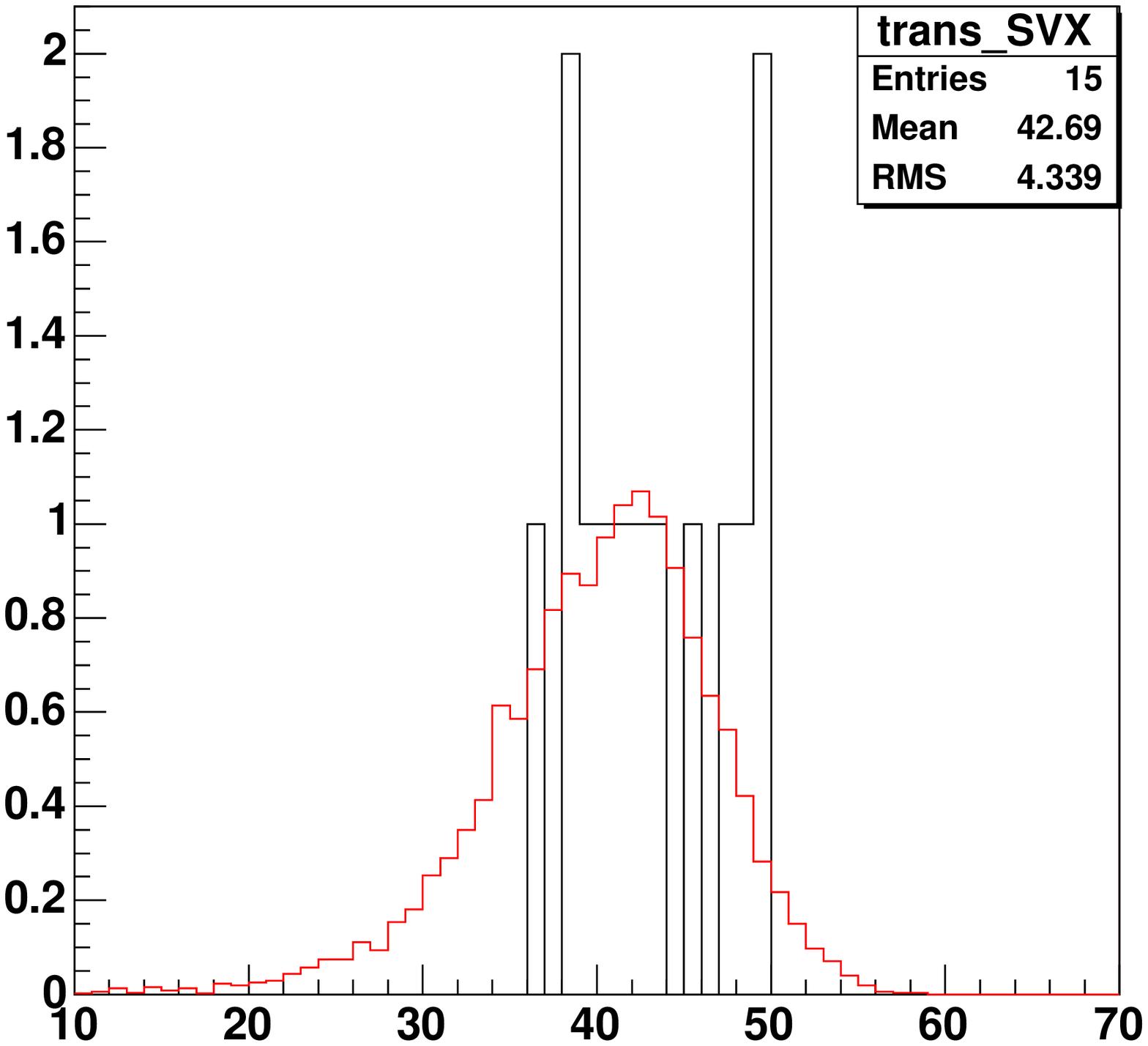,width=1.9in}} \quad
\hspace{-0.4in}
\subfigure
[$\mugmet$: \newline $\mu$ tracks without SVX hits]
{\epsfig{file=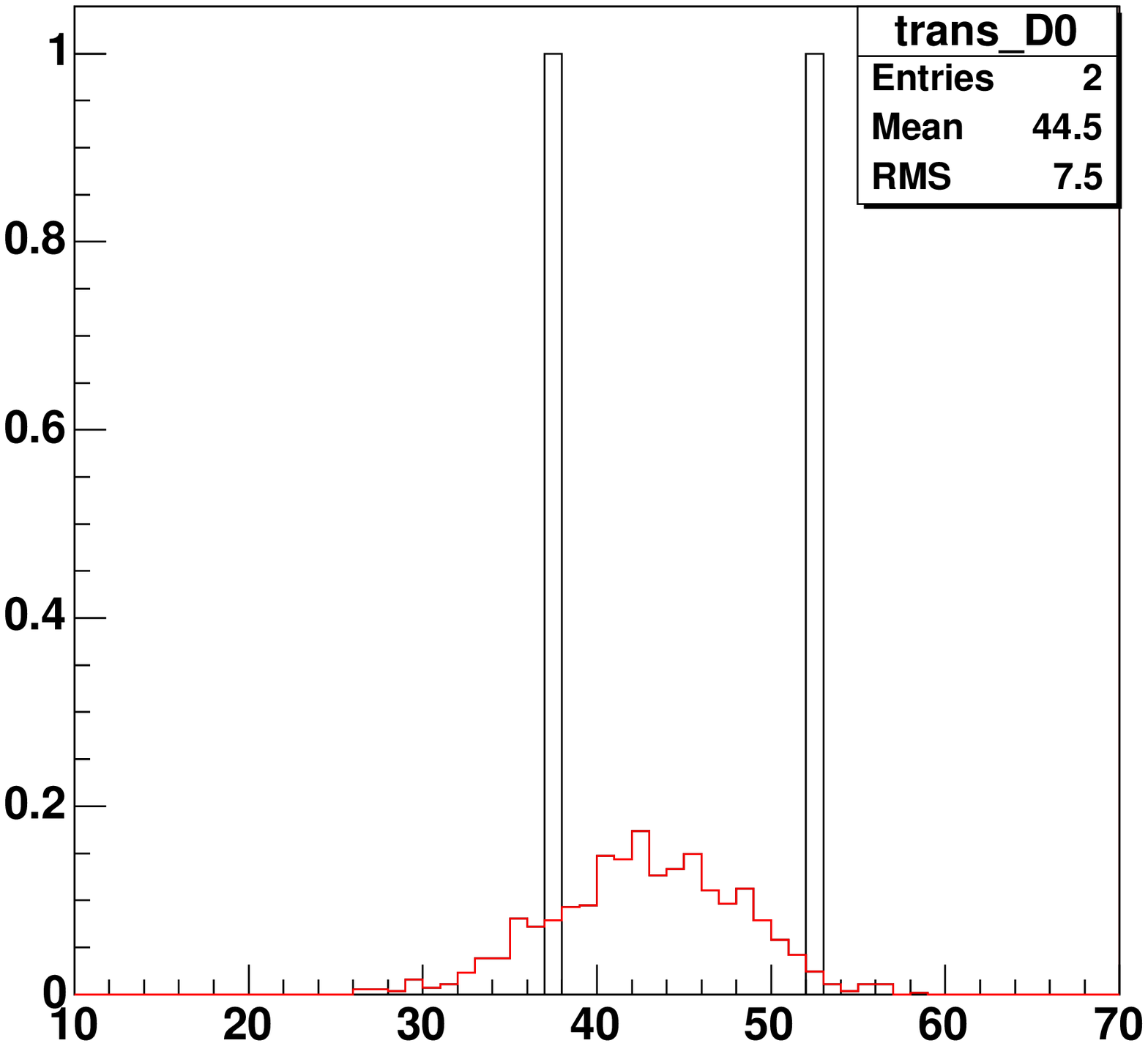,width=1.9in}} \quad
      }
\caption 
[Number of transitions for $\mumug$ and $\mugmet$ samples compared to $\Zmumu$]
{Number of transitions for $\mumug$ (a) $\mu$ tracks with SVX
hits, (b)$\mu$ tracks without SVX hits; $\mugmet$ (c) $\mu$
tracks with SVX hits, (d) $\mu$ tracks without SVX
hits. Muons from $\Zmumu$ are shown as red histogram. }
\label{dif_mumug_mugmet.figure}
  \end{center}
\end{figure}

Figure~\ref{ztight_mu_transitions_good_mu_transitions} shows the
number of transitions in muons in the $\Zmumu$ control sample, and in
a sample enriched in hadron decays by selecting events with a large
$\met > 25~\GeV$, at least one jet and muon that have large impact
parameter $d_0>0.2~$cm. 
Figure~\ref{dif_mumug_mugmet.figure} shows the number of transitions
for muon tracks with and without silicon hits. The red curve is the
distribution for the muons from $\Zmumu$ sample. Decays-in-flight have
a distribution that peaks much lower, with few events above 30
transitions. We see no evidence that any of these tracks are DIF
muons.

\subsubsection{$\mumug$, $\mugmet$ Cosmics Background}

For this we invert cosmic cut, and require the event to be tagged as
cosmic~\cite{cosmics}. We processed the unstripped high-$\Pt$ muon
sample with this inverted requirement and found no $\mumug$ or
$\mugmet$ candidate events.

In addition we scanned our $\mugmet$ and $\mumug$ candidate events
with CDF Run II Event Display (Section~\ref{evd}) and made sure that
none of them look like beam halo or cosmics events.

\clearpage
\bibliography{thesis}
\end{document}